\documentclass{jfm}
\usepackage{rotating,graphicx}
\usepackage{graphicx}
\usepackage{epstopdf, epsfig}

\usepackage{amsmath}
\usepackage{amssymb}

\usepackage{lscape}

\usepackage[utf8]{inputenc}
\usepackage[T1]{fontenc}
\usepackage{mathptmx}
\usepackage{soul}

\usepackage{microtype}

\usepackage{booktabs} 

\usepackage{graphicx}
\usepackage{caption}
\usepackage{hyperref}
\usepackage{subcaption}

\usepackage{algpseudocode}
\usepackage{algorithm}
\usepackage{amsmath}
\let\Psi\varPsi
\usepackage{placeins}
\usepackage{listings}
\usepackage[dvipsnames]{xcolor}
\usepackage{soul}
\usepackage{placeins} 

\usepackage[toc,page]{appendix}

\usepackage{cleveref}
\newcommand\myshade{85}
\colorlet{mylinkcolor}{violet}
\colorlet{mycitecolor}{Aquamarine}
\colorlet{myurlcolor}{Aquamarine}

\hypersetup{
  linkcolor  = mylinkcolor!\myshade!black,
  citecolor  = mycitecolor!\myshade!black,
  urlcolor   = myurlcolor!\myshade!black,
  colorlinks = true,
}

\definecolor{codegreen}{rgb}{0,0.6,0}
\definecolor{codegray}{rgb}{0.5,0.5,0.5}
\definecolor{codepurple}{rgb}{0.58,0,0.82}
\definecolor{backcolour}{rgb}{0.95,0.95,0.92}

\lstdefinestyle{mystyle}{
    backgroundcolor=\color{backcolour},   
    commentstyle=\color{codegreen},
    keywordstyle=\color{magenta},
    numberstyle=\tiny\color{codegray},
    stringstyle=\color{codepurple},
    basicstyle=\ttfamily\footnotesize,
    breakatwhitespace=false,         
    breaklines=true,                 
    captionpos=b,                    
    keepspaces=true,                 
    numbers=left,                    
    numbersep=5pt,                  
    showspaces=false,                
    showstringspaces=false,
    showtabs=false,                  
    tabsize=2
}

\newcommand\R{\mbox{\text{Re}}}
\newcommand\Ca{\mbox{\text{Ca}}}
\newcommand\We{\mbox{\text{$Ca^{-1}$}}}
\newcommand\Ka{\mbox{\text{Ka}}}
\newcommand\F{\mbox{\text{Fr}}}

\usepackage{amsmath,accents}
\usepackage{comment}
\usepackage{tcolorbox}
\usepackage{mathtools}
\usepackage{float}

\shorttitle{Linear stability of a liquid film over a moving substrate}
\shortauthor{F. Pino, M.A. Mendez and B. Scheid}

\title{Linear stability analysis of a vertical liquid film over a moving substrate}

\author{Fabio Pino\aff{1,2}
  \corresp{\email{fabio.pino@vki.ac.be}},
 Miguel A. Mendez\aff{1}
 \and Benoit Scheid \aff{2}}

\affiliation{\aff{1}The von Karman Institute for Fluid Dynamics, EA Department, Sint Genesius Rode, Belgium
\aff{2} Transfers, Interfaces and Processes (TIPs), Université libre de Bruxelles, 1050 Brussels, Belgium
}

\begin{document}

\maketitle

\begin{abstract}
The stability of liquid film flows are important in many industrial applications. In the dip-coating process, a liquid film is formed over a substrate extracted at a constant speed from a liquid bath. We studied the linear stability of this film considering different thicknesses $\hat{h}$ for four liquids, spanning a large range of Kapitza numbers ($\Ka$). By solving the Orr-Sommerfeld eigenvalue problem with the Chebyshev-Tau spectral method, we calculated the neutral curves, investigated the instability mechanism and computed the absolute/convective threshold. 
The instability mechanism was studied through the analysis of vorticity distribution and the kinetic energy balance of the perturbations. It was found that liquids with low $\Ka$ (e.g. corn oil, $\Ka$ = 4) have a smaller area of stability than a liquid at high $\Ka$ (e.g. Liquid Zinc, $\Ka$ = 11525). Surface tension has both a stabilizing and a destabilizing effect, especially for large $\Ka$. For long waves, it curves the vorticity lines near the substrate, reducing the flow under the crests. For short waves, it fosters vorticity production at the interface and creates a region of intense vorticity near the substrate. In addition, we discovered that the surface tension contributes to both the production and dissipation of perturbation's energy depending on the $\Ka$ number.
In terms of absolute/convective threshold, we found a window of absolute instability in the $\R-\hat{h}$ space, showing that the Landau-Levich-Derjaguin solution ($\hat{h}=0.945 \R^{1/9}\Ka^{-1/6}$) is always convectively unstable. Moreover, we show that for $\Ka<17$, the Derjaguin's solution ($\hat{h}=1$) is always convectively unstable.
\end{abstract}


\section{Introduction}
The linear stability analysis of film flow solutions predicts whether small disturbances grow or decay in the long term. This is particularly important in the dip-coating industrial process, where growing disturbances make the coating layer uneven, reducing the quality of the final product \citep{scriven1988physics}.
The dip coating process consists in applying a thin film of protective material over a solid substrate \citep{weinstein2004coating}, with applications ranging from food industry \citep{suhag2020film}, e.g. coating composed of hydrophilic polymers dissolved in water \citep{jose2020advances}, to corrosion-protection material, e.g. zinc coating in the hot-dip galvanization \cite[Chapter~2]{kuklik2016hot}. The substrate is coated by dipping and then withdrawing it from a liquid bath. Thereby, a liquid film forms on the substrate surface, which then solidifies in a protective layer. The thickness of this liquid film $\Bar{h}$ depends on the withdrawal velocity $U_p$ and the action of external control actuators.

In the uncontrolled case, known as free-coating or drag-out problem \citep{wilson1982drag}, $\Bar{h}$ is given by the Landau-Levich-Derjaguin (LLD) solution for small capillary numbers ($\Ca\ll 1$) \citep{derjaguin_thickdipcoating,landau1988dragging} and by the \citet{derjaguin1993thickness}'s solution for large capillary numbers ($\Ca\gg 1$). Both solutions define a monotonically non-decreasing relation between $\Bar{h}$ and $U_p$ with thicker films for faster substrates. This contrasts with industrial needs aiming at thin films and fast substrates. To this end, in industrial lines, external actuators are used to remove the liquid excess from the film, e.g. impinging gas jets in the hot-dip galvanization \citep{gosset2007jet,buchlin1997modeling}, allowing the control of the coating thickness regardless of the substrate velocity \citep{mendez2021dynamics}.

\cite{tu1986stability} and \cite{Gosset2007-ew} studied the stability properties of the controlled liquid film, finding that the cut-off wavenumber is $\propto \Bar{h}^2\R^{1/2}\Ca^{1/2}$. They solved the Orr-Sommerfeld eigenvalue problem with an asymptotic long-wave expansion at order $O(k)$, which is valid only for long-wave and small Reynolds numbers \citep{benjamin1957wave,yih1991stability}. Other authors relied on integral boundary layer models to extend the analysis to larger $\Re$ and short wavelengths. \cite{ivanova2022evolution} found that the cut-off wavenumber is $\propto \Bar{h}^{3/2}\R^{1/2}\Ca^{1/6}$, while \cite{BarreiroVillaverde2023a} showed that the film is more stable to 3D than to 2D disturbances. The disagreement between the OS approximated long-wave solution and integral models suggests that a more thorough analysis of the full OS eigenvalue problem is required, which is still missing in the literature. 

An important aspect of unstable perturbations is the physical mechanism leading to their growth. For a falling liquid film, asymptotic \citep{smith1990mechanism}, vorticity and energy arguments \citep{kelly1989mechanism} showed that the motion induced by a small perturbation of the liquid film's free surface feeds the perturbation's energy by the work of shear stresses at the free surface. The extracted energy is stored in the disturbance's kinetic and potential surface tension energy. Understanding the growth mechanism in dip-coating conditions would shed some light on the role of substrate motion in the instability mechanism.
Another important result given by the linear stability analysis is the threshold between absolutely and convectively unstable film flows \citep[Subsection~7.1.2]{kalliadasis2011falling}. Knowing this threshold is important for the design of control actions, which can affect the whole liquid film or only a part of it \citep{pier2003open}. The liquid film's impulse response can produce waves propagating along the flow direction (convectively unstable flow) or everywhere in the domain (absolutely unstable flow). Experiments \citep{liu1993measurements}, analytical and numerical works \citep{brevdo1999linear} proved that the falling film is convectively unstable, and the suspended film is absolutely unstable \citep{sterman2017rayleigh}. Between these two extremes, a critical inclination angle defines the threshold between absolute and convective film flows  \citep{brun2015rayleigh,scheid2016critical,pino2024absolute}. Likewise, a critical liquid film thickness in the dip-coating defines the Absolute-Convective (AC) threshold. For small coating thicknesses, the entrainment due to the substrate motion sweeps the instabilities upwards; for large coating thicknesses, the gravitational effects push the instability downwards. For intermediate coating thicknesses, gravity and viscous entrainment should compensate each other, resulting in a window of absolute instability. Knowing the position of this window in the parameters space is essential for the design of control laws, which can affect the liquid film downstream, upstream or everywhere. Despite its implications, this remains an open question that has not been answered yet, and which we address theoretically in this paper.

This work studies the stability of controlled and uncontrolled liquid film flow solutions over a substrate moving against gravity. Given the wide variety of coating liquids, going from vegetable oils \citep{sharmin2015recent} to liquid metals, this analysis investigates four liquids with a ratio of surface tension forces to inertial forces (Kapitza number $\Ka$) in the range between $\Ka\sim O(1)$ and $\Ka\sim O(10^4)$. We solve the OS eigenvalue problem via the Chebyshev-Tau spectral method \citep[Chapter~VII]{ortiz1969tau,johnson1996chebyshev,lanczos1988applied} and compare the neutral stability conditions for different values of nondimensional film thickness. Based on the OS solution, we calculate the components of the energy balance equations, and compute the absolute/convective threshold. The analysis focuses on two-dimensional (2D) disturbances since the stability to three-dimensional (3D) disturbances can be brought back to that of an equivalent 2D one at a lower Reynolds number using the reduction law found by  \cite{yih1955stability}.

The rest of the article is organized as follows: the problem setup is described in Section~\ref{sec:prob_description}, and a description of the scaling quantities is given in Section~\ref{sec:scaling}. Section~\ref{sec:formulation} reports the governing equations, the eigenvalue problem formulation, and the instability energy balance equation. The numerical implementation is reported in Section~\ref{sec:methodology} with methods used to solve the OS problem, calculate the energy balance equation and search the absolute/convective threshold. Results are presented in Section~\ref{sec:Results} with the stability curves, the perturbation's energy budget, and the absolute/convective instability threshold. Conclusions and perspectives are given in Section~\ref{Conclusion}.
\begin{figure}
\centering
  \begin{subfigure}[b]{0.31\textwidth}
  \centering
    \includegraphics[width=1\textwidth]{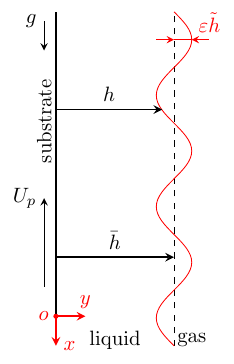}
    \caption{}
    \label{fig:problem_description_a}
  \end{subfigure}
  \hfill
  \begin{subfigure}[b]{0.31\textwidth}
  \centering
    \includegraphics[width=0.67\textwidth]{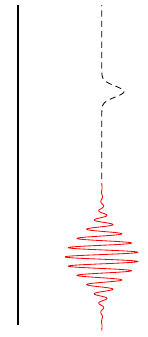}
    \caption{}
  \end{subfigure}
  \hfill
  \begin{subfigure}[b]{0.31\textwidth}
  \centering
    \includegraphics[width=0.67\textwidth]{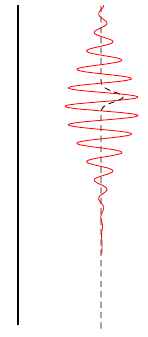}
    \caption{}
  \end{subfigure}
  \caption{(a) schematic of the investigated configuration with a vertical liquid film over a substrate moving with velocity $U_p$, where the liquid film height $h$ is decomposed into a base state $\Bar{h}$ and a harmonic perturbation $\Tilde{h}$ of O(1) with $\epsilon\ll 1$. Liquid film response (red continuous curve) of a (b) convectively and (c) absolutely unstable base state solution to an initial pulse (black dashed line).}
  \label{fig:problem_description} 
\end{figure}
\section{Problem Description}
\label{sec:prob_description}
We consider a 2D liquid film with density $\rho$, dynamic viscosity $\mu$, kinematic viscosity $\nu$, and surface tension $\sigma$, over a vertical substrate moving against gravity $g$ at constant speed $U_{\!p}$. Table~\ref{tab:liquid_prop} reports the physical properties of the four liquids considered in this work: liquid zinc, water, water-glycerol solution (glycerol concentration 45\% by volume), and corn oil. These cover a broad range of conditions encountered in dip or slot coating \cite{Gosset2019b,BarreiroVillaverde2023a}. Figure~\ref{fig:problem_description_a} shows the fixed reference system ($\mathcal{O}xy$) with $x$ aligned with the plate and pointing in the direction of the gravitational acceleration, $y$ along the wall-normal direction towards the free-surface, with origin $\mathcal{O}$ on an arbitrary point of the substrate, given the translational invariance of the problem.
\begin{table}
\centering
\begin{tabular}{c@{\hspace{0.6cm}}c@{\hspace{0.5cm}}c@{\hspace{0.5cm}}c@{\hspace{0.5cm}}c@{\hspace{0.5cm}}c@{\hspace{0.5cm}}c@{\hspace{0.5cm}}}
\toprule
            & $\mathbf{\rho}$ & $\mathbf{\mu}$  & $\mathbf{\sigma}$ &$\Ka$ & $\R$ \\ 
            & (kg/m$^3$) & (mPa\,s) & (mN/m) && \\
            \midrule
Corn Oil    & 1023 & 87.5 & 32& 4 & $1 - 180$ \\
Water-glycerol solution  & 1120 & 8.1 & 65.4 &195 & $4-616$ \\
Water       & 1000 & 1 & 72.8 & 3400 & $10 - 1659$ \\
Liquid Zinc & 6570 & 3.5  & 700 & 11525 & $14 - 2272$\\
\bottomrule
\end{tabular}
\caption{Liquid properties (density, dynamic viscosity, and surface tension from left to right), Kapitza number ($\Ka$) and range of Reynolds numbers ($\R$) for substrate velocity $U_{\!p}\in[0.1,3] \,\rm{m/s}$ for the four liquids considered in the analysis.}
\label{tab:liquid_prop}
\end{table}
\FloatBarrier
\subsection{Scaling quantities}
\label{sec:scaling}
For a given liquid, the two control parameters of the systems are the substrate velocity $U_{\!p}$ and the liquid film thickness $h$. We define as reference velocity $U_{\!p}$ and as reference length the film thickness resulting from the steady state viscous-gravity balance (see \cite{mendez2021dynamics}):

\begin{equation}
\label{eq:scaling_u_h}
    u_{\rm ref} = U_{\!p}, \qquad\qquad\qquad\qquad h_{\rm ref} = \sqrt{\frac{\nu U_{\!p}}{g}}\,,
\end{equation}
where the subscript $\textit{\rm `ref'}$ denotes reference quantities.
We define the capillary length $\ell_{c}$, relating gravity and surface tension, and the viscous length $\ell_{\nu}$, relating gravity and viscosity, as:
\begin{equation}
    \ell_{c} = \sqrt{\frac{\sigma}{\rho g}},\qquad\qquad\qquad\qquad \ell_{\nu}=\sqrt[3]{\frac{\nu^2}{g}}.
\end{equation}
Based on velocity and length scales in \eqref{eq:scaling_u_h}, the dependent and independent variables are scaled accordingly:

\begin{subequations}
\begin{equation}
    (u,v)\rightarrow u_{\rm ref}\,(\hat{u},\hat{v}), \qquad\qquad (x,y)\rightarrow h_{\rm ref}\,(\hat{x},\hat{y}), \qquad\qquad h\rightarrow h_{\rm ref}\,\hat{h},
\end{equation} 
\begin{equation}
t\rightarrow \frac{h_{\rm ref}}{u_{\rm ref}}\,\hat{t},\qquad\qquad p\rightarrow p_{\infty} + \rho g h_{\rm ref}\,\hat{p},
\end{equation}
\end{subequations}
where the hat $\hat{\cdot}$ denotes the non-dimensional quantities and $p_{\infty}$ the atmospheric pressure. To make the nondimensional thickness $\hat{h}$ independent of $U_{\!p}$ and to have a clearer understanding of the role of the control parameters, we introduce another nondimensional film thickness $\Breve{h}$ based on $\ell_{\nu}$:
\begin{equation}
\label{eq:h_breve}
    h\rightarrow \ell_{\nu}\breve{h}.
\end{equation}

The non-dimensional groups arising from our scaling are the Reynolds number:
\begin{equation}
    \R = \frac{u_{\rm ref}h_{\rm ref}}{\nu} = \sqrt{\frac{U_{\!p}^3}{g\nu}}\,,
\end{equation}
which is equivalent to the Froude number defined as:
\begin{equation}
    \F = \frac{u_{\rm ref}^2}{g h_{\rm ref}} = \sqrt{\frac{U_{\!p}^3}{g\nu}}.
\end{equation}
This means that $\R$ represent the ratio of inertia over viscous forces and the ratio between inertia and gravitational forces. The other nondimensional group is the inverse of the capillary number defined as the ratio between surface tension and gravity:
\begin{equation}
    \We = \frac{\sigma}{u_{\rm ref} \mu} = \frac{\Ka}{\R^{2/3}}\, ,\label{eq:We}
\end{equation}
where $\Ka$ is the Kapitza number defined as:
\begin{equation}
    \Ka = \frac{\sigma}{\rho g^{1/3} \nu^{4/3}},
\end{equation}
Based on these nondimensional groups, the nondimensional capillary and viscous wave numbers read:
\begin{equation}
\label{eq:non_dim_cap_wave_num}
    k_{\ell_{c}} = \frac{2\pi}{\hat{\ell}_{c}} = \frac{2\pi\R^{1/3}}{\Ka^{1/2}}, \qquad\qquad\qquad\qquad k_{\ell_{\nu}} = \frac{2\pi}{\hat{\ell}_{\nu}} = 2\pi\R^{1/3}.
\end{equation}
Using the scaling in \eqref{eq:scaling_u_h}, the \citet{derjaguin1993thickness}'s flat film solution corresponds to $\hat{h}=1$ and the Landau-Levich-Derjaguin (LLD) 
solution \citep{derjaguin_thickdipcoating,landau1988dragging,snoeijer_plate_withdrawn} corresponds to:
\begin{equation}
\label{eq:LLD_sol}
    \accentset{\circ}{h} = 0.945\Ca^{1/6}.
\end{equation}

\subsection{Mathematical formulation}
\label{sec:formulation}
The liquid film is governed by the incompressible 2D Navier-Stokes equations, with a velocity vector $\mathbf{v}=(u(x,y),v(x,y))$ and a pressure field $p(x,y)$. The equations are accompanied by the non-slip condition at the substrate $\mathbf{v}(x,y=0)=(-U_p,0)$ and a set of kinematic and dynamic boundary conditions at the free surface ($y=h$), accounting for the continuity of the interface, and the normal and tangential force balance (see \citet[Chapter~2]{kalliadasis2011falling}). The nondimensional steady-state solutions (base states) are given by a flat interface ($\hat{h}=\overline{h}$) and the following velocity and pressure fields:
\begin{equation}
\label{eq:base_state}
\bar{u}(\hat{y}) = -\frac{1}{2}\hat{y}^2 + \Bar{h}\hat{y} - 1,\qquad\qquad
\bar{v}(\hat{y}) = 0,\qquad\qquad
\bar{p}(\hat{y}) = 0,
\end{equation}
where $\Bar{\cdot}$ denotes the base state quantities. The balance between wall shear stress and gravity, along with the imposed velocity at the boundary, produce a non-monotonic relation between non-dimensional flow rate $\hat{q}=q/(u_{\rm ref}h_{\rm ref})$ and $\hat{h}$:
\begin{equation}
    \bar{q} = \frac{1}{3}\bar{h}^3 - \bar{h}.
\end{equation}

\begin{figure}
\centering
  \begin{subfigure}[b]{0.49\textwidth}
    \includegraphics[width=\textwidth]{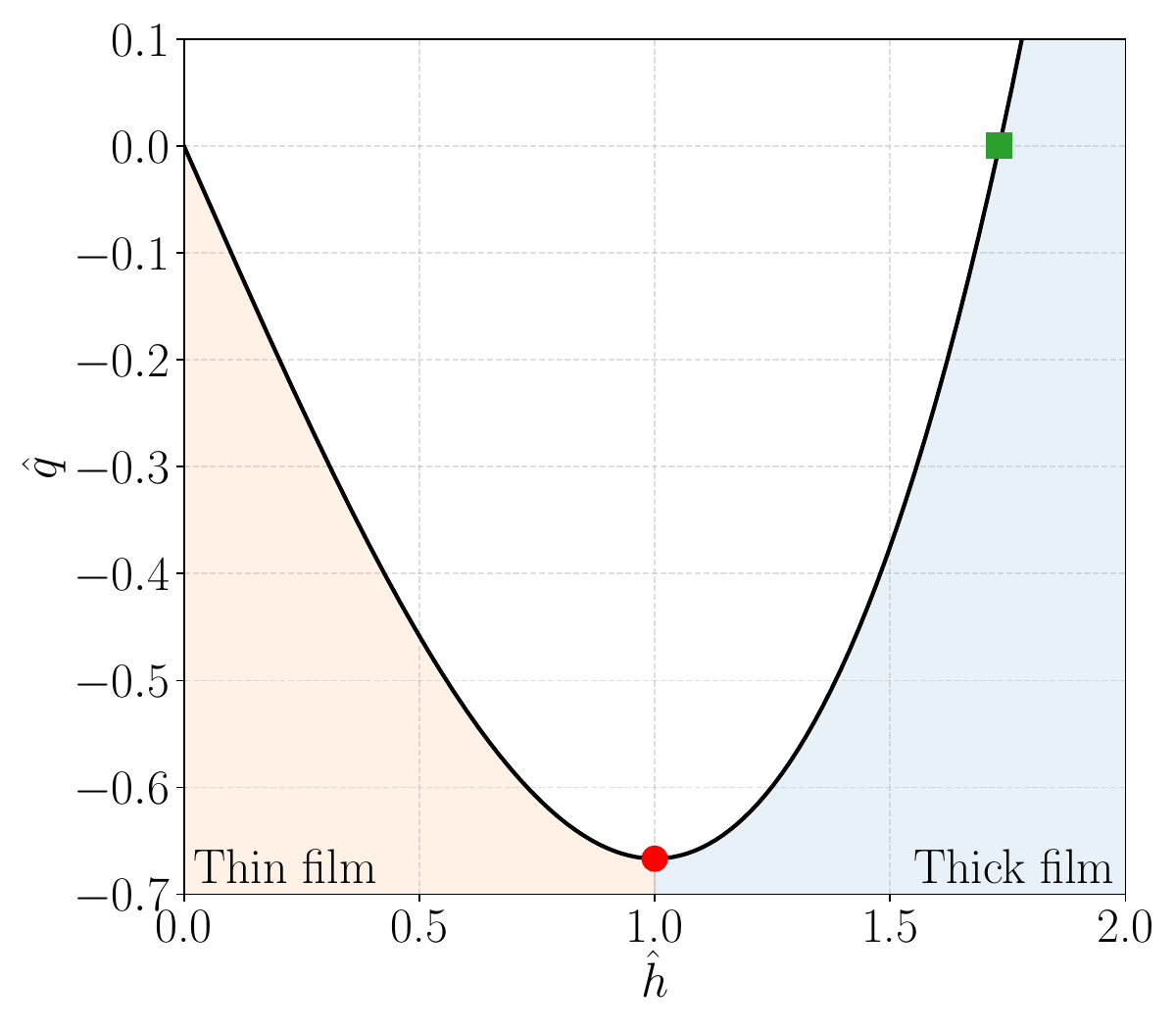}
    \caption{}
    \label{fig:scaling_fig_a}
  \end{subfigure}
  \hfill
  \begin{subfigure}[b]{0.49\textwidth}
    \includegraphics[width=\textwidth]{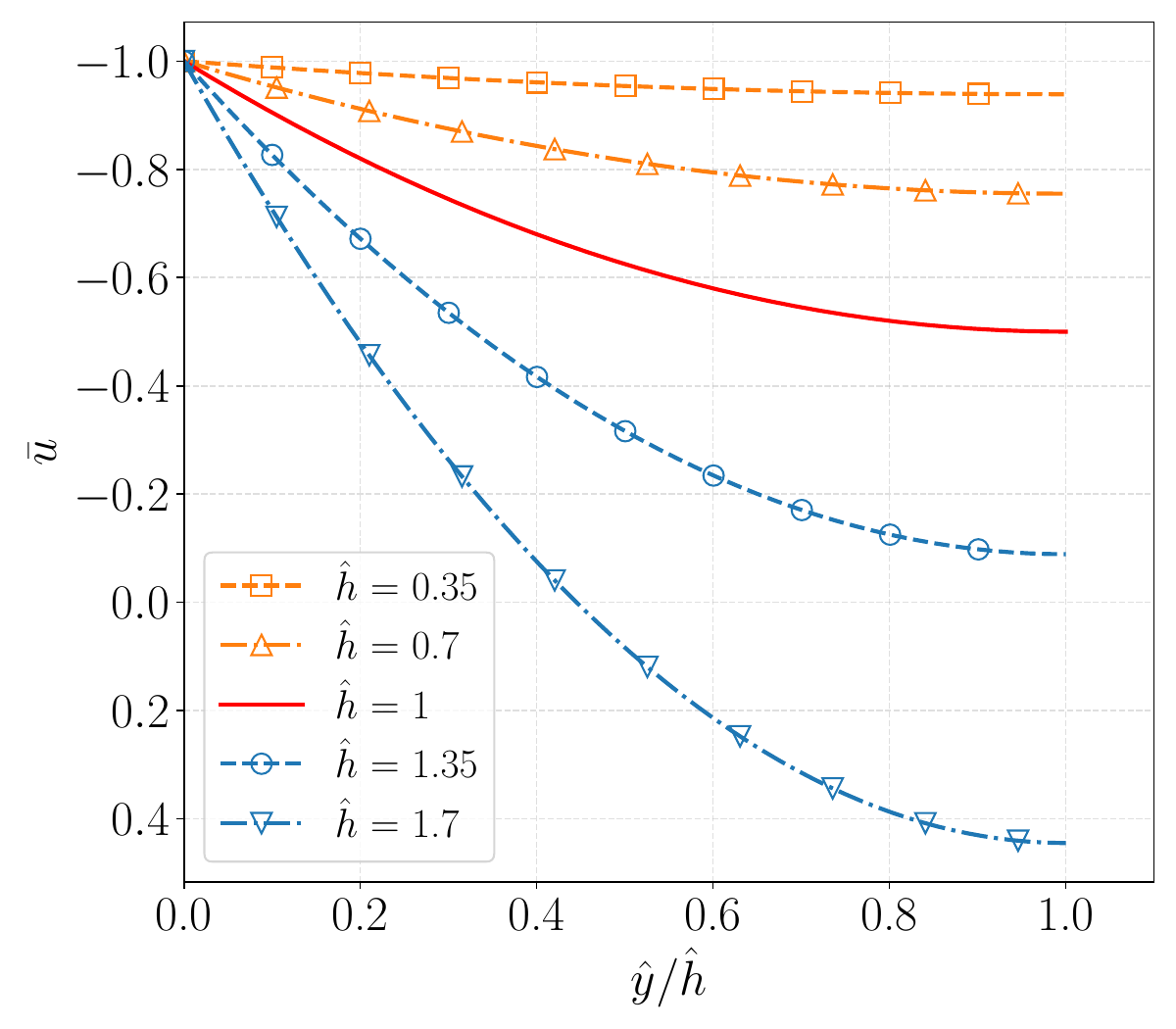}
    \caption{}
    \label{fig:scaling_fig_b}
  \end{subfigure}
    \caption{(a) Relation between the nondimensional liquid film $\hat{h}$ and flow rate $\hat{q}$ with the Derjaguin's solution (red circle), the maximum thickness $\hat h=\sqrt{3}$ (green square) and the thin film (orange shadowed) and thick film (blue shadowed) regions with (b) the associated velocity profiles.}
    \label{fig:scaling_fig}
\end{figure}

As shown in figure~\ref{fig:scaling_fig_a}, this relation entails two branches of steady-state solutions: a thin one ($\bar{h} < 1$) (orange area) and a thick one ($\bar{h} > 1$) (blue area) with the Derjaguin's solution (red dot) and the solution $\bar{h}=\sqrt{3}$ (green square) defining the limit of zero net flow rate, above which we enter the falling film regime, i.e. for $\bar{h}>\sqrt{3}$. Figure~\ref{fig:scaling_fig_b} shows the nondimensional stream-wise velocity profile $\Bar{u}(\hat{y})$ associated with Dejarguin's solution (continuous red line) and the thin (orange) and thick (blue) film solution (lines with markers).

\newpage
To analyse the stability of the base states given by \eqref{eq:base_state}, we decompose the dependent variables as follows:
\begin{equation}
\label{eq:decom_varia}
    \hat{u} = \bar{u} + \varepsilon\tilde{u},\qquad\qquad \hat{v}= \bar{v}  + \varepsilon\tilde{v},\qquad\qquad \hat{p} = \bar{p}  + \varepsilon\tilde{p},\qquad\qquad \hat{h} = \bar{h}  + \varepsilon\tilde{h},
\end{equation}
where $\tilde{\cdot}$ represents the perturbations with order \textit{O}(1) and $\varepsilon\ll1$ a small parameter. Injecting \eqref{eq:decom_varia} into the governing equations with the kinematic and dynamic boundary conditions and collecting the term at \textit{O}$(\varepsilon)$, yields the linearized perturbation (Navier-Stokes) equations:
\begin{subequations}\label{eq:linearized_eqs}
\begin{gather} 
    \partial_{\hat{x}}\tilde{u} + \partial_{\hat{y}}\tilde{v} = 0\label{linearize_eq:1}, \\
    \R\Big(\partial_{\hat{t}}\tilde{u} + \Bar{u}\partial_{\hat{x}}\tilde{u} + \tilde{v} D\Bar{u}\Big) = -\partial_{\hat{x}}\tilde{p} + \nabla^2\tilde{u}\label{linearize_eq:2},\\
    \R\Big(\partial_{\hat{t}}\tilde{v} + \Bar{u}\partial_{\hat{x}}\tilde{v}\Big) = -\partial_{\hat{y}}\tilde{p} + \nabla^2\tilde{v}\label{linearize_eq:3},
\end{gather}   
\end{subequations}
where $D(\cdot)=\partial_{\hat{y}}(\cdot)$ is the wall-normal differential operator. The boundary conditions at the substrate ($\hat{y}=0$) reads:
\begin{equation}
    \tilde{u}=\tilde{v}=0,\label{linearize_eq:bc1}
\end{equation}
and at the free surface ($\hat{y}=\hat{h}$):
\begin{subequations}
\label{eq:bc_fs_linearized}
\begin{gather}
\tilde{v} = \partial_{\hat{t}}\tilde{h} + \Bar{u}\partial_{\hat{x}}\tilde{u}\label{linearize_eq:bc2},\\
\tilde{p} = 2\partial_{\hat{y}}\tilde{v} - \We\partial_{\hat{x}\hat{x}}\tilde{h}\label{linearize_eq:bc3},\\
\tilde{h} = \partial_{\hat{y}}\tilde{u} + \partial_{\hat{x}}\tilde{v}\label{linearize_eq:bc4}.
\end{gather}   
\end{subequations}
We concatenate the streamwise $\tilde{u}$ and cross-stream $\tilde{v}$ perturbations, by recasting \eqref{eq:linearized_eqs} in terms of the streamfunction $\Psi$ defined as:
\begin{equation}
\tilde{u} = \partial_{\hat{y}}\Psi, \qquad\qquad\qquad\qquad \tilde{v} = -\partial_{\hat{x}}\Psi,
\end{equation}
and assuming a normal mode solution of the form:

\begin{equation}
\label{eq:def_normal_mod_strf}
    \Psi = \varphi(\hat{y}) \exp(i(k\hat{x} - \omega\hat{t})), \qquad\qquad\qquad \tilde{h} = \eta\exp(i(k\hat{x} - \omega\hat{t})),
\end{equation}
where $\varphi(\hat{y}) = \varphi_r(\hat{y}) + i\varphi_i(\hat{y})$ and $\eta$ are the amplitudes of the streamfunction and the film thickness respectively, $k=k_r + ik_i$ is the wavenumber, $\omega = \omega_r + i\omega_i$ the angular frequency and $c=c_r + ic_i=\omega/k$ is the perturbation's complex phase speed. This yields the following Orr-Sommerfeld eigenvalue problem: 

\begin{equation}
    \label{eq:Orr_Sommerfeld}
    \mathbf{OS}(k,c,\R)\varphi(\hat{y}) = [\mathbf{A}(k,\R) - c\mathbf{B}(k,\R)]\varphi(\hat{y})=0,   
\end{equation}

where $\mathbf{A}(k,\R)$ is given by:
\begin{equation}
\label{eq:Orr_Sommerfeld_A}
    \mathbf{A}(k,\R) = (D^2-k^2)^2 - i\R k[\Bar{u}(D^2-k^2)+1]
\end{equation}
and $\mathbf{B}(k,\R)=\partial_c\mathbf{OS}(k,c,\R)$ is given by:
\begin{equation}
\label{eq:Orr_Sommerfeld_B}
    \mathbf{B}(k,\R) = -i\R k(D^2 -k^2)
\end{equation}
with boundary conditions $\mathbf{OS}_{BC}\varphi|_{0,\hat{h}}=0$ defined as:
\begin{subequations}
\label{eq:bcs}
\begin{equation}
\varphi(0)=D\varphi(0)= 0,\\
\end{equation}
\begin{equation}
\eta = \varphi(\hat{h})/(c-a),\label{kinematic_con_eq}
\end{equation}
\begin{equation}
[(D^2 - 3k^2) + i \,\R \,k (c -a)]D\varphi(\hat{h}) - i\eta \We \,k^3 = 0,\label{eq:normal_stress_cond}
\end{equation}
\begin{equation}
(D^2 + k^2)\varphi(\hat{h}) - \eta = 0, \label{eq:bc_OS_4}
\end{equation}
\end{subequations}
where $a = \Bar{u}(\Bar{h}) = (\Bar{h}^2/2-1)$ is the base-state velocity at the interface.
 
The solution of the eigenvalue problem sets a relation between the wavenumber $k$ and angular frequency $\omega$. This is known as the dispersion relation and depends on three parameters:
\begin{equation}
\label{eq:dispersion_relation}
    \mathcal{D}(\omega,k; \R, \Ka, \Bar{h})= 0.
\end{equation} 

These solutions are linked to the system's response to a local impulse, known as Green's function \citep[pp. 270-71]{brevdo1999linear,schmid2002stability}. It can be shown that these are the system's poles \cite[pp. 97]{charru2011hydrodynamic} and thus control whether a disturbance grows or vanishes.

The analysis of poles for varying $\omega$ with a fixed $k$ is known as a temporal analysis, while the analysis of poles for varying $k$ and fixed $\omega$ is known as a spatial analysis. The images of any straight lines $k_i=\mathit{C}$, where $\mathit{C}$ is an arbitrary constant, are called temporal branches in the complex frequency space, while the images of any straight line $\omega_i=\mathit{E}$, where $\mathit{E}$ is an arbitrary constant, are called spatial branches in the complex wavenumber space \citep{kupfer1987cusp}. The points of intersection of spatial branches are called spatial branch points.

For a real $k$, a base state is classified as temporally stable or unstable, depending on the value of the growth rate ($\omega_i$): a state is unstable if $\omega_i>0$, as this results in the unbounded temporal growth of infinitesimal perturbation, while it is stable if $\omega_i<0$, as this results in the return of the perturbed liquid film to its steady state equilibrium conditions for all $k$. The locus of points with zero growth rate ($\omega_i=0$) corresponds to the neutral curve, with neither amplified nor damped perturbations.

A base state is absolutely unstable if the solution with zero group velocity $c_g=\partial\omega/\partial k=0$  has $\omega_i>0$ \citep{gaster1968growth,charru2011hydrodynamic}. The condition $c_g=0$ is equivalent to $\partial \omega_r/\partial k_r=\partial \omega_i/\partial k_r=0$ or, because of the Cauchy-Riemann relation, to $\partial \omega_i/\partial k_i=\partial \omega_r/\partial k_i=0$. Since both $\omega_r$ and $\omega_i$ must also satisfy the Laplace equation in the wavenumber domain $k$ for a differentiable (holomorphic) $\omega(k)$, this implies that the condition $c_g=\partial\omega/\partial k=0$ corresponds to a saddle point in the wavenumber domain $k$. However, not all saddle points are admissible; we return to this point in section~\ref{subsec:numerical_method_zero_group_velocity}.

\section{Methodology}
\label{sec:methodology}

\subsection{Numerical solution of the eigenvalue problem}
\label{subsec:numerical_OS}
Before presenting the numerical methods used to solve the OS problem, we get rid of $\eta$ from the boundary conditions \eqref{eq:bcs}, substituting the kinematic condition \eqref{kinematic_con_eq} into the shear stress balance \eqref{eq:bc_OS_4} and the original shear stress balance \eqref{eq:bc_OS_4} into the normal stress balance \eqref{eq:normal_stress_cond} \citep{pelisson2018numerical}. The OS eigenvalue problem \eqref{eq:Orr_Sommerfeld} with the modified boundary conditions is solved using the Chebyshev-Tau spectral method \citep[Section. 3.1]{johnson1996chebyshev,canuto2012spectral}. The eigenfunction $\varphi(\hat{y})$ is approximated with a combination of $N+1$ Chebyshev polynomials of the first kind, with the coefficients $a_i$ being collected in the vector $\mathbf{\phi}=[a_0,a_1,\cdots,a_N]^T$. The approximated eigenfunction is introduced in \eqref{eq:Orr_Sommerfeld}, and the residual is projected onto another base of Chebyshev polynomials of the first kind. This results in an algebraic system of $N+1$ equations representing a generalized eigenvalue problem with eigenvector $\phi$ and eigenvalue $c$:
\begin{equation}
\label{eq:algeb_system}
    \Big(\mathit{A}(Re,k) - c\mathit{B}(Re,k)\Big)\phi = 0,
\end{equation}
with matrices $\mathit{A}$ and $\mathit{B}$ representing the discretization of the operators $\mathbf{A}$ in \eqref{eq:Orr_Sommerfeld_A} and $\mathbf{B}$ in \eqref{eq:Orr_Sommerfeld_B}. Replacing the last four rows of the system in \eqref{eq:algeb_system} with the modified boundary conditions \citep[Section~6.4]{boyd2001chebyshev} leads to the generalized eigenvalue problem:
\begin{equation}
\label{eq:discrete_OS}
    \hat{A}(Re,k)\phi = c \hat{B}(Re,k)\phi,
\end{equation}
with $\hat{A}$ and $\hat{B}$ representing matrices $\mathit{A}$ and $\mathit{B}$ in \eqref{eq:algeb_system} with the enforced boundary conditions. The eigenvalue problem is solved using Python's function \textit{numpy.linalg.eig}. The approximation error is given by the magnitude of four $\tau$ coefficients \cite[Chapter~7, Section~12]{lanczos1988applied}. The smaller the magnitude of these complex coefficients, the better the approximation.

To cope with the spurious eigenvalues \citep{bourne2003hydrodynamic,dawkins1998origin}, we solve \eqref{eq:discrete_OS} for $N=20$ and $N=80$, retaining the eigenvalues whose difference in magnitude, using the Euclidean norm ($||\cdot||$), is below 0.1 \citep{gardner1989modified}. Among these, the one with the largest growth rate $\omega_{i1}=c_{i1}/k$ (most unstable) is selected. To further improve the computational accuracy of the associated eigenvector $\phi_1^{\{1\}}$, we run ten iterations of the inverse power method \cite[Lecture~27]{trefethen2022numerical}:
\begin{equation}
\label{eq:inverse_power_iteration}
    \phi_1^{\{j+1\}} = \frac{(C - I/c_1)^{-1}\phi^{\{j\}}_{1}}{||(C - I/c_1)^{-1}\phi_1^{\{j\}}||}\,,
\end{equation}
where $j\in\{1,2,\cdots,9\}$ is the iteration count and $C=\hat{A}^{-1}\hat{B}$. We defined $C$ this way, with the inverse of $\hat{A}$ rather than the inverse of $\hat{B}$, because $\hat{B}$ is always singular

Given $\Ka$ and $\hat{h}$, to compute the neutral stability curves in the $k-\R$ space with $k\in\mathbb{R}$, we start from the long-wave most unstable mode ($(\phi^{\{1\}},c^{\{1\}})$ with the largest $c_i$), with $k^{\{1\}}=3.5\times 10^{-6}$, and we march it along the discrete $k$ axis running twenty iterations of Rayleigh quotient iteration at every point $k^{\{j\}}$ and collecting the values of $c^{\{j\}}$:
\begin{equation}
\label{eq:rayleigh_quotient}
    \phi^{\{j+1\}} = \frac{(C - I/c^{(j)})\phi^{\{j\}}}{||(C - 1/c^{\{j\}})^{-1}\phi^{\{j\}}||}, \qquad\qquad\qquad I/c^{\{j+1\}} = \frac{\phi^{*\{j+1\}}C\phi^{\{j+1\}}}{\phi^{*\{j+1\}}\phi^{\{j+1\}}}\,,
\end{equation}
where $\cdot^*$ stands for complex conjugate. Thereby, converting $c_i$ into $\omega_i$, we find the growth rate as a function of $k$. The zero of this curve $k_f$ ($k\in\mathbb{R}:\omega_i(k)=0\wedge k\neq 0$), also known as the cut-off frequency, belongs to the neutral curve. 
The neutral curve is constructed by repeating this procedure varying $\R$.
\subsection{Numerical method for zero group-velocity perturbation}
\label{subsec:numerical_method_zero_group_velocity}

To assess if a base state is absolutely or convectively unstable, we follow the Brigg's method (outlined in \citet[Subsection~7.2.2]{schmid2002stability}). This consists in mapping the complex $k$ space into the complex $\omega$ space, identifying the saddle points, and checking \textit{a posteriori} if these respect the causality condition using the collision criterion \citep{briggs1964electron,huerre1985absolute,avanci2019geometrical,thual2013absolute}. According to this criterion, the only valid saddle points arise as spatial branch points, pinching two spatial branches coming from the positive and negative half-planes of the complex wavenumber space. Depending on the value of the growth rate in these points, the base state is convectively ($\omega_i<0$) or absolutely ($\omega_i>0$) unstable.

Given a set of parameters ($\R$, $\Ka$, $\hat{h}$), we explore a portion of the complex wavenumber space $k_i\in[k_{i_{min}},k_{i_{max}}]$ and $k_r\in[k_{r_{min}},k_{r_{max}}]$, discretized with a uniform mesh of $M\times M$ elements with spacing $\Delta k_i$ and $\Delta k_r$. 
Starting from the most unstable mode ($\phi^{\{1\}},c^{\{1\}}$), obtained solving the OS problem \eqref{eq:discrete_OS} for a given pair ($k_r^{\{1\}},k_i^{\{1\}}$), we march over the discretized wavenumber space running twenty iterations of Rayleigh quotient iteration \eqref{eq:rayleigh_quotient} at every point ($k_r^{\{j\}},k_i^{\{j\}}$) and collecting the values of $c^{\{j\}}$. The solution at one point of the grid serves as a starting point for the computation of the solution in the adjacent one. We march over the wavenumber space with a spiral matrix algorithm to avoid two consecutive steps falling outside the iterative method's convergence region. Given the computed mapping $\omega_r(k_r,k_i),\omega_i(k_r,k_i)$, we calculate the saddle point location using numerical differentiation and checking the collision condition.

\begin{figure}
\centering
  \begin{subfigure}[b]{0.45\textwidth}
      \centering
    \includegraphics[width=\textwidth]{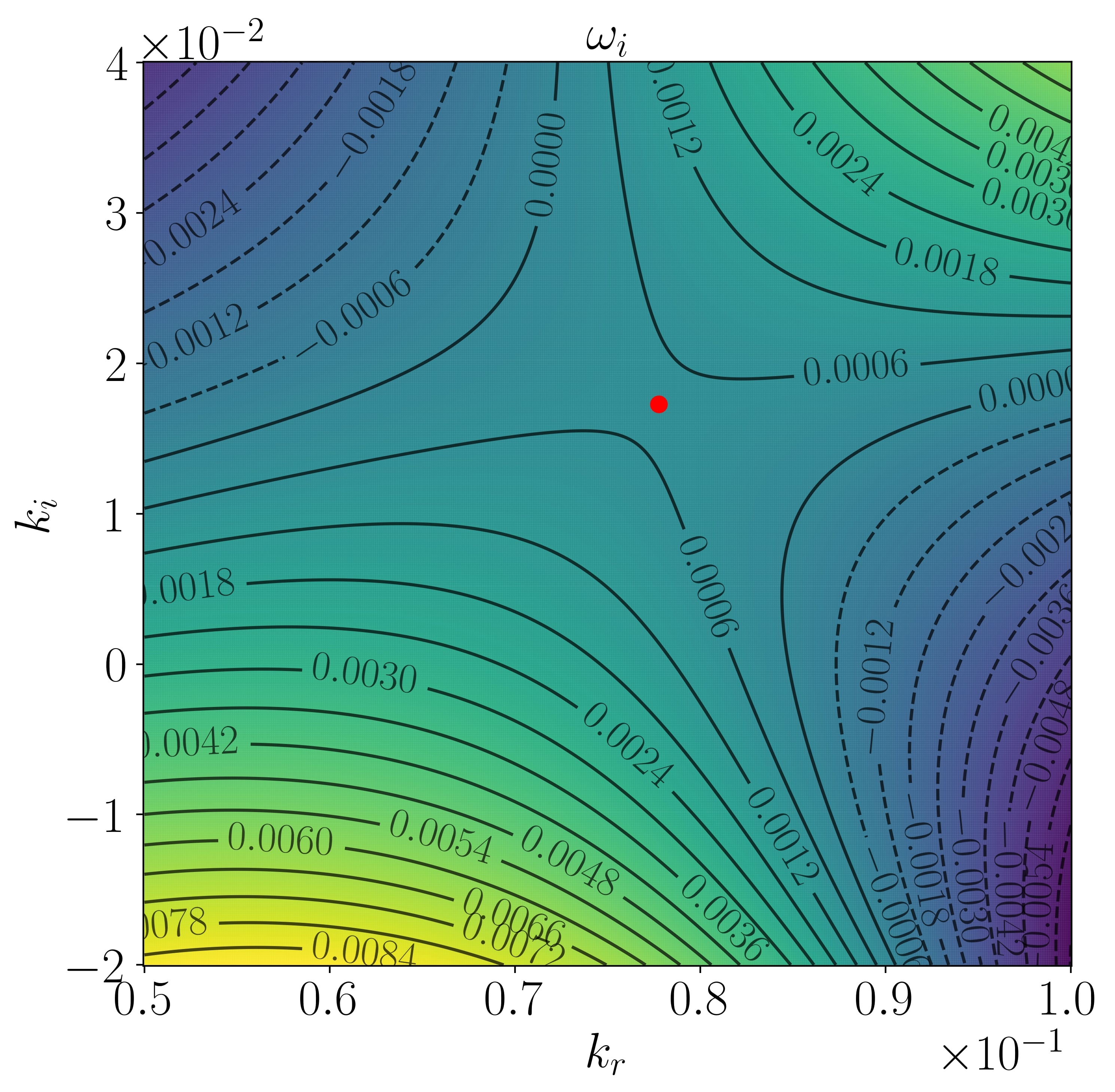}
    \caption{}
    \label{fig:saddle_point_causality_a}
  \end{subfigure}
  \hfill
  \begin{subfigure}[b]{0.45\textwidth}
    \centering
    \includegraphics[width=\textwidth]{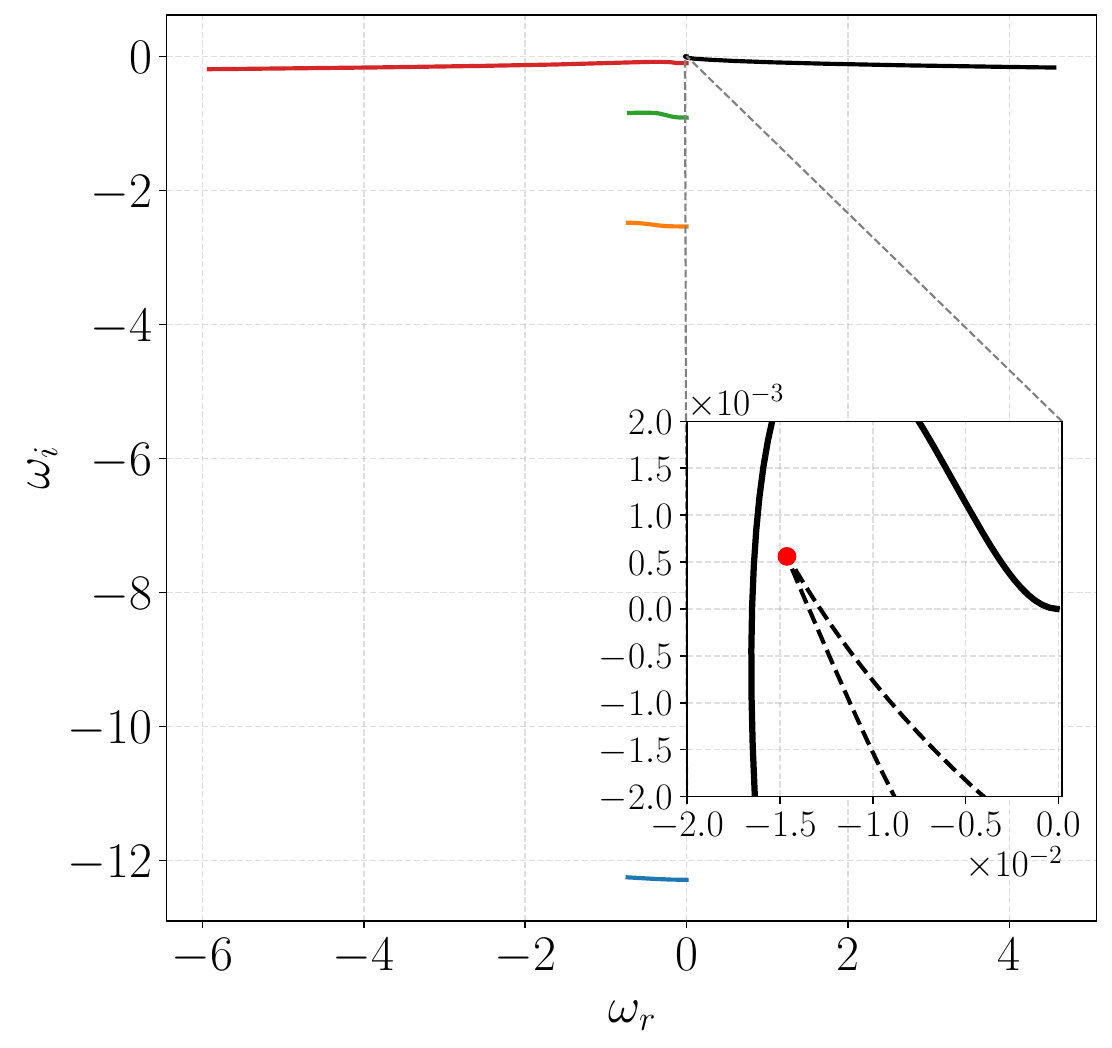}
    \caption{}
    \label{fig:saddle_point_causality_b}
  \end{subfigure}
    \caption{Position of the saddle point (red dot) for the zinc with $\hat{h}$=0.8885, Re=30 (a) in the complex $k$ space with the colourmap of $\omega_i$ and (b) in the complex frequency space with the cusp (black dashed line) and the spatial branches of the first four modes (coloured lines), sorted by ascending values of $\omega_i$, obtained solving the OS problem along the path $k_i=0$ in the complex wavenumber space}
    \label{fig:saddle_point_causality}
\end{figure}

Figure \ref{fig:saddle_point_causality_a} presents an example of the mapping, with $\omega_i$'s colour plot and contour map as a function of $k$, along with the position of the saddle point (red point). To check if the collision criterion is satisfied, we check the position of the saddle point in the complex $\omega$ space. Figure~\ref{fig:saddle_point_causality_b} shows the saddle point position (red dot) in the complex $\omega$ space and the temporal branches of the first five modes (coloured lines), sorted by ascending values of $\omega_i$, along the path $k_i=0$. The collision criterion is respected if the saddle point is surmounted by an odd number of spatial branches \citep{kupfer1987cusp,suslov2006numerical}. 

To determine the absolute/convective instability threshold (saddle point with  $\omega_i=0$.), we explore the parameter spaces ($\Ka-\R$) and ($\hat{h}-\R$). We fix $\R$ number and run a line search along the other parameter. A detailed description of the pseudocode is reported in Appendix~\ref{appxsubsec:pseudocode_AC_threshold}.
\section{Energy balance and mechanism of the unstable perturbation}
\label{sec:energy_balance}
In this section, we present the long-wave mechanism of instability using an asymptotic expansion of the linearized Navier-Stokes equations (Subsection~\ref{subsec:mechanism_long_wave_inst}) and the derivation of the perturbation's energy balance equation (Subsection~\ref{subsec:energy_balance}).
\subsection{Mechanism of long-wave instability in a moving reference frame}
\label{subsec:mechanism_long_wave_inst}
We consider a solution of \eqref{eq:linearized_eqs} with boundary conditions \eqref{linearize_eq:bc1} and \eqref{eq:bc_fs_linearized} in the form of normal modes given by:
\begin{subequations}
\label{eq:expansion_normal_modes_Smith}
\begin{equation}
    \tilde{u} = \acute{u}(\hat{y})\exp(ik(\hat{x} - c\hat{t})), \qquad\qquad\qquad \tilde{v} = \acute{v}(\hat{y})\exp(ik(\hat{x} - c\hat{t})),
\end{equation}
\begin{equation}
    \tilde{h} = \eta\exp(ik(\hat{x} - c\hat{t})), \qquad\qquad\qquad \tilde{p} = \acute{p}(\hat{y})\exp(ik(\hat{x} - c\hat{t})),
\end{equation}
\end{subequations}
where the $\acute{\cdot}$ denotes the amplitudes. We consider a reference system moving with the substrate velocity $U_p$ ($\hat{u}=-1$), with the change of variables:
\begin{equation}
    \acute{u}_m(\hat{y}) = \acute{u}(\hat{y}) + 1, \qquad\qquad c_m = c + 1, \qquad\qquad \Bar{u}_m(\hat{y}) = \Bar{u}(\hat{y}) + 1 = -\frac{y^2}{2} + \hat{h}\hat{y},
\end{equation}
where $\acute{u}_m$, $c_m$ and $\Bar{u}_m$ are the streamwise velocity amplitude, the phase speed and the base state velocity in the moving reference frame. We seek an approximated solution via long-wave expansion of the amplitudes and the phase speed up to $O(k)$ \citep{yih1963stability,benjamin1957wave,smith1990mechanism}:

\begin{subequations}
    \begin{equation}
    \acute{u}_m = \acute{u}_{m0} + \acute{u}_{m1}k + O(k^2), \quad\quad \acute{v} = \acute{v}_0 + \acute{v}_1k + O(k^2),\quad\quad\acute{p} = \acute{p}_0 + \acute{p}_1k + O(k^2),
    \end{equation}
    \begin{equation}
     \eta = \eta_0 + \eta_1k + O(k^2),\qquad\qquad c_m = c_{m0} + c_{m1}k + O(k^2),    
    \end{equation}
\end{subequations}
with the normalization $\eta_0 = 1$ and $\eta_1=0$. Collecting the terms at $O(1)$ gives the system:
\begin{equation}
    \label{eq:leading_order_smith_like}
        D\acute{v}_0(\hat{y}) = 0, \qquad\qquad  \acute{v}_0(\hat{y})D\Bar{u}_m(\hat{y})\R = D^2 \acute{u}_{m0}(\hat{y}), \qquad\qquad D\acute{p}_0(\hat{y}) = D^2 \acute{v}_0(\hat{y}),
\end{equation}
with boundary conditions:
\begin{subequations}
\label{eq:leading_order_smith_like_bc}
    \begin{equation}
        \acute{u}_{m0}(0) = 0, \qquad  \acute{v}_0(0) = 0,
    \end{equation}
    \begin{equation}
        \acute{p}_0(\hat{h}) = 2\acute{v}_0(\hat{h}), \qquad \acute{v}_0(\hat{h}) = 0 \qquad D\acute{u}_{m0}(\hat{h})=1.
    \end{equation}
\end{subequations}

At order $O(k)$, we obtain the system:
\begin{subequations}
\label{eq:first_order_smith_like}
\begin{equation}
\label{eq:asym_exp_cont_k_Smith}
    \acute{v}_1(\hat{y})+i (\acute{u}_{m0}(\hat{y})-1) = 0, 
\end{equation}
\begin{equation}
\label{eq:Ok_moving_syst}
    -\R(\acute{u}_{m0}(\hat{y})-1) (c_{m0}-\Bar{u}_m(\hat{y})+1)+\acute{p}_0(\hat{h})-i\R \acute{v}_1(\hat{y}) D\Bar{u}_m(\hat{y})+i D^2\acute{u}_{m1}(\hat{y}) = 0,
\end{equation}
\begin{equation}
    -i c_{m0} \R \acute{v}_0(\hat{y})+D\acute{p}_1(\hat{y})+i\R \Bar{u}_m(\hat{y}) \acute{v}_0(\hat{y})-i\R \acute{v}_0(\hat{y})-D\acute{v}_1(\hat{y}) = 0,
\end{equation}
\end{subequations}
with the boundary conditions:
\begin{subequations}
\label{eq:first_order_smith_like_bc}
    \begin{equation}
        \acute{u}_1(0)=0\qquad \acute{v}_1(0)=0,
    \end{equation}
    \begin{equation}
    \label{eq:kin_exp_ord_k_Smith}
        \acute{v}_1(\hat{h}) = -i (c_{m0} - \Bar{u}_m(\hat{h})+1),
    \end{equation}
    \begin{equation}
        \acute{p}_1(\hat{y})-2 D\acute{v}_1(\hat{h})=0, \qquad\qquad D\acute{u}_{1m}(\hat{h})+i \acute{v}_0(\hat{h}) = 0 .
    \end{equation}
\end{subequations}
The solution of these systems is presented in Subsection \ref{res_susection:energy_balance_inst_mec}.

\subsection{Energy balance of the perturbation}
\label{subsec:energy_balance}
To study the instability mechanism, we analyse the contributions to the perturbation's kinetic energy equation as in \cite{kelly1989mechanism} and \cite{lin1970roles}. This equation is obtained by summing up the product of \eqref{linearize_eq:2} by $\tilde{u}$ and of \eqref{linearize_eq:3} by $\tilde{v}$, averaging over a wavelength $\lambda$, integrating over the liquid film thickness $\hat{h}$ and then using \eqref{linearize_eq:1} and the boundary conditions (\ref{linearize_eq:bc1},\char`\~\ref{linearize_eq:bc4}) to obtain: 
\begin{subequations}
\begin{equation}
\label{eq:energy_balance}
    RKINE + SURTE = SHEST + REYNS + DISSI.
\end{equation} with 
\begin{equation}
    RKINE = \frac{1}{2\lambda}\frac{d}{dt}\int_{0}^{\lambda}\,\int_{0}^{\hat{h}}\,\Big(\tilde{u}^2 + \tilde{v}^2\Big)\,d\hat{y}d\hat{x},
\end{equation}
\begin{equation}
    SURTE = -\frac{\We}{Re\lambda}\int_{0}^{\lambda}\,\Big[\tilde{v}|_{\hat{h}}(\partial_{\hat{x}\hat{x}}\tilde{h})\Big]\, d\hat{x},
\end{equation}
\begin{equation}
    SHEST = \frac{1}{Re\lambda}\int_{0}^{\lambda}\,\tilde{u}|_{\hat{h}}\Big(\partial_{\hat{y}}\tilde{u}|_{\hat{h}} + \partial_{\hat{x}}\tilde{v}|_{\hat{h}}\Big)\, d\hat{x},
    \label{eq:SHE_definition}
\end{equation}
\begin{equation}
    REYNS = -\frac{1}{\lambda}\int_{0}^{\lambda}\,\int_{0}^{\hat{h}}\,\tilde{u}\tilde{v}D\Bar{u}\,d\hat{y}d\hat{x},
\end{equation}
\begin{equation}
\label{eq:DISSIs}
    DISSI = DISSI_1 + DISSI_2 + DISSI_3.
\end{equation} 
\end{subequations}

The terms on the left-hand side represent the energy stored in the perturbation in the form of kinetic (RKINE) and surface tension potential (SURTE) energies. The terms on the right-hand side represent the energy extracted from the base state through the work of shear stress at the interface (SHEST), Reynolds stress (REYNS), and dissipative viscous effects (DISSI). DISSI encompass the contribution of extensional ($DISSI_1$ and $DISSI_3$) and shear terms ($DISSI_2$), given by:
\begin{subequations}
\label{eq:shear_s_def}
    \begin{equation}
        DISSI_1 = -\frac{1}{Re\lambda}\int_{0}^{\lambda}\,\int_{0}^{\hat{h}}\, 2(\partial_{\hat{x}}\tilde{u})^2 d\hat{y}d\hat{x},
    \end{equation}
    \begin{equation}
        DISSI_2 = -\frac{1}{Re\lambda}\int_{0}^{\lambda}\,\int_{0}^{\hat{h}}\, \upsilon^2 d\hat{y}d\hat{x},
    \end{equation}
    \begin{equation}
        DISSI_3 = -\frac{1}{Re\lambda}\int_{0}^{\lambda}\,\int_{0}^{\hat{h}}\, 2(\partial_{\hat{y}}\tilde{v})^2d\hat{y}d\hat{x},
    \end{equation}
\end{subequations}
where $\upsilon = \partial_{\hat{y}}\tilde{u} + \partial_{\hat{x}}\tilde{v}$ is proportional to the strain rate.
The perturbation quantities are given by: \begin{subequations}
\begin{equation}
  \tilde{u} = (D\varphi_r\cos(\theta) - D\varphi_i\sin(\theta))E,
\end{equation}
\begin{equation}
    \tilde{v} = k(\varphi_i\cos(\theta) + \varphi_r\sin(\theta))E,
\end{equation}
\begin{equation}
    (\partial_{\hat{y}}\tilde{u} + \partial_{\hat{x}}\tilde{v}) = \Big[(D^2 + k^2)(\varphi_r\cos(\theta) - \varphi_i\sin(\theta)\Big]E,
    \label{eq:ref}
\end{equation}
\begin{equation}
\label{eq:vorticiy_expression}
    \tilde{\omega} = [(D^2 -k^2)(\varphi_r\cos(\theta) - \varphi_i\sin(\theta)]E,
\end{equation}
\begin{equation}
    \tilde{h} = \Big(\eta_r\cos(\theta) - \eta_i\sin(\theta)\Big)E,
\end{equation}  
\end{subequations}
with:
\begin{equation}
    \theta = k(x - c_rt), \quad\quad E=\exp(kc_it), \quad\quad \tilde{\omega} = \partial_{\hat{y}}\tilde{u} - \partial_{\hat{x}}\tilde{v},
\end{equation}
where $\tilde{\omega}$ is the perturbation vorticity.

Using the kinematic condition \eqref{kinematic_con_eq}, the real $\eta_r$ and imaginary $\eta_i$ parts of  surface deflection are given by:
\begin{subequations}
\begin{gather}
\label{eq:TTT}
    \eta_r = \Big(\varphi_r(\hat{h})\hat{c} + \varphi_i(\hat{h})c_i\Big)/(\hat{c}^2 + c_i^2),\\
    \eta_i = \Big(\varphi_i(\hat{h})\hat{c} - \varphi_r(\hat{h})c_i\Big)/(\hat{c}^2 + c_i^2),
\end{gather}
\end{subequations}
where $\hat{c} = c_r - a$. To have an accurate computation of the terms in \eqref{eq:energy_balance}, we calculate the integral along $\hat{x}$ analytically and along $\hat{y}$ numerically, using the Simpson’s rule over a grid of $10^4$ equispaced point in the range $[0,\hat{h}]$ such that the difference in magnitude between the right-hand side and the left-hand side of \eqref{eq:energy_balance} is below$1\%$ of the kinetic energy (RKINE) \citep{lin1990absolute}.
\section{Results and Discussions}
\label{sec:Results}

In this section, we report the results of the linear stability analysis in terms of neutral curves and growth rates (Subsection~\ref{res_susection:neutral_curves}), instability mechanism (Subsection~\ref{res_susection:energy_balance_inst_mec}) and absolute/convective threshold (Subsection~\ref{res_susection:AC_threshold}) for the fluids in Table~\ref{tab:liquid_prop}. Moreover, we compute the threshold for the Derjaguin's flat film solution ($\hat{h}=1$) in the ($\Ka-\R$) parameter space.

The OS eigenvalue problem \eqref{eq:Orr_Sommerfeld} is solved using $N=20$ Chebyshev polynomials in the approximation of the stream function amplitude, and a grid spacing for the saddle point search of $\Delta k_r=2.5\times 10^{-4}$ and $\Delta k_i=3\times 10^{-4}$ (see Appendix~\ref{subsec:convergence_study}). To verify our implementation, we compare the dispersion relations and the eigenfunctions against a long-wave asymptotic expansion of the OS problem up to the third order in $k$  (see Appendix~\ref{subsec:verification}).

\subsection{Neutral stability curves and growth rates}
\label{res_susection:neutral_curves}
\begin{table}
\centering
\begin{tabular}{c|@{\hspace{0.6cm}}c@{\hspace{0.6cm}}c|@{\hspace{0.6cm}}c@{\hspace{0.6cm}}c|@{\hspace{0.6cm}}c@{\hspace{0.6cm}}c|@{\hspace{0.6cm}}c@{\hspace{0.6cm}}c|@{\hspace{0.6cm}}}
\toprule
 $\Ka$ &\multicolumn{2}{c@{\hspace{1.2cm}}}{4} & \multicolumn{2}{c@{\hspace{1.2cm}}}{195} & \multicolumn{2}{c@{\hspace{1.2cm}}}{3400} & \multicolumn{2}{c@{\hspace{1.2cm}}}{11525} \\
 \midrule
 $\hat{h}$ & $k_{max}$ & $\omega_{i,{\rm max}}$ & $k_{max}$ & $\omega_{i,{\rm max}}$ & $k_{max}$ & $\omega_{i,{\rm max}}$  &  $k_{max}$& $\omega_{i,{\rm max}}$\\
\midrule
0.35 & 0.65 &  0.0015 &  0.11 & 4.6$\times 10^{-5}$ &  0.027 & 2.7$\times 10^{-6}$  & 0.015&  8$\times 10^{-7}$\\
0.7 & 0.69  &  0.033  & 0.25  & 0.011  &  0.075 & 0.0013 & 0.041& 0.0004 \\
1 &  0.57 & 0.049 & 0.26 & 0.029 & 0.1 & 0.011 & 0.061 & 0.0054\\
1.4 & 0.5 & 0.049 & 0.24 & 0.036 & 0.1 & 0.021 & 0.065 & 0.015\\ 
1.7 & 0.53 & 0.049 & 0.24 & 0.037 & 0.095 & 0.023 & 0.063
& 0.018 \\ \bottomrule
\end{tabular}
\caption{Peak of growth rate $\omega_{i,max}$ and the associated wavenumber $k_{max}$ for different liquids and $\hat{h}$ at $\R =30$.}
\label{tab:disp_relation_peaks}
\end{table}

\begin{figure}
  \begin{subfigure}[b]{0.49\textwidth}
    \includegraphics[width=\textwidth]{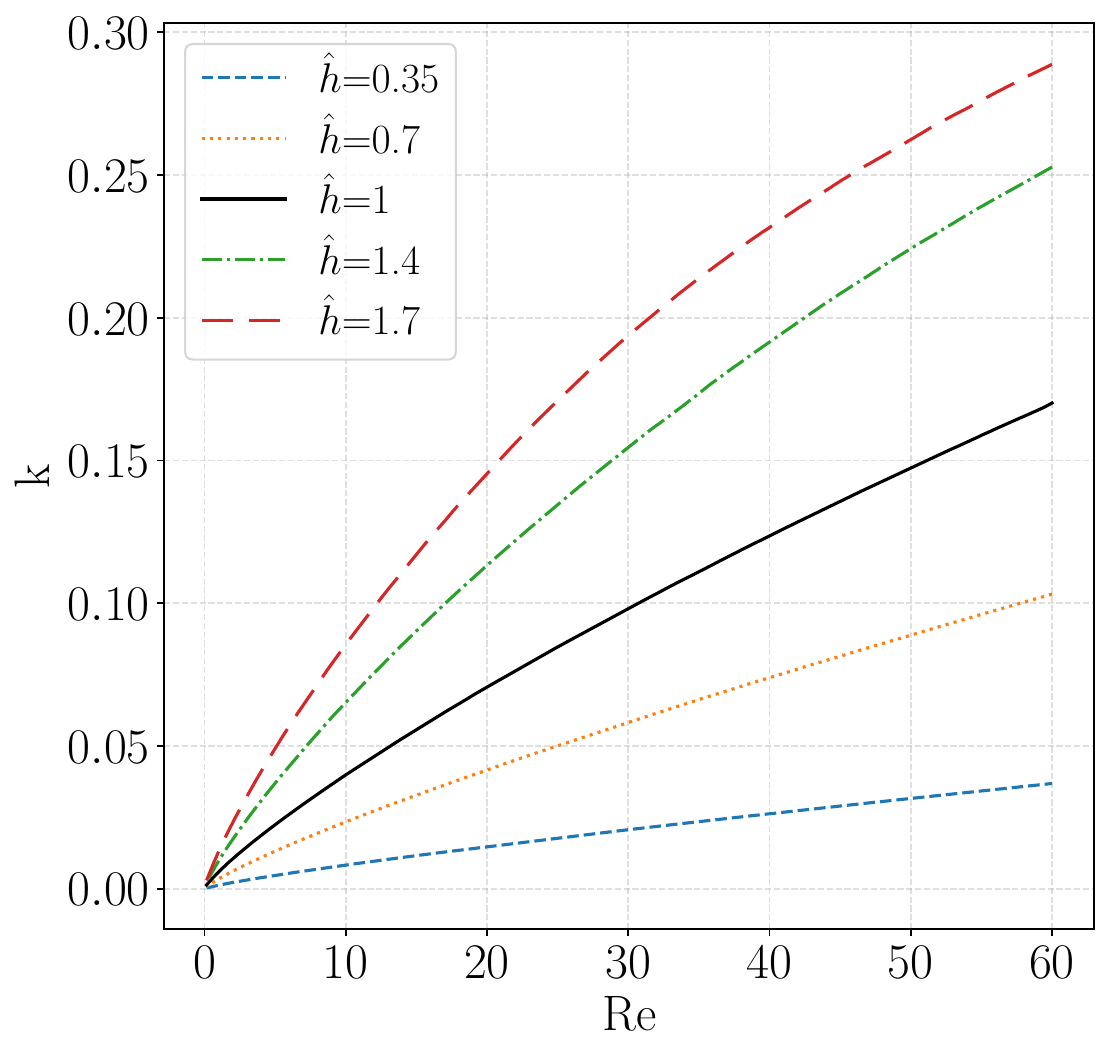}
    \caption{}
  \end{subfigure}
  \hfill
  \begin{subfigure}[b]{0.49\textwidth}
    \includegraphics[width=\textwidth]{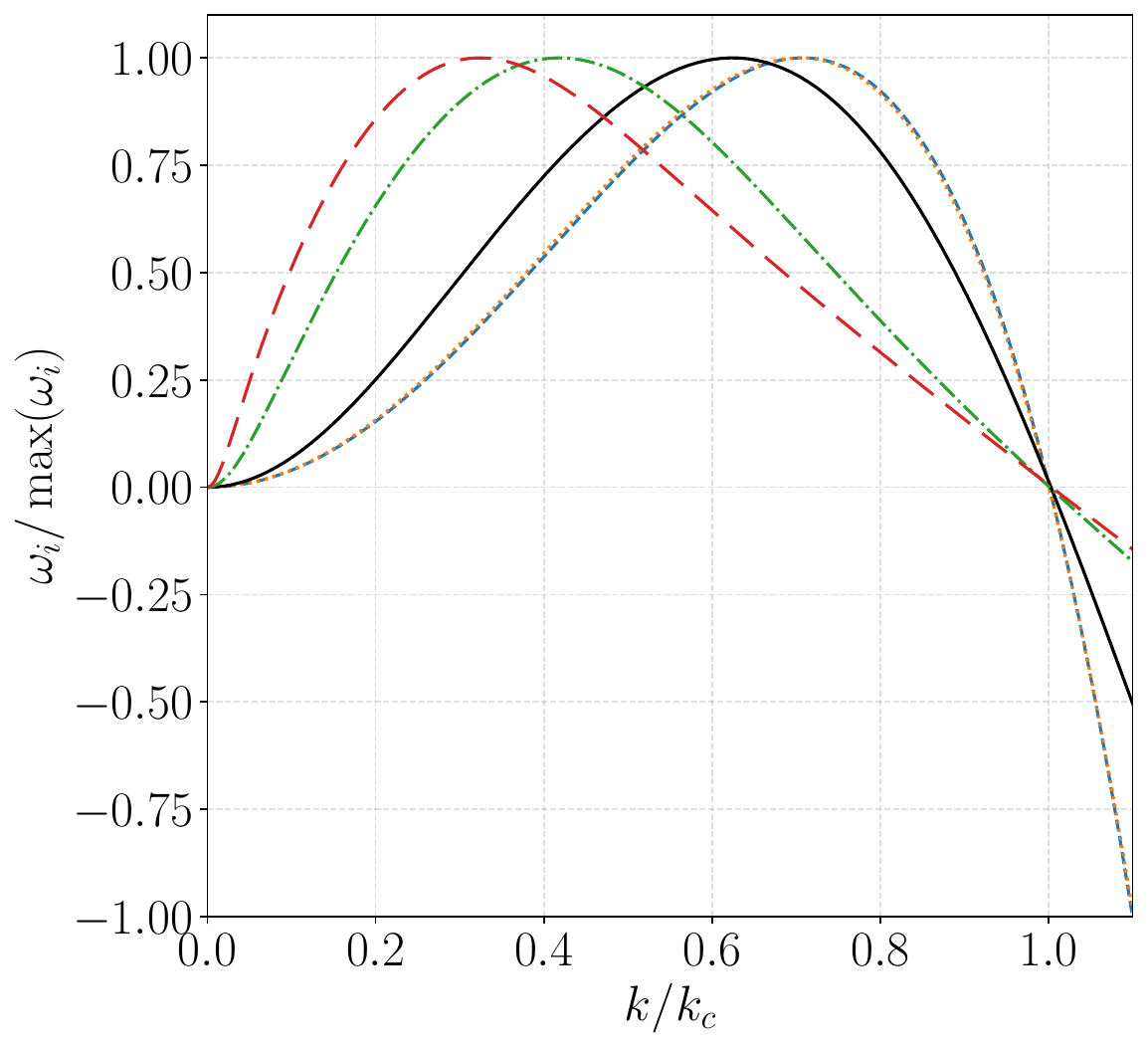}
    \caption{}
  \end{subfigure}
    \begin{subfigure}[b]{0.49\textwidth}
    \includegraphics[width=\textwidth]{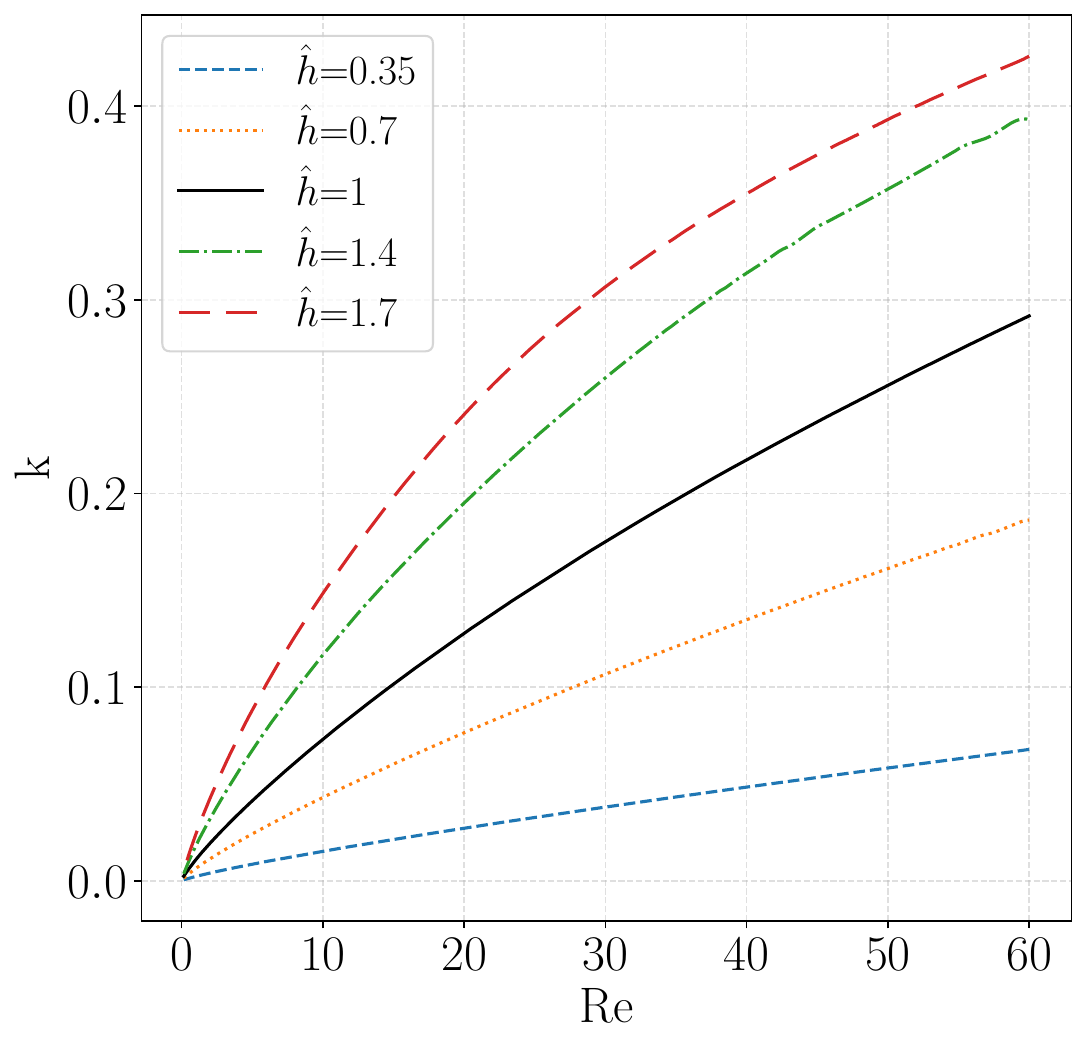}
    \caption{}
  \end{subfigure}
  \hfill
  \begin{subfigure}[b]{0.49\textwidth}
    \includegraphics[width=\textwidth]{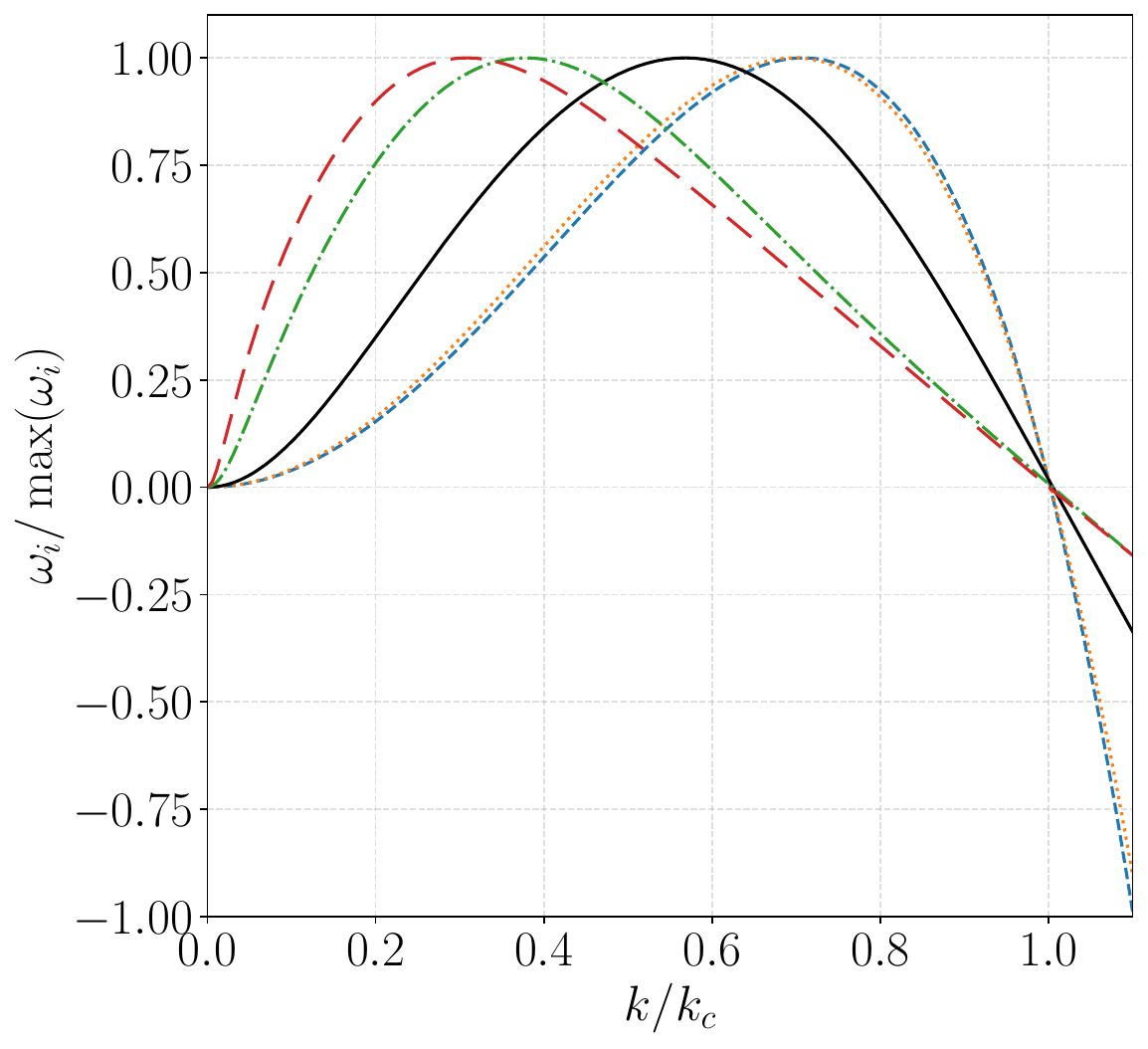}
    \caption{}
  \end{subfigure}
  \caption{Neutral curves for (a and c) different values of the non-dimensional liquid film thickness $\hat{h}$, (b and d) dispersion relations normalized with $\max(\omega_i)$ and the cut-off frequency $k_c$ for $\R=30$  for (a and b) liquid zinc and (c and d) water.}
  \label{fig:neutral_curves_zinc_water}
\end{figure}

\begin{figure}
  \begin{subfigure}[b]{0.49\textwidth}
    \includegraphics[width=\textwidth]{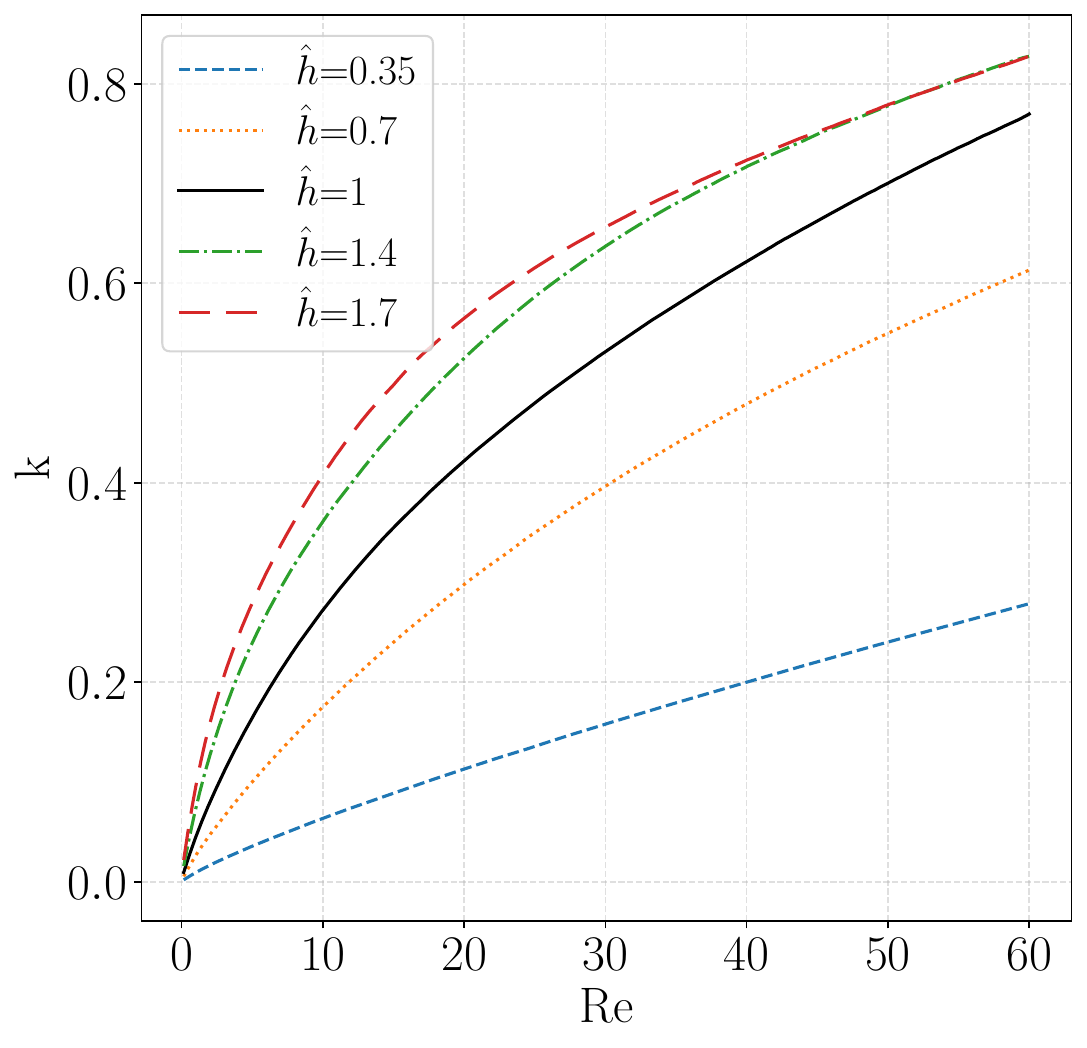}
    \caption{}
  \end{subfigure}
  \hfill
  \begin{subfigure}[b]{0.49\textwidth}
    \includegraphics[width=\textwidth]{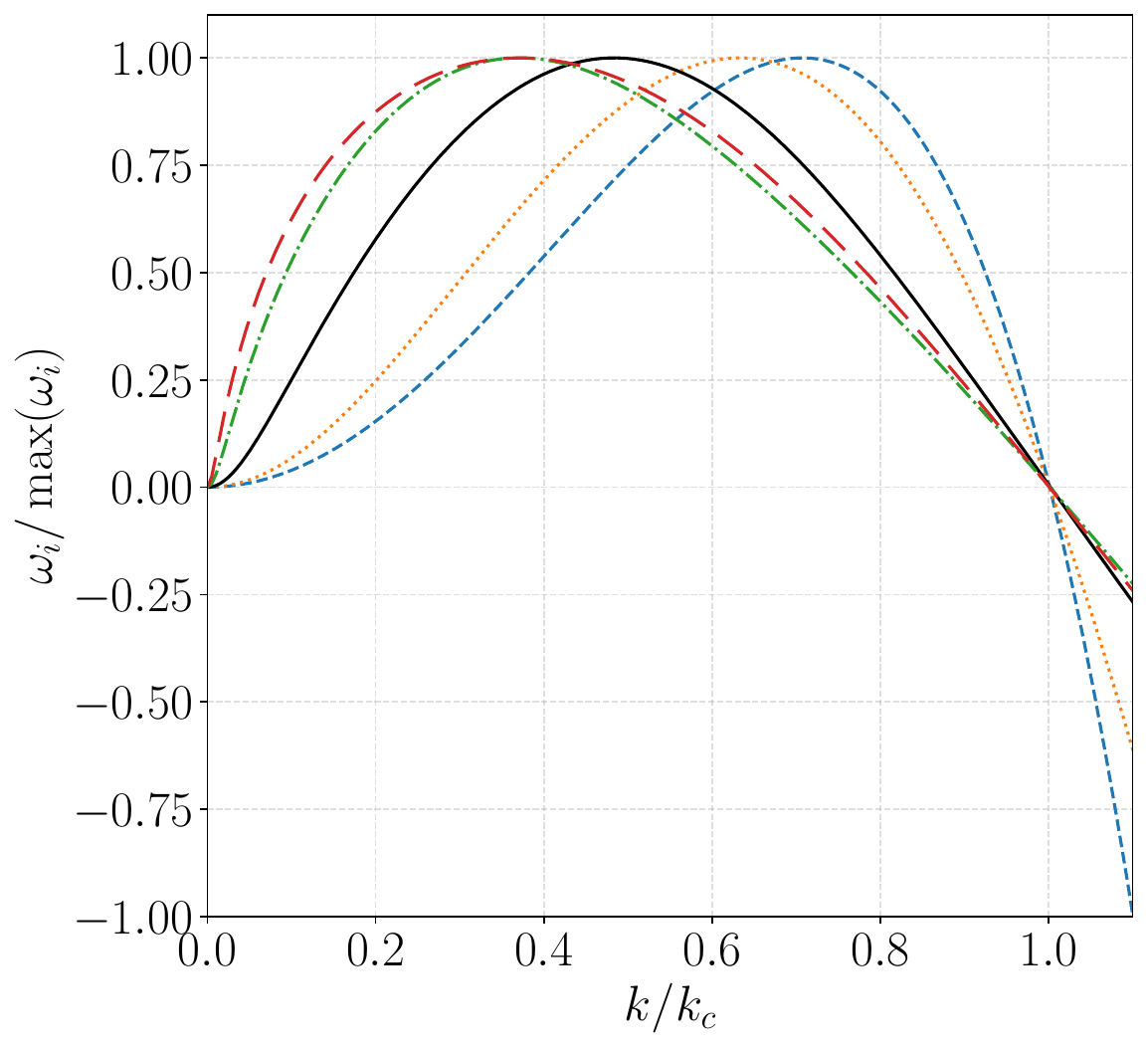}
    \caption{}
  \end{subfigure}
    \begin{subfigure}[b]{0.49\textwidth}
    \includegraphics[width=\textwidth]{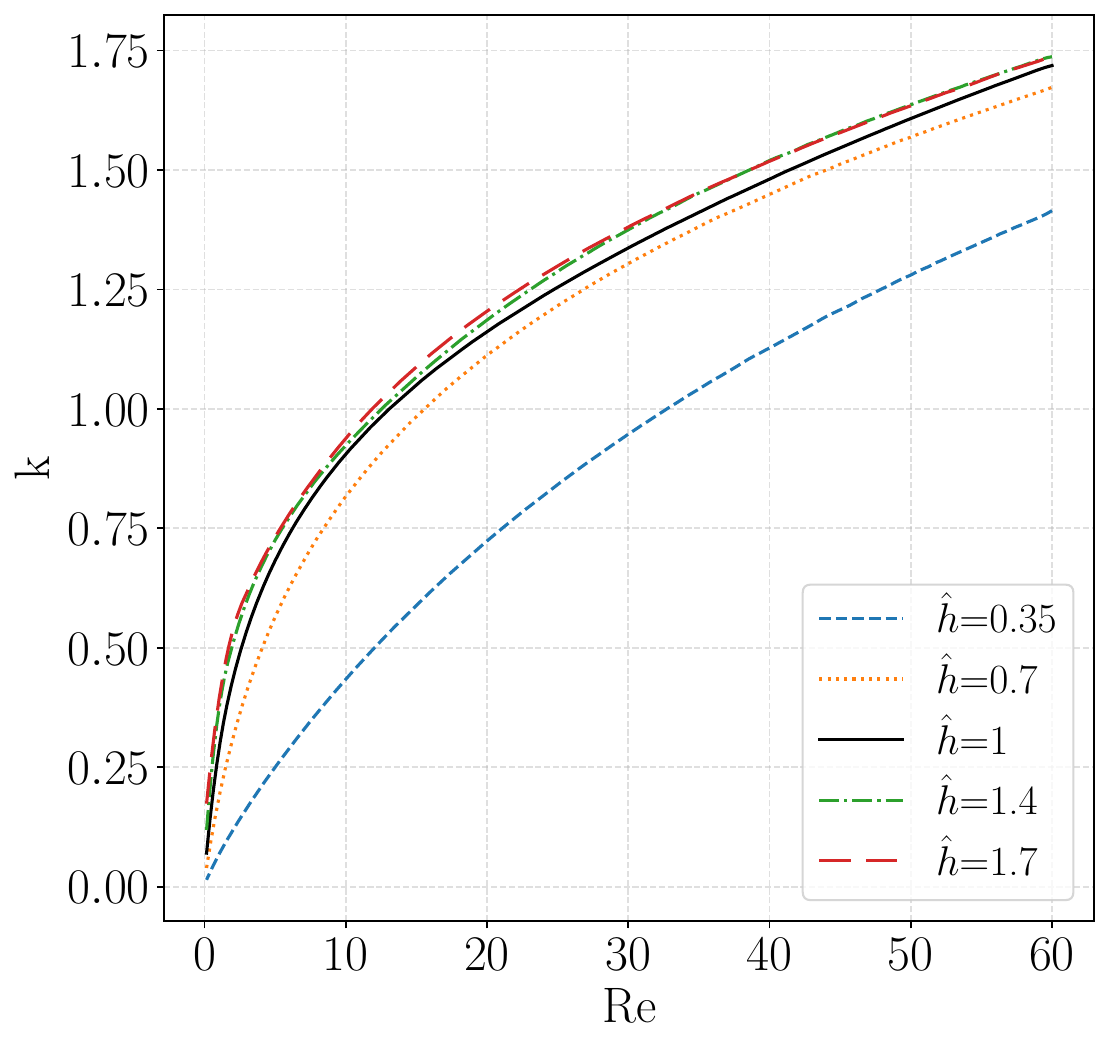}
    \caption{}
  \end{subfigure}
  \hfill
  \begin{subfigure}[b]{0.49\textwidth}
    \includegraphics[width=\textwidth]{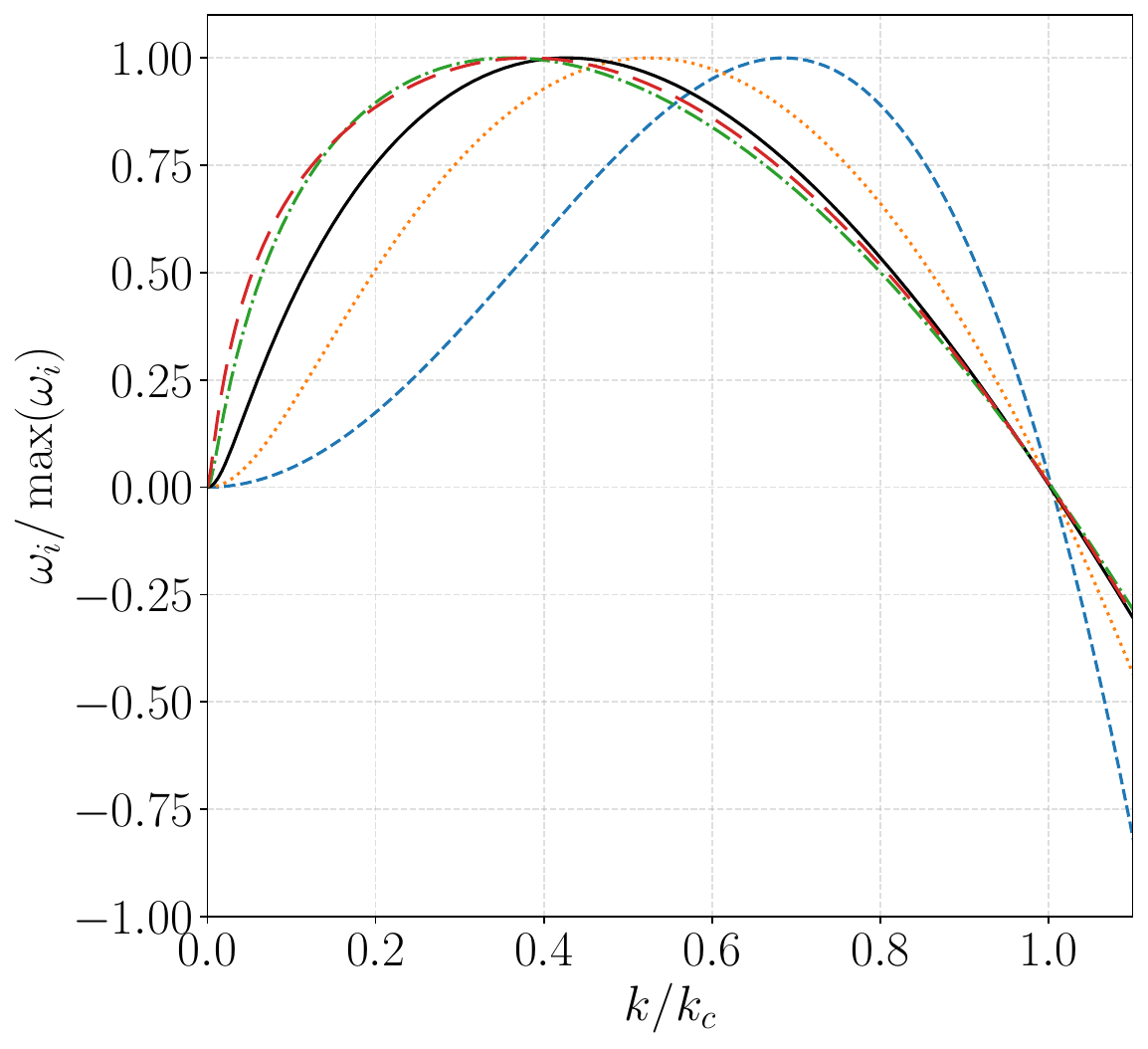}
    \caption{}
  \end{subfigure}
  
  \caption{(a and c) Neutral curve and (b and d) dispersion relations normalized with $\max(\omega_i)$ and the cut-off frequency $k_c$, for $\R=30$ and different values of $\hat{h}$, for (a and b) liquid water-glycerol solution and (c and d) corn oil.}
  \label{fig:neutral_curves_water_glycerol_corn_oil}
\end{figure}
\begin{figure}
  \begin{subfigure}[b]{0.49\textwidth}
    \includegraphics[width=\textwidth]{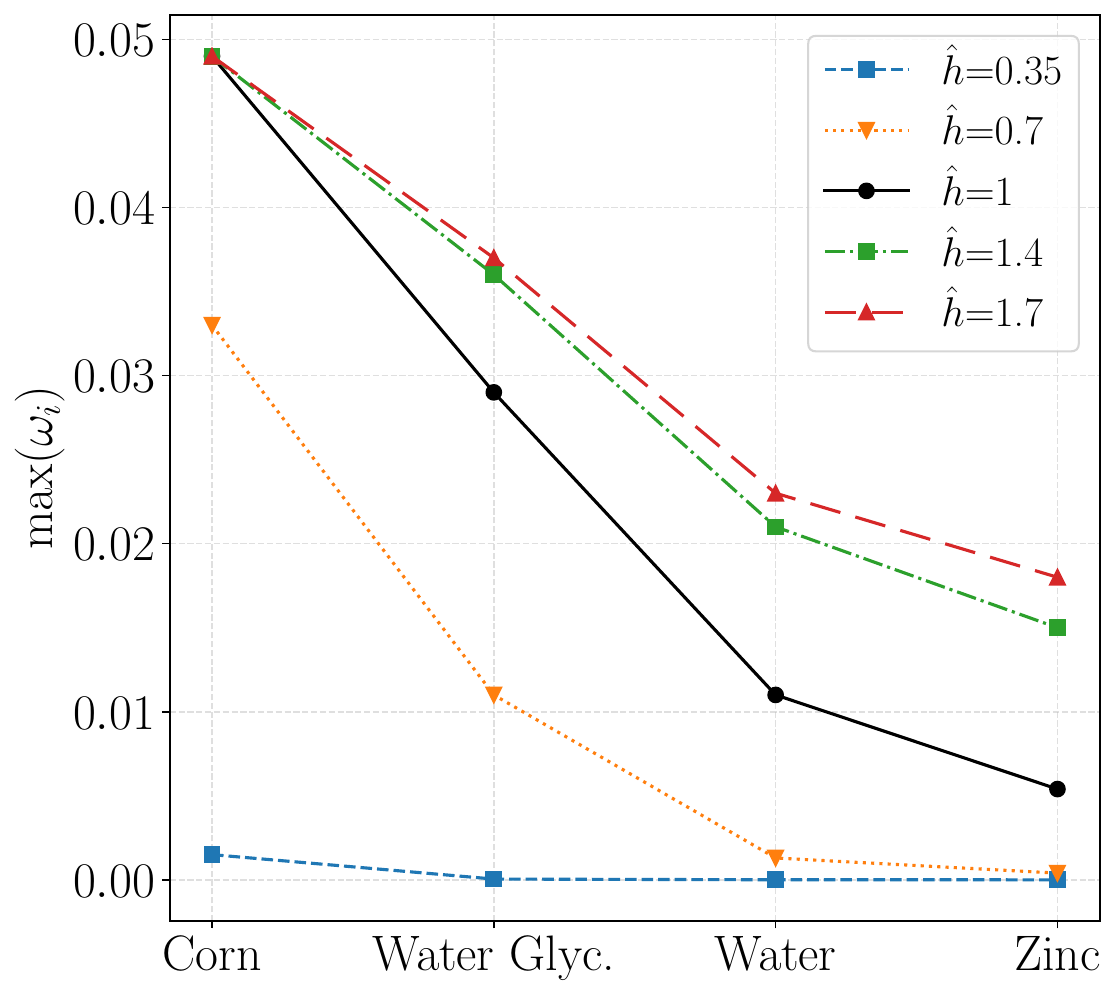}
    \caption{}
  \end{subfigure}
  \hfill
  \begin{subfigure}[b]{0.49\textwidth}
    \includegraphics[width=\textwidth]{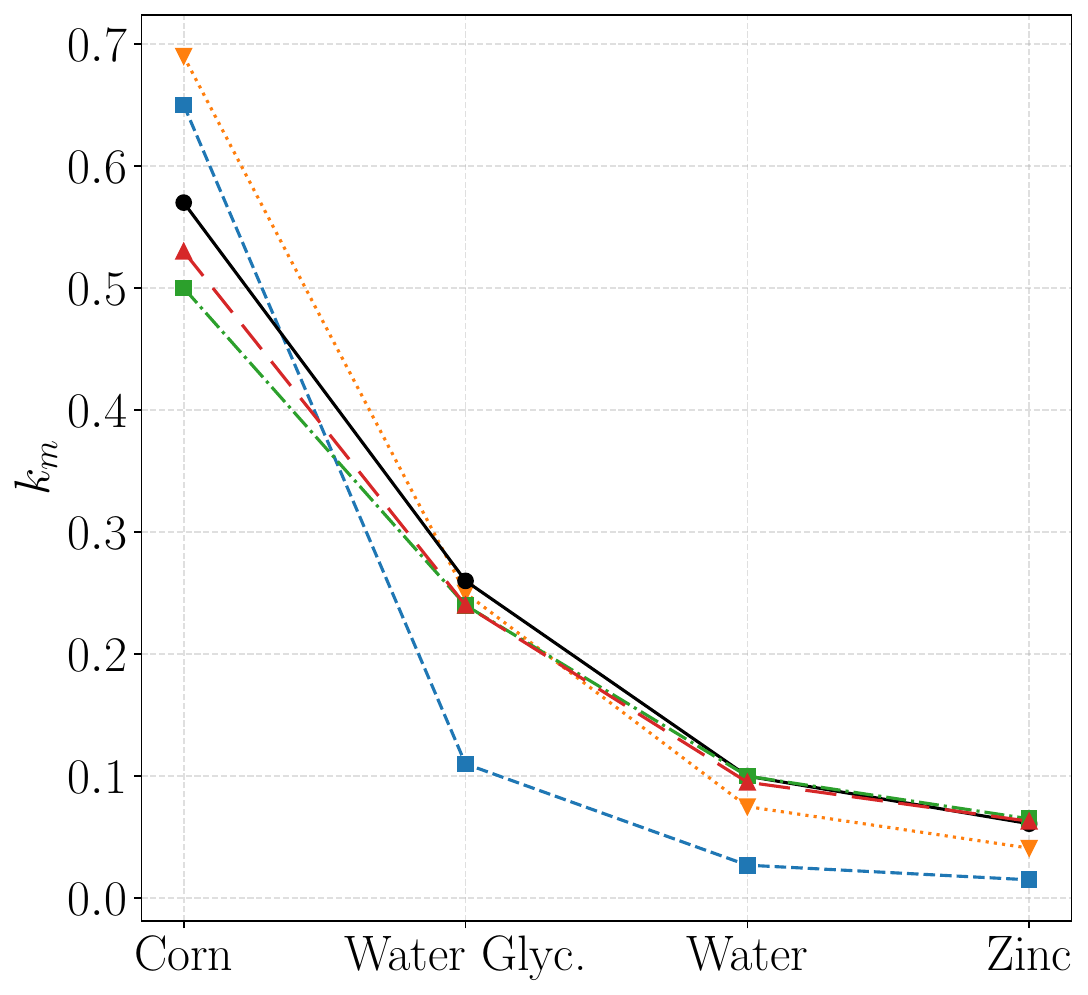}
    \caption{}
  \end{subfigure}
  \caption{Values of (a) the maximum growth rate $\text{max}(\omega_i)$ and (b) the associated wavenumber $k_m$ for different $\hat{h}$ and liquids at Re=30}
  \label{fig:position_n_value_max_growth_rate}
\end{figure}
\begin{figure}
  \begin{subfigure}[b]{0.49\textwidth}
    \includegraphics[width=\textwidth]{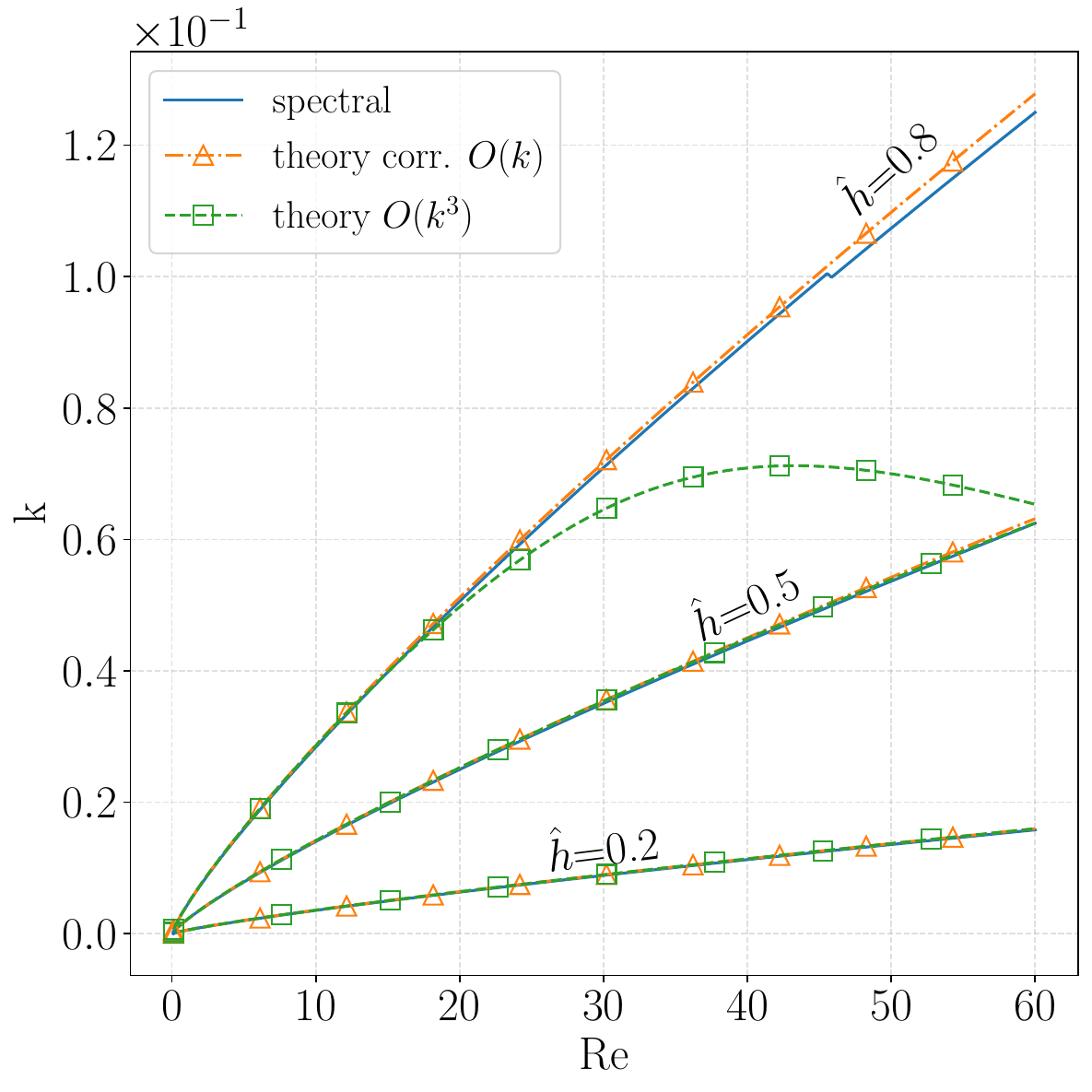}
    \caption{}
  \end{subfigure}
  \hfill
  \begin{subfigure}[b]{0.49\textwidth}
    \includegraphics[width=\textwidth]{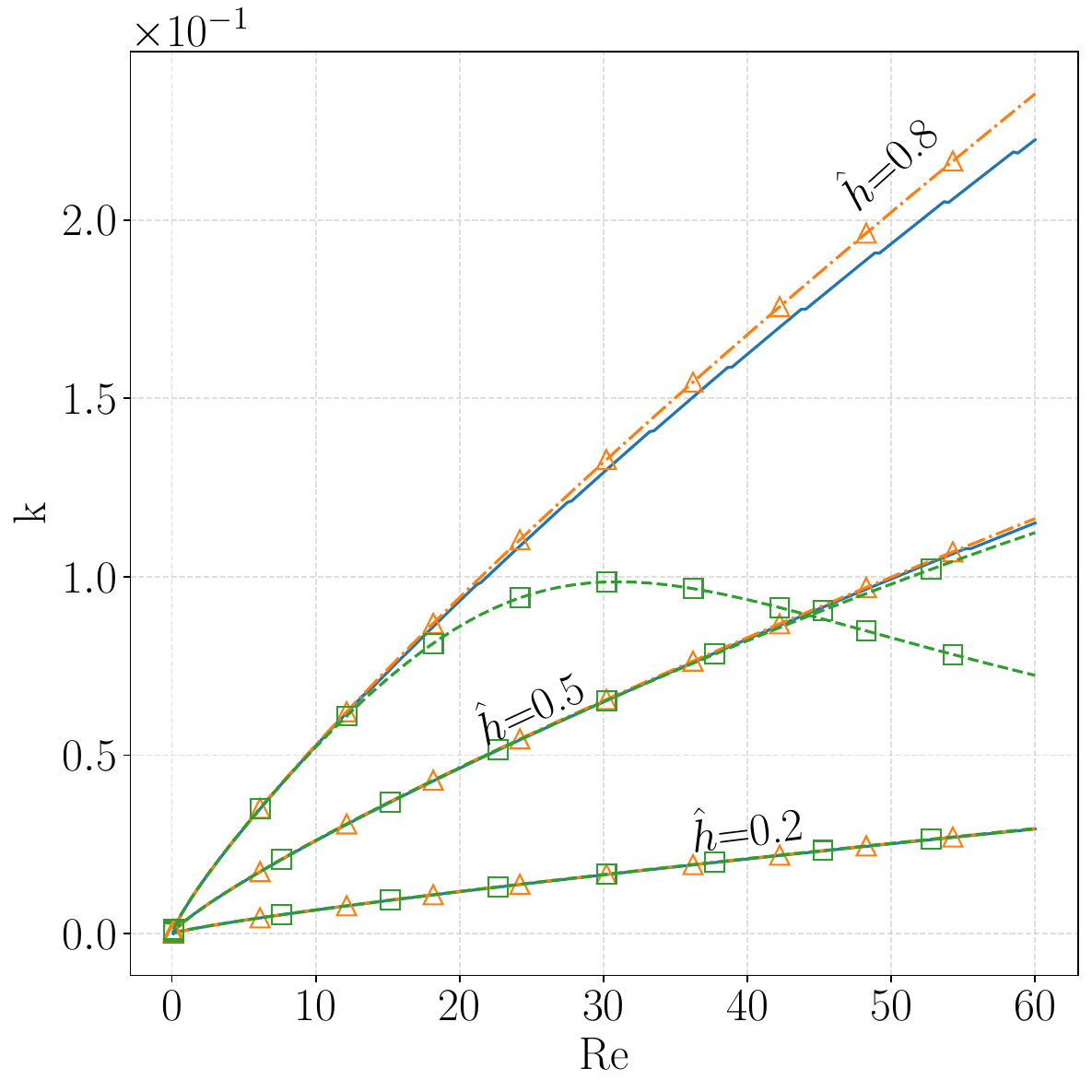}
    \caption{}
  \end{subfigure}
  \caption{Neutral curves separating the region of stable (above) and unstable (below) wavenumbers, based on asymptotic expansions with surface tension correction up to $\text{\textit{O}}(k)$ (orange dash-dotted line with triangles) and without correction up to $\text{\textit{O}}(k^3)$ (green dashed line with squares), against the numerical ones (continuous blue line), for $\hat{h}\in[0.2,0.5,0.8]$ with (a) zinc and (b) water}
  \label{fig:neutral_curves_comp_theory_num}
\end{figure}

First, we compare the neutral stability curves given by the numerical solution of the OS problem and by analytical representation. These representations are obtained by calculating the zeros of the long-wave approximation obtained in Appendix~\ref{subsec:convergence_study}, {\it i.e.} cancelling $c_1^\star$ in (\ref{exp:order_k_w_Ca}), leading to the expression:
\begin{equation}
    k = \sqrt{\frac{2}{5}\frac{\hat{h}^3}{Ka}\R^{5/3}},
\end{equation}
for an expansion up to $\text{\textit{O}}(k)$ with the surface tension correction at leading order ($\We k^2 = \text{\textit{O}}(1)$), and to the expression
\begin{equation}
    k = \frac{24 \sqrt{15015}\hat{h}^{3/2} \sqrt{\R}}{\sqrt{2427904 \hat{h}^{11} \R^3+45463275 \hat{h}^5\R+21621600 \frac{\Ka}{\R^{2/3}}}}\,,
\end{equation}
for an expansion up to order $\text{\textit{O}}(k^3)$ without correction.
Figure~\ref{fig:neutral_curves_comp_theory_num} shows the approximated neutral curves and the one obtained with the spectral methods for $\hat{h}\in[0.2,0.5,0.8]$ considering (a) zinc and (b) water. The instability region lies between the curve and the $k=0$ axis. The analytical expressions agree with the numerics for small values of $\hat{h}$. The solution with correction agrees better than the full third order one, which highlights the validity of the assumption $Ca^{-1}\times k^2=\text{\textit{O}}(1)$ for large $\Ka$ and small $\hat{h}$.

Going more in-depth in the analysis of the neutral curves and the dispersion relations, Figures~\ref{fig:neutral_curves_zinc_water} and \ref{fig:neutral_curves_water_glycerol_corn_oil} show the numerical neutral curves for different values of $\hat{h}$ and the dispersion relations with axes normalized with the maximum value of $\omega_i$ ($\max(\omega_i)$) and the cut-off wavenumber $k_m$ at $\R=30$ for the four liquids. The wavenumber $k_{max}$ and the magnitudes $\omega_{i,max}$ of the growth rate peaks are reported in Table~\ref{tab:disp_relation_peaks}. For $\hat{h}\leq 1$, $\omega_i$ gradually increases with $k$, up to the peak, then sharply decreases towards $k_m$. For $\hat{h}>1$, $\omega_i$ reaches the peak after a steep increase, and then it gently gets to $k_m$.

As for the growth rate, $\hat{h}$ also influences the neutral stability curves. For the four fluids, the instability region gets larger for large $\hat{h}$ and small $\Ka$, with curves progressively gathering around the same wave numbers for thick film conditions ($\hat{h}>1$), with the limit case of corn oil, where the neutral curves are almost superimposed. Similar behaviour is also visible in the dispersion relations. As $\Ka$  decreases, the relative distance between the peaks' positions for $\hat{h}> 0.35$ shrinks, with the limit case of corn oil where the curves for $\hat{h}=0.4$ and $\hat{h}=1$ also change shapes, becoming similar to those for $\hat{h}>1$.

This highlights stabilizing mechanisms given by the balance of inertia, viscous and gravitational forces without the effect of surface tension. As the $\Ka$ decreases, the stabilizing effects of the surface tension diminish compared to the viscous effects. Even increasing $\hat{h}$, the instability region does not expand much. This suggests that neutral modes with $k=\text{\textit{O}}(1)$ arise as an equilibrium of mostly viscous and gravitational forces, which is not linked to the velocity of the plate. Indeed, increasing $\R$ does not change the relative position of the neutral curves; it just brings this equilibrium point to larger $k$.

Figure~\ref{fig:position_n_value_max_growth_rate} shows the evolution of (a) the $\omega_i$'s peak magnitude $\max(\omega_i)$ and (b) the wavenumber $k_{max}$ varying $\hat{h}$ for the different fluids. As $\Ka$ decreases, the maximum growth rate increases. Corn oil has the largest growth rates of all the $\hat{h}$ with its peaks overlapping for $\hat{h}\geq 1$.

Interestingly, the peaks' locations do not have a monotonic behaviour with $\hat{h}$. For the corn oil, the peak position advances towards smaller wavelengths for $\hat{h}=0.7$, and then it goes to longer wavelengths for higher values of $\hat{h}$. As $\Ka$ increases, the peak is located at smaller wave numbers, and it moves to $\hat{h}=1.7$, passing through $\hat{h}=1$ for the water-glycerol solution.

\begin{figure}
  \begin{subfigure}[b]{0.32\textwidth}
    \includegraphics[width=\textwidth]{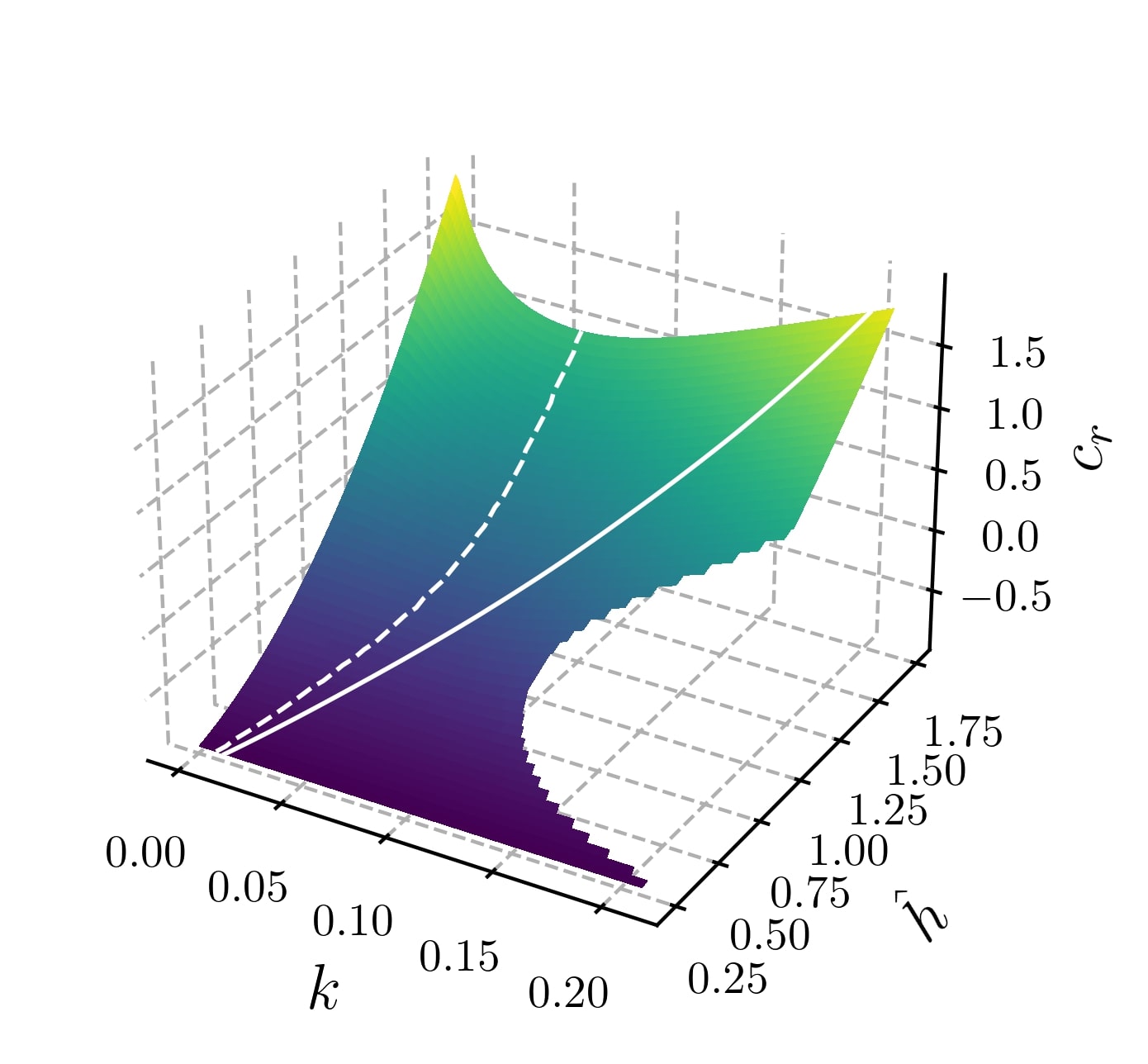}
    \caption{}
  \end{subfigure}
  \hfill
  \begin{subfigure}[b]{0.32\textwidth}
    \includegraphics[width=\textwidth]{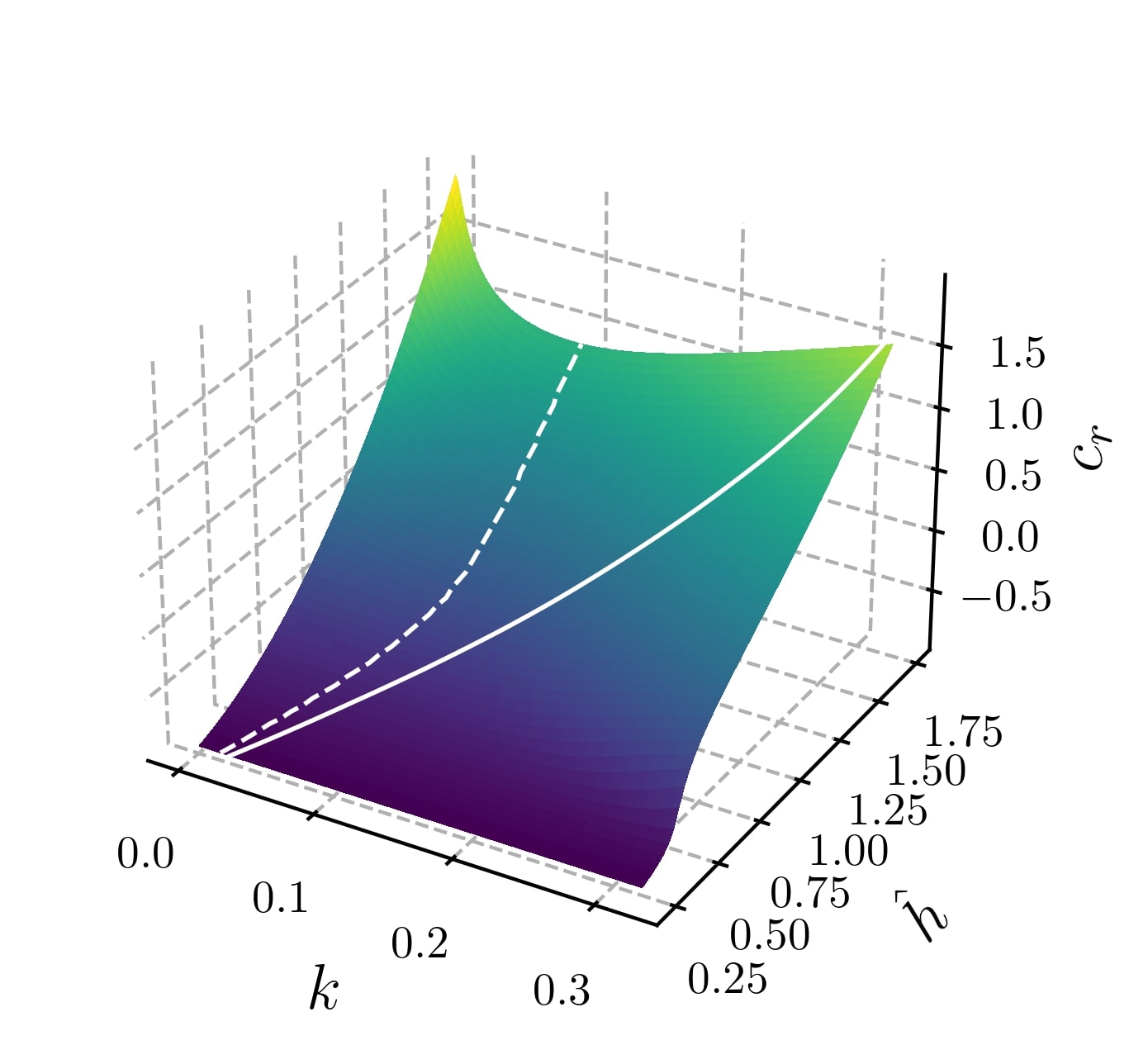}
    \caption{}
  \end{subfigure}
  \begin{subfigure}[b]{0.32\textwidth}
    \includegraphics[width=\textwidth]{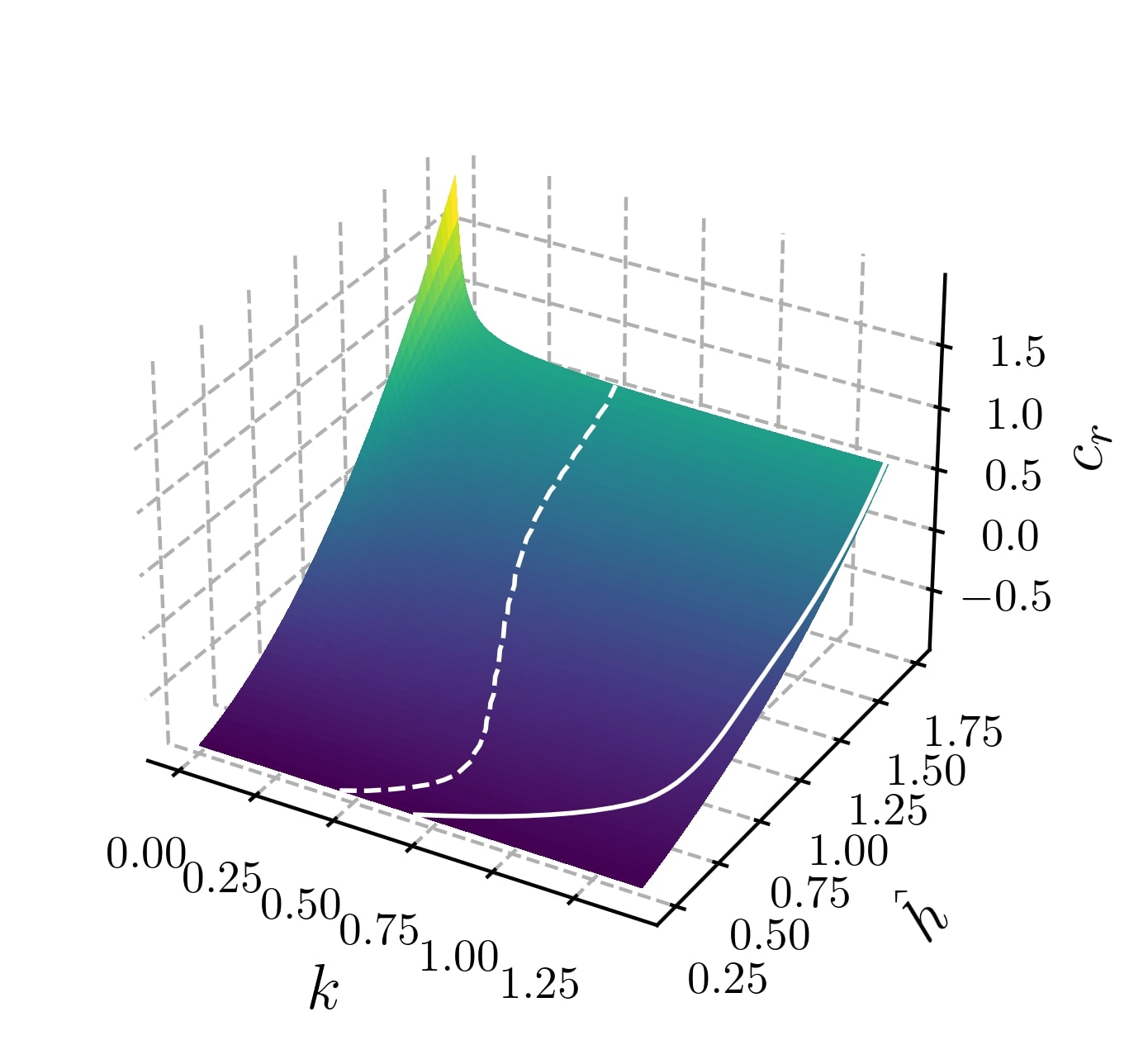}
    \caption{}
  \end{subfigure}
\caption{Coloured surface showing the real part of the phase speed $c_r$ as a function of the wavenumber $k$ and the flat liquid film height $\hat{h}$ at $\R=30$ with the neutral curve (white continuous line) and maximum growth rate (white dashed line) for (a) liquid zinc, (b) water and (c) corn oil.}
    \label{fig:variation_phase_speed}
\end{figure}

The liquid film height $\hat{h}$ and the Kapitza number also play a role in the phase speed of the unstable perturbations. Figure~\ref{fig:variation_phase_speed} shows how the phase speed $c_r$ as a function of the wavenumber $k$ and the liquid film height $\hat{h}$ at $\R=30$ with a highlight on the neutral curve (continuous white line) and the positions of the maximum growth rate (white dashed line) for (a) liquid zinc, (b) water and (c) corn oil. As we move along $k$ for a fixed $\hat{h}$, the phase speed varies first linearly (for small $\hat{h}$) and then non-linearly (for large $\hat{h}$). For zinc and water, this relation is quadratic, with the minimum at the same wavenumber of the maximum growth rate. In contrast, for the corn oil, after a steep variation for $k\rightarrow 0$, it becomes linear again with a minimum not coinciding with the peak of growth rate. As we move along $\hat{h}$ for a fixed $k$, the phase speed of long waves varies quadratically with $\hat{h}$, in accordance with the leading order approximation of the asymptotic expansion \eqref{eq:leading_order_phi_exp} ($c_r = \hat{h}^2-1$). As we increase $k$, this approximation losesvalidity, with waves propagating slower. This relation tends to become linear as we move to shorter wavelengths. The slope of this relationship changes with $k$ and \Ka, becoming almost insensitive to $k$, in the case of corn oil.
A high \Ka, the most unstable mode, travels slower than any other unstable mode.

\subsection{Long-wave mechanism and energy balance of the unstable perturbation}
\label{res_susection:energy_balance_inst_mec}
In this subsection, we describe the mechanism behind the growth of unstable perturbations using momentum and vorticity arguments. Moreover, we investigate how the perturbation's energy is extracted and stored depending on the values of the nondimensional groups and the liquid film height, highlighting the role of viscous terms.
In the falling liquid film case, \cite{kelly1989mechanism}, with the energy-based approach, and \cite{smith1990mechanism}, with asymptotic expansions, showed that the long-wave instability is fed by a streamwise flow field resulting from the base state's shear stress deficiency at the free-surface.

\subsubsection{Long-wave instability mechanism}

Considering the harmonic disturbance $\Tilde{h}$ over a flat interface $\Bar{h}$ (see Section~\ref{subsec:mechanism_long_wave_inst}), the shear stress at the perturbed interface is given by a base state and a perturbation contribution. Expanding $D \hat{u}$ around the base state thickness $\Bar{h}$, we obtain at first order:
\begin{equation}
\label{eq:shear_stress_new_interface_0}
D \hat{u}(\hat{h}) = D\Bar{u}(\Bar{h}) + D\Tilde{u}(\Bar{h}) + D^2\Bar{u}(\Bar{h})\Tilde{h} =0.    
\end{equation}

Since the base state is shear-free at the interface ($D\Bar{u}(\Bar{h})=0$) and knowing that $D^2\Bar{u}(\Bar{h})=-1$, \eqref{eq:shear_stress_new_interface_0} reduces to:
\begin{equation}
\label{eq:shear_stress_new_interface}
D\Tilde{u}(\Bar{h})  - \Tilde{h}=0.    
\end{equation}

The equilibrium of forces implies that the disturbance has to generate a positive shear stress $D\Tilde{u}(\Bar{h})$ to compensate for $\Tilde{h}$. To analyse this mechanism, we considered a reference frame moving at the substrate velocity $U_p$ ($\hat{u}=-1$), and we expand the streamwise velocity amplitude $\acute{u}$ in a power series of $k$ assuming long-wave conditions \eqref{eq:expansion_normal_modes_Smith} \citep{smith1990mechanism}. The solution at $\text{\textit{O}}(1)$ of the equations \eqref{eq:leading_order_smith_like} with boundary conditions \eqref{eq:leading_order_smith_like_bc} is given by a linear velocity amplitude $\acute{u}_{m0}(\hat{y})$ and a positive phase speed $c_{m0}$:
\begin{equation}
\label{eq:asym_ex_leading_order_Smith}
    \acute{u}_{m0}(\hat{y}) = \hat{y} + 1 \qquad\qquad\qquad\qquad c_{m0} = \Bar{h}^2.
\end{equation}
The behaviour of $c_{m0}$ is solely determined by gravity. The thicker the base state, the more gravity is important, and the more the wave travels faster downwards along the $\hat{x}$ direction. 

\begin{figure}
  \begin{subfigure}[b]{0.49\textwidth}
    \includegraphics[width=\textwidth]{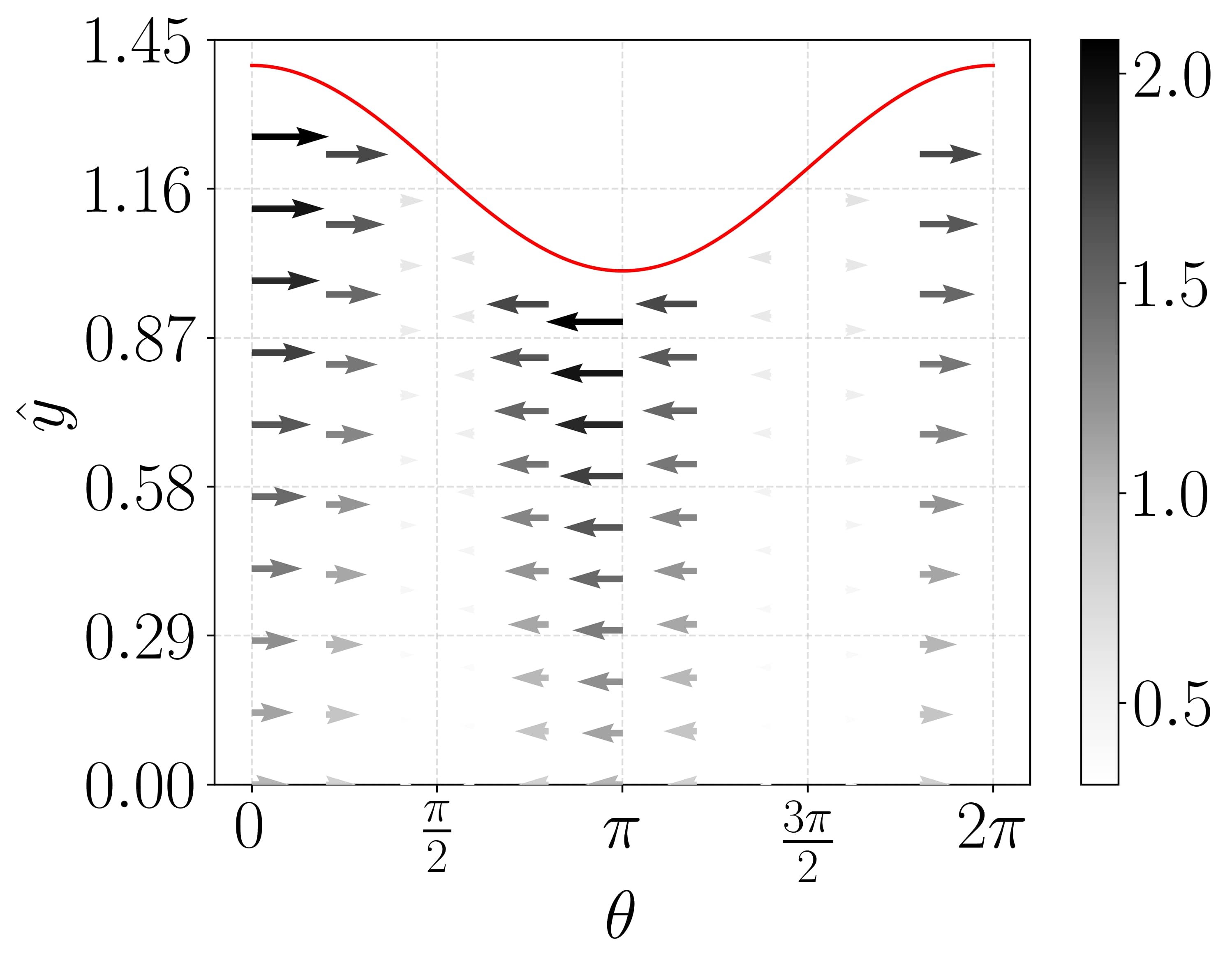}
    \caption{}
    \label{fig:asym_sol_smith_a}
  \end{subfigure}
  \hfill
  \begin{subfigure}[b]{0.49\textwidth}
    \includegraphics[width=\textwidth]{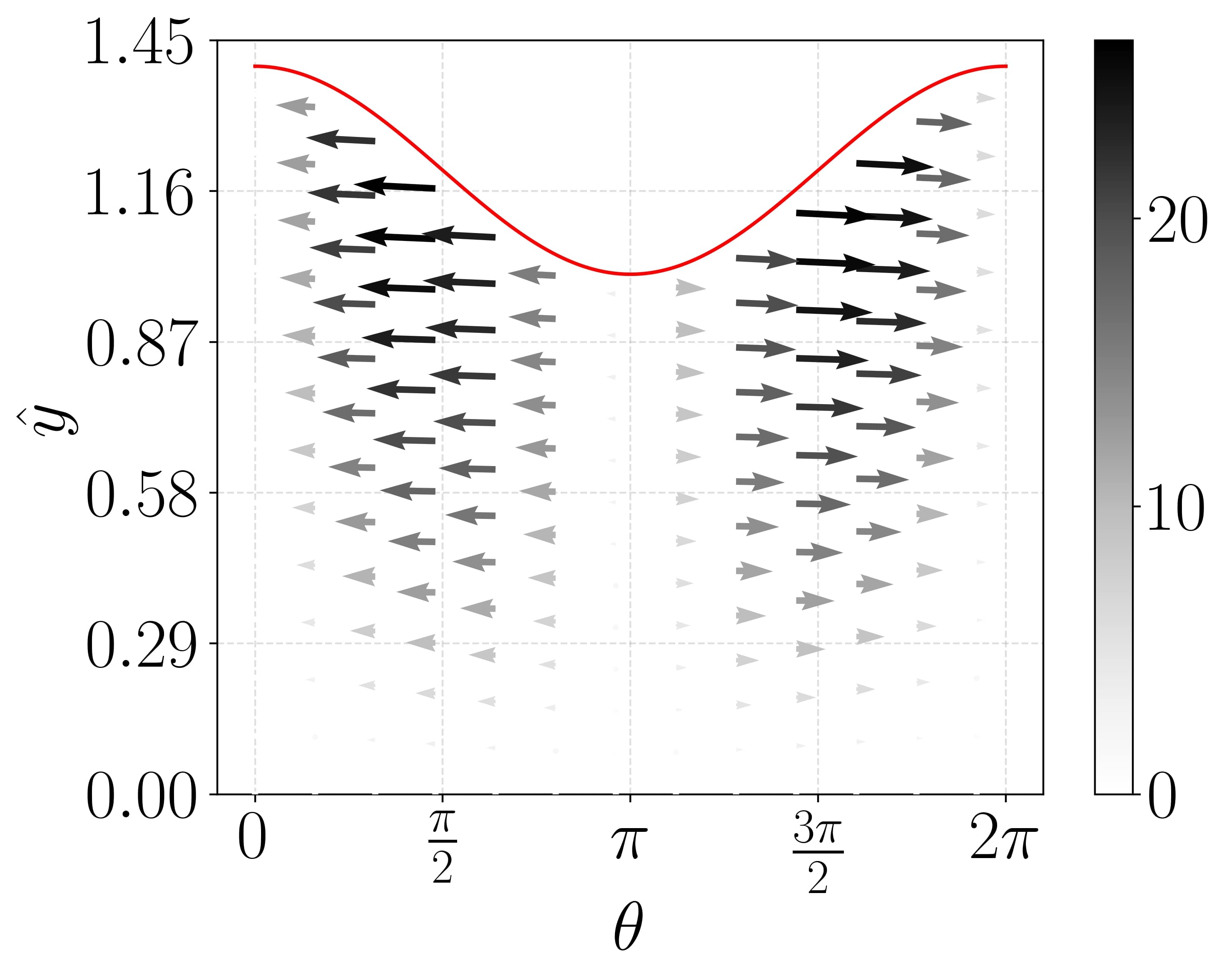}
    \caption{}
    \label{fig:asym_sol_smith_b}
  \end{subfigure}
  \caption{Magnitude (greyscale) and vector field (arrows) of the perturbation's velocity field (a) at leading order and (b) at first order in the wavenumber k}
  \label{fig:asym_sol_smith}
\end{figure}

Figure~\ref{fig:asym_sol_smith_a} shows the flow field at leading order as a consequence of a harmonic displacement of the interface (red line). Since the flow field is in phase with the liquid film displacement, the streamwise velocity is maximum at the peak and minimum at the trough. Considering a control volume in the range $\theta\in[0,\pi/2]$, enclosed between a peak and a node at the interface, this has a positive net flow rate. The velocity field pushes liquid to the right at the crest, with a zero flow rate at the node. In accordance with the continuity equation, this implies a positive displacement of the film interface to accommodate this accumulation of mass, leading to a travelling wave in the positive $\hat{x}$ direction. The solution of \eqref{eq:asym_exp_cont_k_Smith} gives the normal velocity at order $\text{\textit{O}}(k)$:
\begin{equation}
    v_{1} = -i\hat{y}^2/2.
\end{equation}
The link between the phase speed and the flow rate is given by the kinematic boundary condition at the interface \eqref{eq:kin_exp_ord_k_Smith}:
\begin{equation}
    c_{m0} - \Bar{u}_m(\hat{h}) = v_{1} =\int_{0}^{\hat{h}}\,u_{m0}-1\,d\hat{y} = \hat{h}^2/2.
\end{equation}

This initiating mechanism affects the development of the flow field at order $\text{\textit{O}}(k)$ through two \textit{inertial stresses} in \eqref{eq:Ok_moving_syst}:
\begin{equation}
-i\R(u_{m0}(\hat{y}) -1)(\Bar{u}_m(\hat{y}) - c_{m0} - 1) \qquad \mbox{and} \qquad \R v_{1}(\hat{y})D\Bar{u}_m(\hat{y}).
\end{equation}

The first term corresponds to the advection of the reduced leading order solution by the base state velocity with respect to the reduced leading order wave speed. The second term corresponds to the advection of the base state velocity by the first order wall-normal velocity $v_1$. Since the phase speed $c_{m0}$ is larger than any values in $\Bar{u}_m(\hat{y})$, and being $D\Bar{u}_m(\hat{y})=-\hat{y} + \hat{h}$ positive $\forall \hat{y}\in[0,\hat{h}]$, these two terms are both negative. This leads to a destabilizing flow field given by the solution of the first order equations \eqref{eq:first_order_smith_like} with boundary conditions \eqref{eq:first_order_smith_like_bc}:
\begin{equation}
\label{eq:first_order_Smith_sol}
    u_{m1}(\hat{y}) = \frac{\R}{24} i \left(8\hat{h}^4\hat{y} - 4\hat{h}^2\hat{y}^3 + 12\hat{h}\hat{y} + \hat{h}\hat{y}^4 - 4\hat{y}^3\right).
\end{equation}

Figure~\ref{fig:asym_sol_smith_b} shows the flow field calculated with \eqref{eq:first_order_Smith_sol}.
Since the velocity amplitude $u_{m1}(\hat{y})$ is imaginary and positive, the flow has a phase shift of $\pi/2$ with respect to the wave displacement. Consequently, the velocity pushes fluid from the troughs to the peaks, sustaining the growth of the perturbation. This renders a growth mechanism based on the extraction of energy contained in the leading order solution, which, in turn, extracts energy from the base state through the work done by the shear stress at the interface. 

\subsubsection{Vorticity perturbation at the free surface and in the bulk}
In the previous subsection, we studied the general structure of the instability mechanism. The first order solution in $k$, fed by leading order \textit{inertial stresses}, generates a destabilizing flow from the troughs to the peaks due to a $\pi/2$ phase shift to the free-surface displacement. In this section, we go into more detail, analysing how the phase shift and the magnitude of the perturbation change with the wavenumber and the flat liquid film height. To this end, we look at the growth of the perturbation in terms of the vorticity field. The perturbation's shear stress correction in \eqref{eq:shear_stress_new_interface} can be seen as a source of vorticity at the free surface $\omega_{FS}$. We consider a reference frame moving with the wave speed $c_r$, i.e. $\theta=k(\hat{x}-c_r\hat{t})$, and we assume a sinusoidal displacement of the free surface:
\begin{equation}
    \tilde{h} = \sin{(\theta)},
\end{equation}
with $\theta=0$ at 0, this implies through \eqref{eq:TTT}, that:
\begin{equation}
    \eta_r = 0,\qquad\qquad\qquad\qquad \eta_i = -1.
\end{equation}
By means of the kinematic boundary condition \eqref{kinematic_con_eq}, we scale the eigenfunction such that:
\begin{equation}
    \phi_r(\hat{h}) = c_i, \qquad\qquad\qquad\qquad \phi_i(\hat{h}) = -\hat{c}.
\end{equation}
The shear stress condition \eqref{eq:bc_OS_4} imposes that:
\begin{equation}
    D^2\phi_r(\hat{h}) = -k^2 c_i, \qquad\qquad\qquad\qquad D^2\phi_i(\hat{h}) = -2 + k^2\hat{c}.
\end{equation}
By replacing these expressions in \eqref{eq:vorticiy_expression}, we obtain the vorticity at the free surface:
\begin{equation}
    \omega_{FS} = \gamma \sin{(\theta - \xi)},
\end{equation}
with the amplitude $\gamma$ and the phase shift $\xi$ given by:

\begin{equation}
\label{eq:magn_phase_vorti_inter}
\gamma = 2\sqrt{(1-k^2\hat{c})^2 + (k^2c_i)^2]},
\qquad\qquad\qquad\qquad \xi = \arctan{(k^2c_i/(1-k^2\hat{c})}.
\end{equation}
Depending on $\xi$, $\omega_{FS}$ stabilizes ($\xi<0$) or destabilizes ($\xi>0$) the free surface displacement \citep{kelly1989mechanism}.

\begin{figure}
  \begin{subfigure}[b]{0.32\textwidth}
    \includegraphics[width=\textwidth]{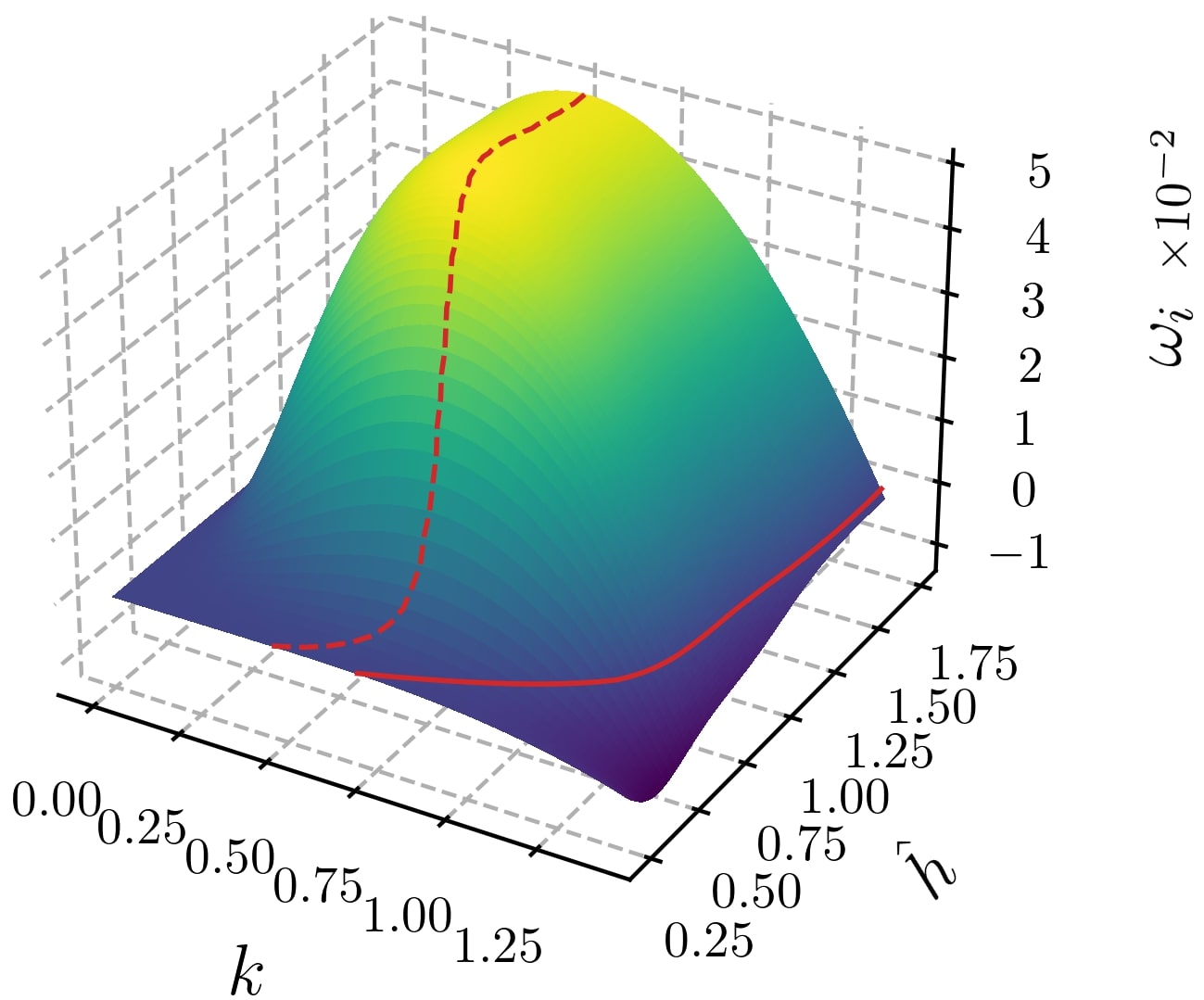}
    \caption{}
  \end{subfigure}
  \hfill
  \begin{subfigure}[b]{0.32\textwidth}
    \includegraphics[width=\textwidth]{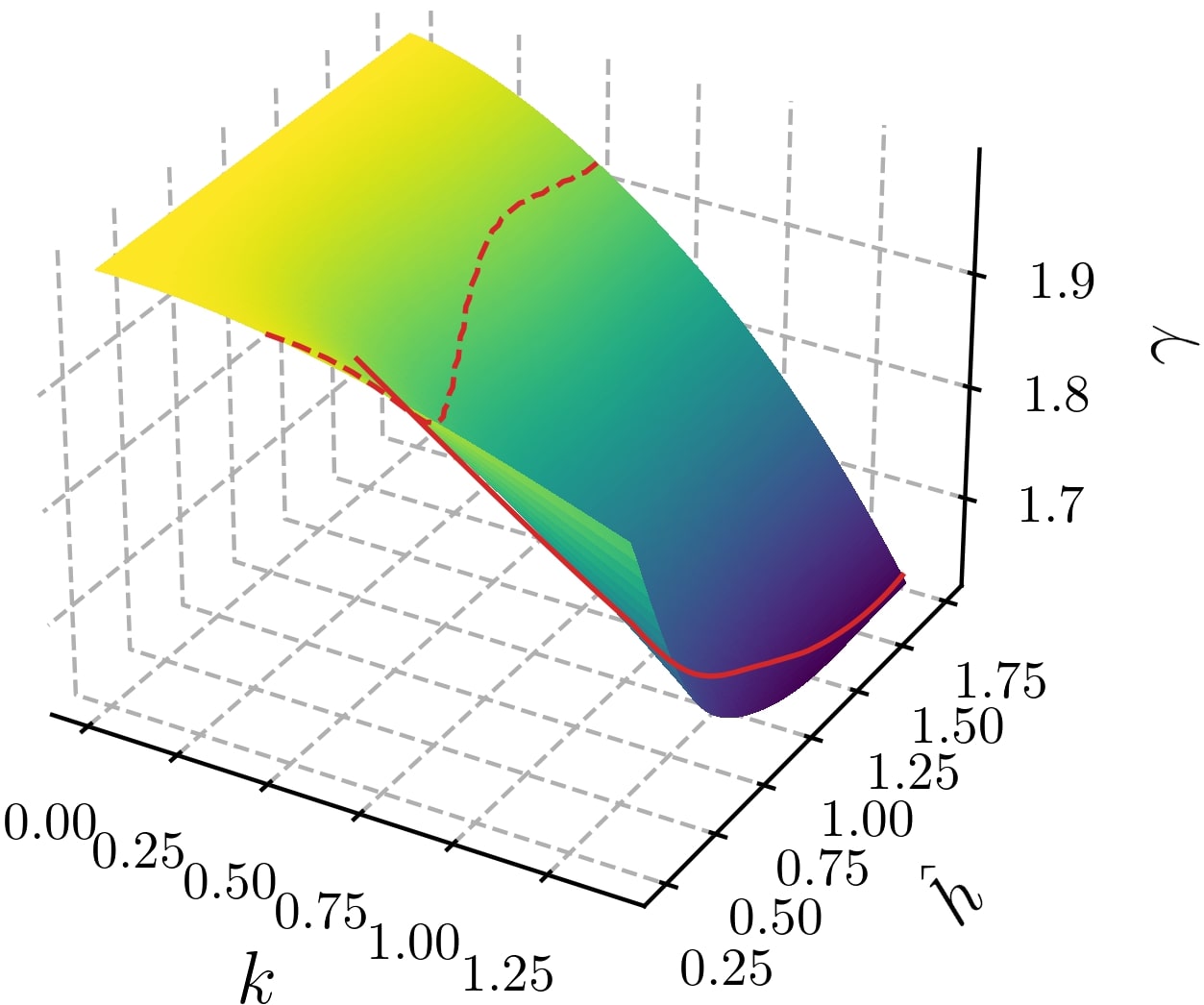}
    \caption{}
  \end{subfigure}
  \hfill
  \begin{subfigure}[b]{0.32\textwidth}
    \includegraphics[width=\textwidth]{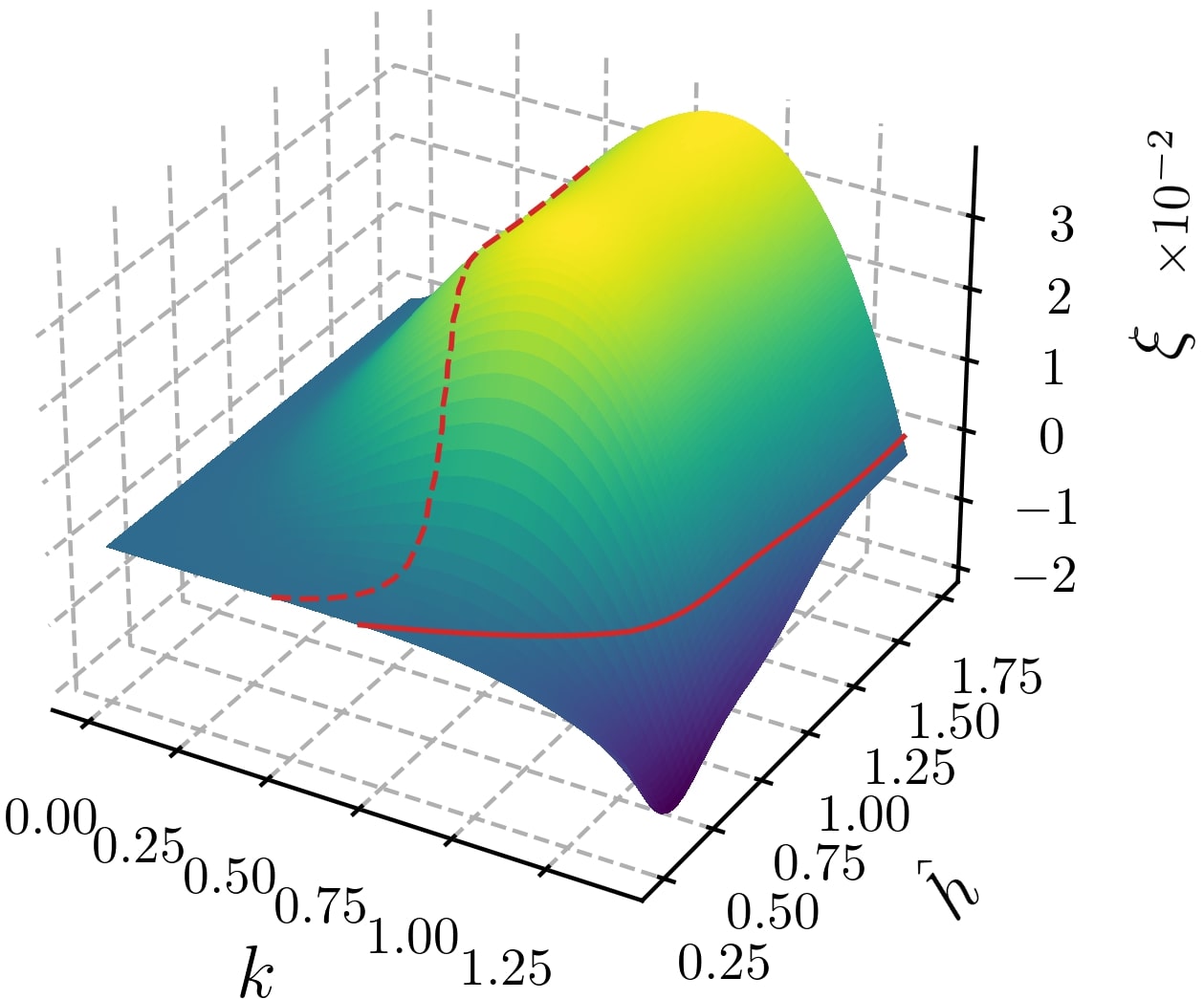}
    \caption{}
  \end{subfigure}
  \begin{subfigure}[b]{0.32\textwidth}
    \includegraphics[width=\textwidth]{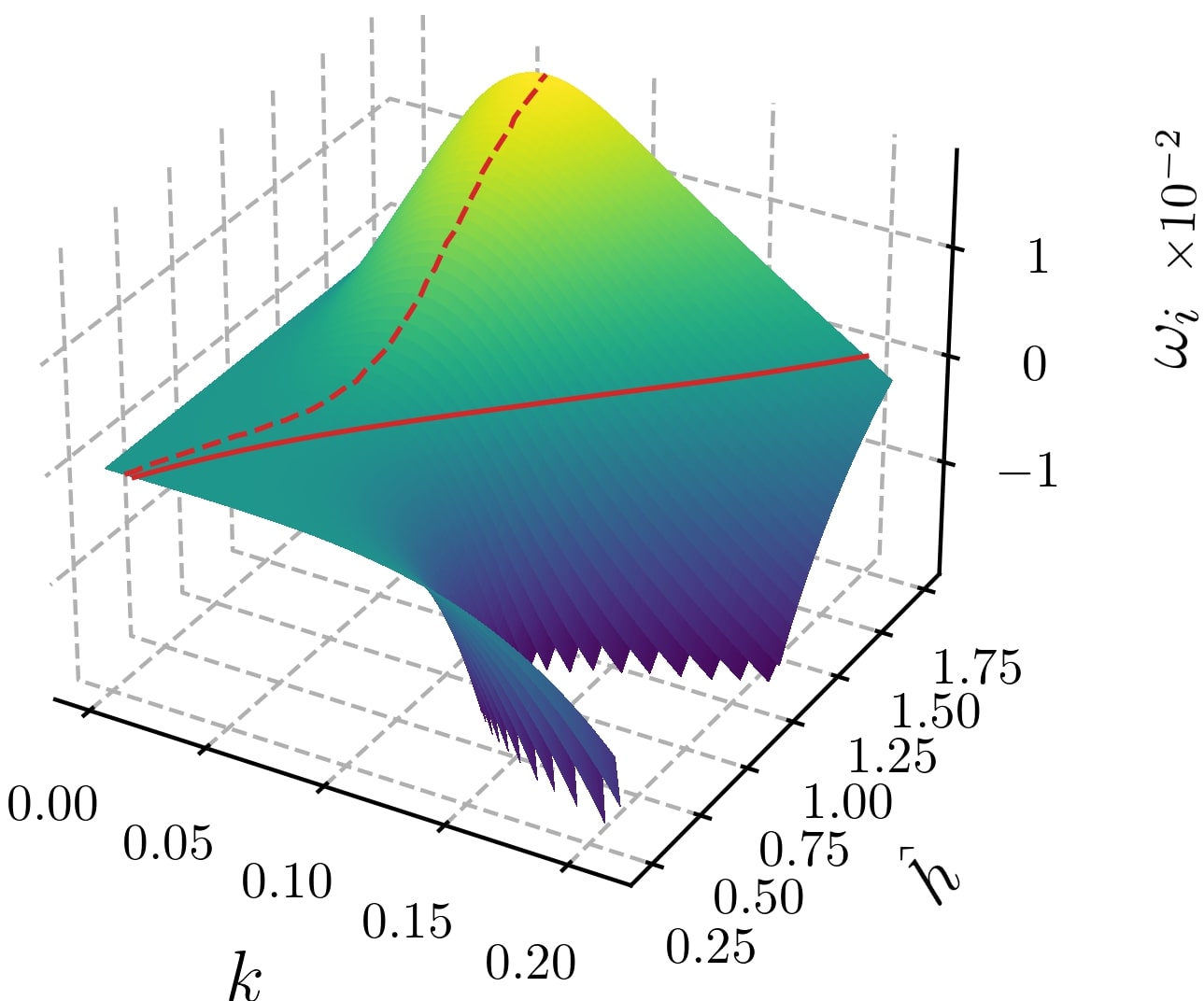}
    \caption{}
  \end{subfigure}
  \hfill
  \begin{subfigure}[b]{0.32\textwidth}
    \includegraphics[width=\textwidth]{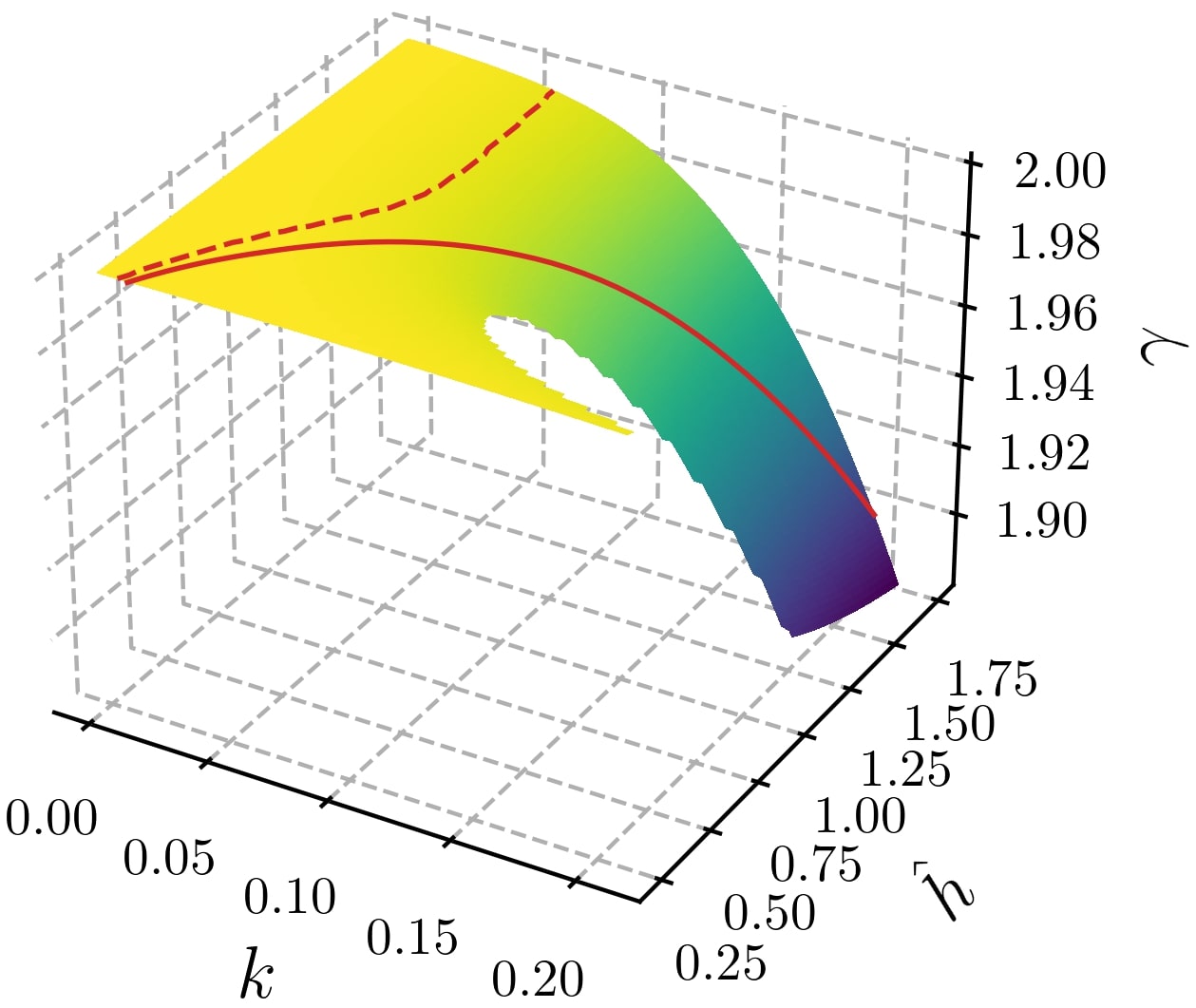}
    \caption{}
  \end{subfigure}
  \hfill
  \begin{subfigure}[b]{0.32\textwidth}
    \includegraphics[width=\textwidth]{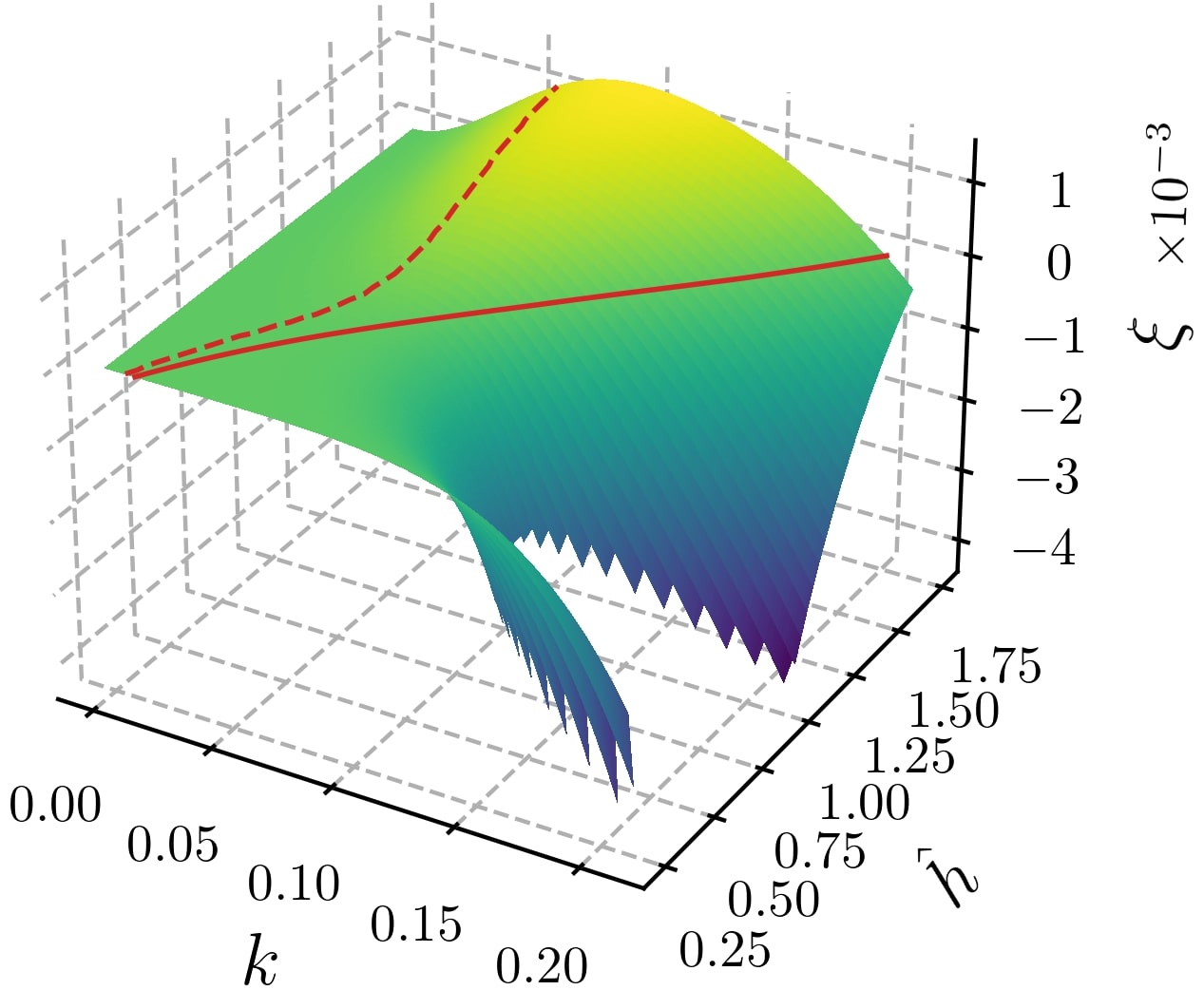}
    \caption{}
  \end{subfigure}
  \hfill
  \caption{(a and d) Growth rate, (b and e) free-surface vorticity amplitude and (c and f) phase shift as a function of $k$ and $\hat{h}$ with the neutral curve (continuous red line) and the positions of maximum growth rates (red dashed line) for $\R=30$, considering (a, b and c) corn oil and (d, e, and f) zinc.}
  \label{fig:vorticiy_fs_amp_phase}
\end{figure}

Figure~\ref{fig:vorticiy_fs_amp_phase} shows (a and d) the growth rate $\omega_i$, (b and e) the amplitude $\gamma$ and (c and f) the phase shift $\xi$ of the free-surface vorticity as a function of the wavenumber $k$ and the liquid film thickness $\hat{h}$ at $\R=30$, with the neutral curve (continuous red line) and the maximum growth rate (dashed red line) for (a, b and c) the corn oil and (d, e, and f) the liquid zinc. For a fixed k, $\omega_i$ gently grows with $\hat{h}$ driven by a vorticity with a large amplitude and a very small positive phase shift. For $\hat{h}=1$, $\omega_i$ grows more sharply, accompanied by an increase in the phase shift and a decrease in the vorticity amplitude. In the corn oil case, the growth rate and the phase shift reach a plateau for large $\hat{h}$, in line with the saturating mechanism introduced in subsection~\ref{res_susection:neutral_curves}.
The phase shift $\xi$ peaks at higher wave numbers than the growth rate $\omega_i$. The decay of the vorticity amplitude $\gamma$ with $k$ compensates for the destabilizing effect of larger phase shifts, decreasing the growth rate. This suggests that the instability mechanism is also driven by the vorticity in the bulk, thus confirming the results in the previous subsections.

\begin{figure}
  \begin{subfigure}[b]{0.32\textwidth}
    \includegraphics[width=\textwidth]{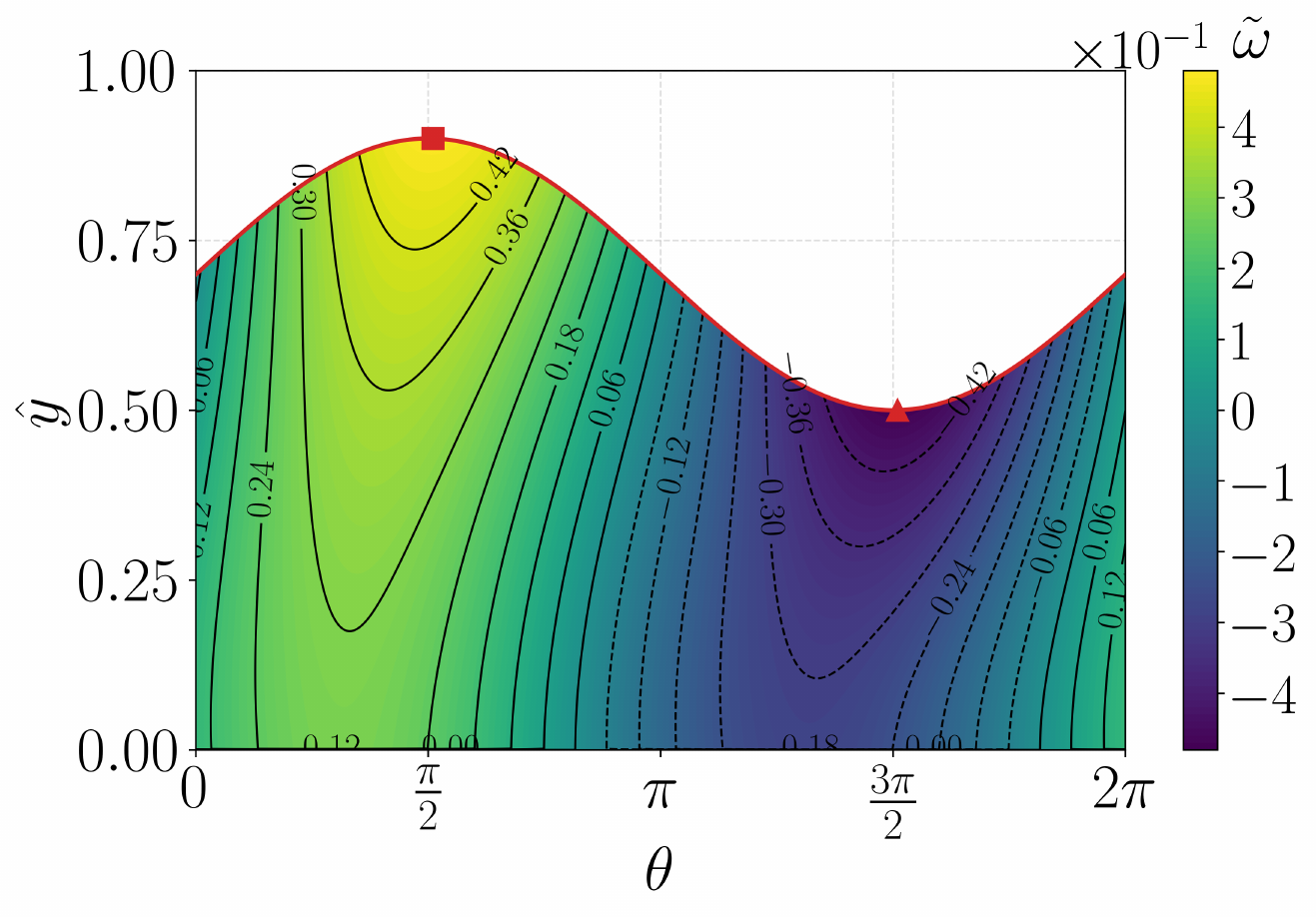}
    \caption{}
  \end{subfigure}
  \hfill
  \begin{subfigure}[b]{0.32\textwidth}
    \includegraphics[width=\textwidth]{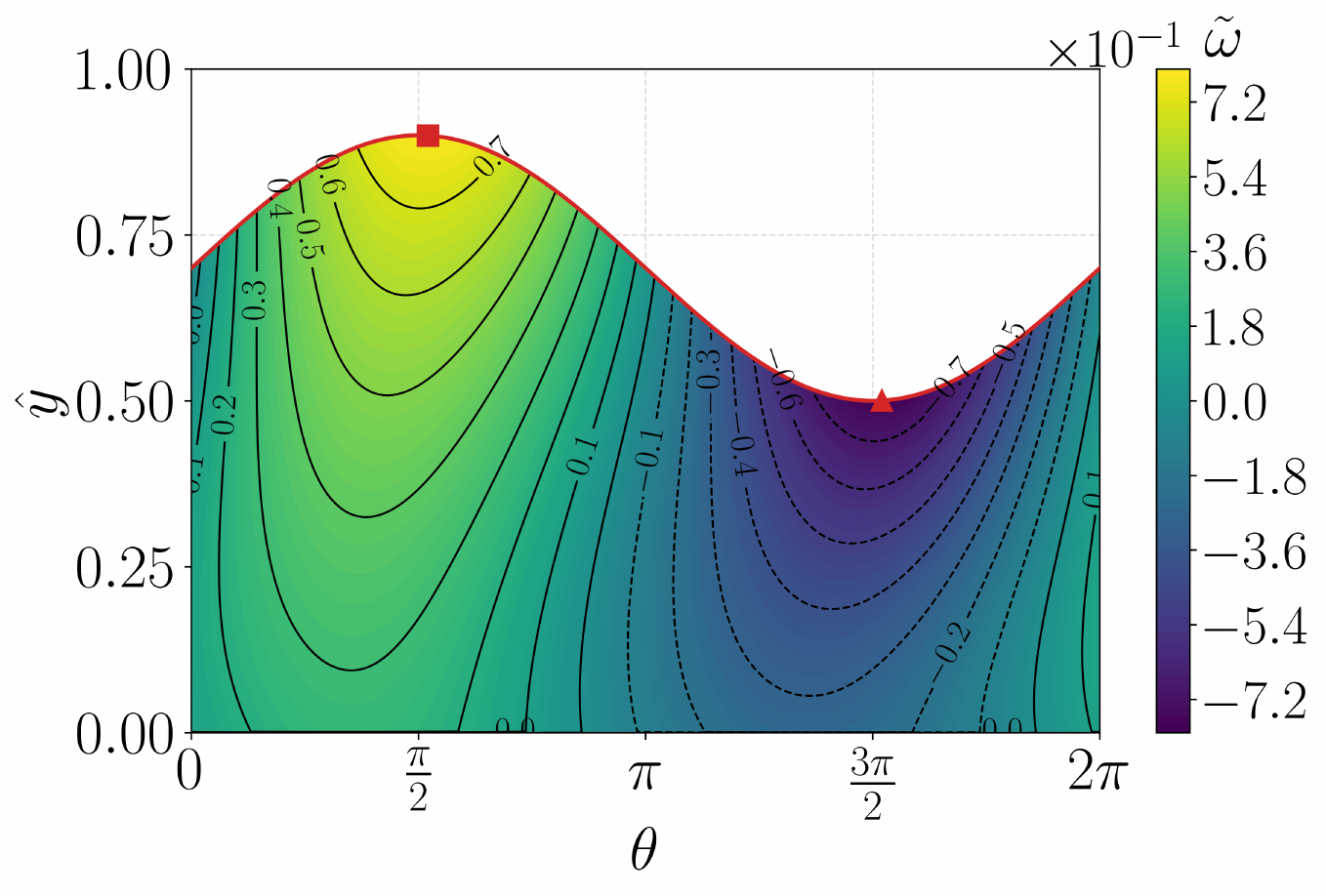}
    \caption{}
  \end{subfigure}
  \hfill
  \begin{subfigure}[b]{0.32\textwidth}
    \includegraphics[width=\textwidth]{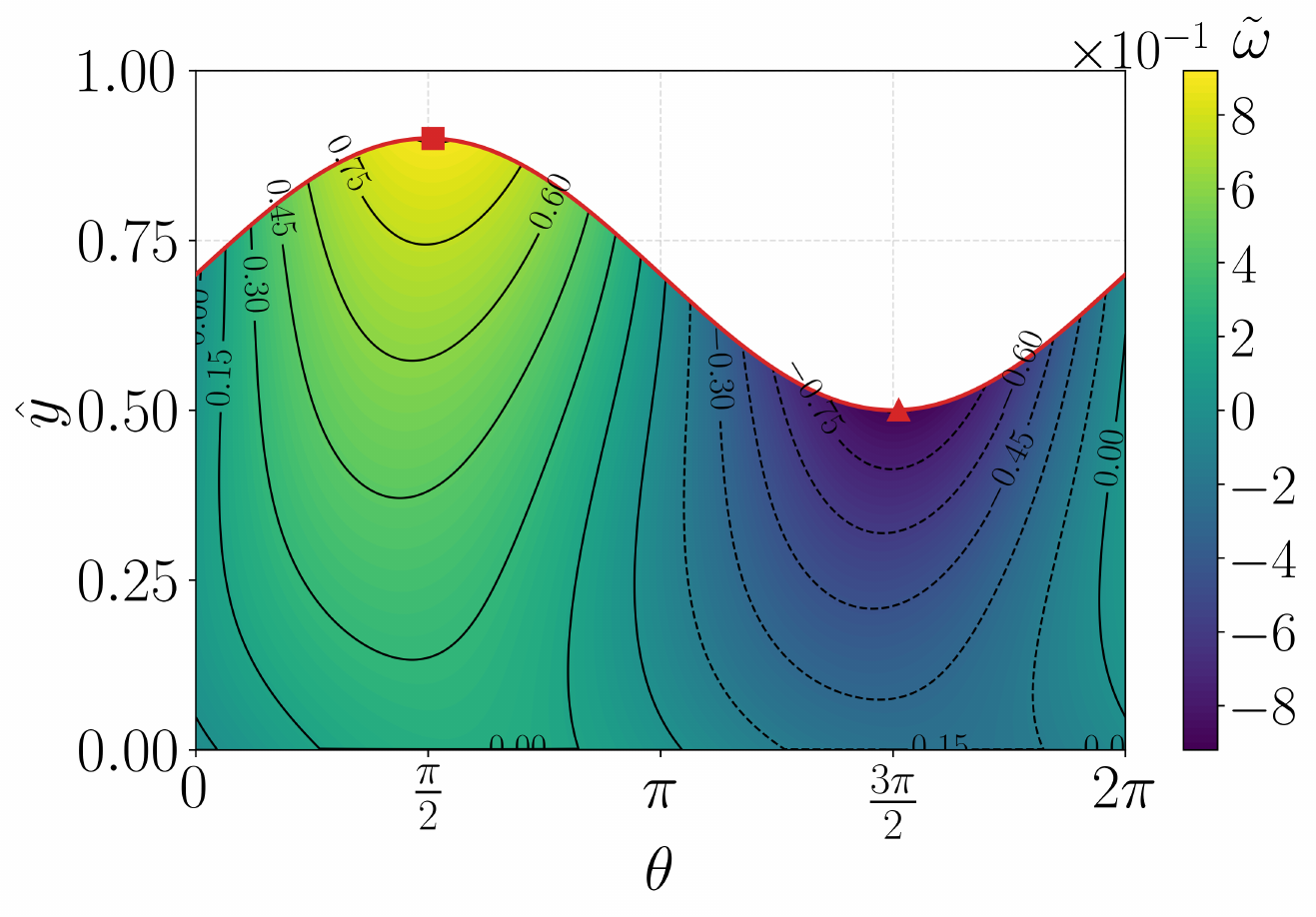}
    \caption{}
  \end{subfigure}
  \hfill
  \begin{subfigure}[b]{0.32\textwidth}
    \includegraphics[width=\textwidth]{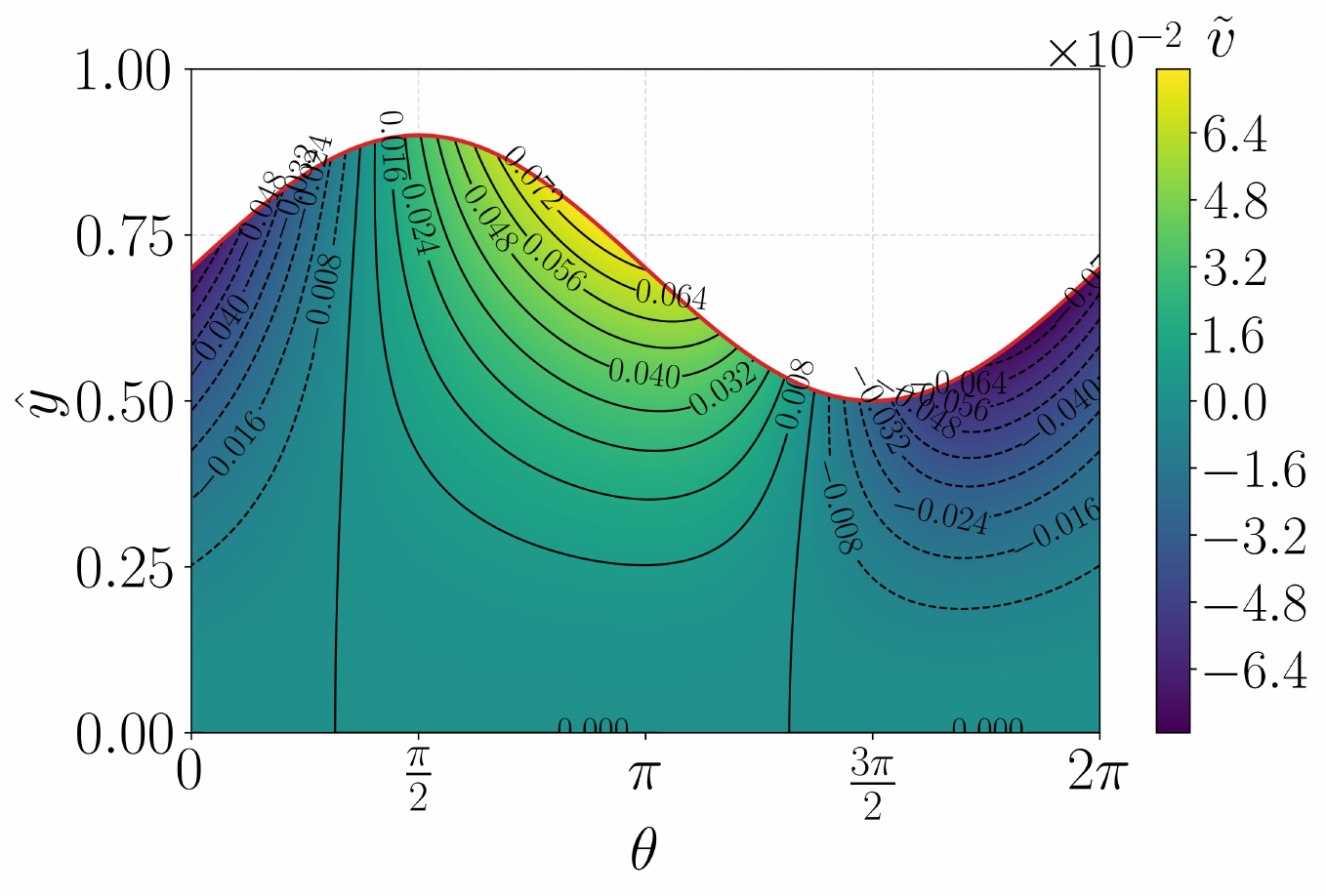}
    \caption{}
  \end{subfigure}
  \hfill
  \begin{subfigure}[b]{0.32\textwidth}
    \includegraphics[width=\textwidth]{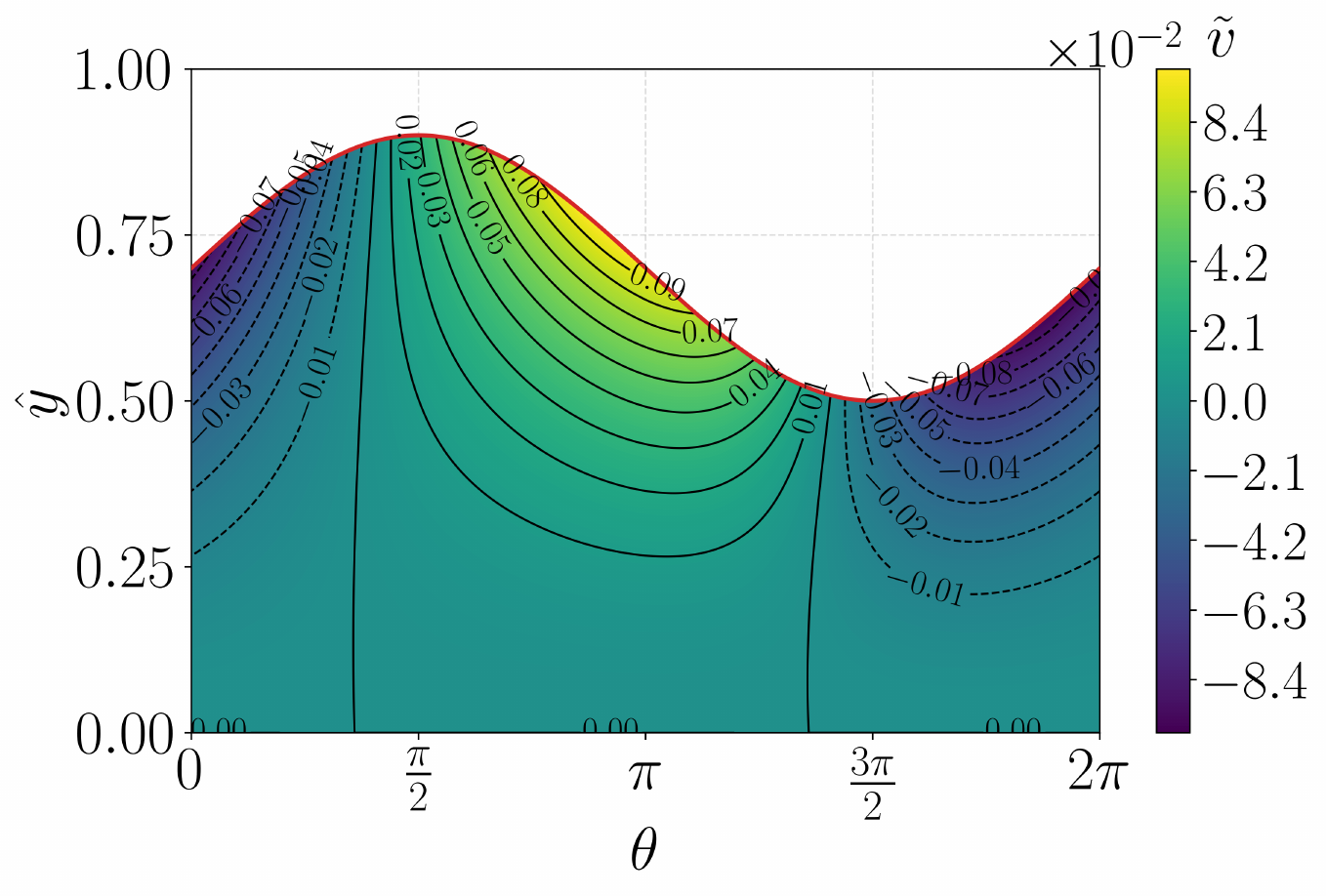}
    \caption{}
  \end{subfigure}
  \hfill
  \begin{subfigure}[b]{0.32\textwidth}
    \includegraphics[width=\textwidth]{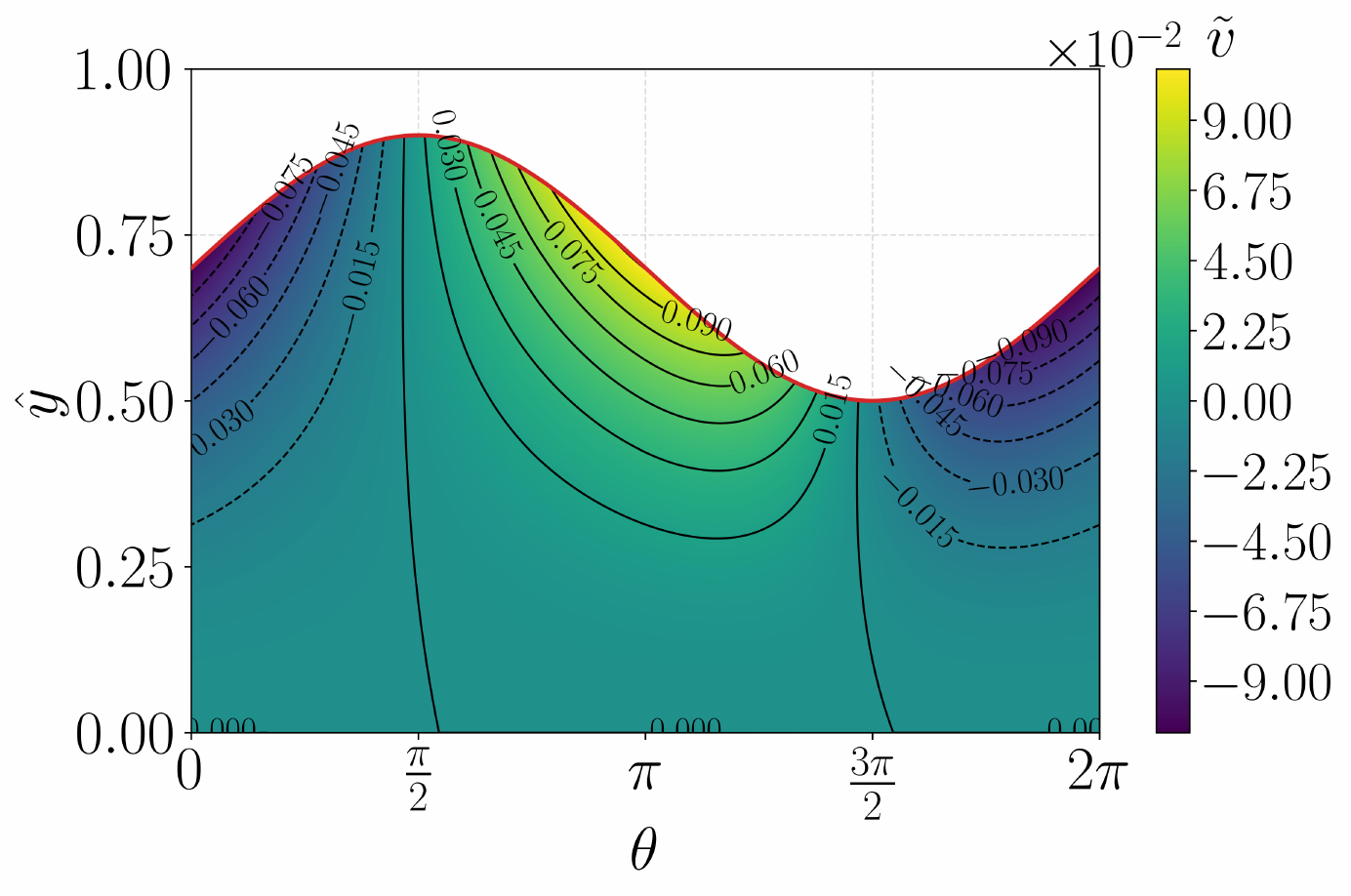}
    \caption{}
  \end{subfigure}
  \begin{subfigure}[b]{0.32\textwidth}
    \includegraphics[width=\textwidth]{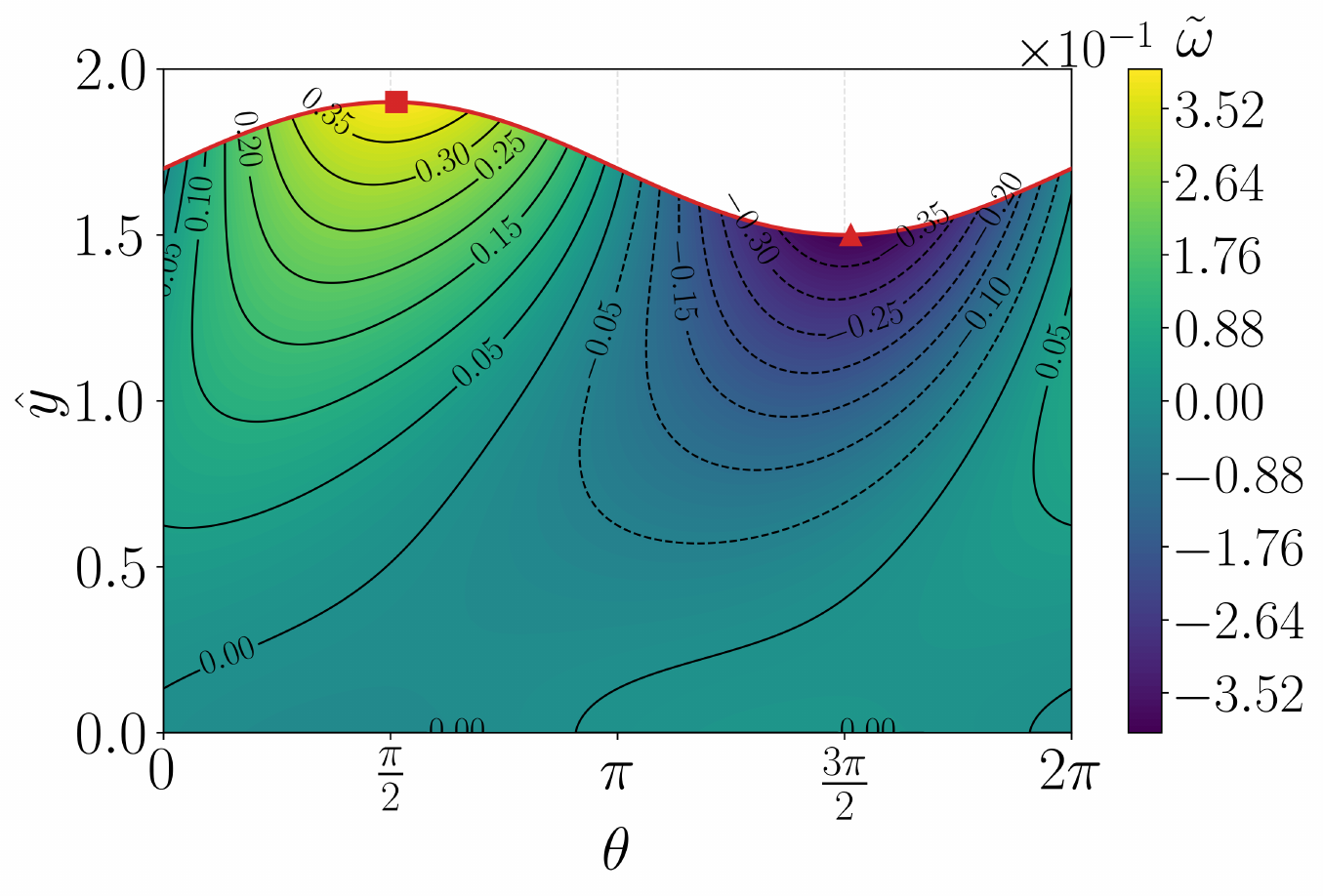}
    \caption{}
  \end{subfigure}
  \hfill
  \begin{subfigure}[b]{0.32\textwidth}
    \includegraphics[width=\textwidth]{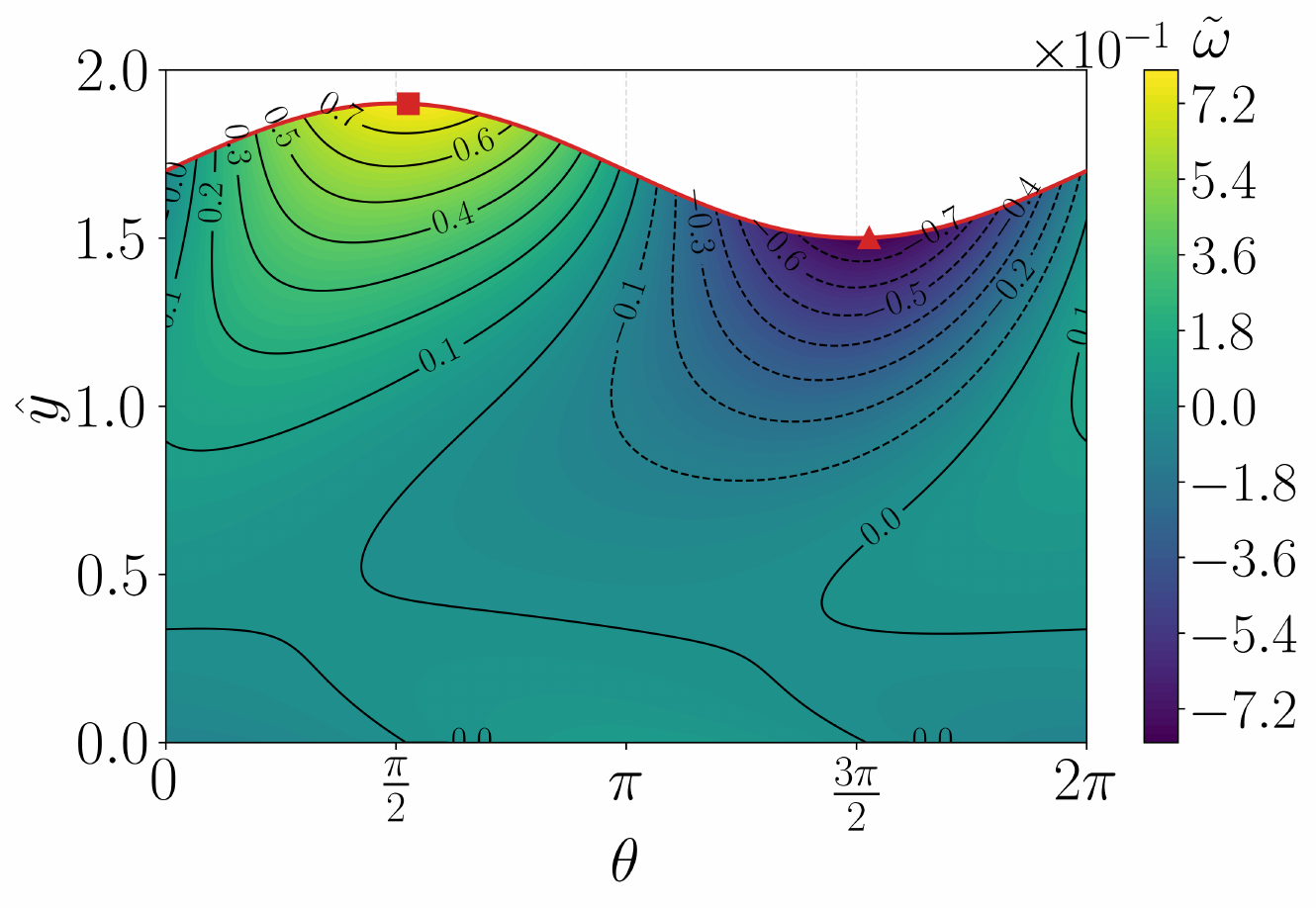}
    \caption{}
  \end{subfigure}
  \hfill
  \begin{subfigure}[b]{0.32\textwidth}
    \includegraphics[width=\textwidth]{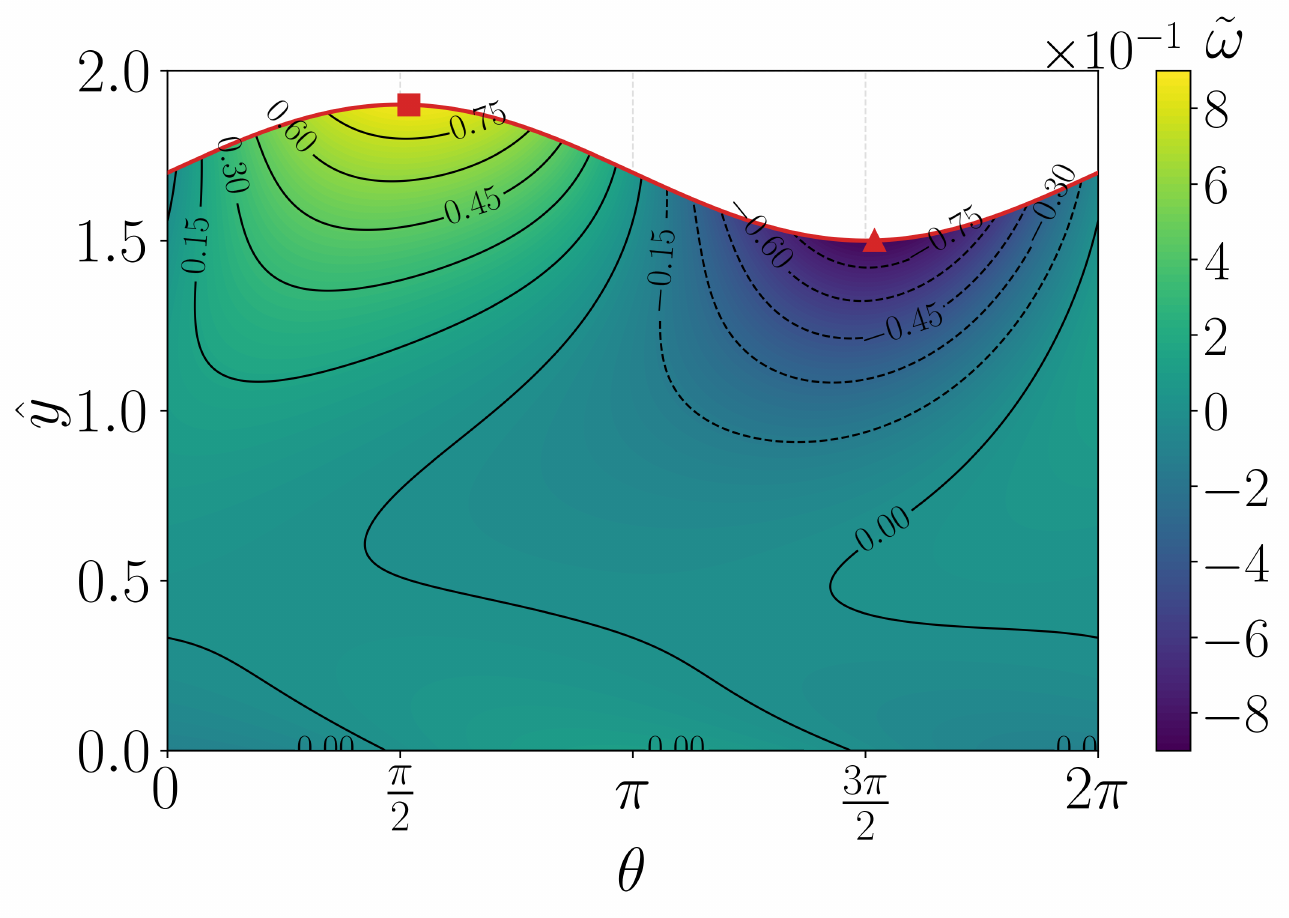}
    \caption{}
  \end{subfigure}
  \hfill
  \begin{subfigure}[b]{0.32\textwidth}
    \includegraphics[width=\textwidth]{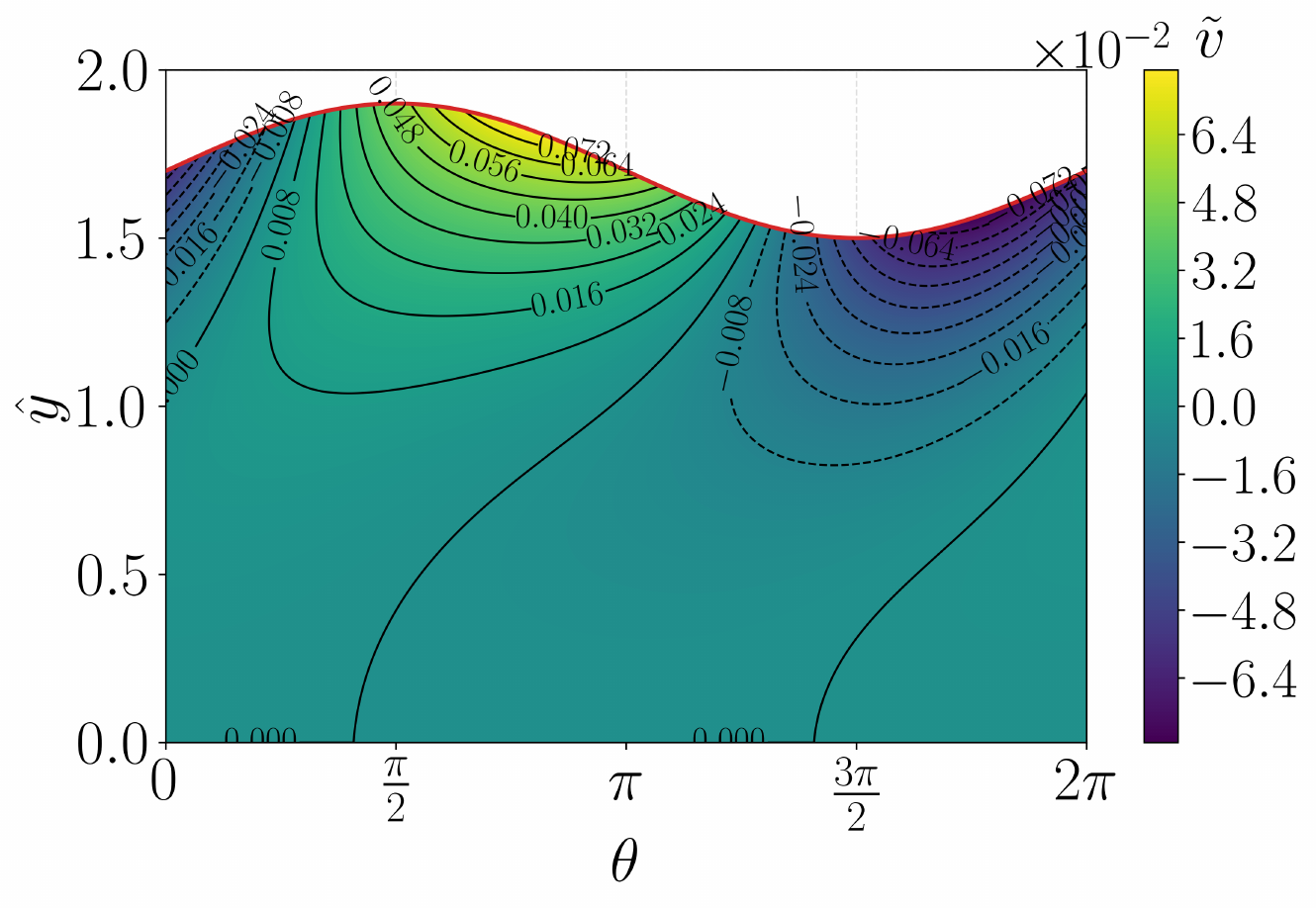}
    \caption{}
  \end{subfigure}
  \hfill
  \begin{subfigure}[b]{0.32\textwidth}
    \includegraphics[width=\textwidth]{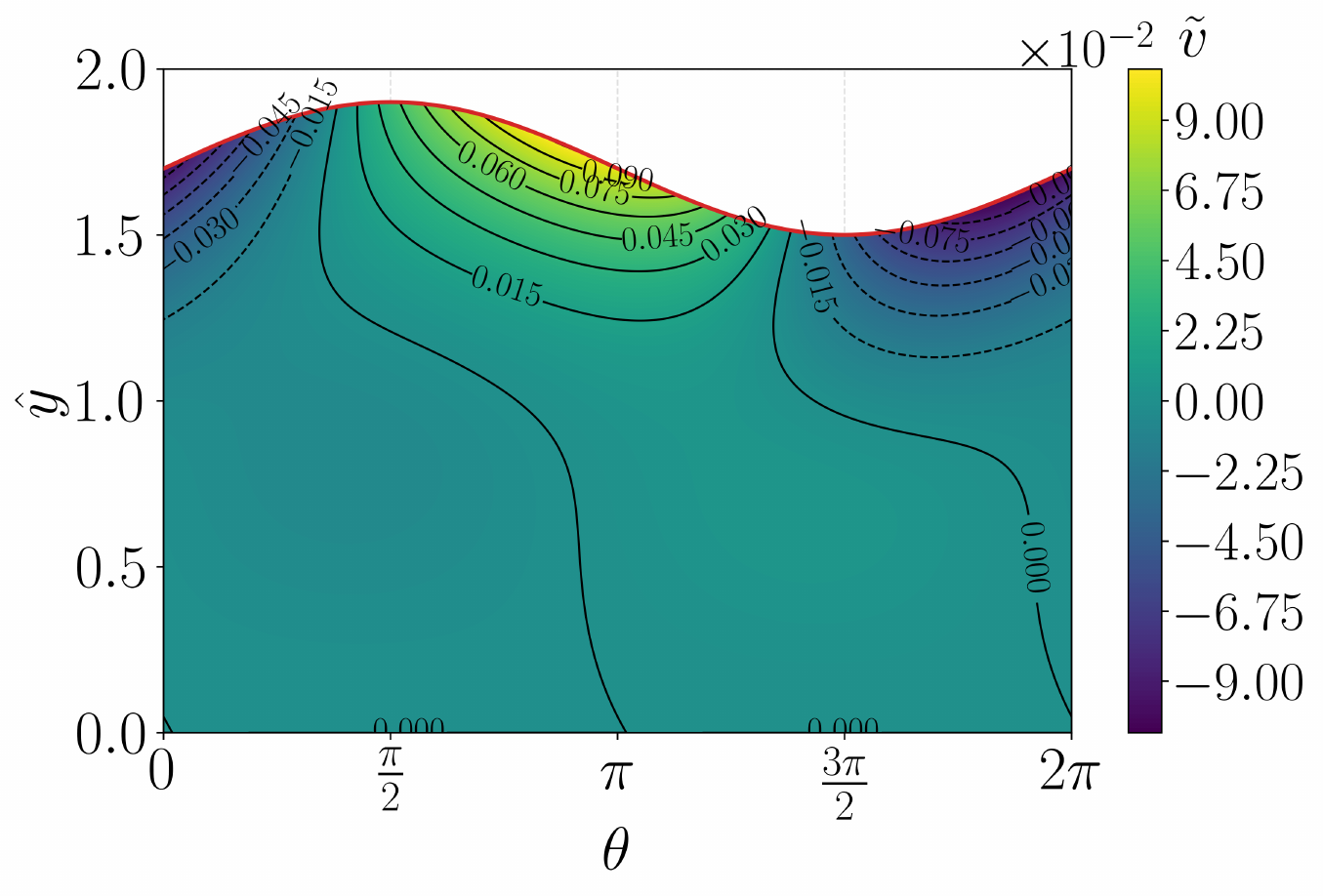}
    \caption{}
  \end{subfigure}
  \hfill
  \begin{subfigure}[b]{0.32\textwidth}
    \includegraphics[width=\textwidth]{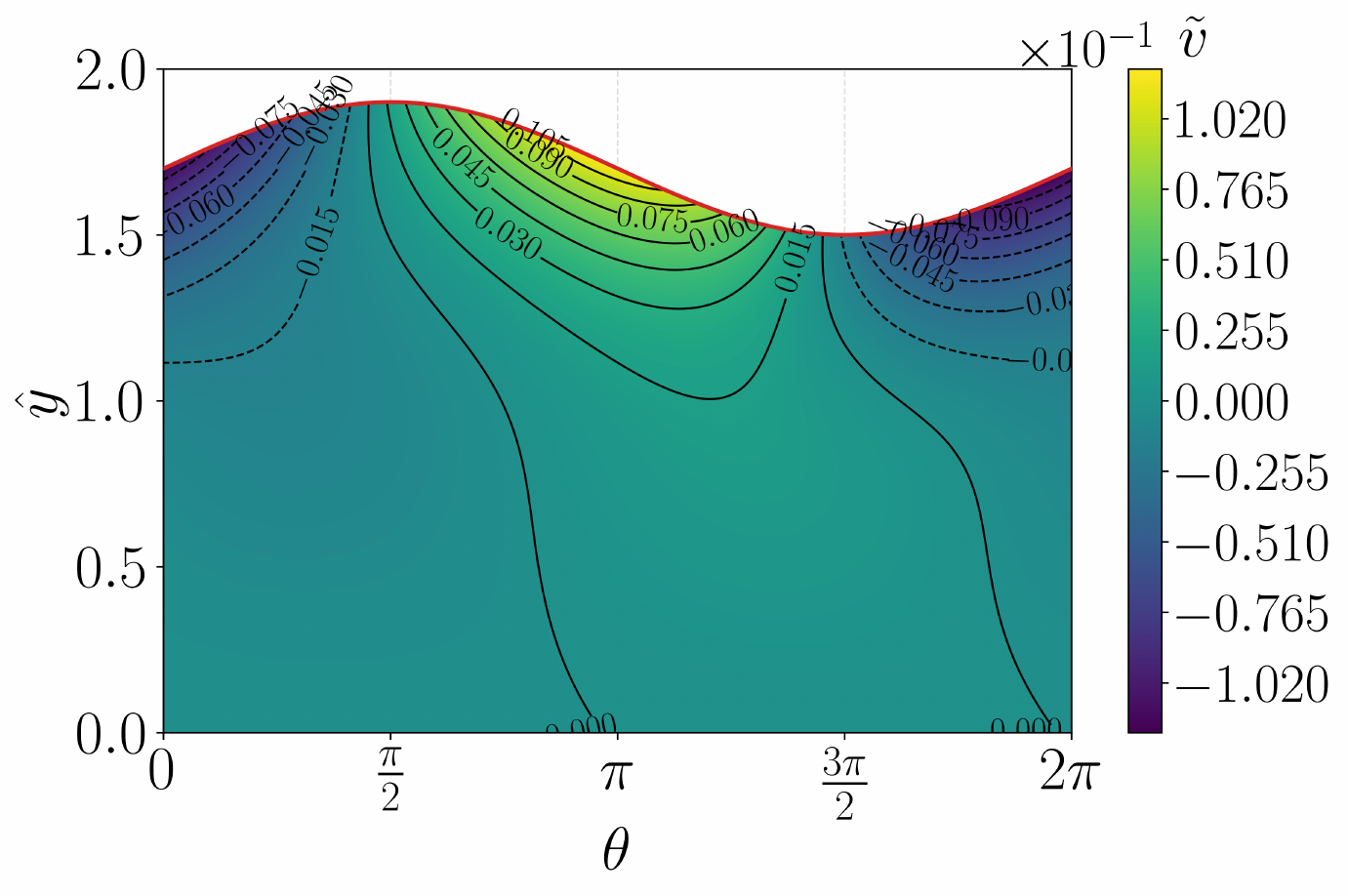}
    \caption{}
  \end{subfigure}
  \caption{Perturbation (a, b, c, g, h and i) vorticity and (d, e, f, j, k and l) normal velocity fields at (a, b, c, d, e and f) $\hat{h}=0.7$  and (g, h, i, j, k and l) $\hat{h}=1.7$ for the corn oil with $k=$0.5 (a, d, g and j), 0.93 (b, e, h and k) and 1.2 (c, f, i and l) with the maximum (red square) and minimum (red triangle) of vorticity at the interface with $\R=30$}
  \label{fig:vorticiy_plots_corn}
\end{figure}

To investigate this behaviour, we analysed the vorticity perturbation $\tilde{\omega}(\theta,\hat{y})=\partial_{\hat{x}}\Tilde{v} - \partial_{\hat{y}}\Tilde{u}$ within the liquid film. The vorticity equation is obtained by taking the curl of the momentum equations \eqref{linearize_eq:2} and \eqref{linearize_eq:3}, considering the continuity equations \eqref{linearize_eq:1} and the base state \eqref{eq:base_state}, which leads to: 
\begin{equation}
    \partial_{\hat{t}}\Tilde{\omega} + \Bar{u}\partial_{\hat{x}}\Tilde{\omega} - \Tilde{v}D^2\Bar{u}  =  \frac{1}{Re}(\partial_{\hat{x}\hat{x}}\Tilde{\omega} + D^2\Tilde{\omega}).
\end{equation}
Considering the reference frame moving with the wave speed $c_r$ and knowing that $D^2\Bar{u}=-1$, we obtain:
\begin{equation}
\label{eq:vorticity}
     k(\Bar{u}-c_r)\partial_{\theta}\Tilde{\omega} + \Tilde{v} = \frac{1}{Re}(k^2\partial_{\theta\theta}\Tilde{\omega} + D^2\Tilde{\omega}).
\end{equation}

On the left-hand side, the first term corresponds to the advection relative to the wave velocity, and the second term corresponds to the vorticity perturbation's advection of the base state. The term on the right-hand side corresponds to the viscous diffusion of vorticity. 
Injecting a normal model solution for the vorticity with $k\in\mathbb{R}$ and $c = c_r + ic_i$:
\begin{equation}
    \tilde{\omega} = \acute{\omega}(\hat{y})\exp(ik(\hat{x} - c\hat{t})) = \acute{\omega}(\hat{y})\exp(kc_i\hat{t})\exp(i\theta),
\end{equation}
and for the wall-normal velocity given by \eqref{eq:expansion_normal_modes_Smith} in \eqref{eq:vorticity} leads to the expression for the free surface amplitude $\acute{v}(\hat{y})$:
\begin{equation}
    \acute{v}(\hat{y}) = ik(c_r - \Bar{u}(\hat{y}))\acute{\omega}(\hat{y}) + \frac{1}{Re}(-k^2 + D^2)\acute{\omega}(\hat{y}).
\end{equation}
Note that for $\R\gg 1$, the relation reduces at leading order in $k$ to 
\begin{equation}
\label{eq:approx_v_vorticity}
    \acute{v}(\hat{y}) \approx ik(c_r - \Bar{u}(\hat{y}))\acute{\omega}(\hat{y}) \,,
\end{equation}
which corresponds to neglecting viscous effects.

Approximating $c_r$ with the long-wave asymptotic solution at $O(1)$ in \eqref{eq:leading_order_phi_exp}, the term $c_r -\Bar{u}$ is always positive $\forall \hat{y}\in[0,\hat{h}]$, implying that the vertical velocity amplitude changes linearly with the vorticity amplitude with a phase shift of $\pi/2$ along $\theta$. 

Figure~\ref{fig:vorticiy_plots_corn} show the perturbation (a, b, c, g, h and i) vorticity and (d, e, f, j, k and l) normal velocity fields at (a, b, c, d, e and f) $\hat{h}=0.7$  and (g, h, i, j, k and l) $\hat{h}=1.7$ for the corn oil with $k=$0.5 (a, d, g and j), 0.93 (b, e, h and k) and 1.2 (c, f, i and l) with the maximum (red square) and minimum (red triangle) of vorticity at the interface with $\R=30$.
For $k=0.5$, the vorticity contours deform towards the left, advected by the flow. Since the advection velocity strengthens, this tilting effect is more intense for large $\hat{h}$. This structure recalls the capillary separation eddy in a falling liquid film of finite amplitude in the capillary flow region \citep{dietze2009experimental}. 
The induced normal velocity field changes sign at the peak of vorticity along $\theta$, leading to a positive (negative) net vertical flow rate under the crests (trough), which fosters the growth of the perturbation.
\begin{figure}
  \begin{subfigure}[b]{0.32\textwidth}
    \includegraphics[width=\textwidth]{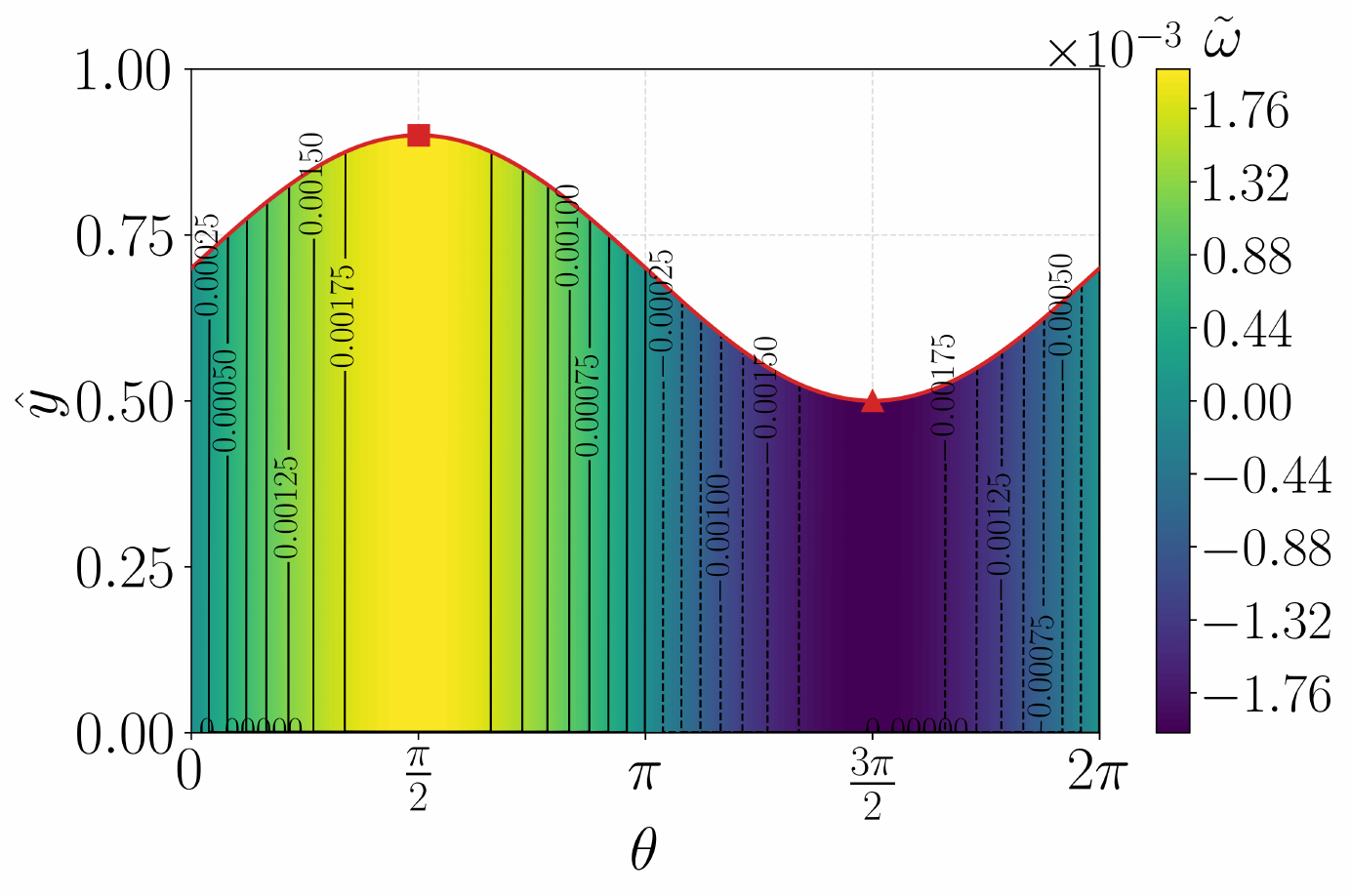}
    \caption{}
  \end{subfigure}
  \hfill
  \begin{subfigure}[b]{0.32\textwidth}
    \includegraphics[width=\textwidth]{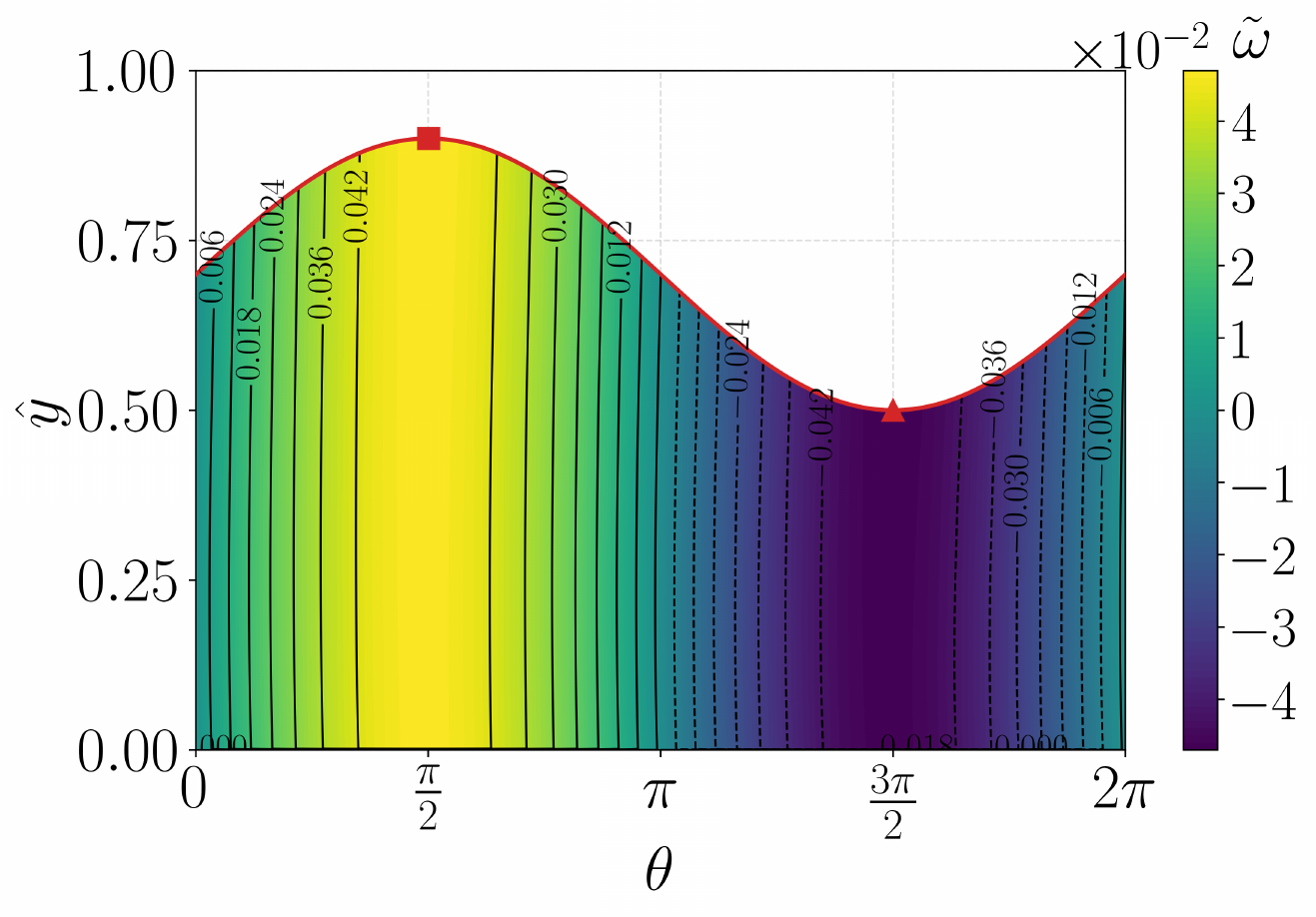}
    \caption{}
  \end{subfigure}
  \hfill
  \begin{subfigure}[b]{0.32\textwidth}
    \includegraphics[width=\textwidth]{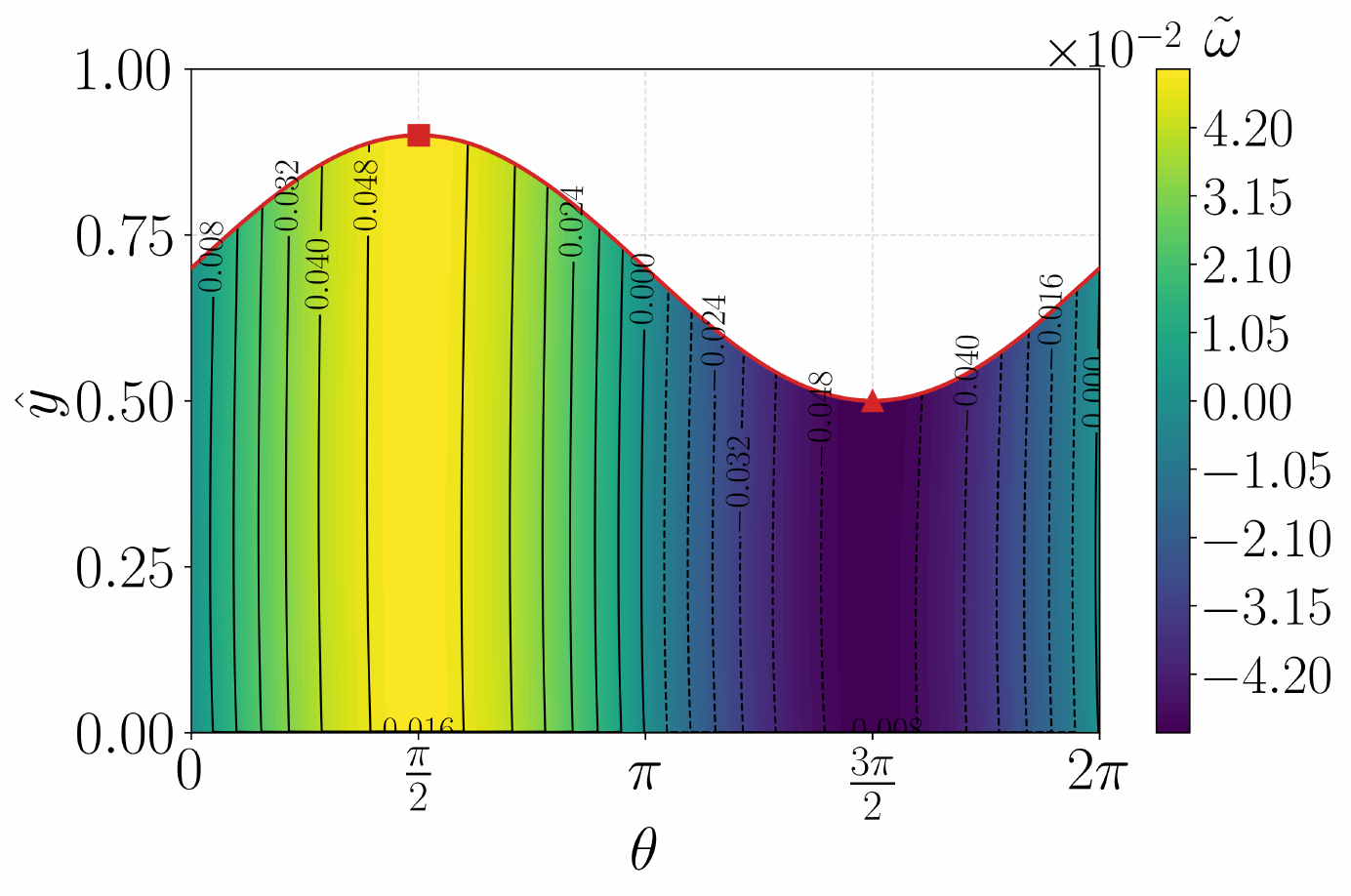}
    \caption{}
  \end{subfigure}
  \hfill
  \begin{subfigure}[b]{0.32\textwidth}
    \includegraphics[width=\textwidth]{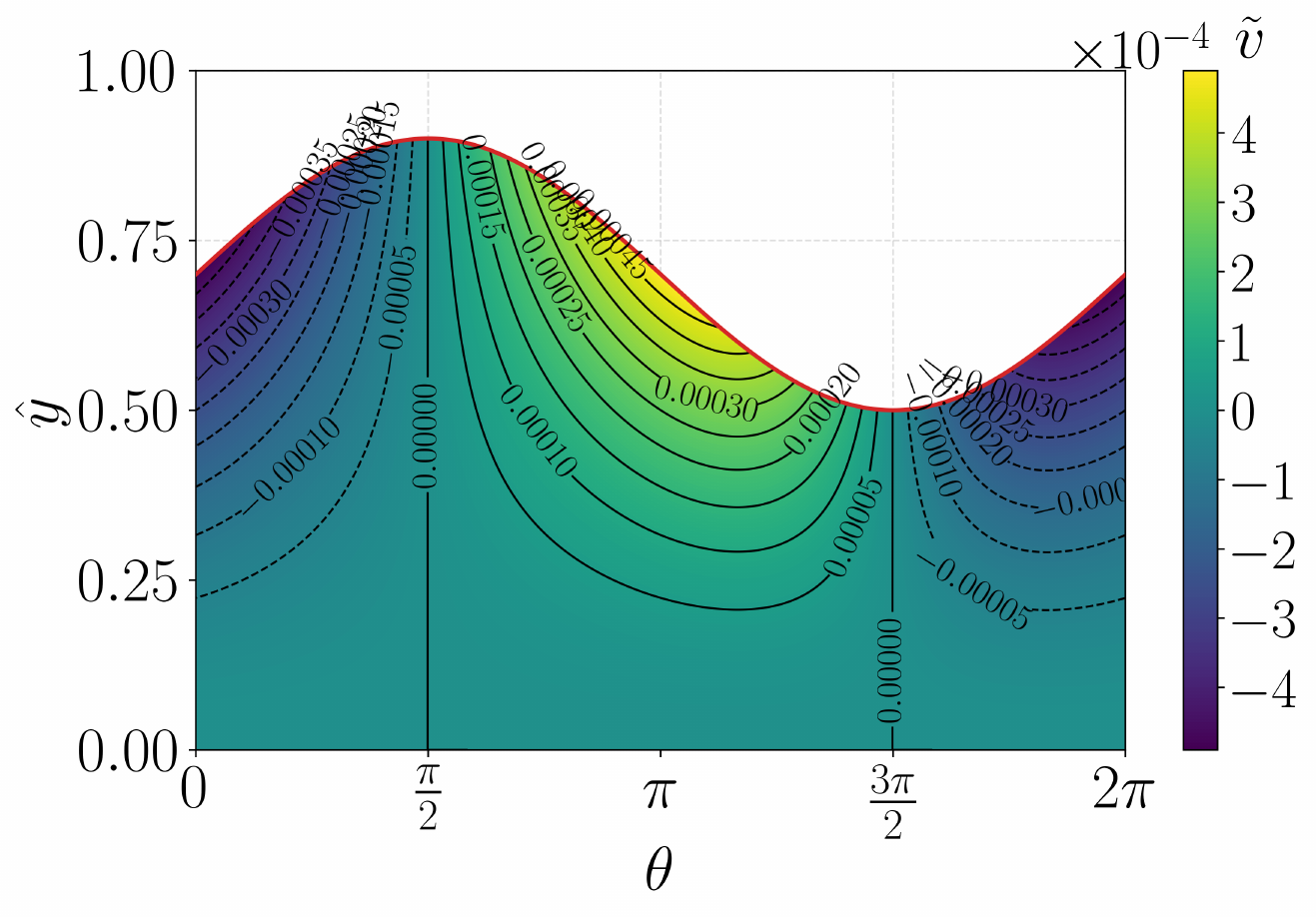}
    \caption{}
  \end{subfigure}
  \hfill
  \begin{subfigure}[b]{0.32\textwidth}
    \includegraphics[width=\textwidth]{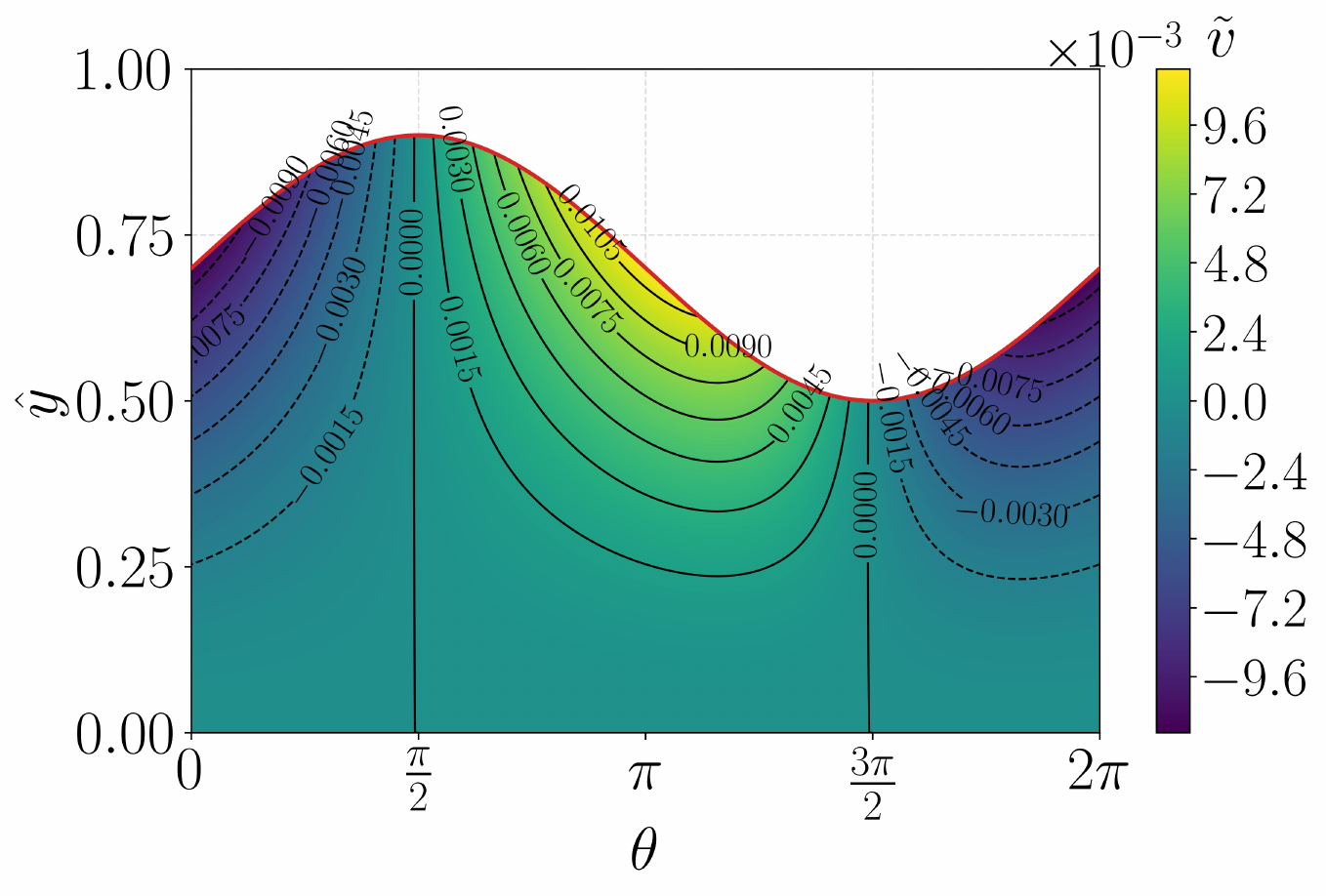}
    \caption{}
  \end{subfigure}
  \hfill
  \begin{subfigure}[b]{0.32\textwidth}
    \includegraphics[width=\textwidth]{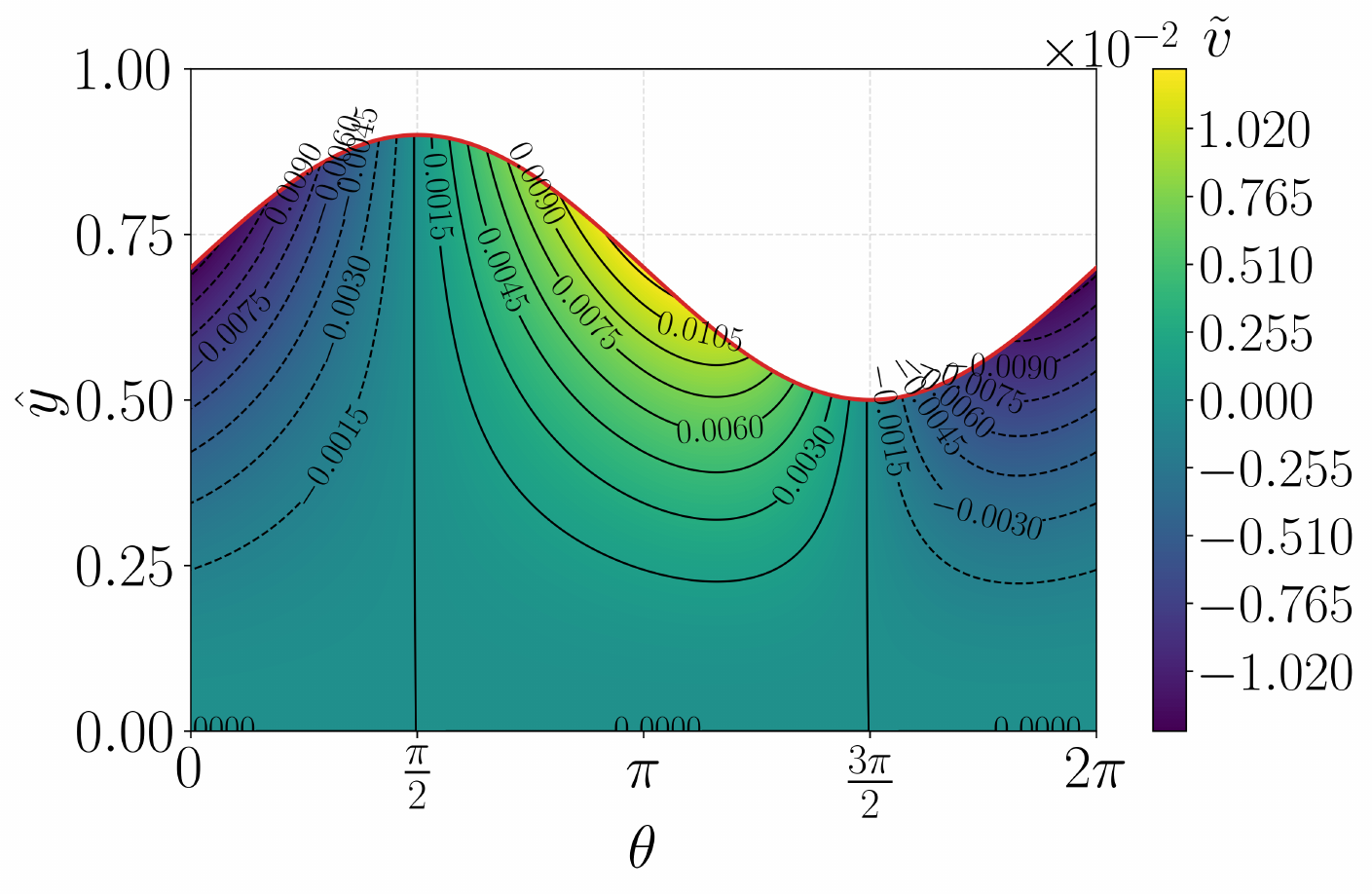}
    \caption{}
  \end{subfigure}
  \begin{subfigure}[b]{0.32\textwidth}
    \includegraphics[width=\textwidth]{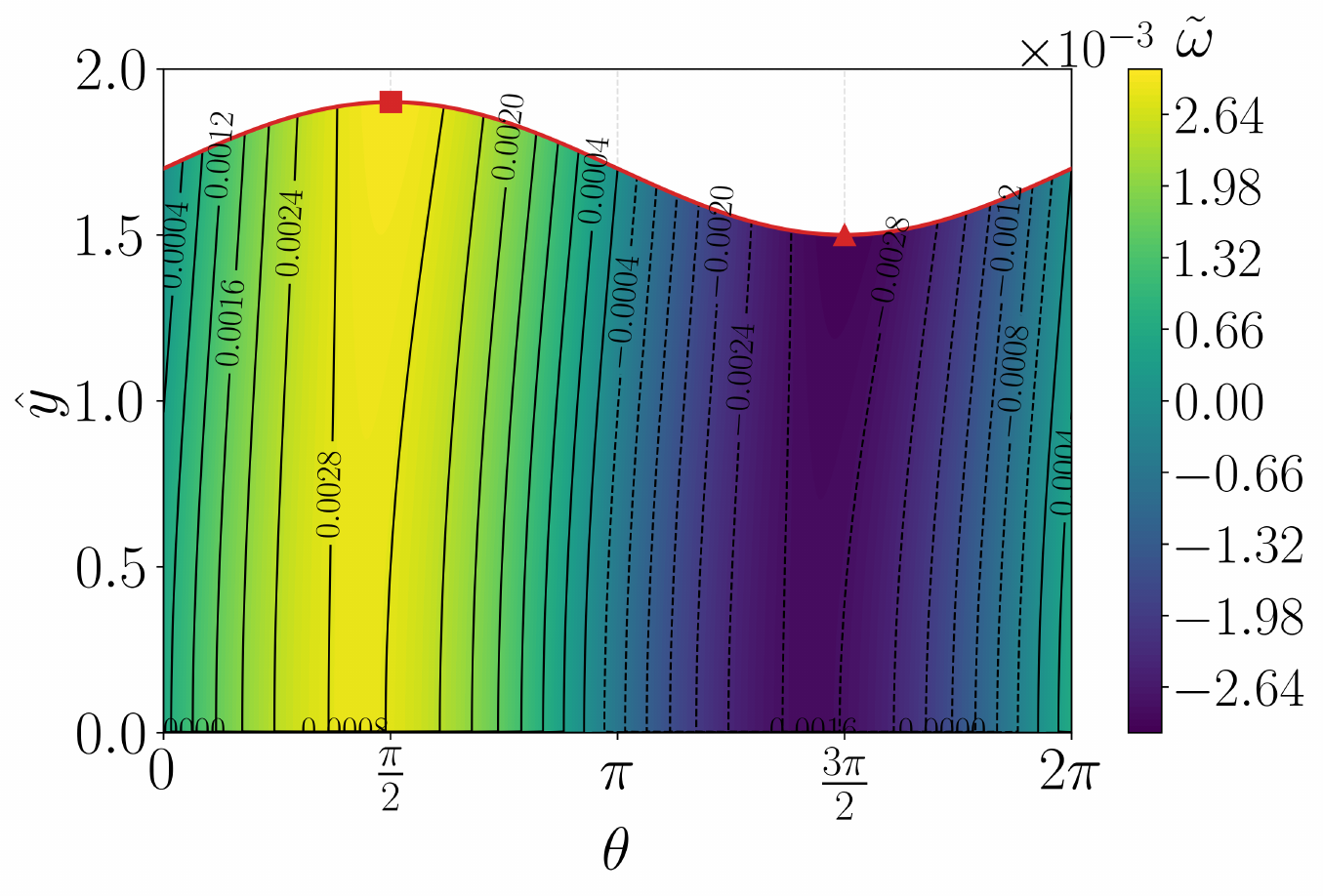}
    \caption{}
  \end{subfigure}
  \hfill
  \begin{subfigure}[b]{0.32\textwidth}
    \includegraphics[width=\textwidth]{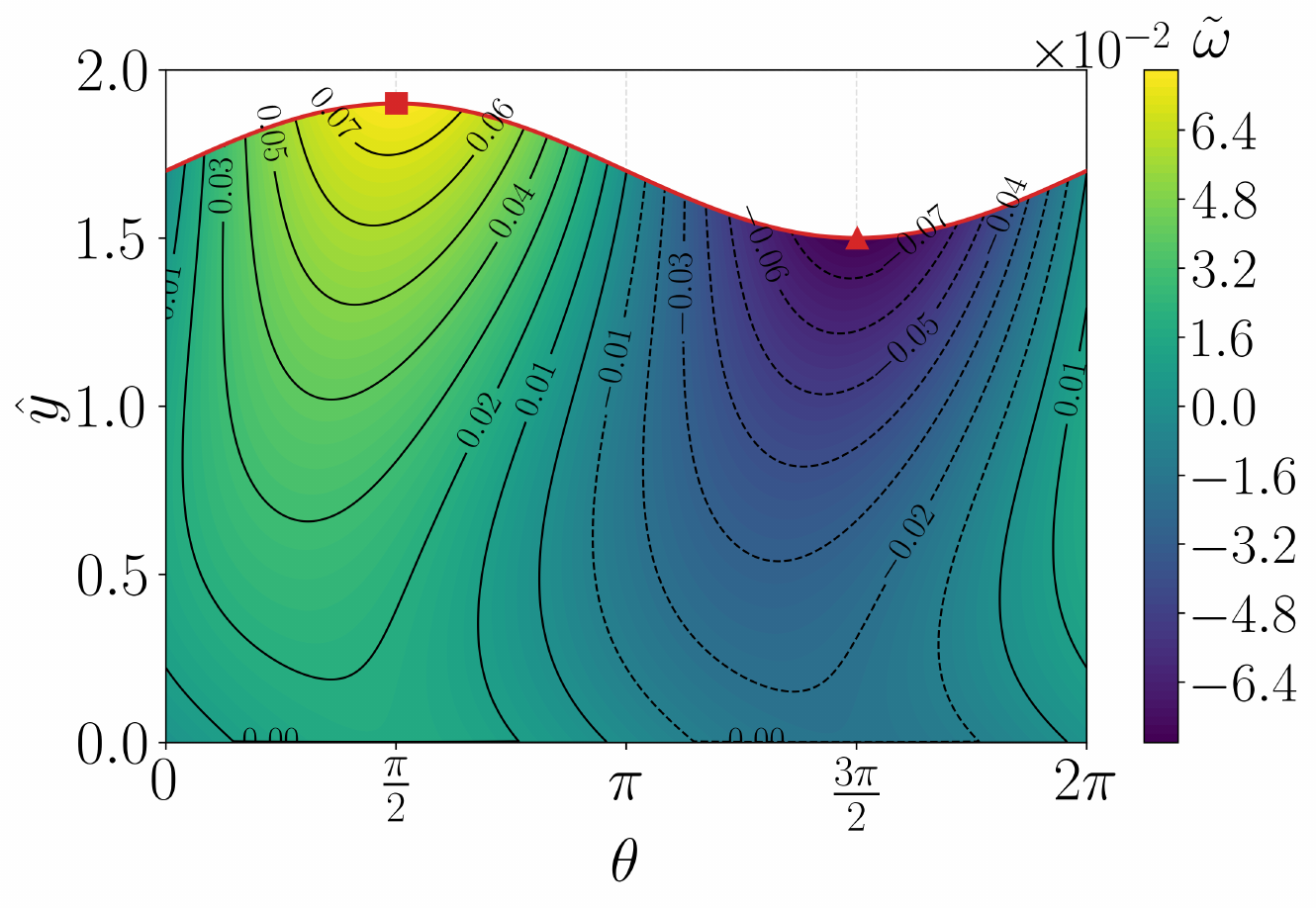}
    \caption{}
  \end{subfigure}
  \hfill
  \begin{subfigure}[b]{0.32\textwidth}
    \includegraphics[width=\textwidth]{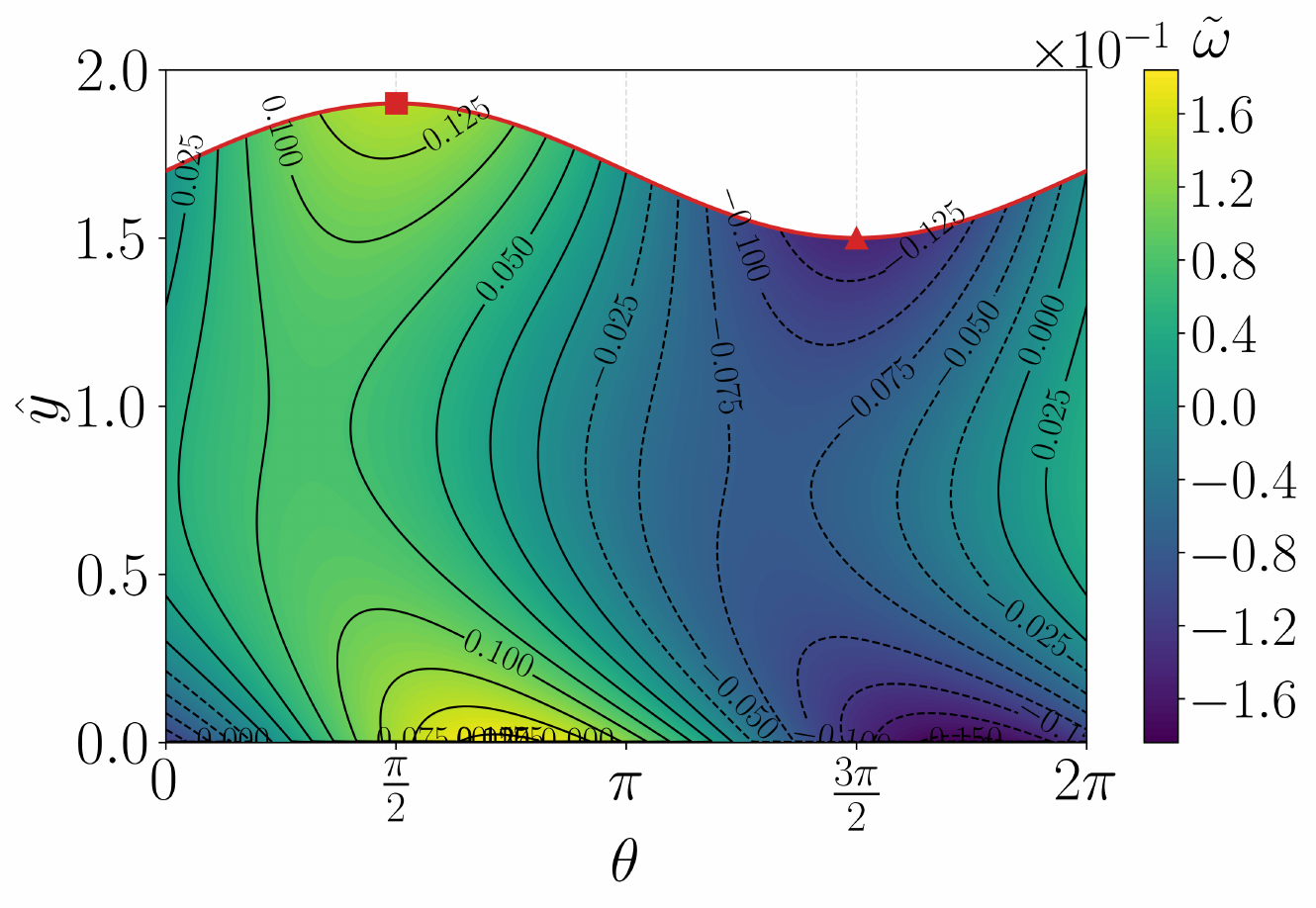}
    \caption{}
  \end{subfigure}
  \hfill
  \begin{subfigure}[b]{0.32\textwidth}
    \includegraphics[width=\textwidth]{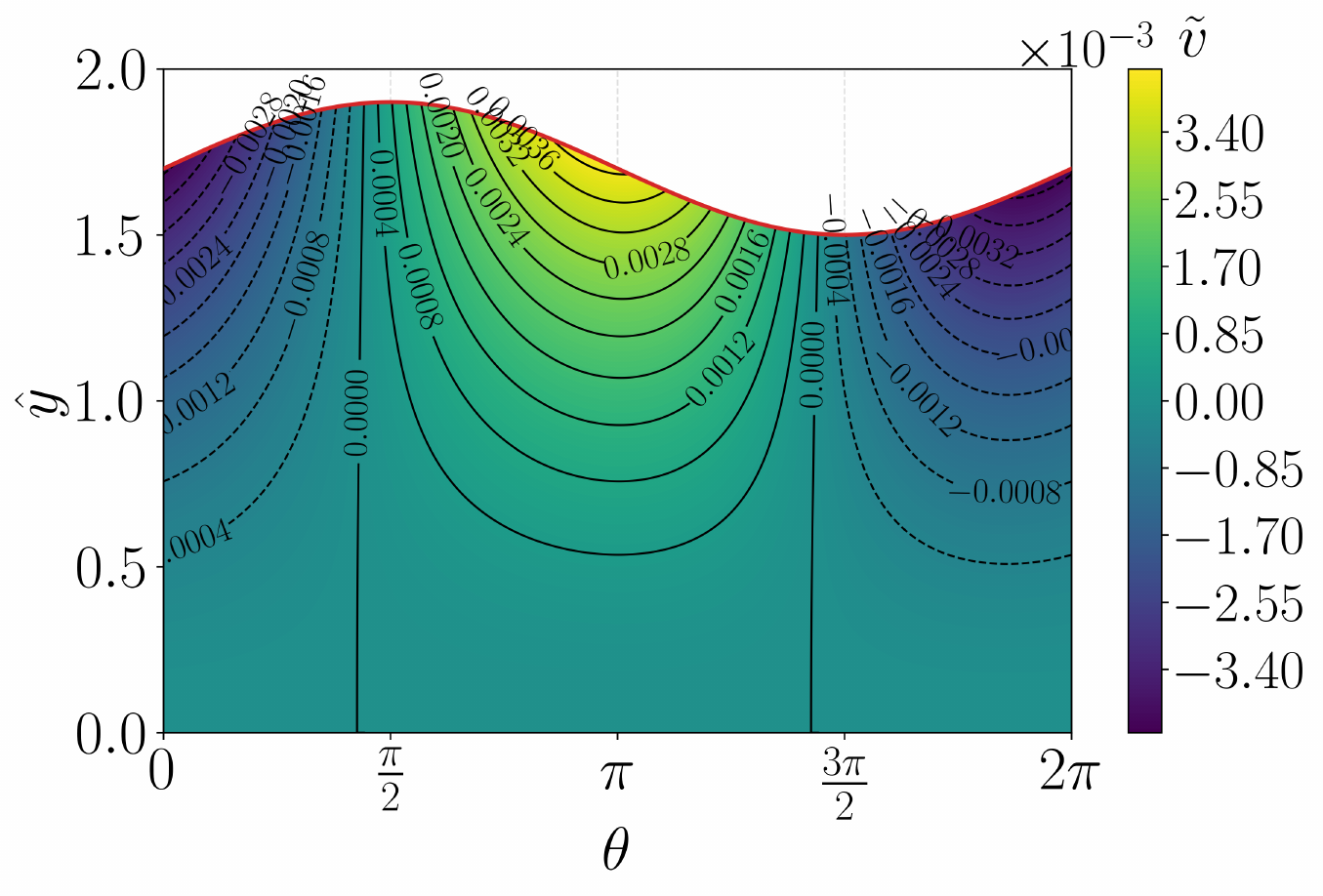}
    \caption{}
  \end{subfigure}
  \hfill
  \begin{subfigure}[b]{0.32\textwidth}
    \includegraphics[width=\textwidth]{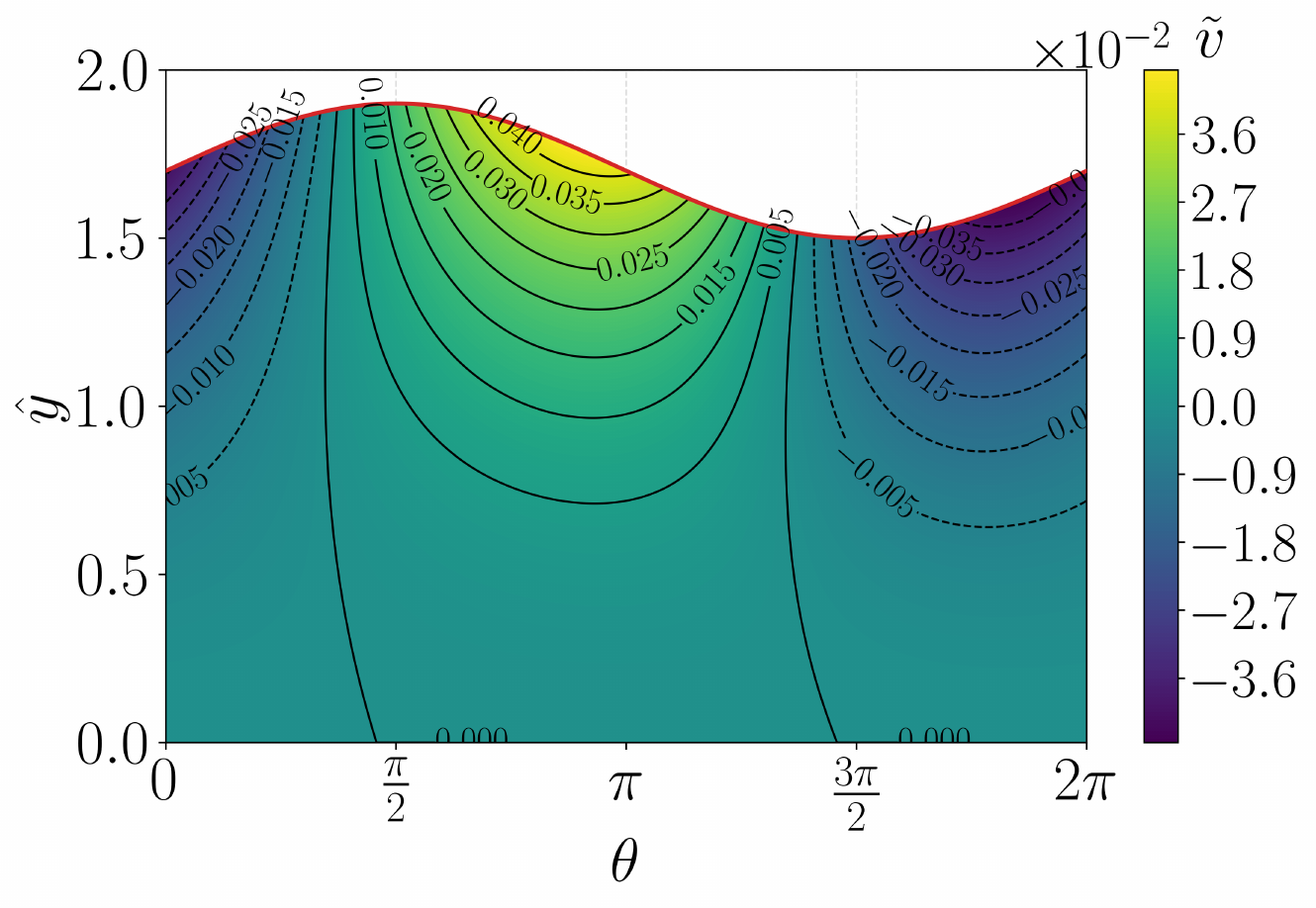}
    \caption{}
  \end{subfigure}
  \hfill
  \begin{subfigure}[b]{0.32\textwidth}
    \includegraphics[width=\textwidth]{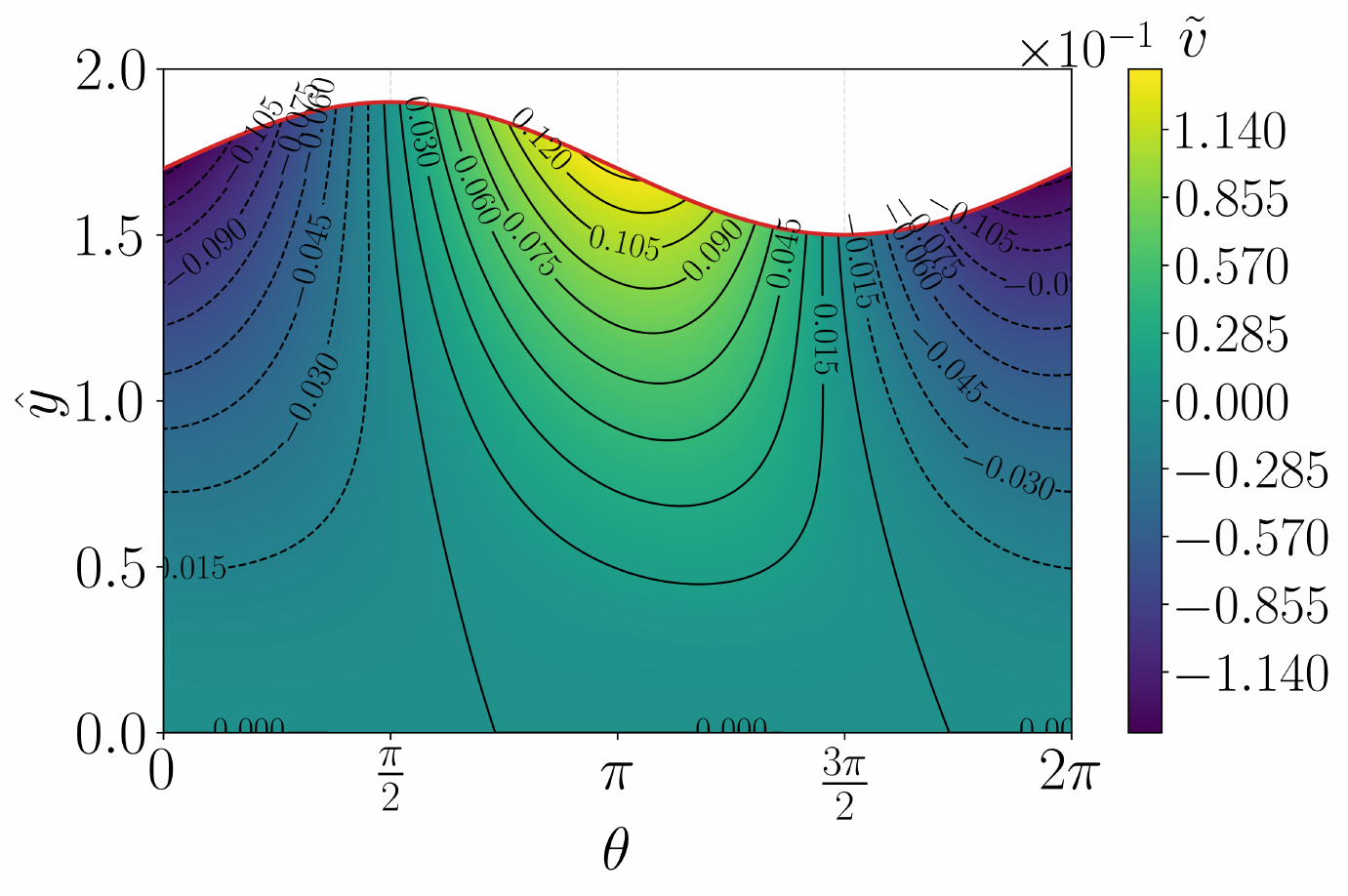}
    \caption{}
  \end{subfigure}
  \caption{Perturbation (a, b, c, g, h and i) vorticity and (d, e, f, j, k and l) normal velocity fields at (a, b, c, d, e and f) $\hat{h}=0.7$ and (g, h, i, j, k and l) $\hat{h}=1.7$ for the liquid zinc with $k$ equal to (a and d) 0.002, (g and j) 0.003, (b and e) 0.047, (c and f) 0.051, (h and k) 0.077 and (i and l) 0.15 with the maximum (red square) and minimum (red triangle) of vorticity at the interface at $\R=30$.}
  \label{fig:vorticiy_plots_zinc}
\end{figure}

As the wavenumber increases, the vorticity lines bend towards the right near the substrate. This curvature stabilizes the perturbation, enlarging the negative (positive) normal velocity area under the peak (trough). At the same time, the vorticity tends to concentrate near the free surface, approaching its maximum value at the interface and forming a boundary layer region with strong shear stresses. Despite the differences in the vorticities and normal velocities' distributions, the stability mechanism at $\hat{h}=0.7$ and $\hat{h}=1.7$ for the corn oil is very similar. As we have seen, the neutral curves and the dispersion relations (figure~\ref{fig:neutral_curves_water_glycerol_corn_oil}), the vorticity at the free surface (figure~\ref{fig:vorticiy_fs_amp_phase}) and the order of magnitude of the vorticity in bulk are almost the same for the two conditions. This means that for small $\Ka$, the instability growth is mainly driven by the vorticity at the free surface and in its proximity.

Things change as we move to larger $\Ka$, Figure~\ref{fig:vorticiy_plots_zinc} shows the perturbation (a, b, c, g, h and i) vorticity and (d, e, f, j, k and l) normal velocity fields at (a, b, c, d, e and f) $\hat{h}=0.7$ and (g, h, i, j, k and l) $\hat{h}=1.7$ for the liquid zinc with $k$ equal to (a and d) 0.002, (g and j) 0.003, (b and e) 0.047, (c and f) 0.051, (h and k) 0.077 and (i and l) 0.15 with the maximum (red square) and minimum (red triangle) of vorticity at the interface at $\R=30$. For $\hat{h}=0.7$, the advection has a very small effect on the vorticity contour's curvature. The viscous effects almost completely dominate the vorticity distribution. As we have seen in figure~\ref{fig:vorticiy_fs_amp_phase}, for $\hat{h}=0.7$, the instability grows mostly due to the vorticity magnitude rather than the phase shift, which is almost zero.

For $\hat{h}=1.7$, the vorticity lines tilt in the advection direction as we increase the wavenumber. Moreover, for $k =0.15$, a region of intense vorticity is created near the substrate, destabilizing the flow, increasing the net flow rate under the crests, and creating a boundary layer near the substrate in addition to the one near the free surface. 

This region is the product of the streamwise pressure gradient induced by the surface tension. The normal vorticity flux at the substrate, also known as \textit{vorticity source strength} \citep{lighthill1963introduction}, is given by \citep{morton1984generation}:
\begin{equation}
    -\partial_{\hat{y}}\omega = \partial_{\hat{x}}\tilde{p} - 1,
\end{equation}
where the right-hand side represents the rate of vorticity production per unit area and comprises the perturbation's streamwise pressure gradient and the gravitational effect. Positive vorticity is generated when the streamwise pressure gradient outweighs the gravitational effect. 

For small $k$, the pressure is constant long $\hat{y}$, with the most sensitive zone to the surface tension close to the substrate. As we have seen in Subsection~\ref{subsec:verification}, for high $\Ka$, the surface tension influences the first-order solution for small k. The first-order streamwise velocity amplitude, derived by differentiating the stream function \eqref{exp:order_k_w_Ca} with respect to the wall-normal coordinate, is given by:
\begin{equation}
    \acute{u}_1 = D \varphi_1^{\star}(\hat{y}) = \frac{i \hat{y}^2}{12}\Big(\hat{h}\R\hat{y}^2 - 4\hat{h}^2\R\hat{y} + 12k^2 Ca \Big).
\end{equation}
Assuming $\hat{h}=\text{\textit{O}}(1)$, the surface tension term is dominant near the substrate for small $\hat{y}$.  

Moreover, the surface tension also increases the vorticity near the free surface (figure e and f). The normal vorticity flux at the interface is given by \citep{dietze2009experimental,wu1995theory}:
\begin{equation}
    -\partial_{\hat{n}}\tilde{\omega} = \frac{\partial_{\hat{x}}\tilde{p}}{\sqrt{1 + (\partial_{\hat{x}}\tilde{h})^2}},
\end{equation}
where $\hat{n}$ is the nondimensional normal vector to the interface pointing toward the liquid. As for the substrate flux, it is dependent on the streamwise pressure gradient. At a peak (trough) location, the pressure gradient in the streamwise direction is negative (positive), so the film produces positive (negative) vorticity at the free surface.

The mechanism leading to the formation of this boundary layer zone is the following: the pressure imposed by the surface tension pushes fluid down (up) at the crests (troughs), which, via the continuity equation, produces a positive (negative) streamwise flow near the substrate. The movement of the substrate increases the shear effects compared to the falling film case, leading to more intense vorticity regions.

These outcomes highlight the importance of $\Ka$, even for long-wave perturbation (small $k$). This means that surface tension is crucial not only for stabilizing short waves and storing perturbation energy but also for the early development of instability and vorticity distribution in liquid film. Moreover, we reveal two mechanisms of instability involving viscous stresses associated with the boundary layer at the free surface for the corn oil and the boundary layer at the substrate for the zinc.
\begin{figure}
  \begin{subfigure}[b]{0.49\textwidth}
    \includegraphics[width=\textwidth]{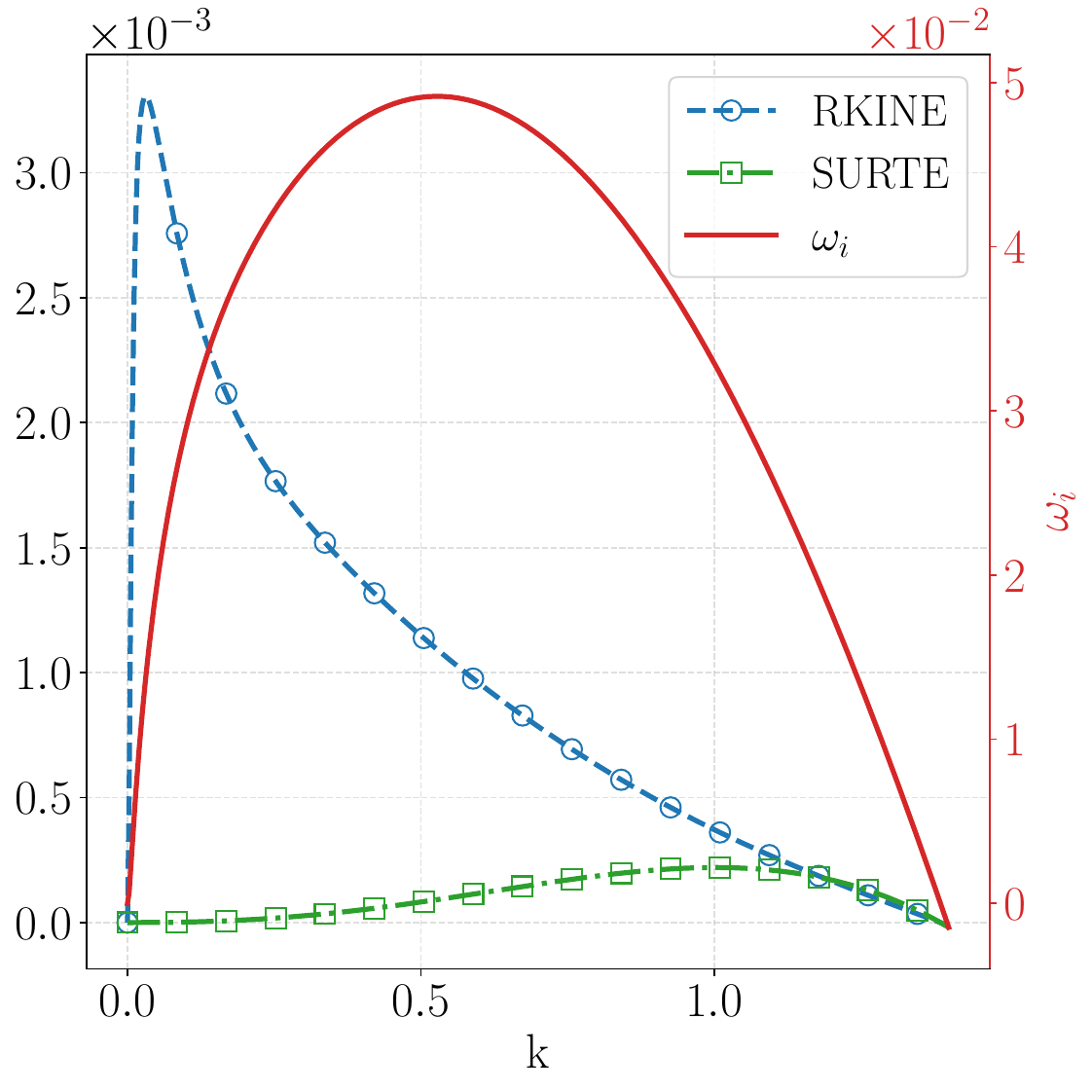}
    \caption{}
  \end{subfigure}
  \hfill
  \begin{subfigure}[b]{0.49\textwidth}
    \includegraphics[width=\textwidth]{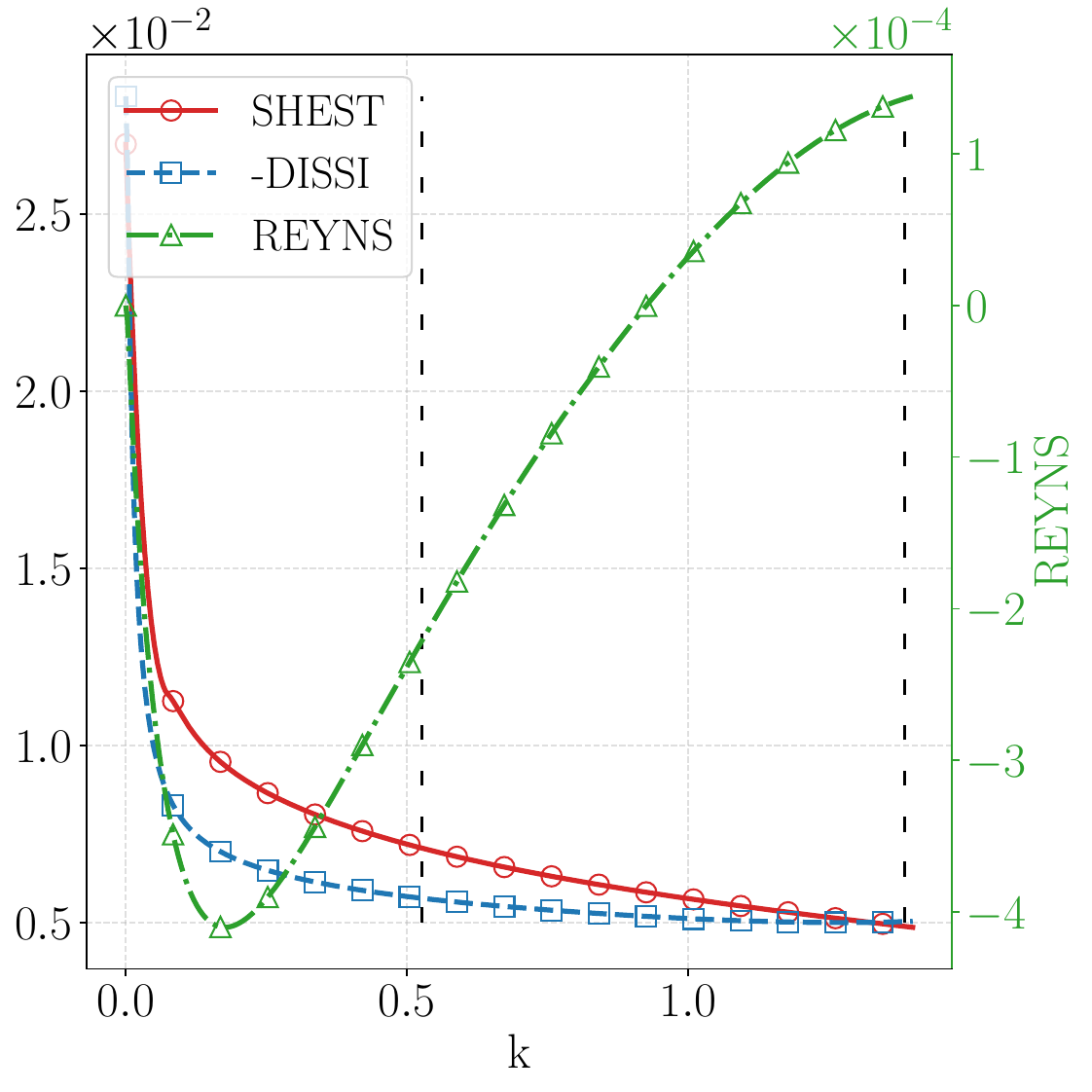}
    \caption{}
  \end{subfigure}
  \hfill
  \begin{subfigure}[b]{0.49\textwidth}
    \includegraphics[width=\textwidth]{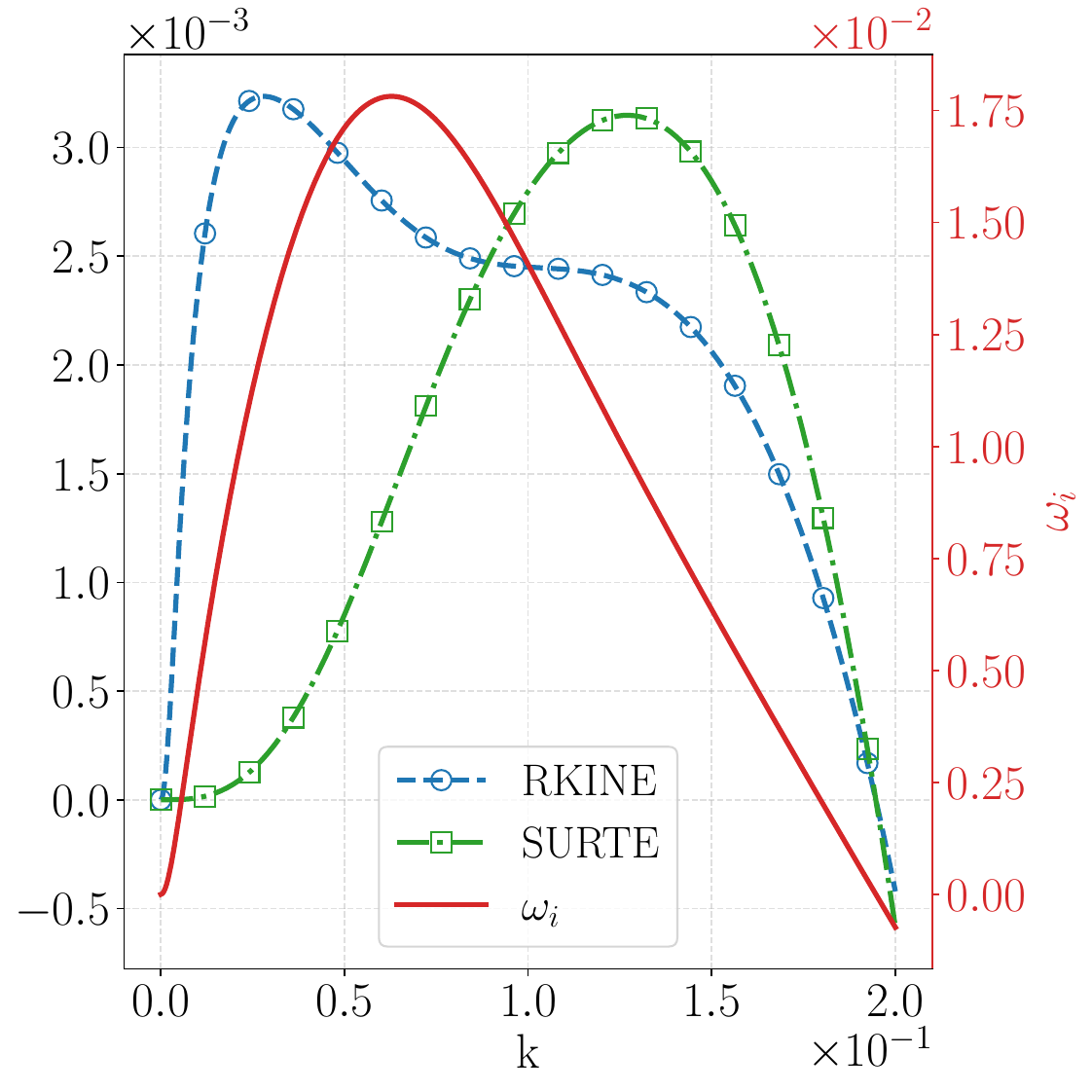}
    \caption{}
  \end{subfigure}
  \hfill
  \begin{subfigure}[b]{0.49\textwidth}
    \includegraphics[width=\textwidth]{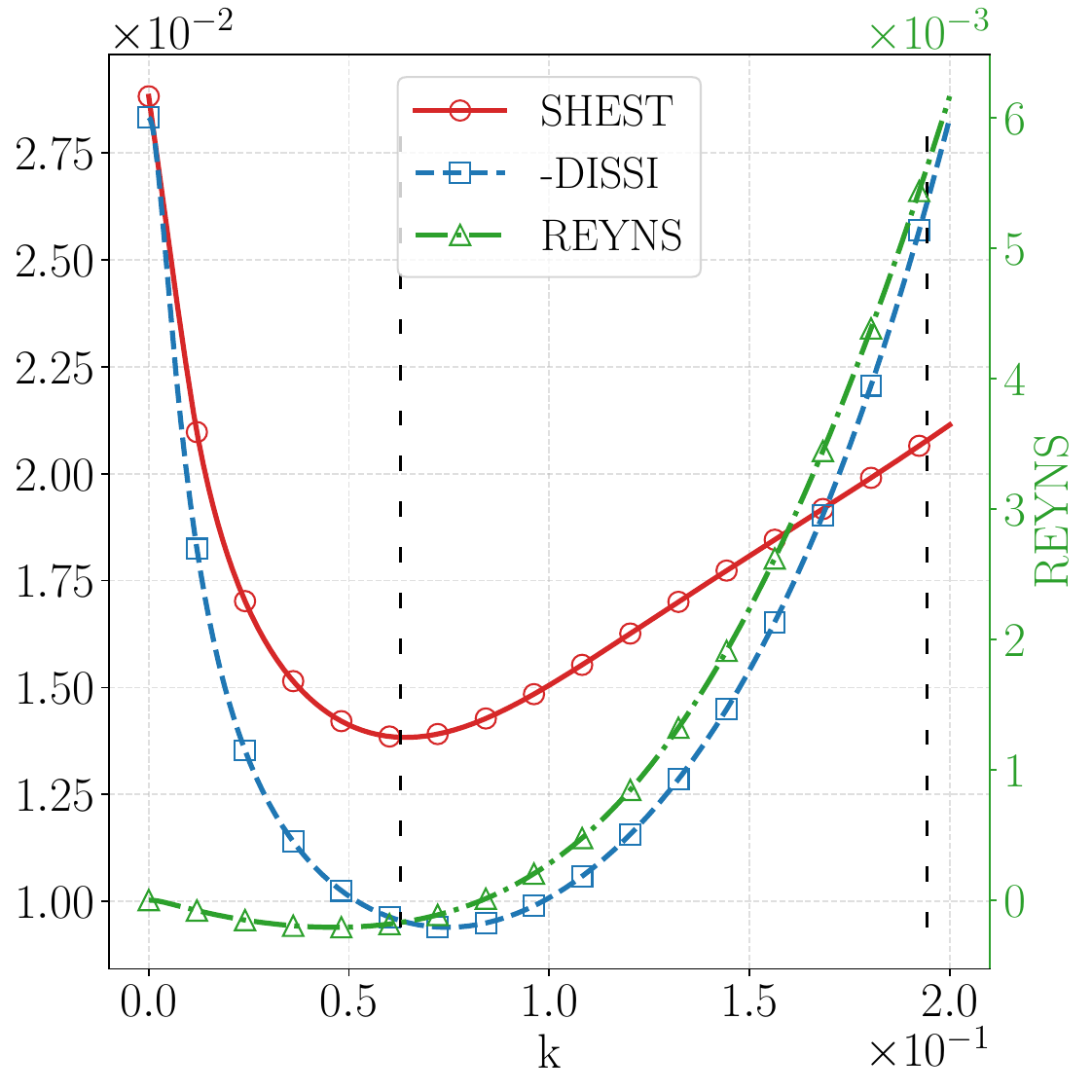}
    \caption{}
  \end{subfigure}
  \caption{Storage (RKINE and SURTE), production (REYNS and SHEST) and dissipation (DISSI) terms in the perturbation's energy balance equation with the growth rate $\omega_i$ (red line) as a function of the wavenumber $k$ with the black loosely dashed lines representing the position of the maximum growth rate and the cut-off wavenumber with $\hat{h}=1.7$ and $\R = 30$ for (a and b) corn and (c and d)liquid zinc.}
  \label{fig:energy_balance}
\end{figure}

\subsubsection{Energy balance of the perturbation}
\begin{figure}
  \begin{subfigure}[b]{0.49\textwidth}
    \includegraphics[width=\textwidth]{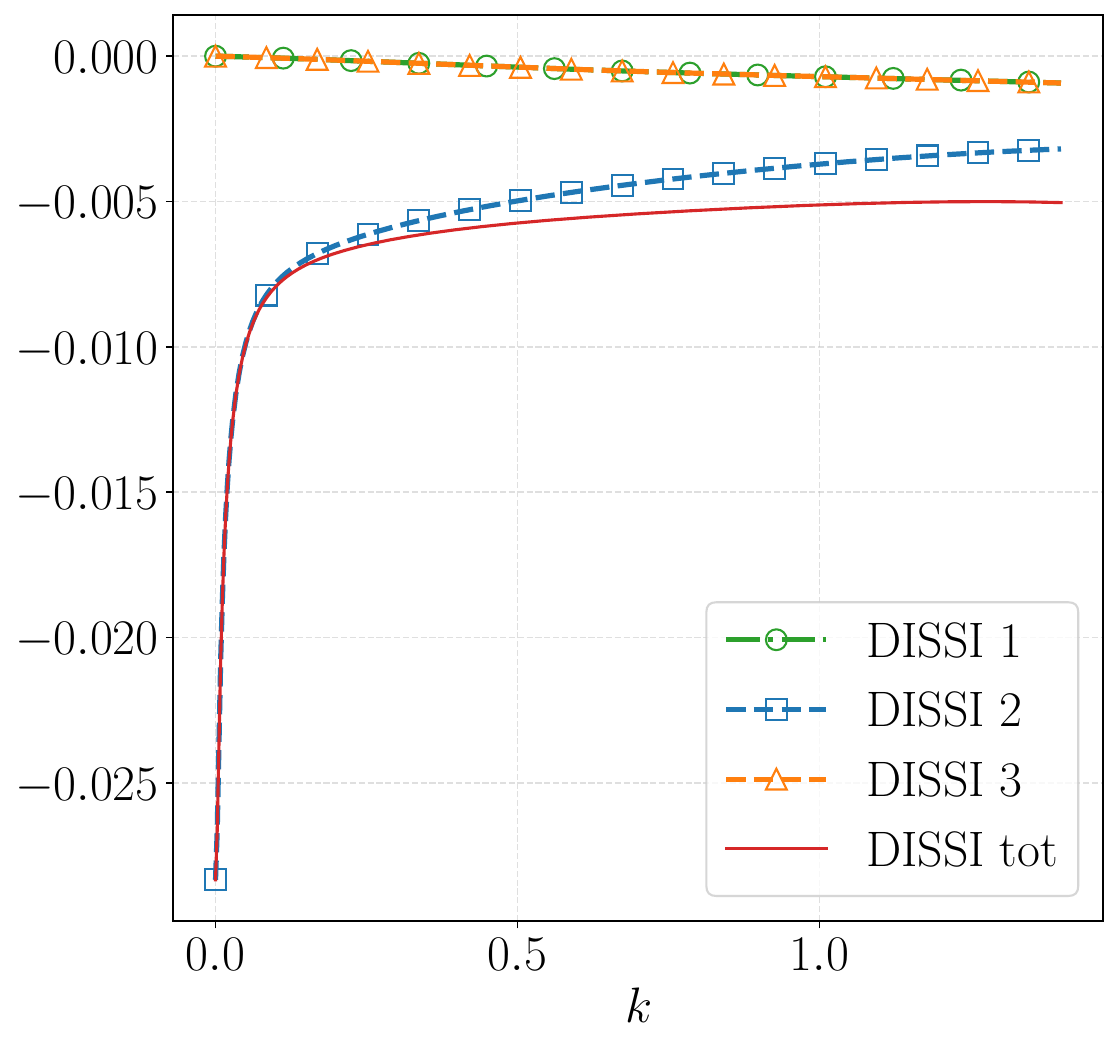}
    \caption{}
  \end{subfigure}
  \hfill
  \begin{subfigure}[b]{0.49\textwidth}
    \includegraphics[width=\textwidth]{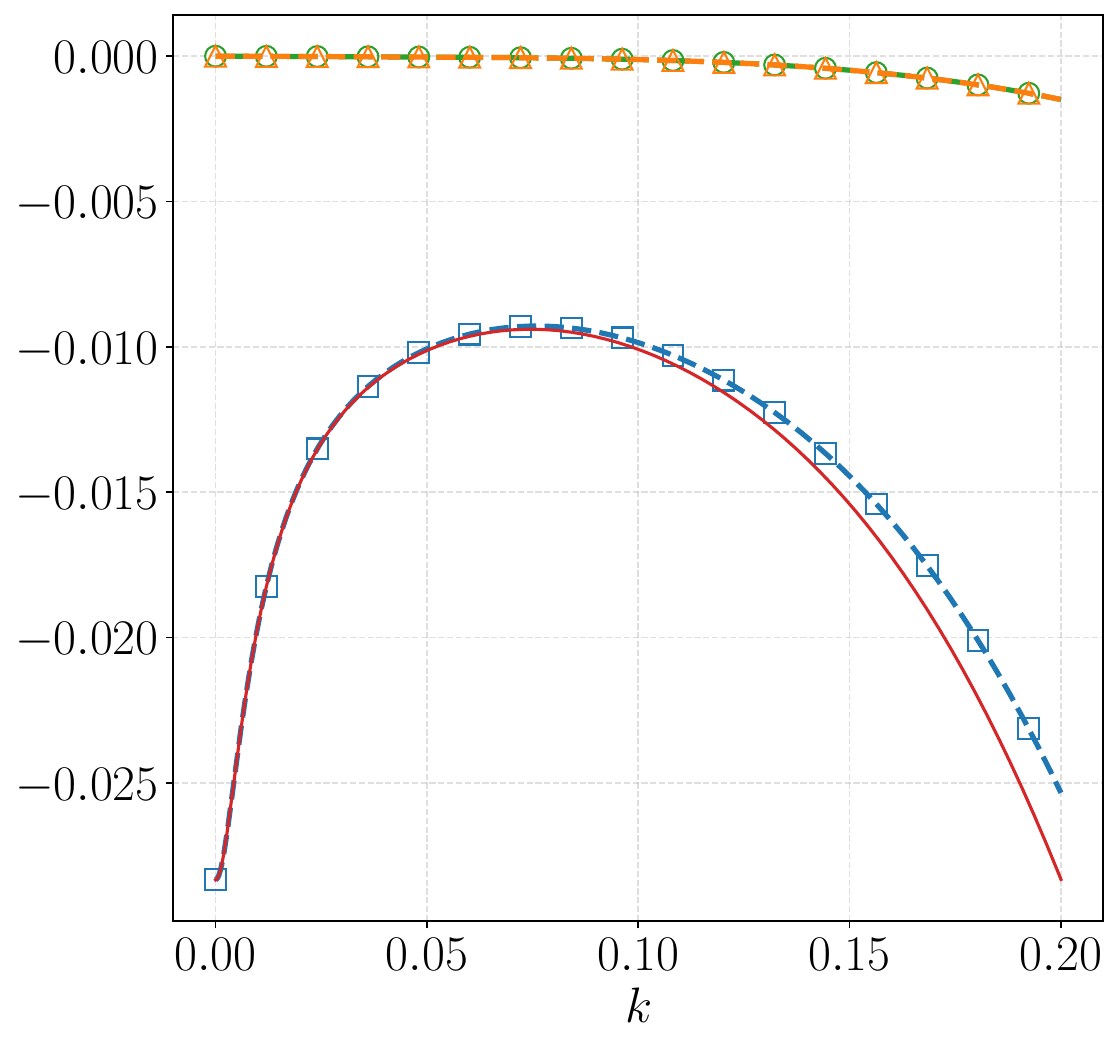}
    \caption{}
  \end{subfigure}
  \caption{Distribution of the extensional stresses $DISSI_1$ and $DISSI_3$ and shear stress $DISSI2$, with $\hat{h}=1.7$ and $\R=30$ for (a) the corn oil and (b) the liquid zinc.}
  \label{fig:distribution_shear_stresses}
\end{figure}

We have analysed the early stages of the long-wave instability mechanism and how the vorticity at the free surface and in bulk fosters the growth of unstable perturbations. Here, we study the terms forming the kinetic energy balance \eqref{eq:energy_balance} of an unstable perturbation, as presented in section~\ref{subsec:energy_balance}.

The vorticity near the substrate also affects the dissipation terms defined in \eqref{eq:DISSIs} and \eqref{eq:shear_s_def}. Figure~\ref{fig:distribution_shear_stresses} shows the elongational ($DISSI_1$ green dashed-dotted line with circles and $DISSI_3$ orange dashed line with triangles) and shear ($DISSI_2$ blue dashed line with squares) stresses contributing to the total dissipative viscous effects ($DISSI_{tot}$ red continuous line) as a function of $k$ at $\hat{h}=1.7$ and $\R=30$ for (a) corn oil and (b) liquid zinc. $DISSI_1$ is equal to $DISSI_3$ due to the continuity equation. For small $k$, the $DISSI_2$, which is a function of the strain rate $\upsilon$, dominates the dissipative effects, with 
$DISSI_1$ and $DISSI_2$ slowly increasing with $k$. For larger $k$, $DISSI_2$ decreases in magnitude. At $\Ka=4$ (corn oil), it decreases monotonically with $k$, whereas for $\Ka=11525$ (zinc), it reaches a minimum and increases again towards the cut-off wavenumber. This is due to the creation of a boundary layer near the substrate. Near the boundary, the strain rate $\upsilon$ is equal to the vorticity, and they are both generated with the same boundary flux densities \citep{morton1984generation}. In addition to the boundary, the bulk is also a source of strain rate. The equation governing the strain rate is given by:
\begin{equation}
    \partial_{\hat{t}}\upsilon = -2 \partial_{\hat{x}\hat{y}}\tilde{p} + \frac{1}{Re}(\partial_{\hat{x}\hat{x}}\upsilon + \partial_{\hat{y}\hat{y}}\upsilon).
\end{equation}
When the surface tension induces a pressure gradient in the liquid film, also the bulk generates a strain rate, feeding $DISSI_2$.

\subsection{Absolute/convective (AC) threshold}
\label{res_susection:AC_threshold}
The previous subsections focused on the instability mechanism using long-wave expansions, vorticity and energy arguments. In this subsection, we further extend the analysis of unstable perturbations, calculating the AC threshold in the $\hat{h}-\R$ and $c_r-\R$ parameters spaces with the LLD solution $\accentset{\circ}{h}$, the real part of the wavenumber $k_r$ (red empty triangles) and the nondimensional capillary length $\hat{\ell}_{c}$ defined in \eqref{eq:non_dim_cap_wave_num} (blue continuous line with empty squares), for the liquids in Table~\ref{tab:liquid_prop}. Moreover, we calculate the threshold also in the $\Ka-\R$ space for the Derjaguin's solution ($\hat{h}=1$). 
The region of absolute instability is depicted as a shadowed area bounded by the AC threshold simulated points (continuous black line with black circles).

In the falling film literature, \cite{brevdo1999linear} showed that a flat liquid film over a vertical substrate is always convectively unstable due to gravitational effects. In the case of an inclined substrate, the hydrostatic effects compensate for the gravity, leading to regions of absolute instability \citep{scheid2016critical,brun2015rayleigh}. Similarly, the leading actors in the moving substrate case are gravity, inertia, viscosity and surface tension. For certain parameters, these compensate each other, leading to regions of absolute instability.
\begin{figure*}
  \begin{subfigure}{0.49\textwidth}
      \centering
    \includegraphics[width=\linewidth]{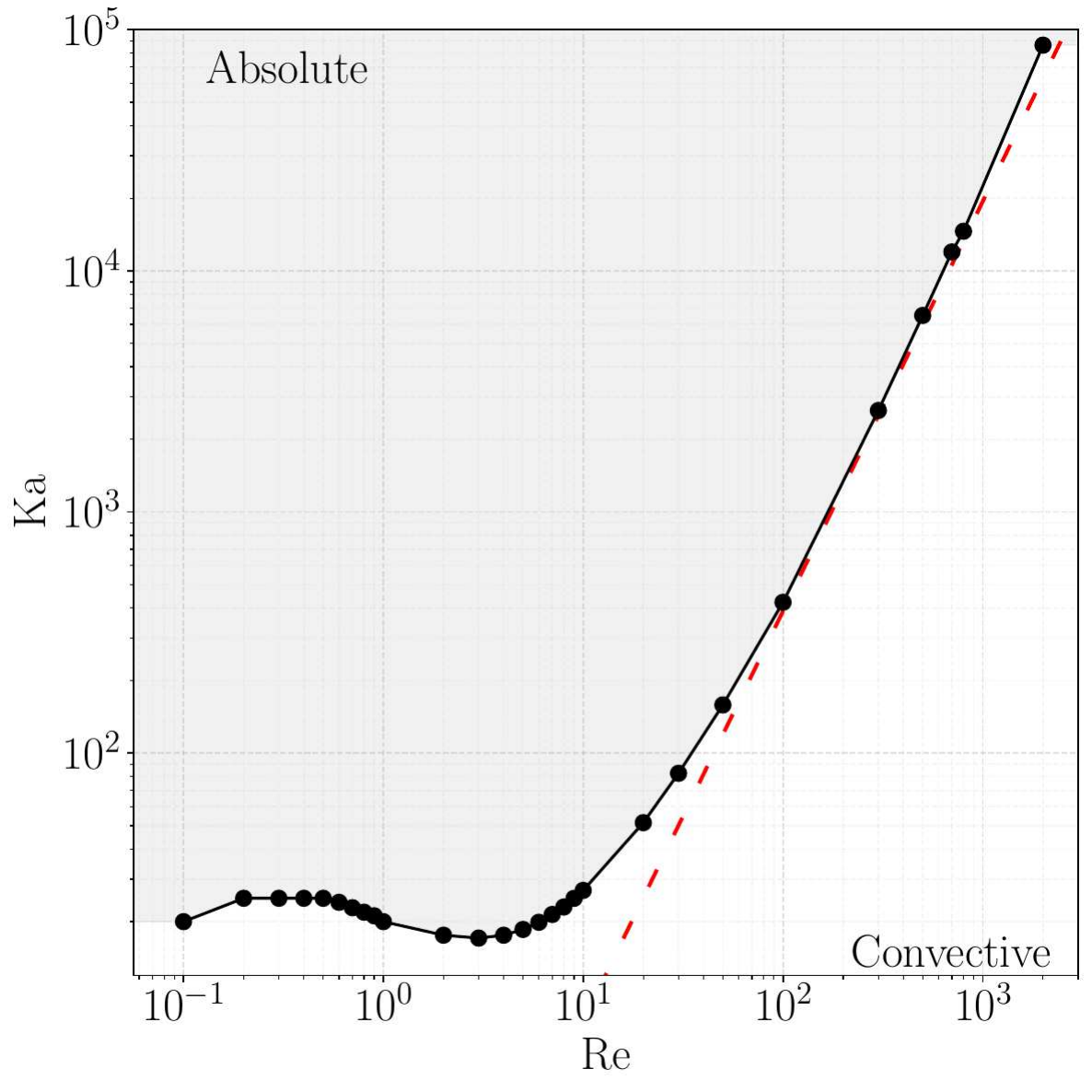}
    \caption{}
    \label{}
  \end{subfigure}
  \hfill
  \begin{subfigure}{0.5\textwidth}
    \centering
    \includegraphics[width=\linewidth]{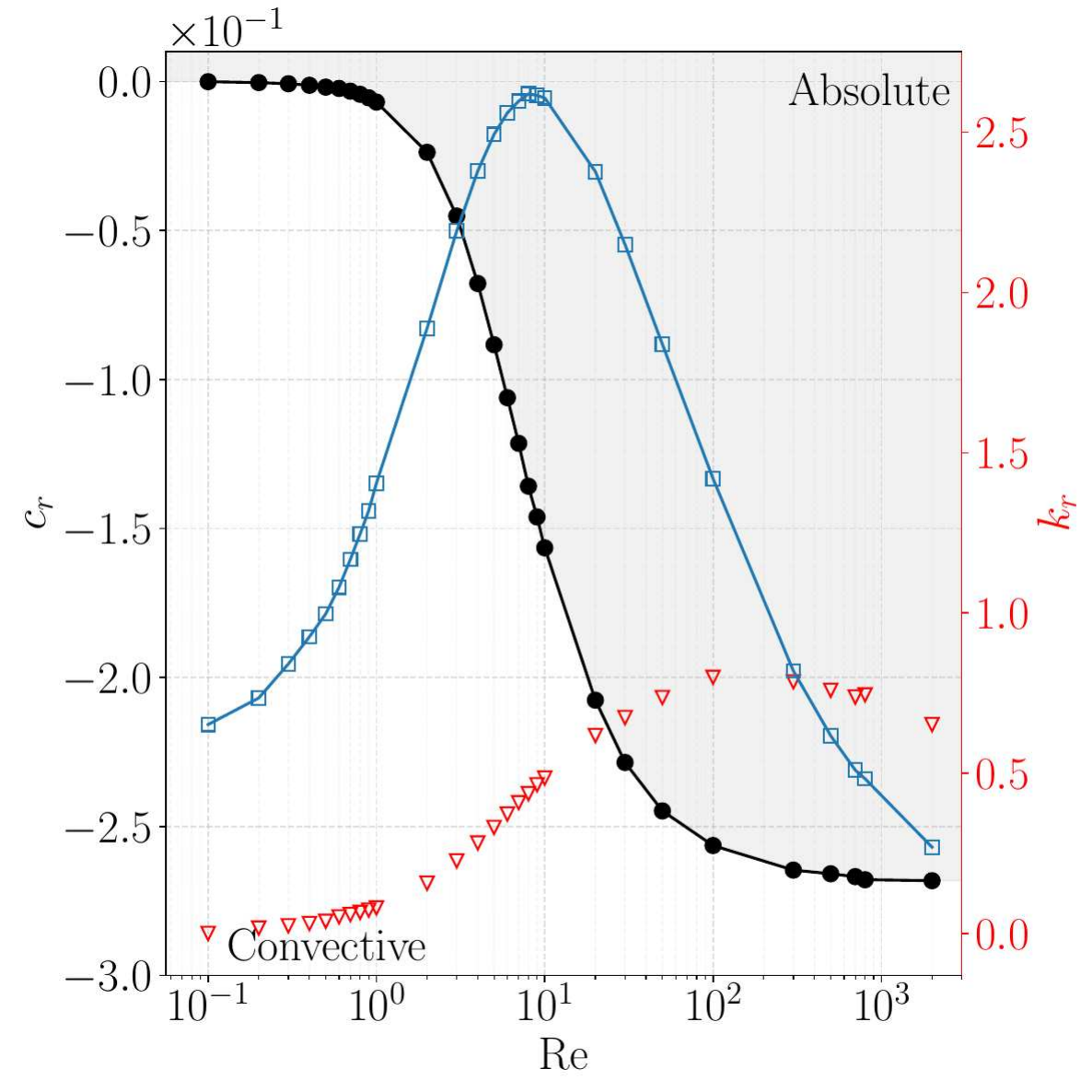}
    \caption{}
    \label{}
  \end{subfigure}
    \caption{Absolute/convective threshold (black dotted line) of the Derjaguin's flat film solution ($\hat{h}=1$) in (a) the $\Ka-\R$ space with the trend line $\Ka=0.15\R^{1.8}$ (red dashed line), and in (b) $c_r-\R$ space with the associated real wavenumber $k_r$ (red triangles), and nondimensional capillary wavenumber $k_{\ell_{c}}$ (continuous blue line with blue squares).}
    \label{fig:ACvaryingKa}
\end{figure*}

\subsubsection{Absolute/convective threshold in the $\Ka-\R$ space with $\hat{h}=1$}
\label{varying_Ka_Re_h_hat_1}
Figure~\ref{fig:ACvaryingKa} shows the AC threshold for the Derjaguin's solution ($\hat{h}=1$) in (a) the $\Ka-\R$ parameter space, with the trend line $\Ka=0.15\R^{1.8}$ (red dashed line) and in (b) the $c_r-\R$ space. For $\Ka<30$, the threshold has a minimum of $\Ka=17$ at $\R=3$ and a maximum of $\Ka=25$ at $\R=0.4$. For $\Ka>30$ and $\R\gtrsim10$, it gets to an asymptote, which goes as $\Ka\sim\R^{1.8}$ or in terms of capillary number as $Ca^{-1}\sim\R^{1.5}$. These results show that for fluids with $\Ka<17$, Derjaguin's solution is always convectively unstable regardless of the Reynolds number.

The neutral perturbations ($\omega_i=0$) associated with the absolute/convective threshold have a negative phase speed for every $\R$. These travel faster than the substrate speed ($|c_r|>|U_p|$) for $\R\gtrsim5.5$. The phase speed has a maximum at $\R=0.1$ and an asymptote at $c_r~2.7$ for $\R>300$. The real wavenumber $k_r$ of the neutral perturbations has a maximum at $k_r\approx0.8$ in the range $100<\R<200$. For $\R>300$, $k_r$ exceeds $\hat{\ell}_{c}$, highlighting how, at these wavelengths, surface tension starts to dominate over gravity.

\subsubsection{Absolute/convective threshold in the $\hat{h}-\R$ space for different $\Ka$}
\begin{table}
\centering
\begin{tabular}{c@{\hspace{0.4cm}}c@{\hspace{0.4cm}}c@{\hspace{0.4cm}}c@{\hspace{0.4cm}}c@{\hspace{0.4cm}}|c@{\hspace{0.4cm}}c@{\hspace{0.4cm}}c@{\hspace{0.4cm}}c@{\hspace{0.4cm}}}
\toprule
$\Ka$ & $4$ & $195$ & $3400$ & $11525$ & $4$ & $195$ & $3400$ & $11525$  \\ \midrule
  & \multicolumn{4}{c}{Minimum} & \multicolumn{4}{|c}{Maximum} \\\midrule
$\hat h_{\rm min}$  & - & 0.904 & 0.838 & 0.824 & 0 & 1.661 & 1.658 & 1.658  \\
$\R_{\rm min}$ & - & 18.28 & 69.08  & 123.46 & 0 & 8.19 & 17.79 & 25 \\\bottomrule
\end{tabular}
\caption{Maximum and minimum of the absolute instability windows in the $\hat{h}-\R$ space for different liquids.}
\label{tab:extremes_AC} 
\end{table}
\begin{figure}
  \begin{subfigure}[b]{0.49\textwidth}
    \includegraphics[width=\textwidth]{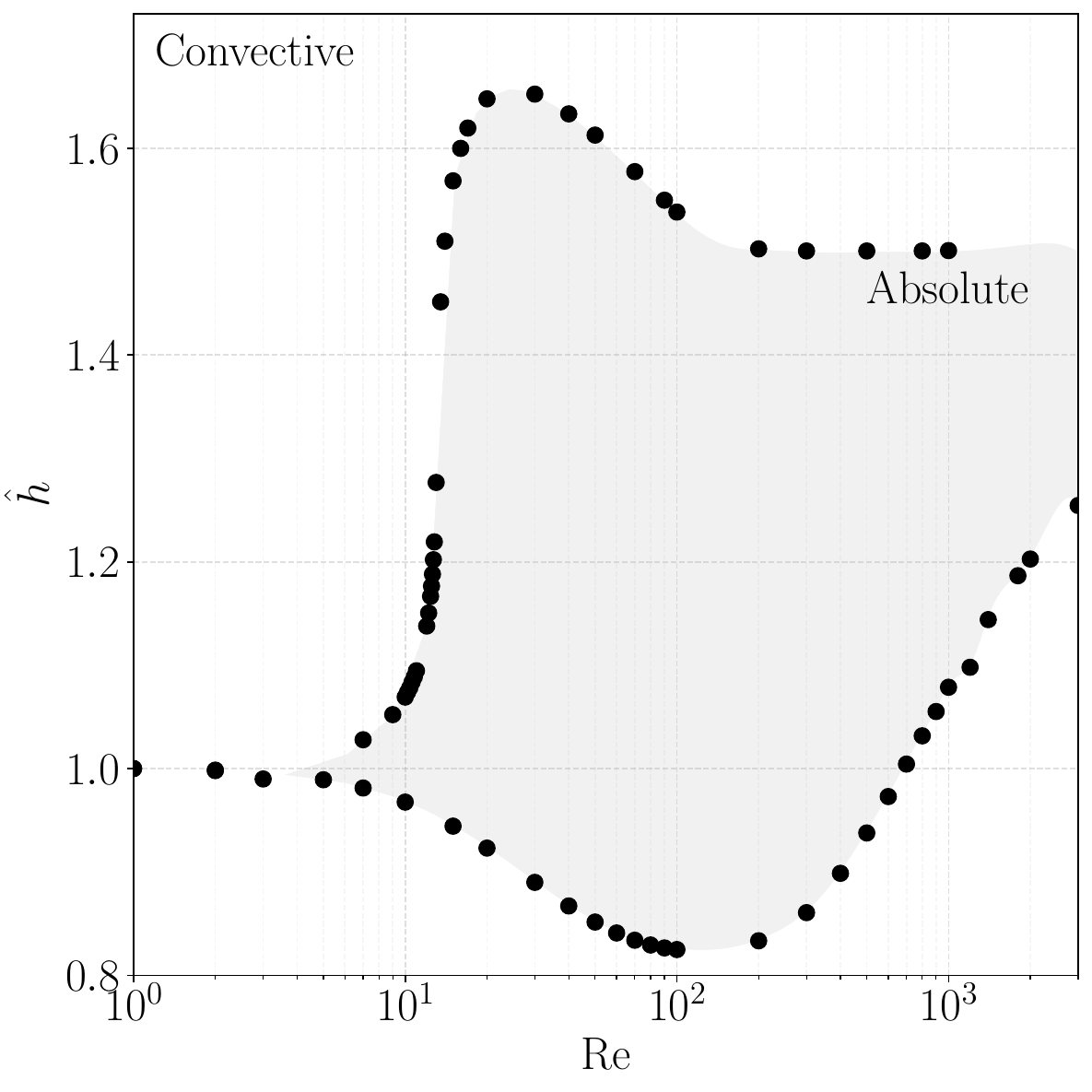}
    \caption{}
  \end{subfigure}
  \hfill
  \begin{subfigure}[b]{0.49\textwidth}
    \includegraphics[width=\textwidth]{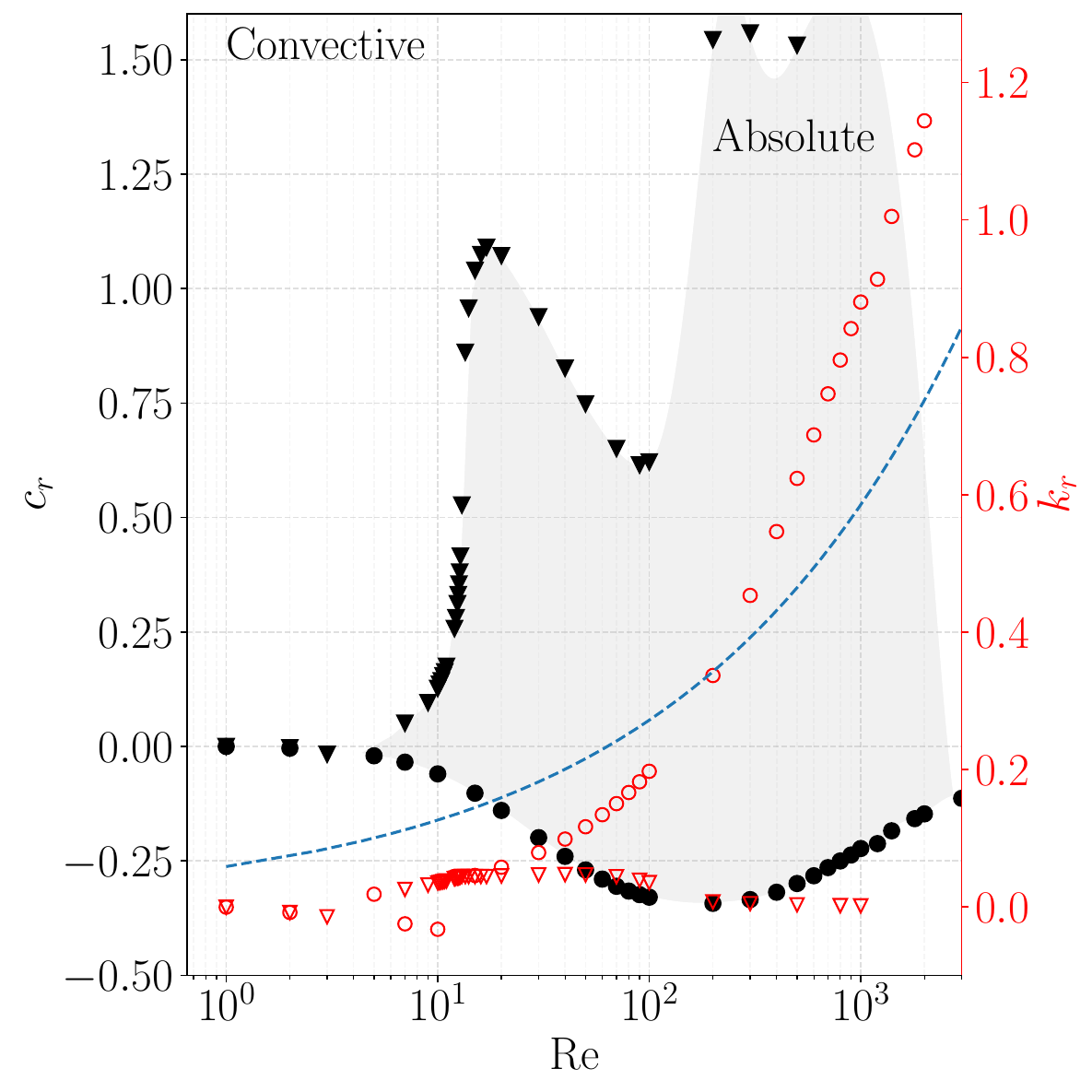}
    \caption{}
  \end{subfigure}
  \begin{subfigure}[b]{0.49\textwidth}
    \includegraphics[width=\textwidth]{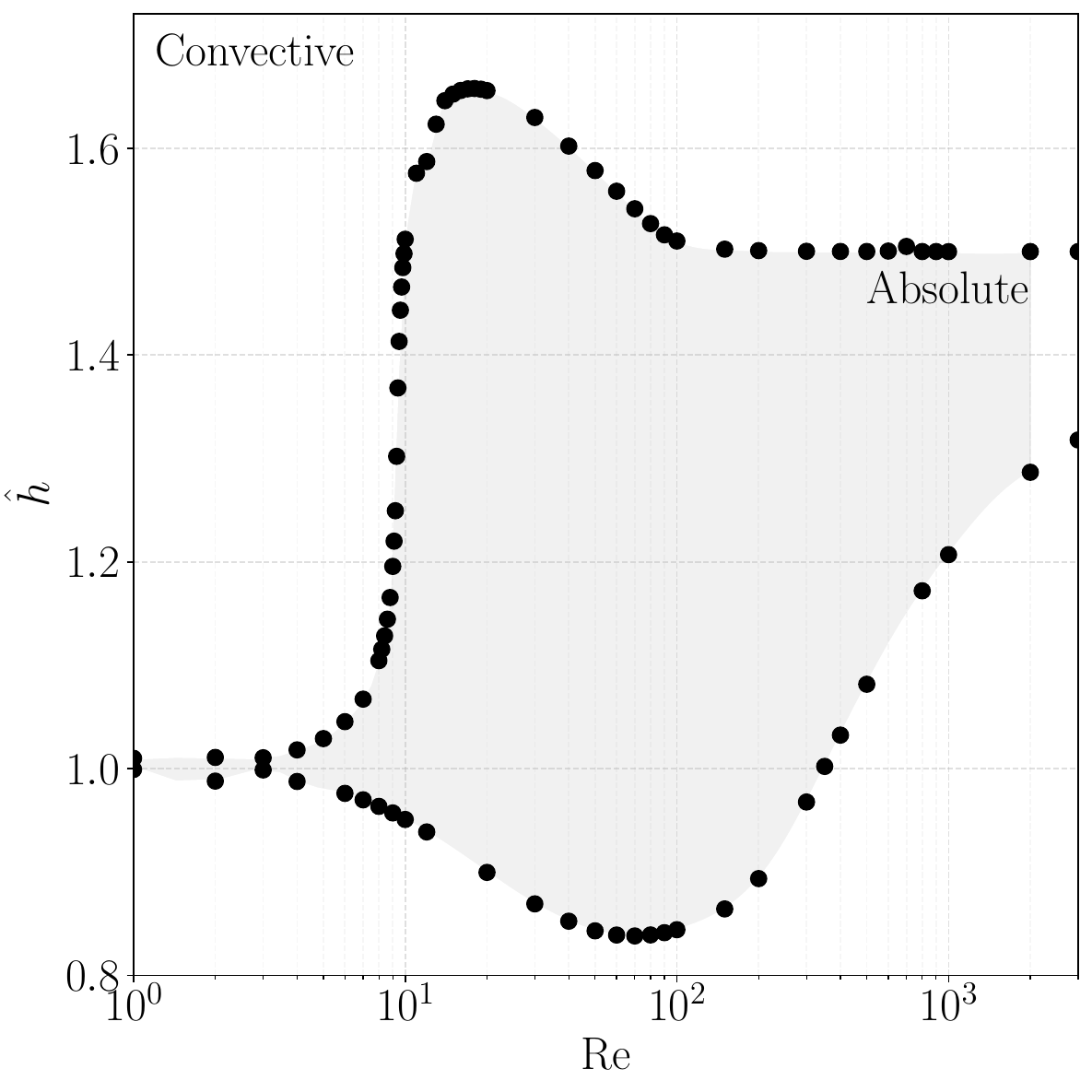}
    \caption{}
  \end{subfigure}
  \hfill
  \begin{subfigure}[b]{0.49\textwidth}
    \includegraphics[width=\textwidth]{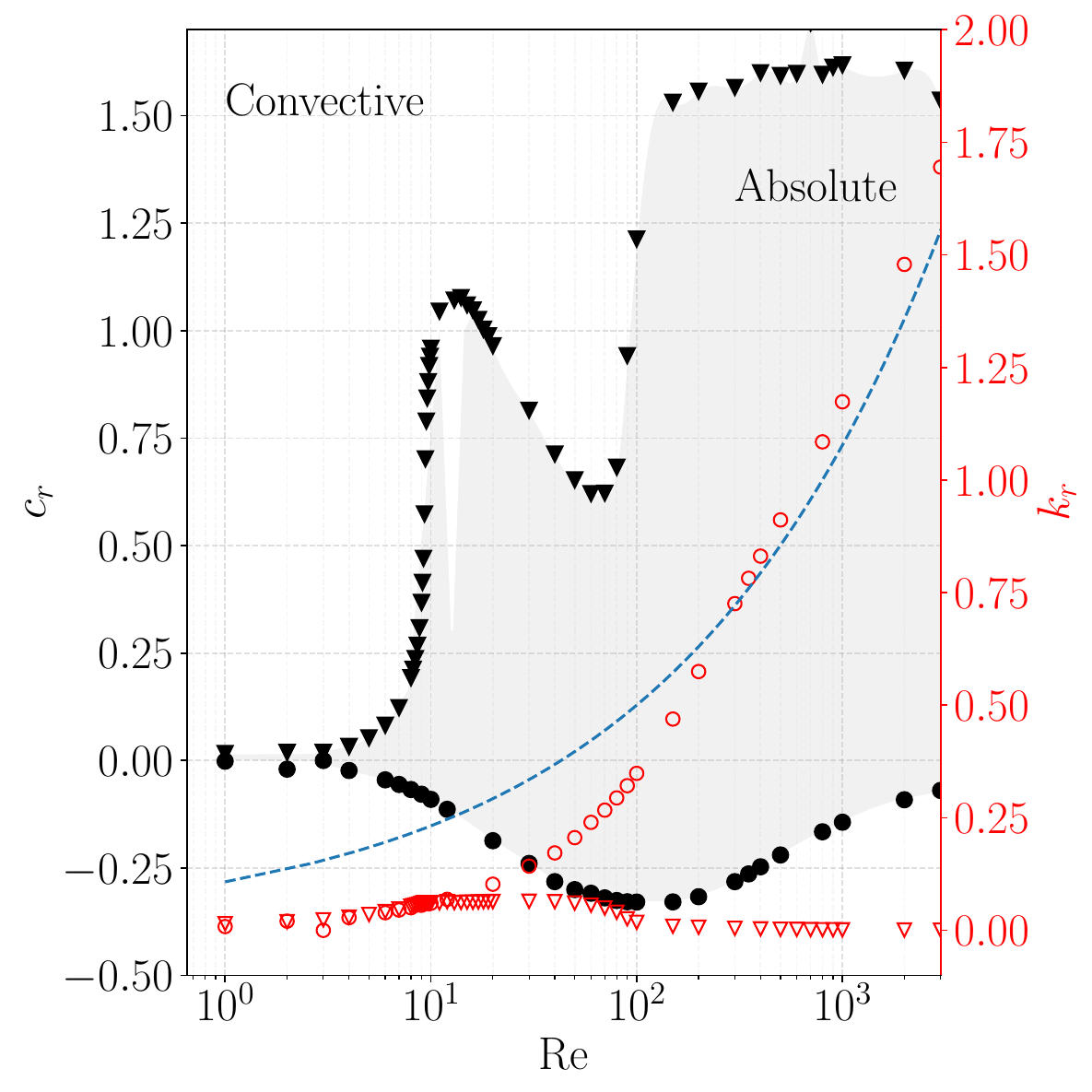}
    \caption{}
  \end{subfigure}
    \caption{Upper and lower branches of the absolute instability region (black dots) in (a and c) the $\hat{h}-\R$ and (b and d) $c_r-\R$ spaces with the associated wavenumber $k_r$ for the lower (circles) and upper (triangles) thresholds and the curve $k_{\ell_c}$ (dashed blue line) for (a and b) liquid zinc ($\Ka$ = 11525) and (c and d) water ($\Ka$ = 3400).}
  \label{fig:ACthreshwater_zinc}
\end{figure}

\begin{figure}
  \begin{subfigure}[b]{0.49\textwidth}
    \includegraphics[width=\textwidth]{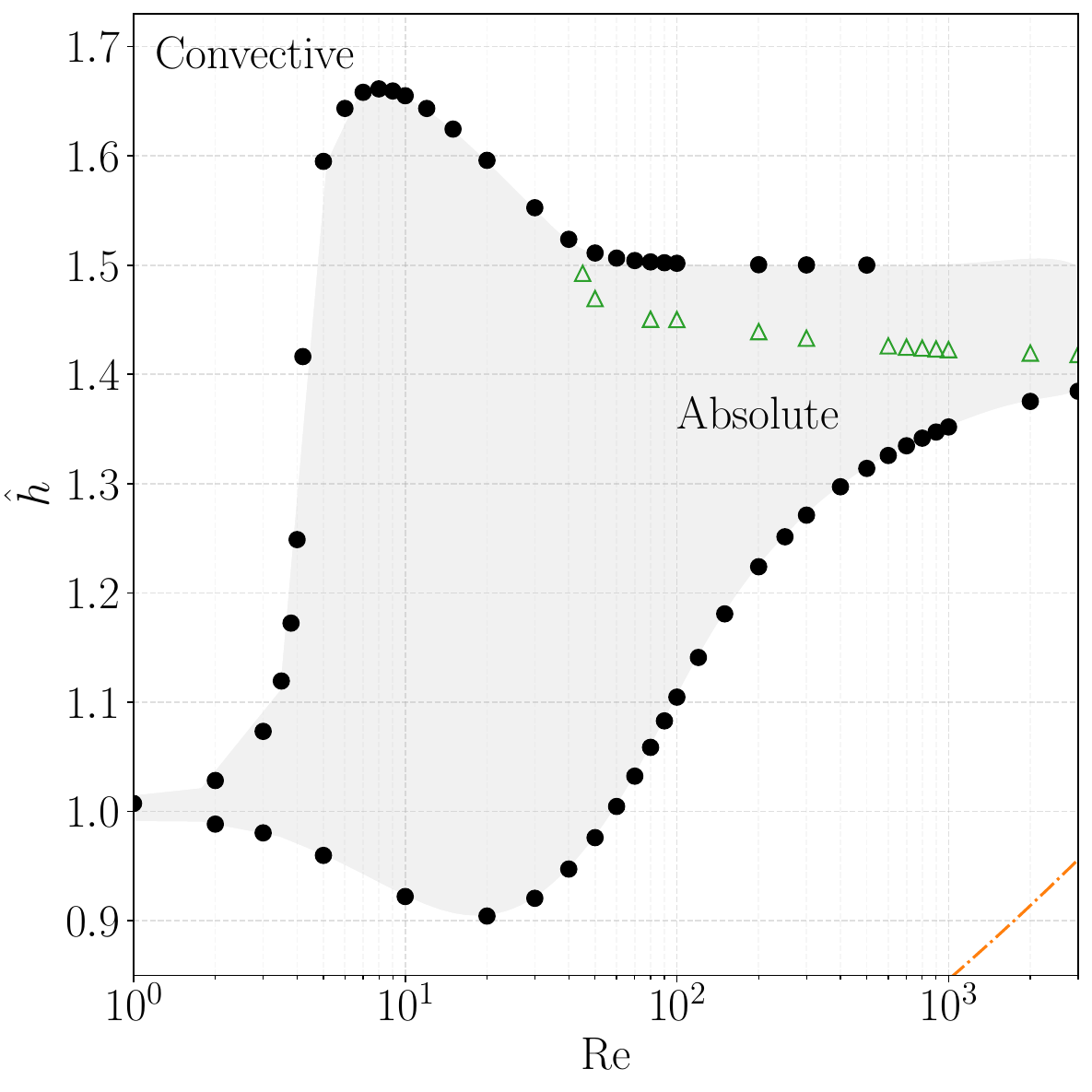}
    \caption{}
    \label{fig:ACthreshwatergly_1}
  \end{subfigure}
  \hfill
  \begin{subfigure}[b]{0.49\textwidth}
    \includegraphics[width=\textwidth]{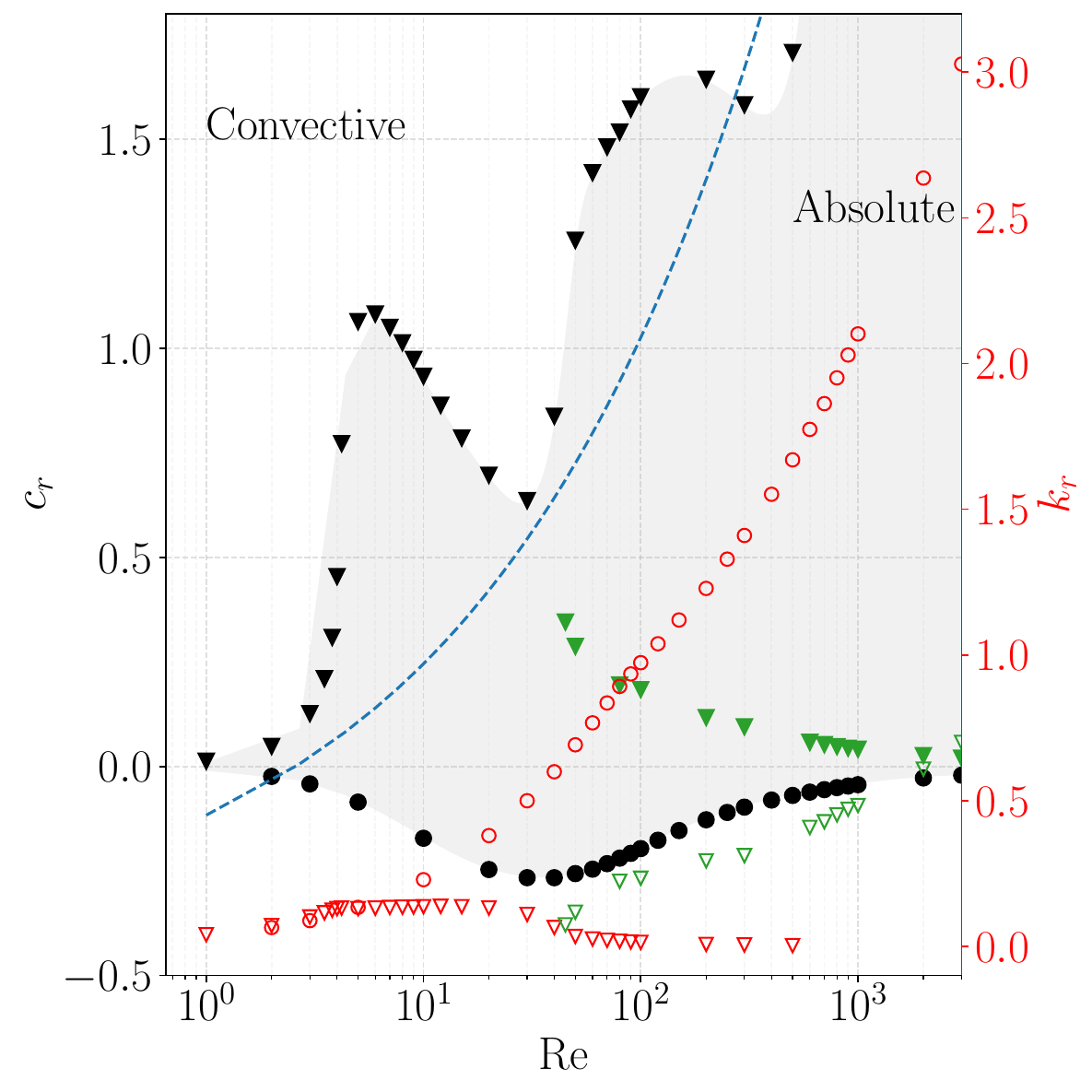}
    \caption{}
    \label{fig:ACthreshwatergly_2}
  \end{subfigure}
  \begin{subfigure}[b]{0.49\textwidth}
    \includegraphics[width=\textwidth]{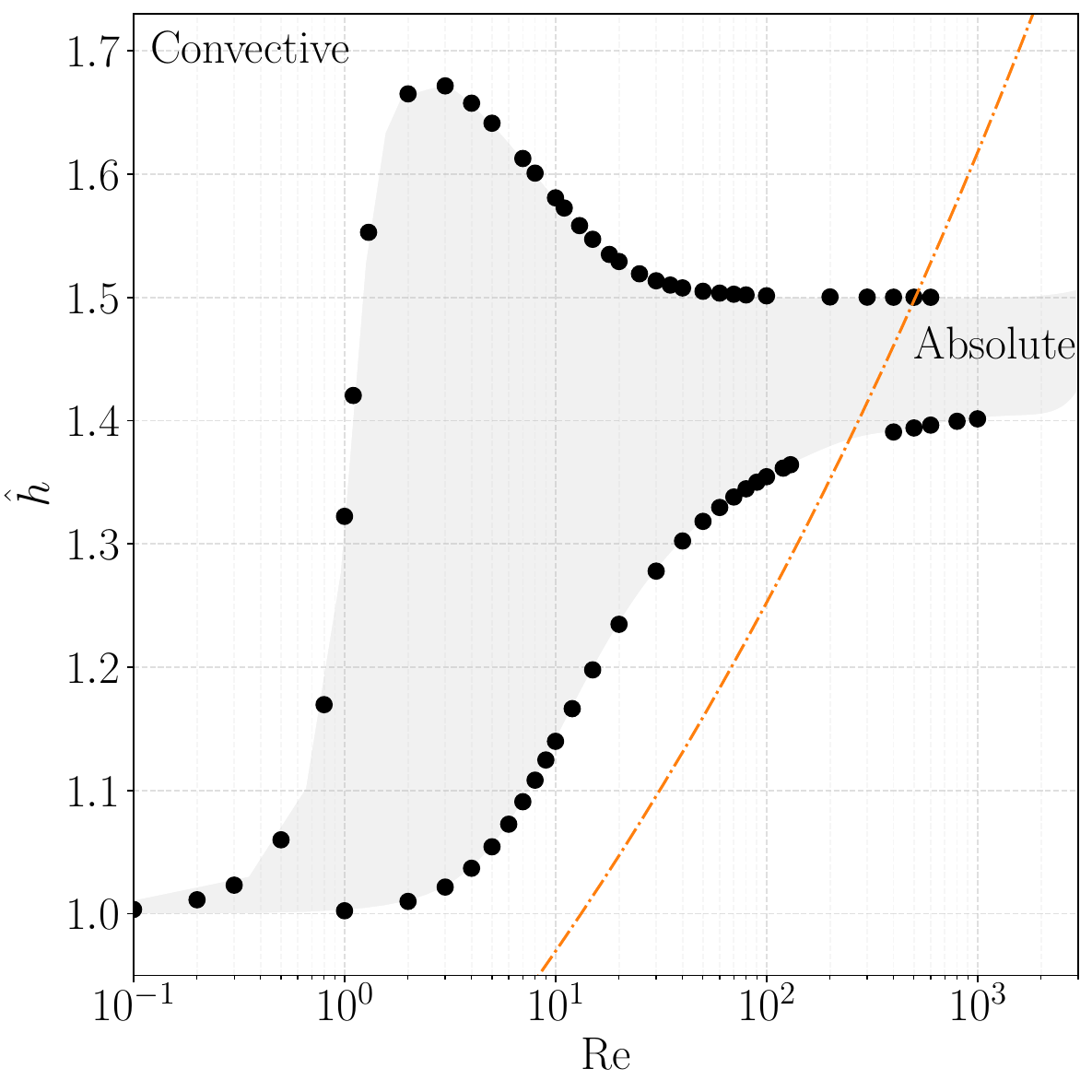}
    \caption{}
    \label{fig:ACthreshcorn_1}
  \end{subfigure}
  \hfill
  \begin{subfigure}[b]{0.49\textwidth}
    \includegraphics[width=\textwidth]{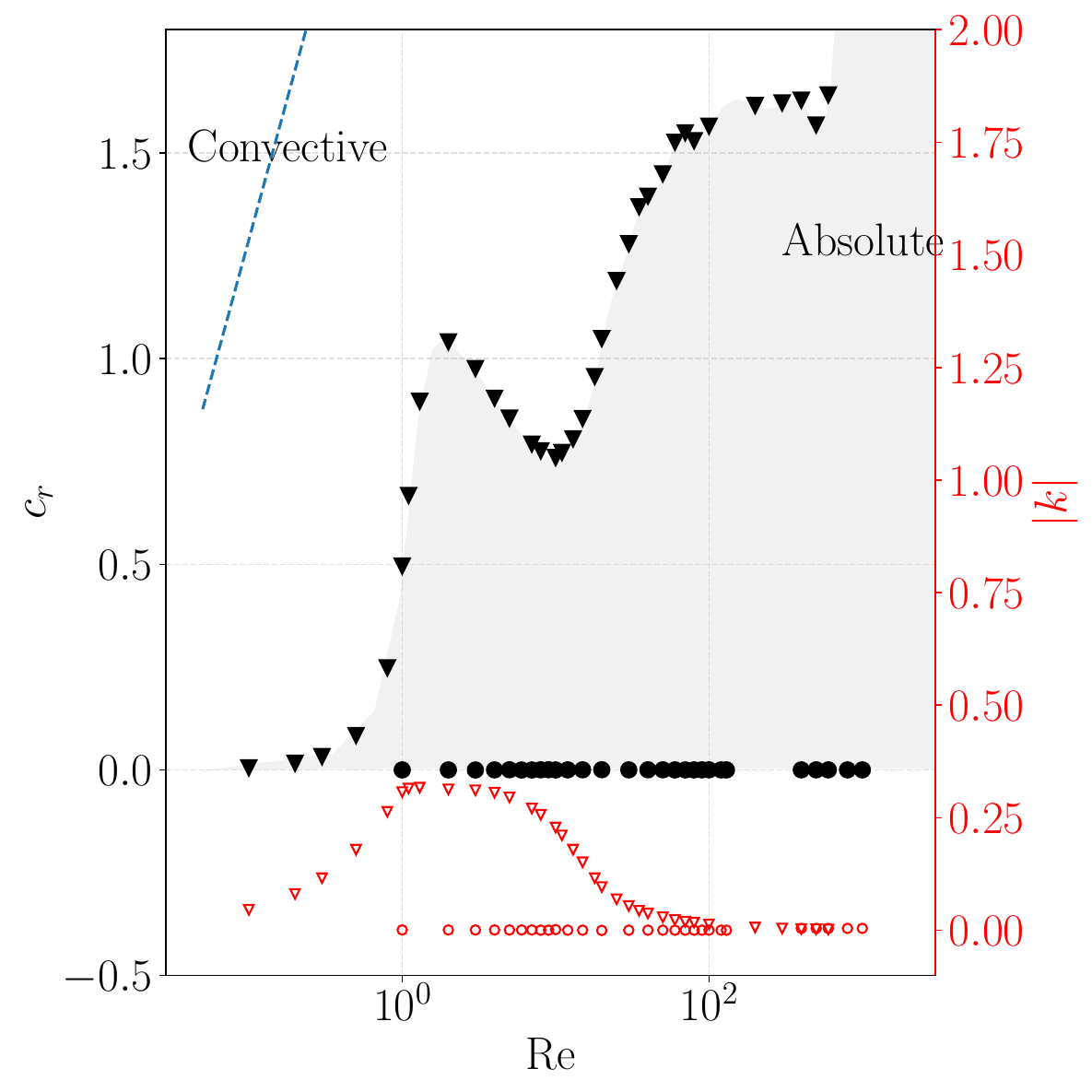}
    \caption{}
    \label{fig:ACthreshcorn_2}
  \end{subfigure}
  \caption{Upper and lower branches of the absolute instability region (black dots) in (a and c) the $\hat{h}-\R$ space with the second branch of the upper threshold (green empty triangles) and the LLD solution (orange dash-dotted line) and in (b and d) the $c_r-\R$ space with the associated wavenumber $k_r$ for the lower (circles) and upper (triangles) thresholds and the curve $k_{\ell_c}$ (dashed blue line) for (a and b) water-glycerol ($\Ka$ = 195) and (c and d) corn oil ($\Ka$ = 4).}
  \label{fig:ACthreshcorn}
\end{figure}
Moving to the AC threshold for the liquids in Table~\ref{tab:liquid_prop}, figures from \ref{fig:ACthreshwater_zinc} to \ref{fig:ACthreshcorn} show the window of absolute instability in (a) the $\hat{h}-\R$ space bounded by a lower and an upper threshold and in (b) the $c_r-\R$ space, with the real wavenumber $k_r$ of the upper (red empty triangular markers) and lower thresholds (red empty circular markers). We also show a second AC upper threshold (green empty markers) and the LLD solution $\accentset{\circ}{h}$ (dash-dotted orange line) for the water glycerol solution and the corn oil. The absolute region's extrema and the upper threshold inflection points are reported in Table~\ref{tab:extremes_AC}.

For the four liquids, the lower threshold stems from $\hat{h}=1$ as $\R\rightarrow 0$ and develops mainly in the thin film domain, with a minimum at intermediate $\R$ values. The threshold extends into the thick film domain, with a plateau at $\hat{h}\approx1.4$ for the water-glycerol and the corn oil. The lower threshold of the corn oil extends solely into the thick film, with thin film base states ($\hat{h}<1$) always convectively unstable. In the $c_r-\R$ space, the neutral waves associated with the lower threshold travel upward against gravity ($c_r<0$). The associated wavenumber $k_r$ monotonically increases with $\R$. $k_r$ overtakes $k_{\ell_{c}}$ for the liquid zinc and water, showing that the surface tension tends to prevail over gravity for large $\R$. Due to surface tension effects, which support the entrainment action against gravity for intermediate $\R$, the minimum of the lower threshold $\hat{h}_{min}$ decreases with $\Ka$, following the relations:
\begin{equation}
    \R_{min} = 1.5496\Ka^{0.4677}, \qquad\qquad\qquad \hat{h}_{min} = - 0.0022\R_{min} + 0.9403,
\end{equation}
where $\R_{min}$ is the $\R$ associated to $\hat{h}_{min}$.

The upper thresholds stem from $\hat{h}=1$ and extend into the thick film domain ($\hat{h}>1$) with a maximum of $\hat{h}\approx1.65$ at $\R\approx 10$ and a plateau at $\hat{h}\approx1.5$ for $\R>100$. In the $c_r-\R$ space, the neutral waves, associated with the upper threshold, travel downwards ($c_r>0$) with a maximum at $\R\approx 1$ for corn oil and $\R\approx 100$ for the other liquids. As for the lower branch, based on the liquid's properties, we define the value of $\hat{h}_{max}$ and the associated $\R$ ($\R_{max}$ as a function of $\Ka$ via the relations:
\begin{equation}
    \R_{max} = 1.9372\Ka^{0.2732}, \qquad\qquad\qquad \hat{h}_{max} = - 0.0008\R_{max} + 1.6663.
\end{equation}

\begin{figure}
\centering
    \begin{subfigure}[b]{0.46\textwidth}
    \centering
    \includegraphics[width=\textwidth]{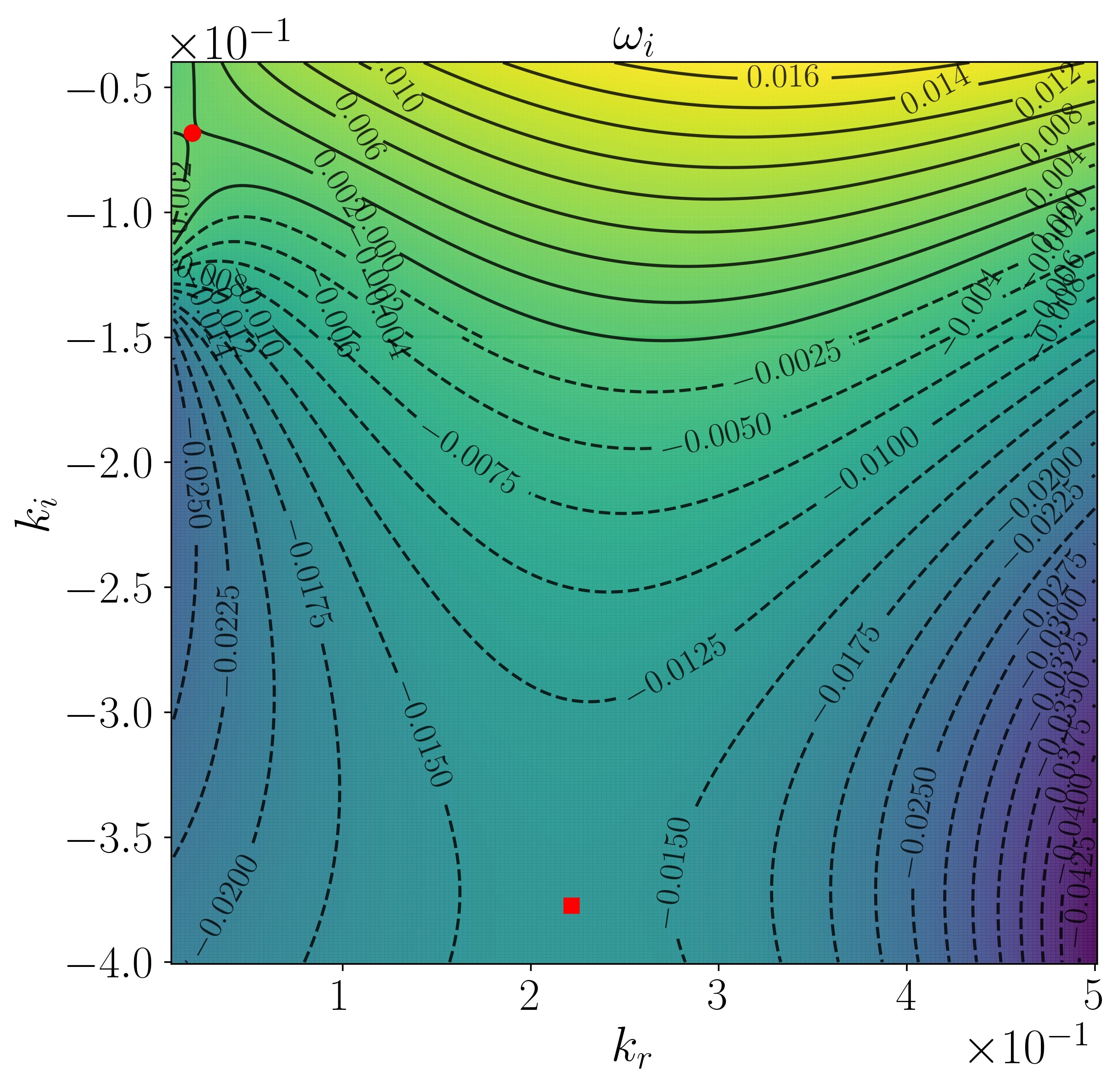}
    \caption{}
  \end{subfigure}
  \hfill
  \begin{subfigure}[b]{0.46\textwidth}
  \centering
    \includegraphics[width=\textwidth]{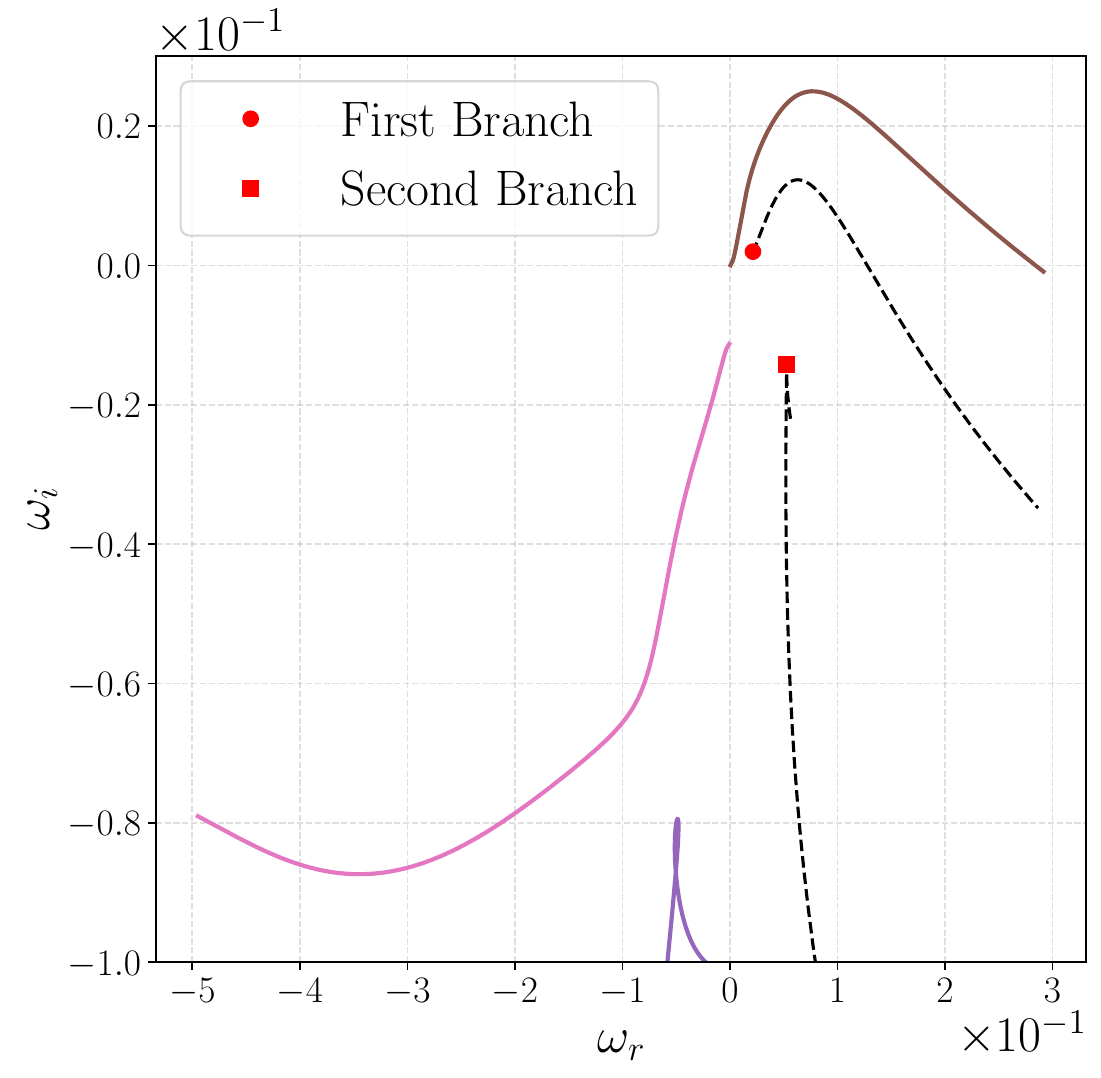}
    \caption{}
  \end{subfigure}
  \caption{Position of two saddle points (red circle and red square) in (a) the $k_r-k_i$ space with the $\omega_i$ colour plot and in (b) the $\omega_r-\omega_i$ space with the temporal branches along the real wavenumber axis with $k_i=0$ (continuous coloured lines) for water-glycerol with $\hat{h}=1.48$ and $\R=100$.}
  \label{fig:multiple_branches}
\end{figure}
\begin{figure}
  \centering
  \begin{subfigure}[b]{0.46\textwidth}
    \centering
    \includegraphics[width=\textwidth]{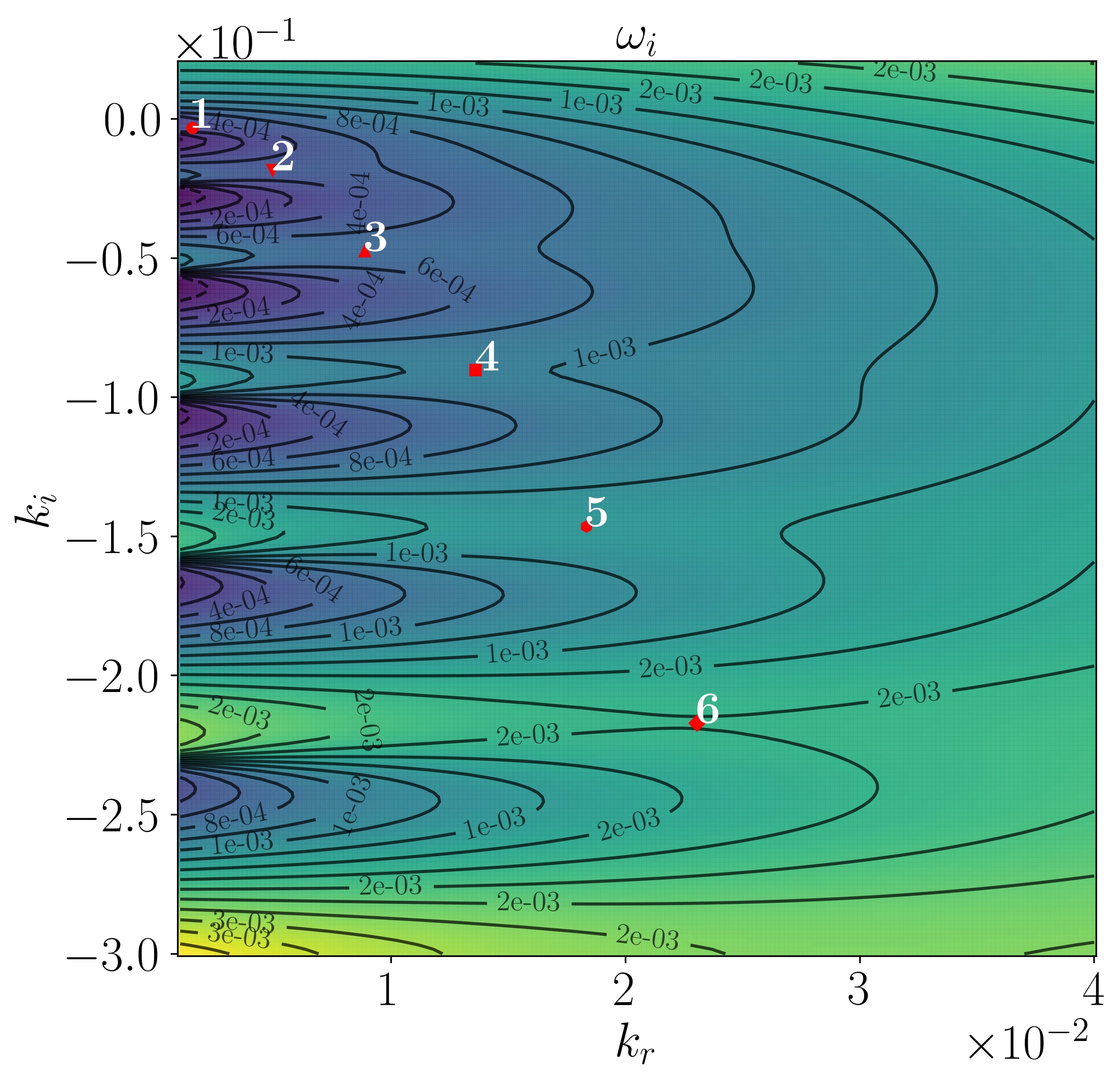}
    \caption{}
    \label{fig:mul_saddl_point_1}
  \end{subfigure}
  \hfill
  \begin{subfigure}[b]{0.46\textwidth}
  \centering
    \includegraphics[width=\textwidth]{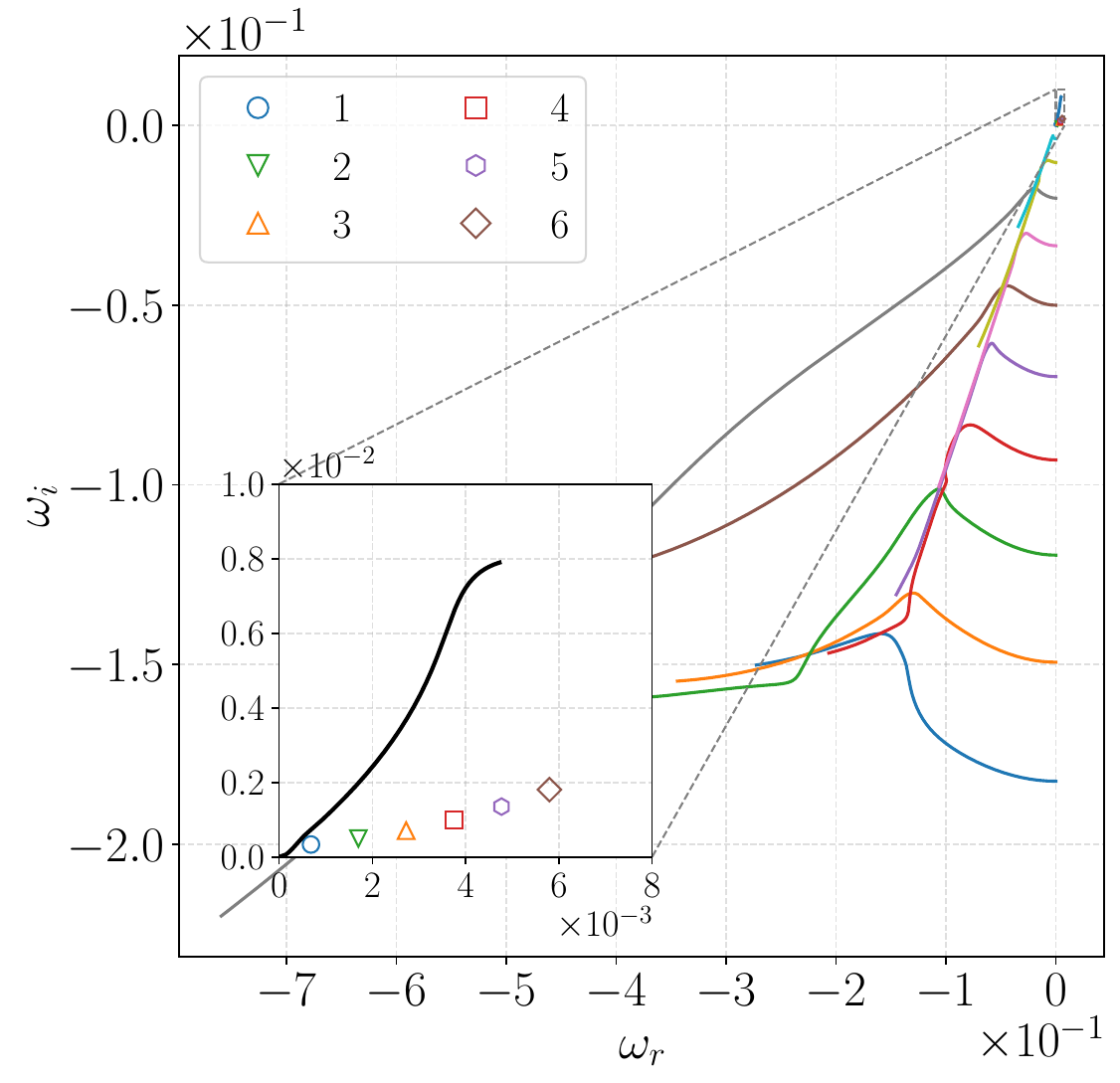}
    \caption{}
    \label{fig:mul_saddl_point_2}
  \end{subfigure}
  \hfill
  \begin{subfigure}[b]{0.46\textwidth}
    \centering
    \includegraphics[width=\textwidth]{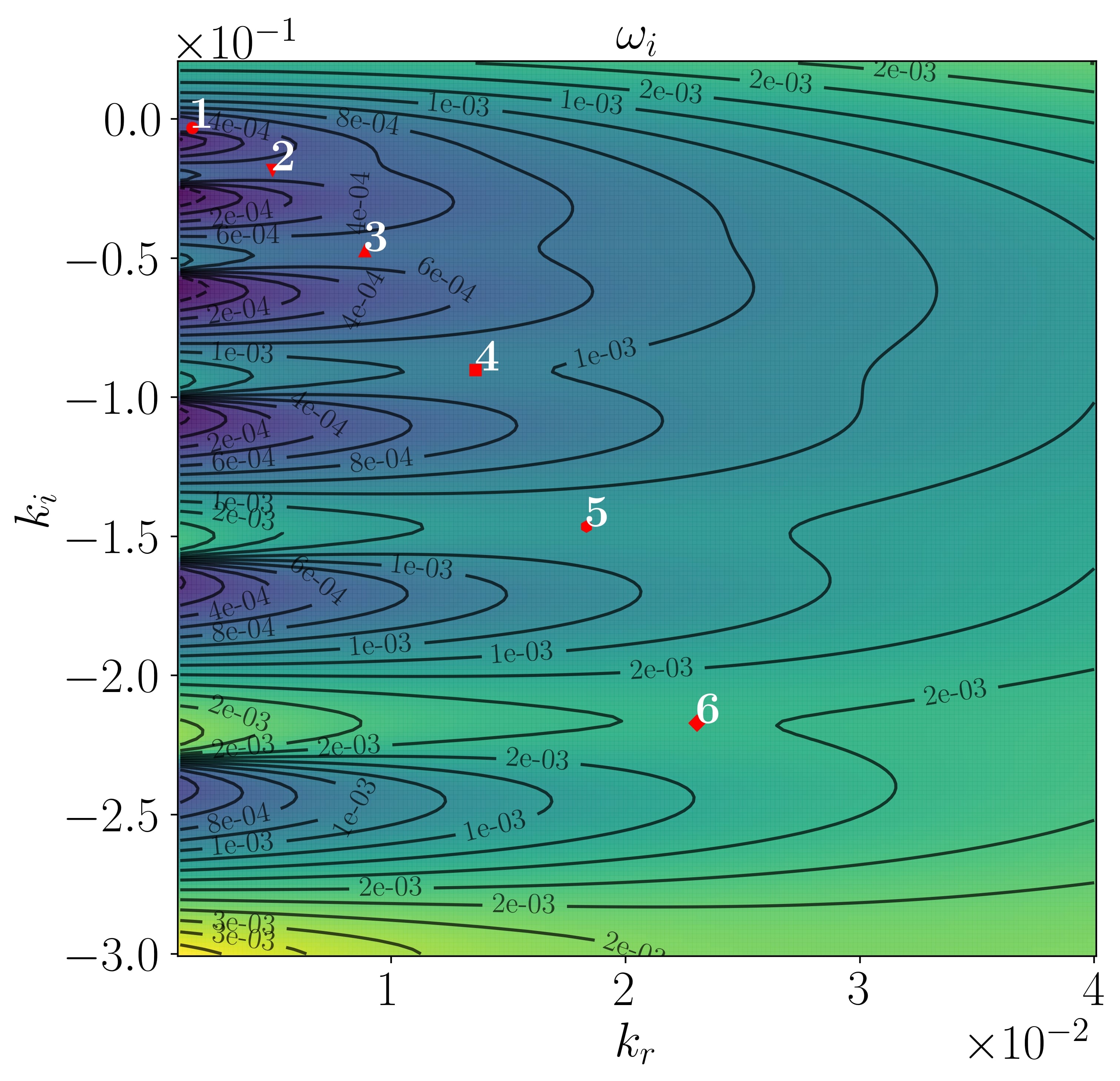}
    \caption{}
    \label{fig:mul_saddl_point_3}
  \end{subfigure}
  \hfill
  \begin{subfigure}[b]{0.46\textwidth}
  \centering
    \includegraphics[width=\textwidth]{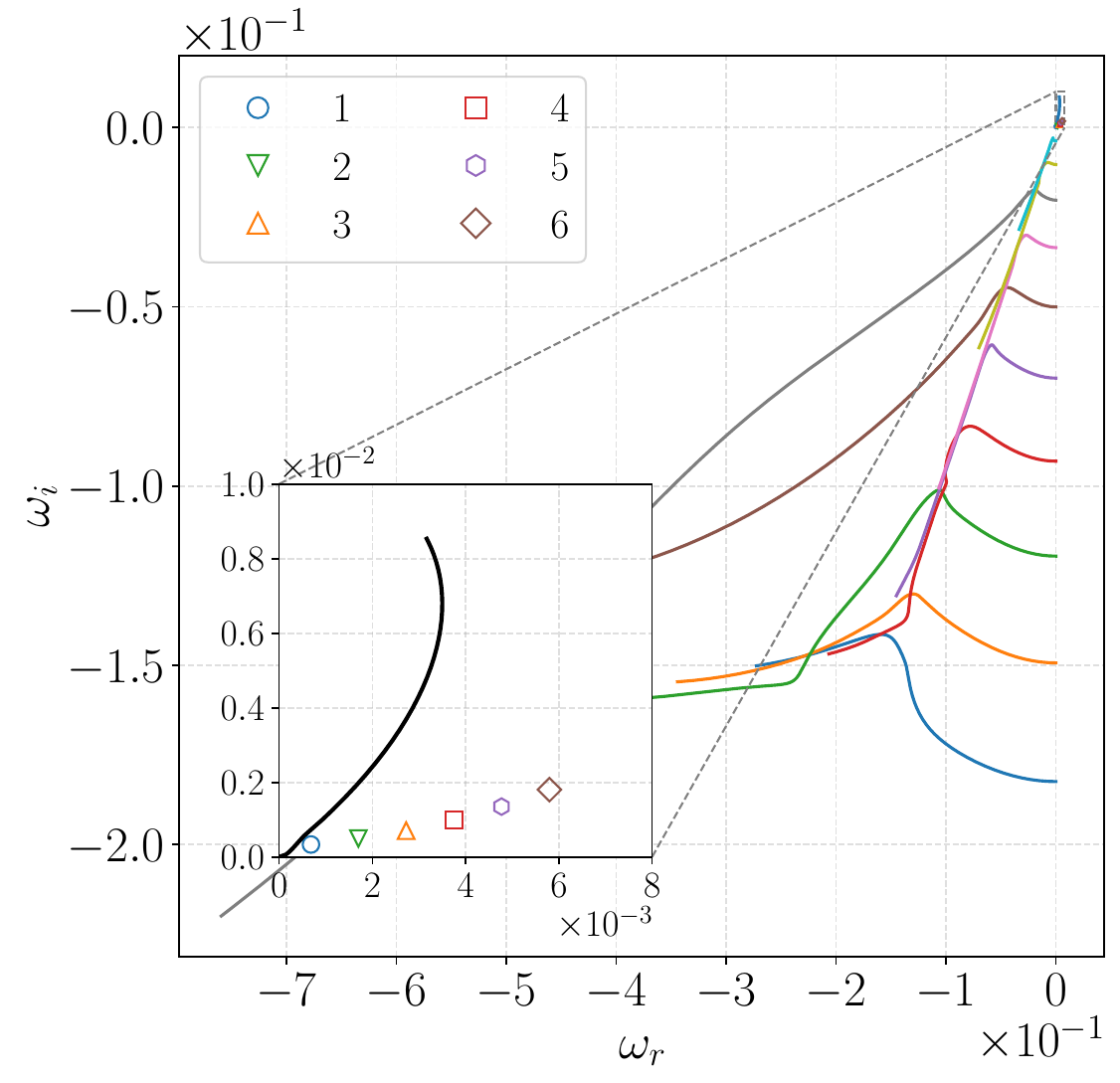}
    \caption{}
    \label{fig:mul_saddl_point_4}
  \end{subfigure}
 \caption{Multiple saddle points in (a and c) the complex wavenumber space $k_r-k_i$ (numbered red markers) and (b and d) in the complex frequency domain $\omega_r-\omega_i$ (empty coloured markers) with the temporal branches along the real wavenumber axis (coloured continuous line) at $\R=3000$ and $\hat{h}=1.41$ for (a and b) water-glycerol ($\Ka=195$) and (c and d) $\Ka=0$.}
  \label{fig:multiple_saddle_points}
\end{figure}

In addition to the saddle point defining the upper branch, we found other valid saddle points for water-glycerol. Figure~\ref{fig:multiple_branches} shows two saddle points (red circle and read square) (a) in the complex wavenumber space $k_r-k_i$ space with the $\omega_i$ colour map and (b) in the complex frequency space $\omega_r-\omega_i$ with the spatial branches along the real $k$ axis with $k_i=0$ for water-glycerol at $\hat{h}=1.48$ and $\R=100$. The two saddle points respect the collision criterion since they are surmounted by an odd number of temporal branches. Based on the second saddle point (red square), which appears around $\R\approx 40$, we traced another AC threshold (green empty triangles in figure~\ref{fig:ACthreshwatergly_1}). Interestingly, this threshold is always below the first one and approaches the lower branch for larger $\R$, closing the window of absolute instability. This is also visible in the $c_r-\R$ space (green triangles in figure~\ref{fig:ACthreshwatergly_2}), where the phase speed of the two branches converges. Moreover, the wave number increases with $\R$ reaching 0.5 for $\R=3000$.
\begin{figure}
  \begin{subfigure}[b]{0.49\textwidth}
    \includegraphics[width=\textwidth]{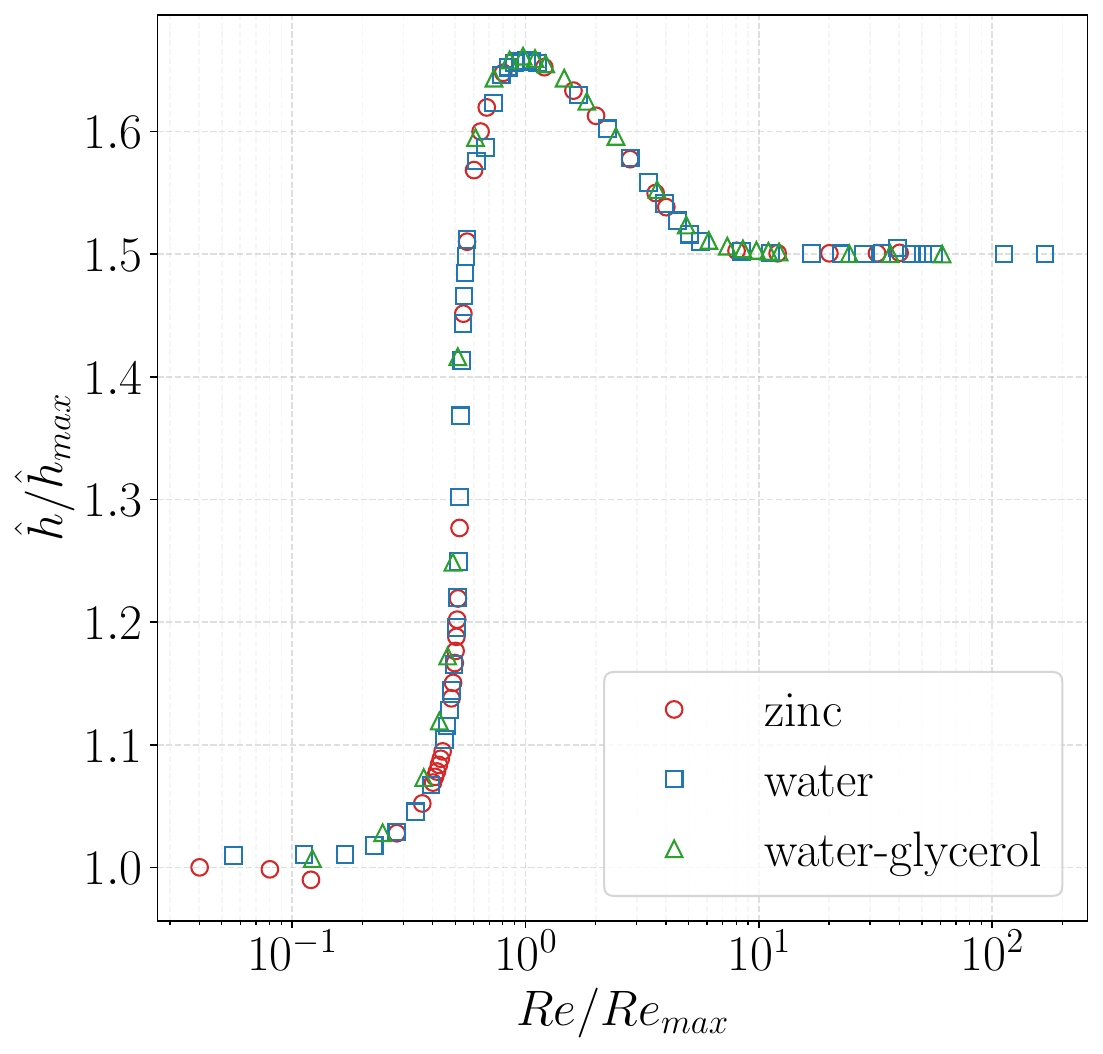}
    \caption{}
  \end{subfigure}
  \hfill
  \begin{subfigure}[b]{0.49\textwidth}
    \includegraphics[width=\textwidth]{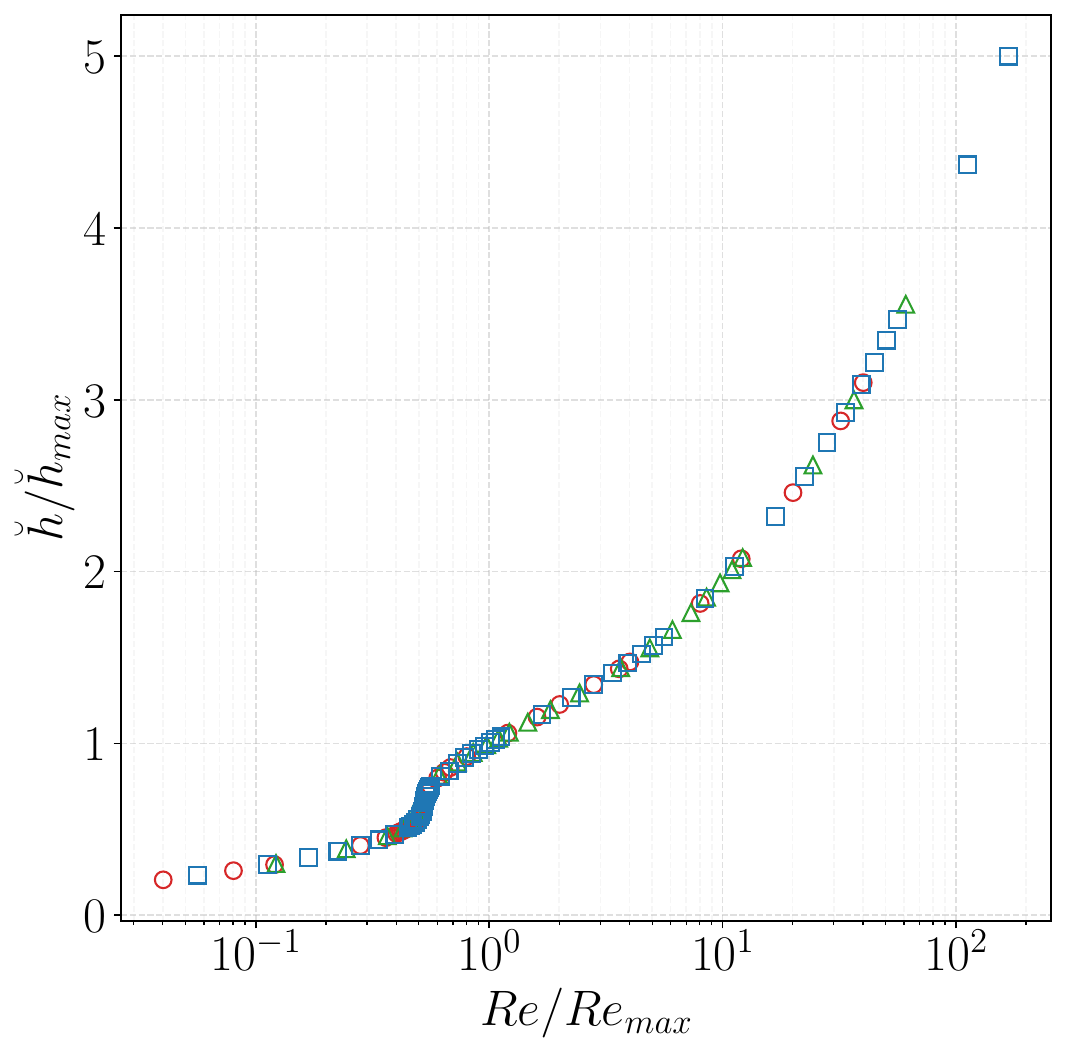}
    \caption{}
  \end{subfigure}
  \hfill
  \begin{subfigure}[b]{0.49\textwidth}
    \includegraphics[width=\textwidth]{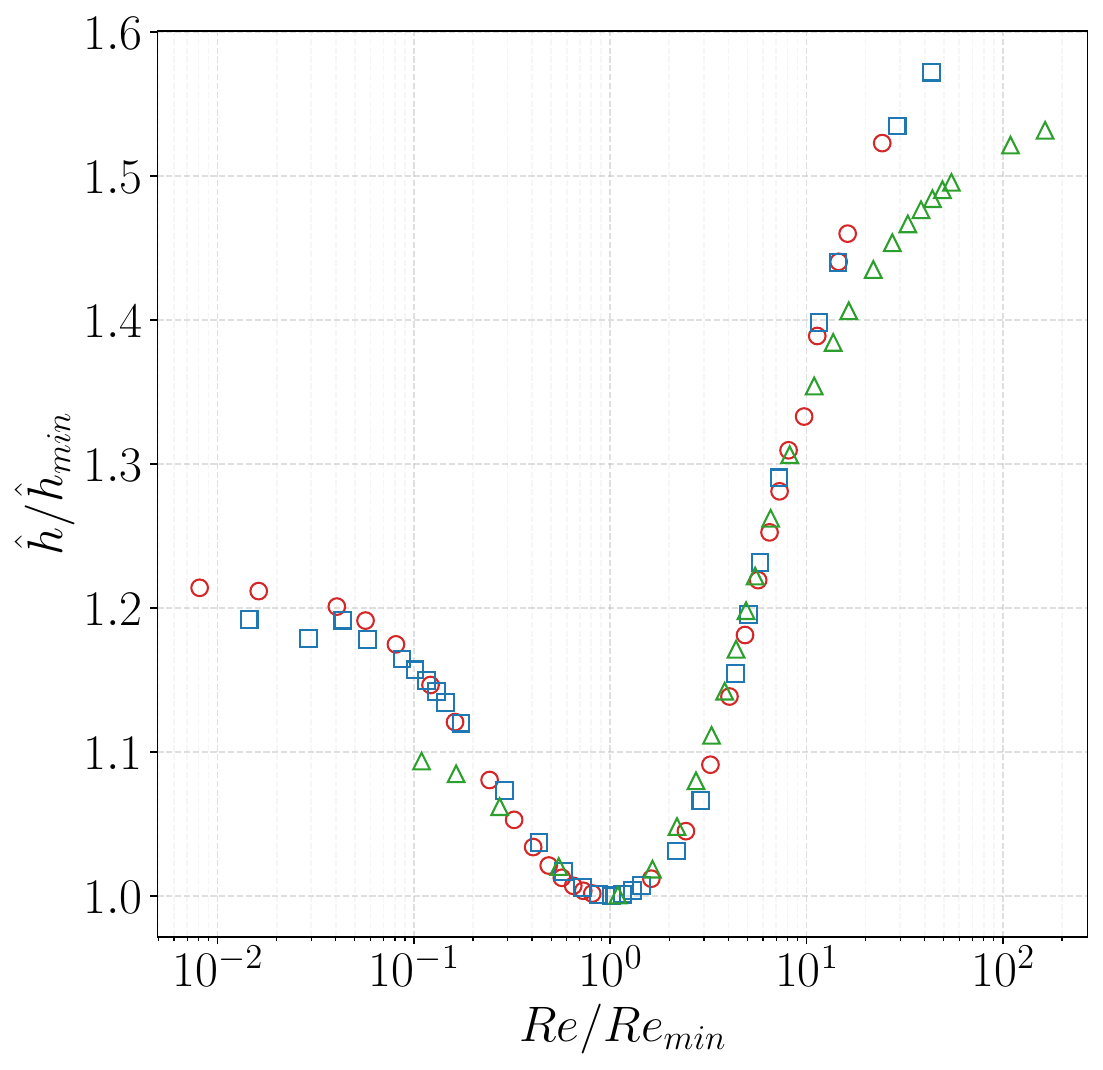}
    \caption{}
  \end{subfigure}
  \hfill
  \begin{subfigure}[b]{0.49\textwidth}
    \includegraphics[width=\textwidth]{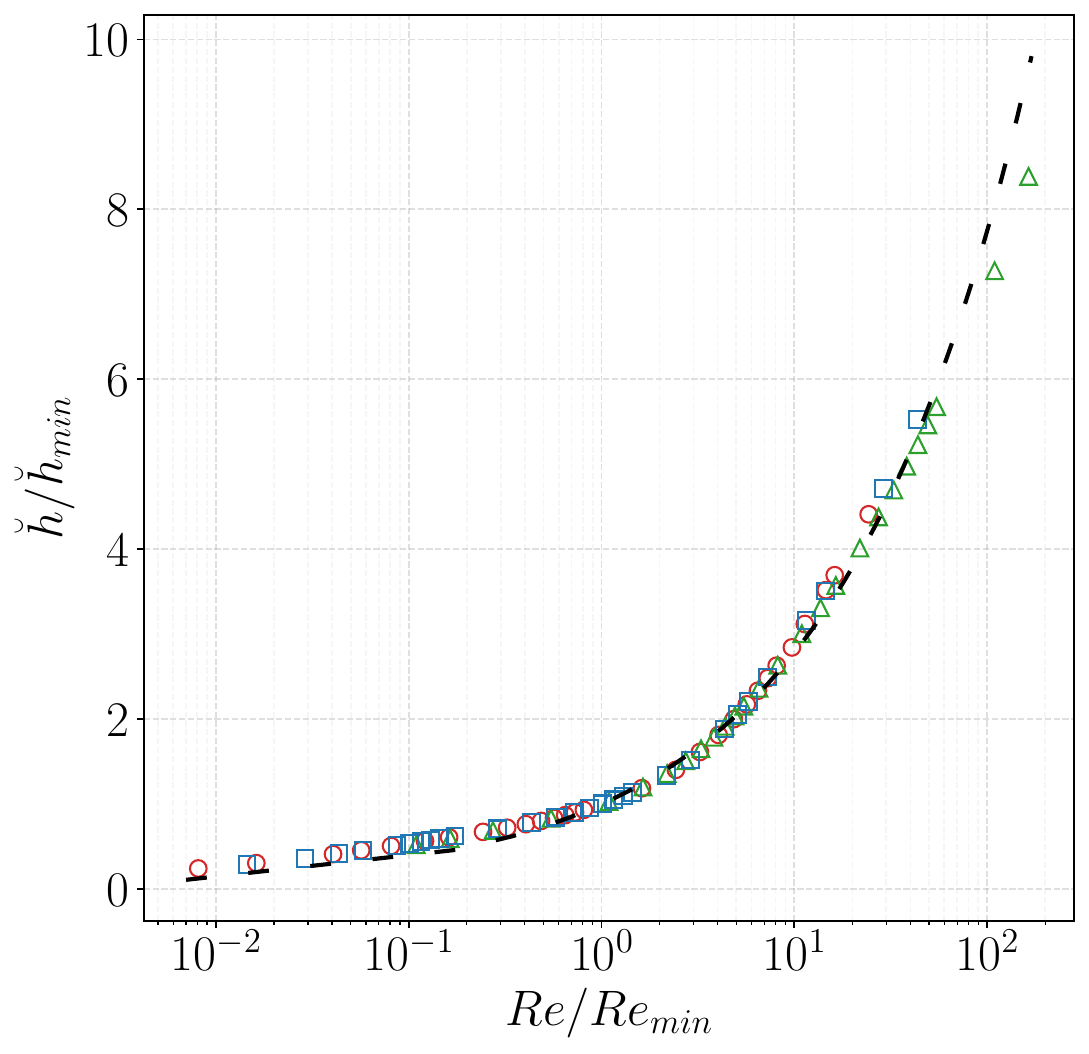}
    \caption{}
    \label{fig:scaled_4}
  \end{subfigure}
  \caption{Scaled absolute instability window for liquid zinc (red circles), water (blue squares) and water-glycerol (green triangles), with the (a and b) upper and (c and d) lower AC thresholds rescaled with their values at the maximum and the minimum in (a and c) the $\hat{h}-\R$ and (b and d) the $\acute{h}-\R$ space with the trend line $\breve{h}/\breve{h}_{min}=(\R/\R_{min})^{4/9}$ (loosely dashed black line).}
  \label{fig:scaled_curves} 
\end{figure}

For $\R=3000$, we discovered a collection of saddle points near the imaginary wave numbers' axis. Figures~\ref{fig:multiple_saddle_points} shows the position of six saddle points with (a and c) red markers in the $k_r-k_i$ space with the $\omega_i$ colour map and with (b and d) coloured markers in the $\omega_r-\omega_i$ space and the temporal branches along the real wavenumber axis (coloured lines) with $\hat{h}=1.41$ and $\R=$ for (a and b) water-glycerol and (c and d) $\Ka=0$. The first saddle point is used to construct the AC threshold described above. For plotting convenience, we do not show the saddle point of the second upper branch because it is at a much larger real wave number. Only the first four of the six saddle points respect the collision criterion, with the first being the last to have a negative growth rate as we increase $\hat{h}$. Since the AC associated with this point represents the upper bound for the region of absolute stability, we did not investigate the AC for the other saddle points. The location of these saddle points is invariant with $\Ka$ since their positions are the same for the case with water glycerol and $\Ka=0$. This implies that this part of the AC upper branch is not affected by the surface tension and is solely given by a balance of gravity and inertia.

The LLD solution $\accentset{\circ}{h}$ (defined in \eqref{eq:LLD_sol}) is always connectively unstable for zinc, water and water-glycerol. For convenience, we did not report this curve for liquid zinc and water because it was at a much smaller $\hat{h}$ compared to the window of absolute instability. For corn oil, $\accentset{\circ}{h}$ crosses the area of absolute instability in the range $100<\R<500$. However, since this intersects the lower branch at a $\R\approx 300$, it is unrealistic to see it in an experiment. In addition, given that the wave number of the lower branch is very small, this would require a very long domain for substrate velocities in the range $U_p\in[0.1-1][m/s]$.

We rescale the upper and lower AC thresholds in the $\hat{h}-\R$ space with ($\hat{h}_{max},\R_{max}$) and ($\hat{h}_{min},\R_{min}$) respectively for liquid zinc, water and water-glycerol. Figure~\ref{fig:scaled_curves} shows the scaled (a and b) upper and (c and d) lower thresholds in (a and c) the $\hat{h}-\R$ and (c and d) the $\breve{h}-\R$ spaces for the liquid zinc (red circle), water (blue squares) and water-glycerol (green triangles), where $\breve{h}$ is the nondimensional film thickness based on the viscous length $\ell_{\nu}$ defined in \ref{eq:h_breve}. The curve matches perfectly apart from the lower bound in the $\hat{h}-\R$ space at small and large $\R$. Moreover, we found a simple approximation to the lower AC threshold. Figure~\ref{fig:scaled_4} shows a trend line (black loosely dashed line) for the lower branch, given by:
\begin{equation}
    \breve{h}/\breve{h}_{min}=\Big(\R/\R_{min}\Big)^{4/9}.
\end{equation}
This approximation fits the simulation's data very well, generalizing the AC threshold to any fluid with $\Ka>195$.

\section{Conclusion and perspectives}
\label{Conclusion}
This study investigated the linear stability of a vertical liquid film over a substrate moving against gravity for four liquids with $\Ka$ numbers ranging from 4 to 11525. Using long-wave asymptotic analysis and numerical solutions to the Orr-Sommerfeld eigenvalue problem, we identified the region of unstable perturbations. For the unstable solutions, the instability mechanism was described via momentum, vorticity, and energy-based arguments, and the threshold between absolute and convective instability was traced.

The neutral curves, growth rates and phase speed converge around the same values for $\hat{h}\gtrsim0.7$, highlighting a stabilizing mechanism where viscous effects balance gravitational effects with minimal influence from surface tension.

For thin films, the instability is driven by the amplitude of the vorticity with a minimal phase shift. For thick films, the amplitude decreases, and the phase shift increases, with a peak shifted at larger $k$. The surface tension strongly affects this mechanism, simultaneously stabilizing and destabilizing the film, especially for $\Ka=11525$. For long waves, this curves the vorticity lines near the substrate, reducing the flow under the crests. For short waves, this enhances vorticity production at the free surface and creates a region of intense vorticity near the substrate.

Intense Reynolds stresses accompany areas of intense vorticity. At the same time, the dissipative terms also grow in magnitude, leading to two instability mechanisms for small and large $\Ka$. For $\Ka=11525$, surface tension induces a larger production of vorticity at the free surface and the wall, resulting in significant Reynolds stresses and intense dissipative effects. In addition, the contribution to the viscous terms also increases. For $\Ka=4$ (corn oil), shear effects mainly influence the viscous term. For $\Ka=11525$ (zinc), the shear effects diminish, and the extensional term becomes more important.

In terms of AC instability threshold, for $\Ka<17$ the LLD solution is always convectively unstable for any $\R$ while for $\R\gtrsim 10$, the threshold follows an asymptote given by $\Ka\approx\R^{23/13}$. In the $\hat{h}-\R$ space, a window of absolute instability arises between the thin and thick film conditions. This window develops solely in the thick film region for $\Ka=4$ (corn oil). Moreover, a bifurcation point is also present at $\R\approx 40$, where two solution branches exist, one of which is independent of $\Ka$. Rescaling the lower (upper) branch of the absolute instability window with its minimum (maximum), the curves for $\Ka>4$ converge in the $\hat{h}-\R$ space. A relation linking the value of $\R$ at the minimum of this curve to the lower bound was also provided.

These findings shed light on the stability properties of the liquid film in the dip-coating process. Our results are important for validating reduced-order models, designing optimal control functions, and understanding the instability mechanism. 

Further analysis could leverage our results to establish whether the perturbation exhibits remnant behaviour, where waves propagate both upwards and downwards. In addition, the transient growth analysis could predict the wave evolution in experimental campaigns. Additionally, a campaign of direct numerical simulations could assess the absolute/convective threshold's validity and spot other unstable saddle points in the complex frequency space.

\section*{Funding and Declaration of Interest}
F.Pino is supported by an F.R.S.-FNRS FRIA grant, and his PhD research project is funded by Arcelor-Mittal. B.Scheid is Research Director at F.R.S.-FNRS.

-The authors report no conflict of interest.

\bibliographystyle{jfm}
\bibliography{Fabio_et_al_2022}

\appendix 
\section{Pseudocode for the search of the absolute/convective threshold}
\label{appxsubsec:pseudocode_AC_threshold}

The pseudocode below reports the main steps of the absolute/convective threshold search. First, we define the liquid properties and the Kapitza number. Then we select a range of Reynolds numbers to calculate the threshold. Inside the for loop, we select one of these values and a guess value for $\hat{h}$, and we search for a valid saddle point in the complex wavenumber space. This procedure is done via trial and error. The wave numbers space is explored in small windows until we find a saddle point. Then, we validate this point by visually inspecting two spatial branches from different half-planes pinch at the saddle point. If the saddle point is validated, we fix the window in the complex wavenumber space and define the limit for the $\hat{h}$ optimization. We define the range of $\hat{h}$, such that the growth rate at the saddle point has a different value at the extremes of the range. Then we pass this information to a scalar optimization algorithm that computes precisely the location of the threshold associated with a saddle point with $|\omega_i|<10^6$. We store this value and move to another $\R$ number. The procedure for the $Ka-\R$ space is the same; instead of searching along $\hat{h}$, we search along $Ka$ for a fixed Reynolds number. 
\begin{algorithm}[!ht]
\caption{Absolute/convective threshold search in the parameter space $\hat{h}-\R$}
\begin{algorithmic}[1]
  \Function{scalar optimization}{$\mathbf{H},\mathbf{K},Re,Ka$}
  \State Select random value of $\hat{h}_r$ from $\mathbf{H}$
  \State Select the most unstable mode at the border or the wavenumber window $\mathbf{K}$ solving the generalized eigenvalue problem
  \State Map the $\mathbf{K}$ into the frequency space with Rayleigh quotient iteration 
  \State Calculate saddle point location with numerical differentiation
  \State Collect value of growth rate at the saddle point $\omega_{i_{SP}}$
  \While{$|\omega_{i_{SP}}|>1e-6$}
  \State Select value of $\hat{h}_g$ from $\mathbf{H}$
  \State Map $\mathbf{K}$ into frequency space
  \State Compute saddle point location with numerical differentiation
  \State Check collision criterion
  \State Collect value of $\omega_i$ at the saddle point ($\omega_{i_{SP}}$)
  \EndWhile
  \State \Return $\hat{h}_g$
  \EndFunction
  \State Define liquid properties and $Ka$ number
  \State Initialize list of $\R$ numbers: $\mathbf{a} = [\R_0,\R_1,\cdots,\R_N]$
  \State Initialize empty list of $\hat{h}$: $\mathbf{b}$
  \For {$n$ in (1,$N$)}
  \State Select Reynolds from the list $\R=\mathbf{a}[n]$
  \State Define guess value for $\hat{h}=\hat{h}_g$ 
  \State Search saddle point in wavenumber space
  \State Define limit of search window: $\mathbf{K}=[k_{r_{min}},k_{i_{min}},k_{r_{max}},k_{r_{max}}]$
  \State Define limit for $\hat{h}$ line search: $\mathbf{H}=[\hat{h}_{min},\hat{h}_{max}]$
  \State Compute $\hat{h}$ at the threshold $\hat{h}_n\leftarrow$\textproc{scalar optimization}($\mathbf{H},\mathbf{K},Re,Ka$)
  \State Store $\mathbf{b}\leftarrow\hat{h}_n$
  \EndFor
  \label{Alg_AC_threshold}
\end{algorithmic}
\end{algorithm}
\FloatBarrier

\section{Convergence study and code verification}
Subsection~\ref{subsec:convergence_study} reports the convergence study used to define a suitable number of Chebyshev polynomials in the solution of the OS eigenvalue problem and the grid spacing of the complex wavenumber space in the AC threshold search. Subsection~\ref{subsec:verification} validate the numerical implementation, showing the eigenvalue spectrum obtained solving the OS problem and the long-wave approximation of the OS solution.
\subsection{Convergence study}
\FloatBarrier
\label{subsec:convergence_study}
\begin{table}
\centering
\begin{tabular}{cccc}
\toprule 
$N$ & $\omega_r$ & $\omega_i$ & accuracy\\ \midrule
10  &  1.61898344 & 0.4042714 & 1.5299$\times 10^{-4}$\\
20  &  1.61898354 & 0.40427141 & 1.5308$\times 10^{-4}$\\
80  &  1.61897484 & 0.40429792 & 1.5322$\times 10^{-4}$ \\
100 &  1.6188345 & 0.40423644 & -\\ \bottomrule
\end{tabular}
\caption{Values of $\omega_i$ and $\omega_r$ of the most unstable eigenvalue (largest $\omega_i$) for $\R=20$, $\hat{h}=1.7$ and $k=10^{-2}$ for four different number of Chebyshev polynomials $N$ with their relative accuracy expresses as the Euclidean norm of the eigenvalue difference with respect to the $N=100$ case.}
\label{tab:convergence_study}
\end{table}
\begin{table}
\centering
\begin{tabular}{ccccc}
\toprule 
$N$  & $\tau_1$ & $\tau_2$ & $\tau_3$ & $\tau_4$\\ \midrule
10  & (3-4i)$\times 10^{-6}$ & (-0.5+i)$\times 10^{-6}$ & (0.5-2i)$\times 10^{-7}$ & (-0.3 + i)$\times 10^{-8}$\\
20  & (-1-2i)$\times 10^{-14}$ & (-1-9i)$\times 10^{-15}$ & (0.8-i)$\times 10^{-15}$ & (-4-6i)$\times 10^{-15}$ \\
80  &  (0.6+i)$\times 10^{-10}$ & (3+i7)$\times 10^{-11}$ & (1-i)$\times 10^{-11}$ & (4+i4)$\times 10^{-12}$ \\
100 & (2+i4)$\times 10^{-10}$& (-5+i0.1)$\times 10^{-7}$ & (0.5+i2)$\times 10^{-7}$ & (0.3+i1.2)$\times 10^{-8}$\\ \bottomrule
\end{tabular}
\caption{Values of the $\tau$ coefficients estimating the approximation error for four different numbers of Chebyshev polynomials $N$ in the case of liquid zinc with for $\R=20$, $\hat{h}=1.7$ and $k=1\times 10^{-2}$.}
\label{tab:convergence_study_tau}
\end{table}
\begin{table}
\centering
\begin{tabular}{cccc}
\toprule 
$M\times M$ & $k_r$ & $k_i$ & $\omega_i$ \\ \midrule
100$\times$ 100 & 0.0778 & 0.0176 & 0.00056017 \\ 
200$\times$ 200 & 0.0776 & 0.0174 & 0.0005594 \\ 
500$\times$ 500 & 0.0778 & 0.0173 & 0.0005599 \\\bottomrule 
\end{tabular}
\caption{Position of the saddle point in the complex wavenumber space ($k_r$,$k_i$) and the associated growth rate $\omega_i$, for three mesh sizes $M$ using liquid film with Re = 30 and $\hat{h}=0.9$}
\label{tab:convergence_study_saddle_point}
\end{table}
The values of the most unstable eigenvalue are compared by varying the numbers of Chebyshev polynomials ($N=10,20,80,100$) in the approximation of the eigenfunction $\varphi(\hat{y})$ for the liquid zinc with $\R=20$, $\hat{h}=1.7$ and $k=10^{-2}$. Table~\ref{tab:convergence_study} reports the real and imaginary part of the most unstable eigenvalue and the difference in magnitude to the $N=100$ case, expressed by the Euclidean norm. Ten polynomials are sufficient to approximate the unstable mode, with minor variation compared to the more accurate $N=100$ case. The solution with $N=10$ guarantees an approximation error of at least $\text{\textit{O}}(10^{-6})$ for both (a) the real and (b) the imaginary parts, also in terms of $\tau$ coefficients (reported in Table~\ref{tab:convergence_study_tau}). To be conservative and limit the linear system's size, we use $N=20$ for the rest of the computations. In case matrix $\hat{A}$ used for the Rayleigh quotient iteration \eqref{eq:rayleigh_quotient} is singular, we use $N=30$ polynomials.
Concerning the domain and grid spacing for the saddle point computation, we calculate the growth rate at the saddle point in $k_r\in[0.05,0.1]$ and $k_i\in[-0.02,0.04]$, testing three different meshes: $M\times M=\{100\times 100,200\times 200,500\times 500\}$. Table~\ref{tab:convergence_study_saddle_point} reports the value of $k_r$ and $k_i$ at the saddle point and the associated $\omega_i$. A grid of $M\times M=200\times 200$ is sufficient to have an accuracy of the growth rate up to the sixth digit, compared to the $M\times M=500\times 500$ case. Therefore, a $M\times M=200\times 200$ grid is a good compromise between results accuracy and computational cost. This grid spacing corresponds to a discretization step of $\Delta k_r=2.5\times 10^{-4}$ and $\Delta k_i=3\times 10^{-4}$, which are used for the saddle point search.
\subsection{Verification}
\label{subsec:verification}
To verify the numerical implementation, we compare the growth rate and the eigenfunction obtained with the spectral method (with $N=20$) against a long-wave asymptotic expansion, obtained approximating the solution ($\varphi(\hat{y})$, $c$) with a power series up to third order of $k\in\mathbb{R}$   \citep{yih1963stability}:
\begin{subequations}
\label{eq:approx_long_wave}
\begin{equation}
    \varphi(\hat{y}) = \varphi_0(\hat{y}) + \varphi_1(\hat{y})k + \varphi_2(\hat{y})k^2 + \varphi_3(\hat{y})k^3 + \text{\textit{O}}(k^4),
\end{equation}
\begin{equation}
    c= c_0 + c_1k + c_2k^2 + c_3k^3 + \text{\textit{O}}(k^4),
\end{equation}
\end{subequations}
with the long-wave assumption:
\begin{equation} 
    k\ll1, \qquad \text{and}\qquad \R = \text{\textit{O}}(1).
\end{equation}
Injecting \eqref{eq:approx_long_wave} in \eqref{eq:Orr_Sommerfeld} and \eqref{eq:bcs} and solving for $O(1)$ leads to the leading order solution:
\begin{equation}
\label{eq:leading_order_phi_exp}
    \varphi_0(\hat{y}) = \hat{y}^2, \qquad\qquad\qquad c_0 = (\hat{h}^2 - 1).
\end{equation}
In the calculation, we assumed for convenience that the constant associated with $\hat{y}^2$ is equal to unity \citep[Subsection~3.5.3]{kalliadasis2011falling}, which is equivalent to setting the liquid film's displacement amplitude to two ($\eta=2$). This solution corresponds to the displacement of the film thickness associated with a variation of the flow rate with a simply advected perturbation, which is neither amplified nor damped. In the falling liquid film literature, this is known as Goldstone mode \citep{colinet2001nonlinear}. In our case, the magnitude and direction of the phase speed depend on $\hat{h}$. Waves have a zero phase speed for $\hat{h}=1$. For $\hat{h}<1$, waves propagate upwards and for $\hat{h}>1$, waves propagate in the direction of the gravitational acceleration, with waves going faster in magnitude than any other particle inside the base flow ($c>1$) for $\sqrt{2}<\hat{h}<\sqrt{3}$.

\begin{figure}
\centering
  \begin{subfigure}[b]{0.49\textwidth}
    \includegraphics[width=\textwidth]{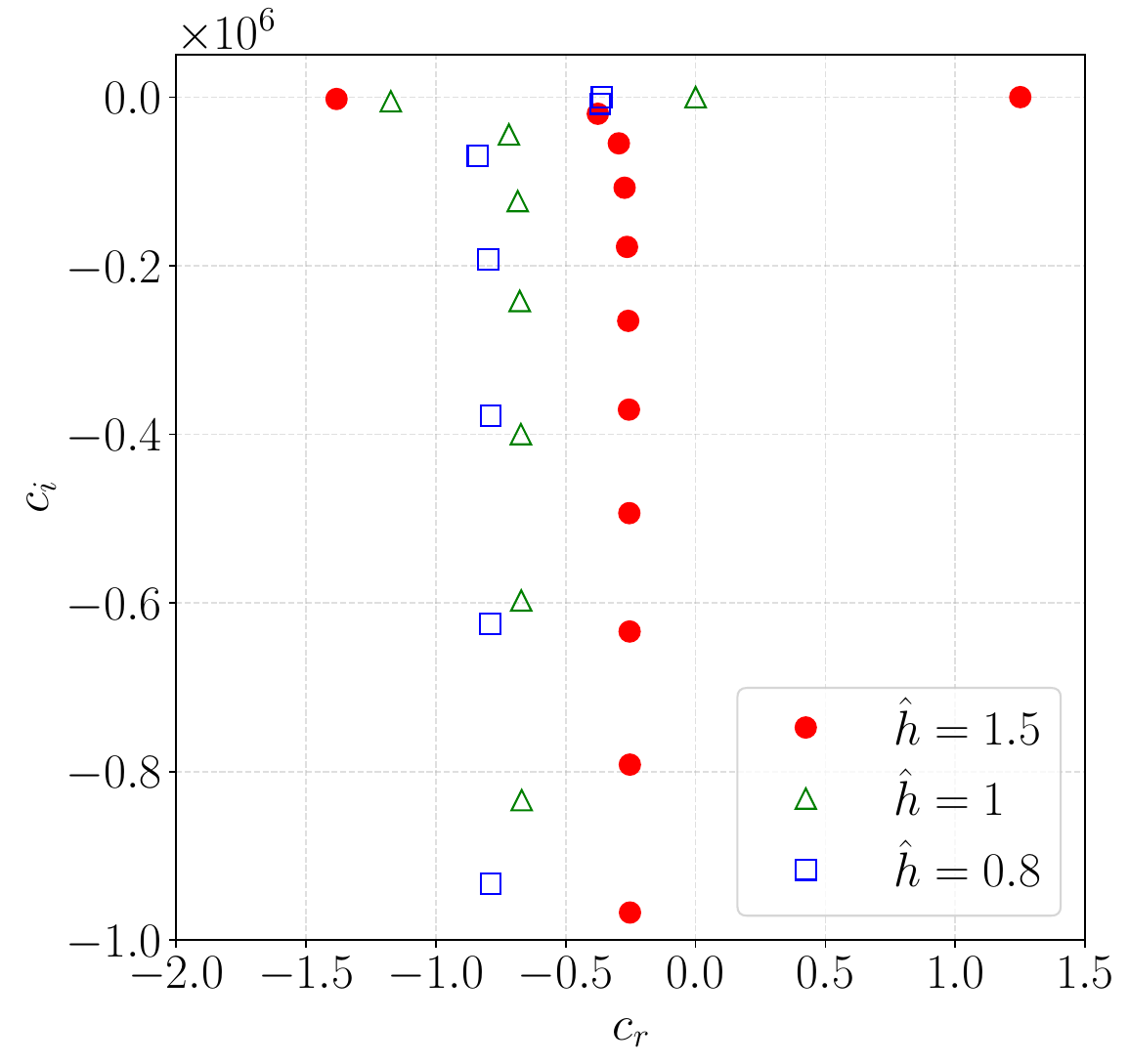}
    \caption{}
  \end{subfigure}
  \hfill
  \begin{subfigure}[b]{0.49\textwidth}
    \includegraphics[width=\textwidth]{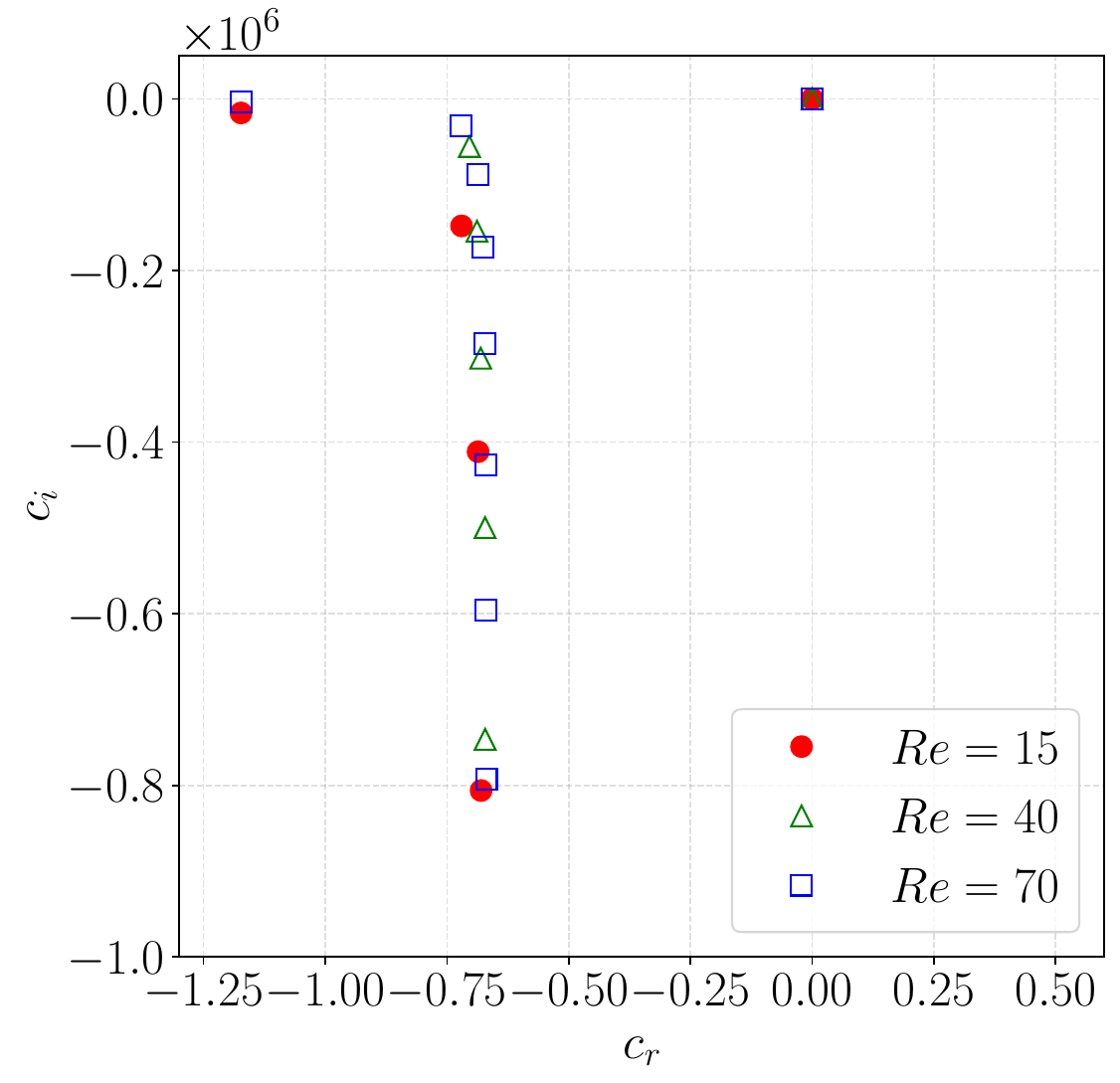}
    \caption{}
  \end{subfigure}
  \caption{Spectrum of the Orr-Sommerfeld operator obtained using 80 Chebyshev polynomials considering liquid zinc and $k=10^{-5}$ (a) for different values of the liquid film height $\hat{h}$ with $\R=50$ and (b) for different values of $\R$ at $\hat{h}=1$.}
   \label{fig:spectrum}
\end{figure}

The spectral method accurately predicts this leading-order solution. Figure~\ref{fig:spectrum} shows the spectrum of the eigenvalue problem for $k=10^{-5}$ (a) for different $\hat{h}$ with $\R=50$ and (b) for different $\R$ with $\hat{h}=1$. The spectrum presents two discrete eigenvalues; the one with the largest $c_r$ corresponds to the Goldstone mode. The spectrum also presents a continuous branch of eigenvalues in the negative $c_r$ plane. As $\hat{h}$ increases, the discrete eigenvalues tend to spread on the real axis, whereas the continuous branch approaches the imaginary axis. As $\R$ decreases, the continuous spectrum spreads along the imaginary axis. While the left eigenvalue tends to spread along the negative real axis, the Goldstone mode does not move, in agreement with the asymptotic expansion.

Moving to higher order terms in the asymptotic expansion, the solution at order $\text{\textit{O}}(k)$ is given by an imaginary streamwise velocity $\varphi_1(\hat{y})$ and an imaginary phase speed $c_1$:
\begin{equation}
\label{exp:order_k}
    \varphi_1(\hat{y}) = i\frac{1}{60} \left(\hat{h}\R \,\hat{y}^5-5\hat{h}^2Re\, \hat{y}^4\right),\qquad\qquad\qquad
    c_1 = i\frac{2}{15}\hat{h}^6\R.
\end{equation}
In the derivation, we set the quadratic term $\hat{y}^2$ to zero such that (\ref{exp:order_k}) represents a pure higher-order polynomial correction to the leading order solution $\varphi_0(\hat{y})$ \citep[Subsection~3.5.3]{kalliadasis2011falling}. In the phase speed, the sixth-power dependence on $\hat{h}$ highlights the important impact of the liquid film height even at small $k$. 
\begin{figure}
  \begin{subfigure}[b]{0.49\textwidth}
    \includegraphics[width=\textwidth]{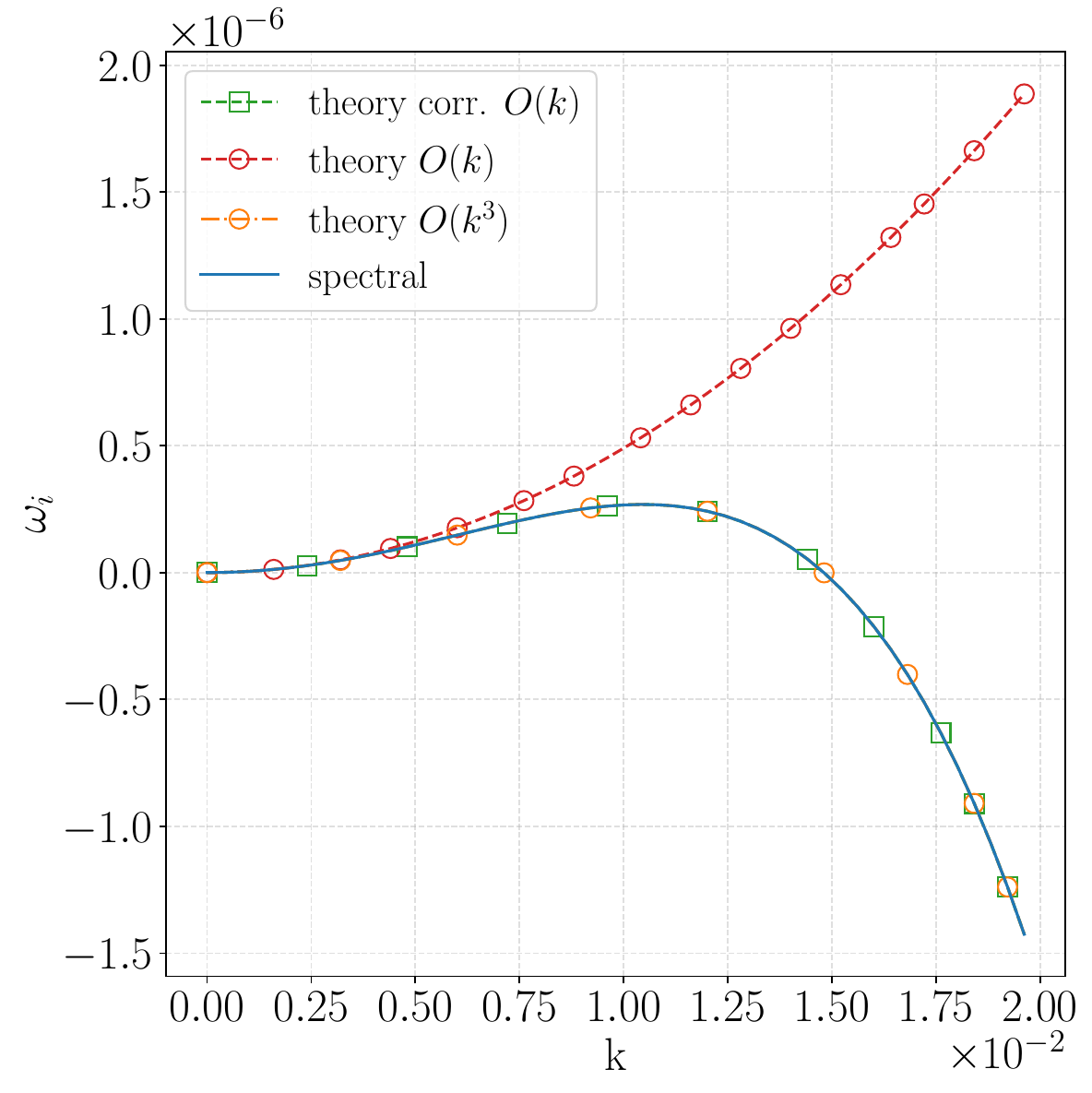}
  \end{subfigure}
  \hfill
  \begin{subfigure}[b]{0.49\textwidth}
    \includegraphics[width=\textwidth]{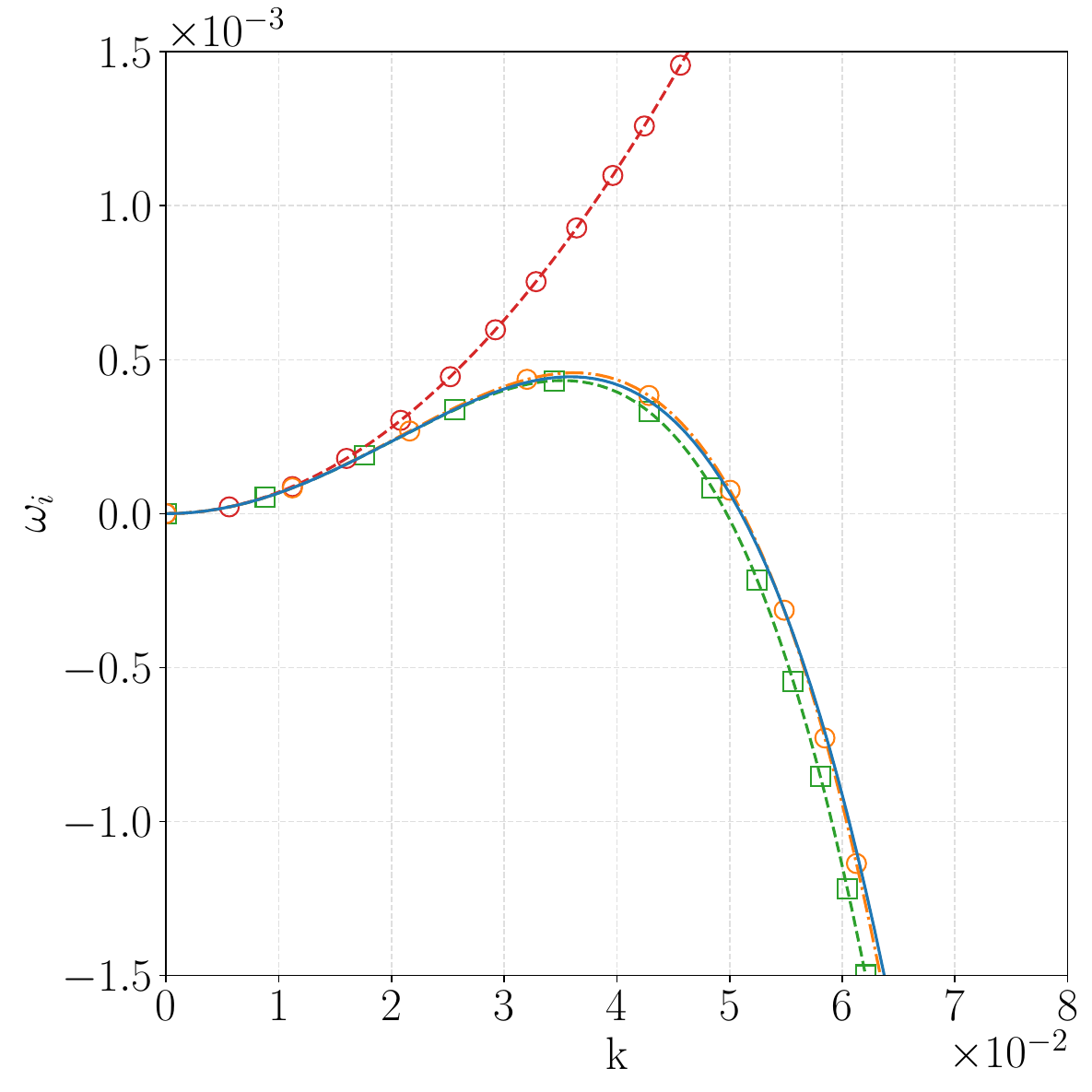}
  \end{subfigure}
  \hfill
  \begin{subfigure}[b]{0.49\textwidth}
    \includegraphics[width=\textwidth]{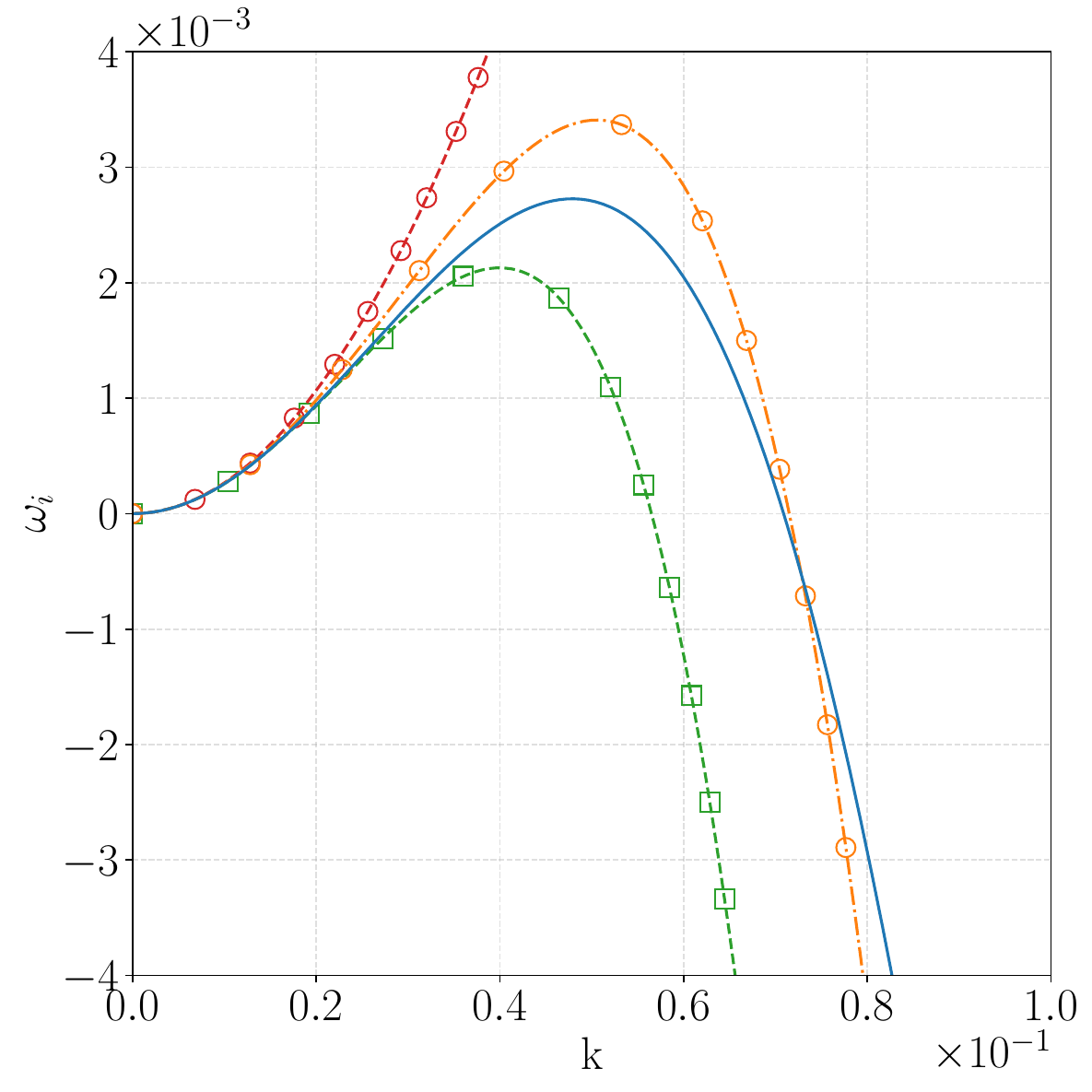}
  \end{subfigure}
  \hfill
  \begin{subfigure}[b]{0.49\textwidth}
    \includegraphics[width=\textwidth]{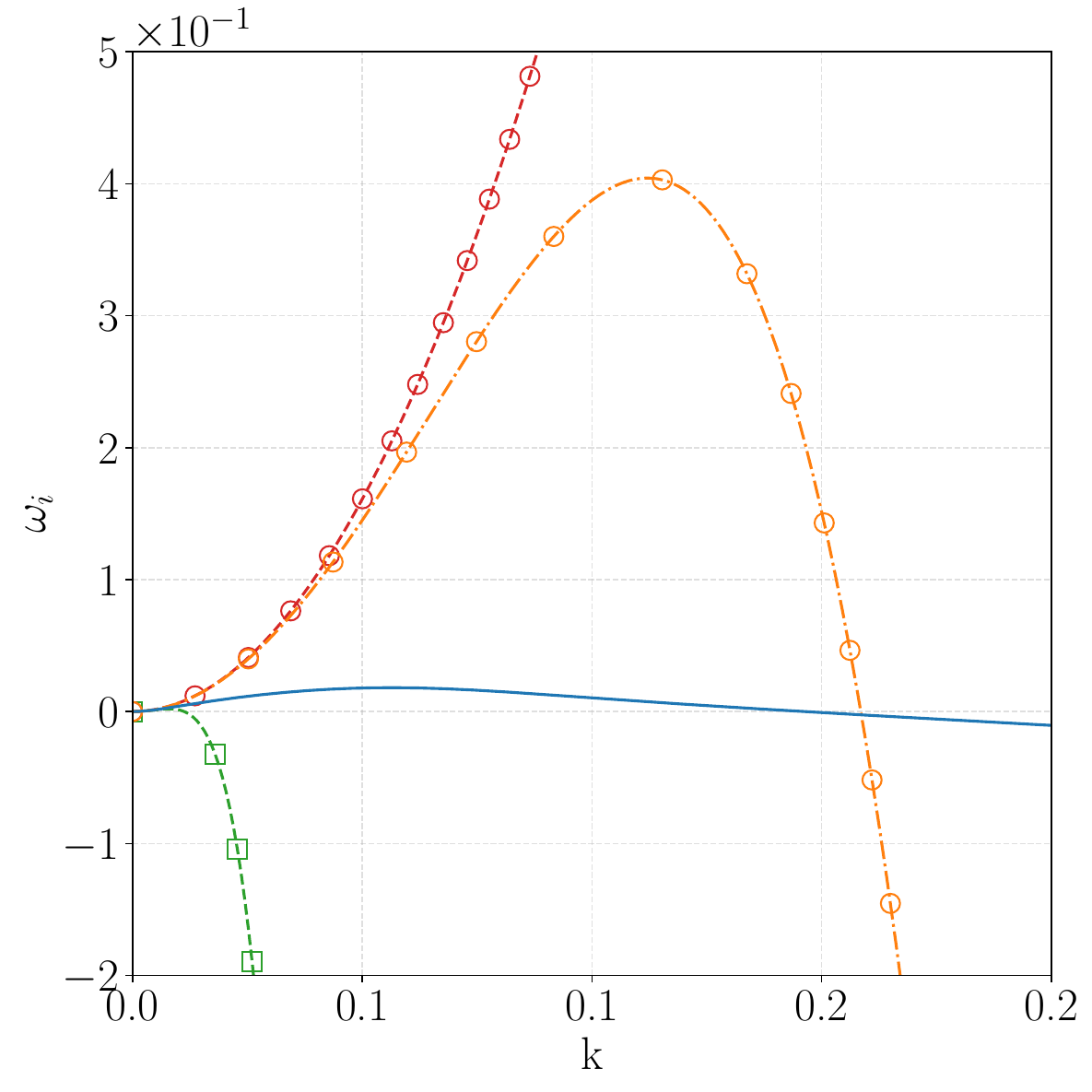}
  \end{subfigure}
  \caption{Comparison of the growth rate $\omega_i$ between long-wave expansion up to $\text{\textit{O}}(k^3)$ (dash-dotted orange line with circles), up to $\text{\textit{O}}(k)$ with (dashed green line with squares) and without (dashed red line with squares) surface tension correction and the result obtained with the spectral method (continuous blue line) considering liquid zinc with $\R = 20$ for (a) $\hat{h}=0.35$, (b) $\hat{h}=0.8$, (c) $\hat{h}=1$ and (d) $\hat{h}=1.7$}
  \label{fig:validation_spectral_eig}
\end{figure}

We can include the surface tension effect at this order, assuming $(\We\times k^2) = \text{\textit{O}}(1)$ \citep{pelisson2018numerical}. This leads to the corrected solution:
\begin{equation}
\label{exp:order_k_w_Ca}
    \varphi_1^{\star}(\hat{y}) = \frac{i}{60}\left(\hat{h}\R \,\hat{y}^5-5\hat{h}^2Re\, \hat{y}^4 +20k^2Ca^{-1}\,\hat{y}^3\right),\qquad\quad
    c_1^{\star} = \frac{i}{15}(2\hat{h}^6 \R - 5\hat{h}^3k^2 \We).
\end{equation}

For the higher order term solutions, we neglect the hypothesis on the surface tension, which then appears starting at $\text{\textit{O}}(k^3)$ \citep{benney1966long}. The solution at order $O(k^2)$ reads:
\begin{subequations}
\begin{equation}
    \varphi_2(\hat{y}) = \frac{\hat{y}^3 \left(\hat{h}\left(\R^2\hat{y}\left(224\hat{h}^5-56\hat{h}^3\hat{y}^2+32\hat{h}^2\hat{y}^3-9\hat{h}\hat{y}^4+\hat{y}^5\right)+6720\right)+3360\hat{y}\right)}{20160},
\end{equation}
\begin{equation}
    c_2 = \frac{\hat{h}^4}{63}\left(4\hat{h}^6\R^2+63\right),
\end{equation}
\end{subequations}
and at order $O(k^3)$ reads:
\begin{subequations}
\begin{equation}
    \begin{split}
        \varphi_3(\hat{y}) = &-\Big(\frac{i\hat{y}^3\left(-17160\R\hat{y}^5 \left(\hat{h}^6\R^2-12\right)-27456\hat{h}\R\hat{y}^4\left(8\hat{h}^6\R^2+45\right)+51480\hat{h}^5\R^3\hat{y}^6\right)}{1037836800}+\\&+ \frac{+i\hat{y}^3\left(5765760\left(13\hat{h}^5\R-60Ca^{-1}\right)-26884\hat{h}^4\R^3\hat{y}^7+7904\hat{h}^3\R^3\hat{y}^8\right)}{1037836800}+\\&+\frac{+i\hat{y}^3\left(17297280\hat{h}^3\R\hat{y}^2-1365\hat{h}^2\R^3\hat{y}^9-1372800\hat{h}^4\R\hat{y}\left(4\hat{h}^6\R^2+63\right)\right)}{1037836800}+\\&+\frac{i\hat{y}^3\left(192192\hat{h}^2\R\hat{y}^3\left(4\hat{h}^6\R^2+15\right)+105\hat{h}\R^3\hat{y}^{10}\right)}{1037836800}\Big),
    \end{split}
\end{equation}
\begin{equation}
    c_3 = -\frac{75872i\hat{h}^{14}\R^3}{2027025}-\frac{157}{224}i\hat{h}^8\R-\frac{1}{3} i\hat{h}^3 Ca^{-1}.
\end{equation}
\end{subequations}

Figure~\ref{fig:validation_spectral_eig} shows a comparison of the growth rates ($\omega_i = k c_i$) of the most unstable mode for the expansion up to $O(k^3)$ (orange dash-dotted line with circles), up to $O(k)$ with (green dashed line with circles) and without (red dashed line with circles) surface tension correction and the one obtained with the spectral method (continuous blue line) at $\R=20$ with liquid zinc's properties for (a) $\hat{h}=0.35$, (b) $\hat{h}=0.8$, (c) $\hat{h}=1$ and (d) $\hat{h}=1.7$. The numerical results agree with the three expansions for $k\rightarrow0$.

For $\hat{h}=0.35$ the solutions at $\text{\textit{O}}(k^3)$, at $\text{\textit{O}}(k)$ with correction and the spectral one match for all $k$ with the solution at $\text{\textit{O}}(k)$ diverging around $k=0.75\times 10^{-2}$. As $\hat{h}$ increases, the curves start to disagree for larger $k$. At $\hat{h}=1$, the solution at $\text{\textit{O}}(k^3)$ predicts very well the cut-off wavenumber (where $\omega_i=0$), despite overpredicting the location and the magnitude of the growth rate peak. At $\hat{h}=1.7$, the asymptotic solutions completely disagree with the OS solution as the range of unstable wave numbers enlarges and the long-wave assumption loses validity. These results verify the numerical implementation and highlight the long-wave nature of the instability for $\hat{h}\rightarrow 0$.
\begin{figure}
  \begin{subfigure}[b]{0.49\textwidth}
    \includegraphics[width=\textwidth]{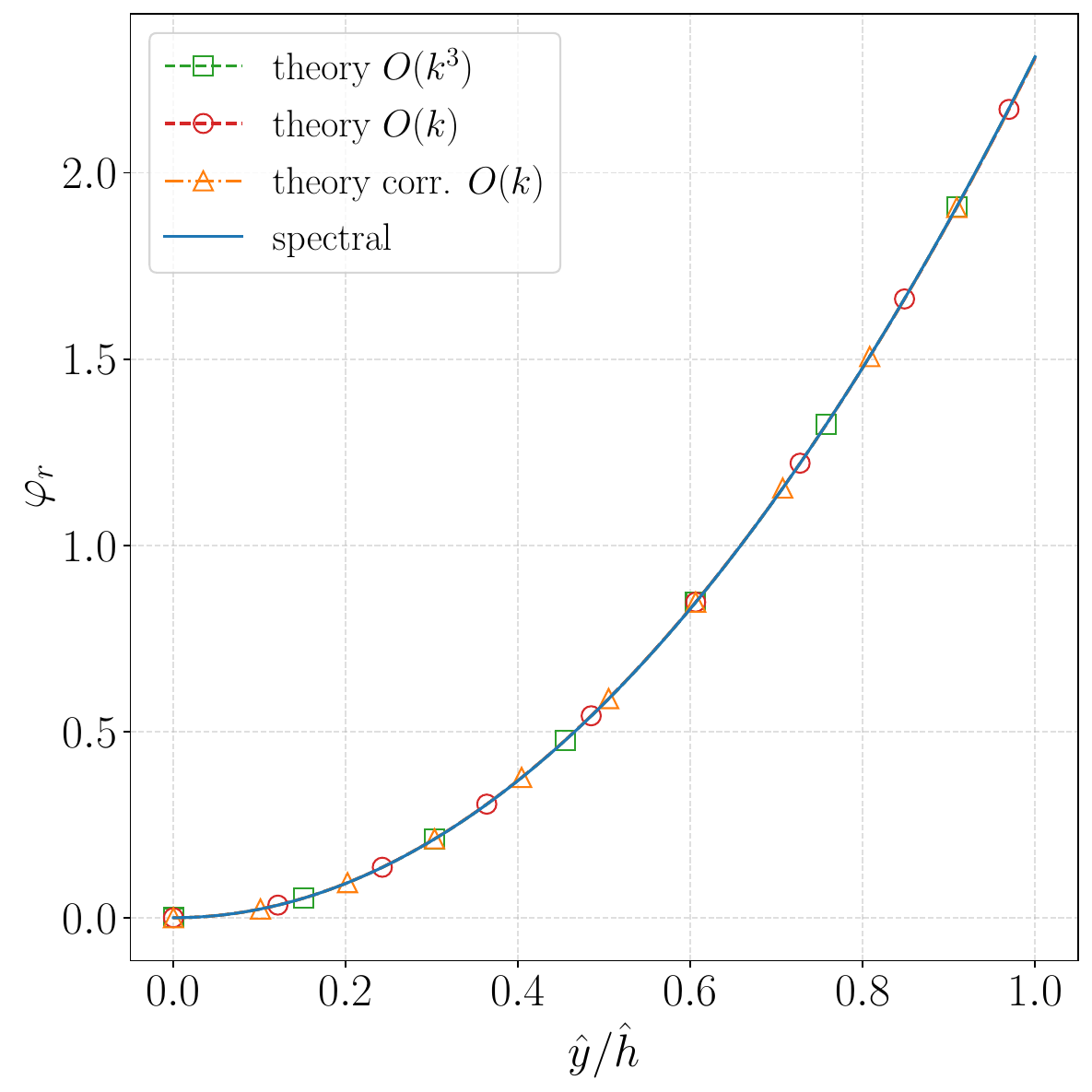}
    \caption{}
  \end{subfigure}
  \hfill
  \begin{subfigure}[b]{0.49\textwidth}
    \includegraphics[width=\textwidth]{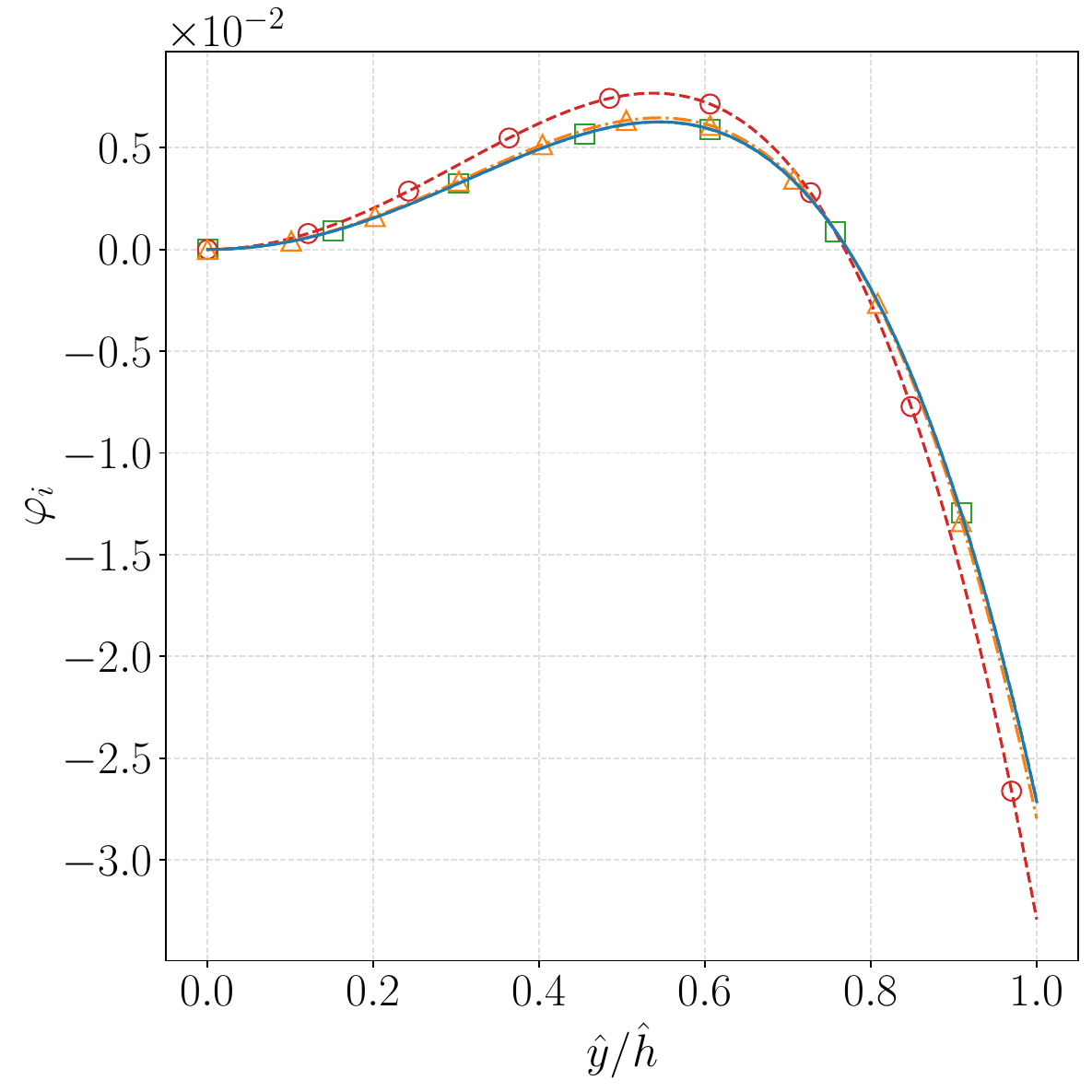}
    \caption{}
  \end{subfigure}
  \caption{Comparison of the eigenfunction $\varphi(\hat{y})$ considering liquid zinc with $\hat{h}$ = 1.3, $\R = 10$ and $k=0.02$ between the long wave expansions up to $\text{\textit{O}}(k)$ (dashed orange line with triangle markers), to $\text{\textit{O}}(k)$ with surface tension correction (dotted green line with square markers) and to $\text{\textit{O}}(k^3)$, against the solution of the OS problem with $N=20$ Chebyshev polynomials (continuous blue line) in terms of the (a) real and (b) imaginary parts.}
   \label{fig:validation_spectral_eigfun}
\end{figure}

The agreement between theory and numerics is also evident in terms of eigenfunctions. Figure~\ref{fig:validation_spectral_eigfun} shows the comparison of the eigenfunction associated with the most unstable mode at $\hat{h}=1.3$, $k=0.02$ and $Re=20$ for both (a) the real and (b) the imaginary parts, scaled with the normalization constraint:
\begin{equation}
\label{eq:int_condition}
\int_{0}^{\hat{h}}\, \varphi(\hat{y})\, d\hat{y} = 1.
\end{equation}

Both the real and the imaginary parts agree perfectly for all the expansions and the spectral results, apart from the solution at $\text{\textit{O}}(k)$, which overpredicts the peak in the imaginary part and a small deviation for $\hat{y}\rightarrow\hat{h}$.

In conclusion, given the agreement of both the growth rates and the eigenfunction, we consider the numerical implementation verified.

\end{document}